\def\OPTIONLoudLabels{0}
\def\OPTIONArxiv{1}

\def\OPTIONArxiv{0}
\def\OPTIONConf{1}
\ifnum\OPTIONConf=1%
  \ifnum\OPTIONArxiv=0%
     \documentclass[nonatbib,reprint]{sigplanconf}
     \setpagenumber{1}   %
  \else
     \documentclass[reprint,nonatbib]{sigplanconf}
  \fi
\else
\documentclass[10pt,leqno]{article}
\fi
\usepackage[overload]{textcase}
\ifnum\OPTIONConf=1%
\else%
   \usepackage[headings]{fullpage}
\fi
\usepackage[hyperref,svgnames]{xcolor}
\usepackage{graphics,pstricks,color}

\usepackage{listings}

\usepackage{pgf,tikz}
\usetikzlibrary{shapes,arrows,matrix,fit,backgrounds,calc,decorations,decorations.pathmorphing}

\usepackage{xspace}
\ifnum\OPTIONConf=1%
\else
    \usepackage{charter}
\fi
\usepackage{euler}
\usepackage{multirow}
\usepackage{stmaryrd}
\usepackage{pifont}
\usepackage{comment}
\usepackage{mathpartir}
\usepackage{framed}
\usepackage{rotating}

\let\MathRightArrow\Rightarrow %
\usepackage{marvosym} %
\def\Rightarrow{\MathRightArrow}

\usepackage{srcltx}
\usepackage{fancyhdr}
\usepackage{enumerate}
\usepackage{relsize}

\usepackage{booktabs}

\usepackage{phaistos}

\usepackage[hyphens]{url}
\usepackage[breaklinks]{hyperref} %
\usepackage{breakurl}
\usepackage{latexsym,stmaryrd}
\usepackage{amsmath,amsthm,amssymb}
\usepackage{thmtools,thm-restate}

\usepackage[authoryear]{natbib}
\bibpunct{(}{)}{;}{a}{}{,}

\def\MathparLineskip{\lineskiplimit=1.0em\lineskip=0.8em plus 0.2em}  %

\usepackage[T1]{fontenc}

\definecolor{dHilite}{rgb}{0.9, 0.9, 0.6}
\definecolor{dRed}{rgb}{0.65, 0.0, 0.0}
\definecolor{dGreen}{rgb}{0.0, 0.65, 0.0}
\definecolor{dDkGreen}{rgb}{0.0, 0.35, 0.0}
\definecolor{dBlue}{rgb}{0.0, 0.0, 0.65}
\definecolor{dPurple}{rgb}{0.65, 0.0, 0.65}
\definecolor{dDigPurple}{rgb}{0.5, 0.0, 0.5}
\definecolor{dFaint}{rgb}{0.7, 0.7, 0.7}
\definecolor{dGray}{rgb}{0.5, 0.5, 0.5}
\definecolor{dDark}{rgb}{0.2, 0.2, 0.2}
\definecolor{dAlmostBlack}{rgb}{0.1, 0.1, 0.1}

\makeatletter
\def\url@MGstyle{%
\def\UrlFont{\tiny\huge\ttfamily}%
\Url@do
}
\makeatother

\newcommand{\marginnote}[1]{\marginparwidth=40pt \marginpar{%
    \raisebox{-2ex}{\parbox{40pt}{\raggedright\scriptsize #1}}}}

\makeatletter
\def\url@vttstyle{%
  \@ifundefined{selectfont}{\def\UrlFont{\tt}}{\def\UrlFont{\normalfont\fontfamily{cmvtt}\selectfont}}}
\makeatother
\newcommand\textvtt[1]{{\normalfont\fontfamily{cmvtt}\selectfont #1}}

\newcommand{\LoudLabel}[1]{\idempotentlabel{#1}%
\ifnum\OPTIONLoudLabels=1%
  \ifnum\OPTIONConf=1%
  \marginnote{\tiny\textvtt{#1}}%
  \else%
  \marginnote{\textvtt{#1}}%
  \fi%
\fi%
}

\newcommand{\idempotentlabel}[1]{%
    \ifcsname IDEMPFLAG#1\endcsname%
      \message{YYY ALREADY DEFINED: #1}
    \else%
      \message{YYZ NOT ALREADY DEFINED: #1}
      \expandafter\gdef\csname IDEMPFLAG#1\endcsname{d}%
      \label{#1}%
    \fi}

\ifnum\OPTIONLoudLabels=1%
\newcommand{\Label}[1]{\LoudLabel{#1}}%
\newcommand{\FLabel}[1]{\idempotentlabel{#1}%
{\tt\scriptsize{#1}}}%
\else%
\newcommand{\Label}[1]{\idempotentlabel{#1}}%
\newcommand{\FLabel}[1]{\idempotentlabel{#1}}%
\fi

\newdimen\zzfontsz
\newcommand{\fontsz}[2]{\zzfontsz=#1%
{\fontsize{\zzfontsz}{1.2\zzfontsz}\selectfont{#2}}}

\newcommand{\mathsz}[2]{\text{\fontsz{#1}{$#2$}}}

\newcommand{\mathcolor}[2]{\text{\textcolor{#1}{\ensuremath{#2}}}}

\newcommand{\smallblacktriangle}{\text{\textscale{0.7}{$\blacktriangleright$}}}

\newcommand{\arr}{\rightarrow}
\def\CompactJudgments{0}
\newcommand{\entails}{\mathrel{\ifnum\CompactJudgments=1%
    \vdash%
  \else%
     \vdash\,%
  \fi}}
\newcommand{\ctxoutsym}{\ifnum\CompactJudgments=1%
    \dashv%
  \else%
     \,\dashv%
  \fi}
\newcommand{\ctxout}[1]{\mathrel{\ctxoutsym}{#1}}

\newcommand{\MonnierCommaSym}{{\smallblacktriangle}}
\newcommand{\MonnierComma}[1]{{\MonnierCommaSym}_{#1}}

\newcommand{\FV}[1]{\mathrm{FV}(#1)}

\newcommand{\beeq}{=_{\beta\eta}}

\newcommand{\emptyctx}{\cdot}

\newcommand{\tyname}[1]{\textsf{\normalfont #1}}
\newcommand{\unitexp}{\text{\normalfont \tt()}}
\newcommand{\unitty}{\tyname{1}}

\newcommand{\subtypingycolor}[1]{\textcolor{dDigPurple}{#1}}
\newcommand{\subtype}{\mathrel{\normalfont\texttt{\subtypingycolor{<:}}}}  %
\newcommand{\declsubtype}{\mathrel{\leq}}

\newcommand{\Figureref}[1]{Figure \ref{#1}}

\newcommand{\Sectionref}[1]{Section \ref{#1}}

\newcommand{\Theoremref}[1]{Theorem \ref{#1}}

\gdef\xxDerivationProofCaseColor{N}

\newcommand{\DerivationProofCase}[3]{%
     \smallskip
     \item %
       \parbox[t]{100ex}{%
       \textbf{Case } \\[-0.5em]
       $~$\hspace{5ex}
       \if\xxDerivationProofCaseColor N%
           \ensuremath{%
              \Infer{#1}{#2}{#3}%
            }
       \else%
           \colorbox{\xxDerivationProofCaseColor}{%
              \ensuremath{%
                \Infer{#1}{#2}{#3}%
              }%
           }%
        \fi%
     }%
     \nopagebreak \\[-0.8ex]
  }

\newcommand{\DoubleDerivationProofCase}[6]{%
     \smallskip
     \item %
       \parbox[t]{100ex}{%
       \textbf{Case } \\[-0.5em]
       $~$\hspace{5ex}
       \if\xxDerivationProofCaseColor N%
           \ensuremath{%
              \Infer{#1}{#2}{#3}%
              ~~~~~
              \Infer{#4}{#5}{#6}%
            }
       \else%
           \colorbox{\xxDerivationProofCaseColor}{%
              \ensuremath{%
                \Infer{#1}{#2}{#3}%
                ~~~~~
                \Infer{#4}{#5}{#6}%
              }%
           }%
        \fi%
     }%
     \nopagebreak \\[-0.8ex]
  }

\newenvironment{displ}{\vspace{1pt} \begin{center} ~\!\!}{\end{center}}
\newenvironment{mathdispl}{\vspace{1pt} \begin{center} ~\!\!\(}{\)\end{center}}

\newcommand{\BeginProof}{\renewcommand{\arraystretch}{1.1} \begin{tabular}[b]{r@{}r @{} l  l}}
\newcommand{\EndProof}{\end{tabular} \renewcommand{\arraystretch}{\mydefaultarraystretch}}

\newenvironment{llproof}{\BeginProof}{\EndProof}

\newcommand{\proofheading}[1]{}  %

\newcommand{\xdom}{\mathsf{dom}}
\newcommand{\dom}[1]{\xdom(#1)}

\newcommand{\xunsolved}{\mathsf{unsolved}}
\newcommand{\unsolved}[1]{\xunsolved(#1)}

\newcommand{\textgraybox}[1]{\boxed{#1}}

\newcommand{\tightcolorbox}[2]{\setlength{\fboxsep}{1pt}\colorbox{#1}{#2}}

\newgray{grayboxgray}{0.85}
\newcommand{\shadebox}[1]{\text{\textshadebox{\ensuremath{#1}}}}

\newcommand{\mathcolorbox}[2]{\text{\tightcolorbox{#1}{$\displaystyle {#2}$}}}

\newcommand{\judgboxfontsize}[1]{%
    \ifnum\OPTIONConf=1%
        \mathsz{11pt}{#1}%
    \else%
        \mathsz{14pt}{#1}%
    \fi}
\newcommand{\judgbox}[2]{%
      {\raggedright \textgraybox{\ensuremath{\judgboxfontsize{#1}}}\!%
        \fontsz{9pt}{\begin{tabular}[c]{l} #2 \end{tabular}} %
}}

\newcommand{\bnfas}{\mathrel{::=}}
\newcommand{\bnfalt}{\mathrel{\mid}}

\newcommand{\AllSym}{\forall}
\newcommand{\xAll}[1]{\AllSym#1}
\newcommand{\All}[1]{\xAll{#1}.\:}

\newcommand{\Infer}[3]{\inferrule*[right={\text{\strut#1}}]{{}#2\mathstrut}{{}#3\mathstrut}}

\newcommand{\keyword}[1]{\textsf{#1}}
\newcommand{\textkw}[1]{\keyword{#1}}
\newcommand{\lam}[1]{\lambda #1.\,}
\newcommand{\fun}[2]{\lam{#1}{#2}}

\newcommand{\Let}[2]{\textkw{let}\;{#1}\,\texttt{=}\,{#2}\;\textkw{in}\;}

\newcommand{\bigprec}{\mathrel{\mathsz{14pt}{\prec}}}

\newcommand{\declsubjudg}[3]{\ensuremath{{#1} \entails {#2} \declsubtype {#3}}}

\newcommand{\subjudg}[4]{\ensuremath{{#1} \entails {#2} \subtype {#3} \ctxout{#4}}}

\newdimen\zzinstsymLTwidth
\newdimen\zzinstsymEQwidth
\newdimen\zzinstsymDiff
\newcommand{\instsymLeq}{%
    \settowidth{\zzinstsymLTwidth}{\text{\normalfont\tt<}}%
    \settowidth{\zzinstsymEQwidth}{\text{\normalfont=}}%
    \setlength{\zzinstsymDiff}{\zzinstsymEQwidth}%
    \addtolength{\zzinstsymDiff}{-\zzinstsymLTwidth}%
    \text{\raisebox{-0.22ex}{\normalfont=}%
    \hspace{-\zzinstsymEQwidth}%
    \hspace{0.5\zzinstsymDiff}%
    \raisebox{0.77ex}{\normalfont\tt<}}}
\newcommand{\instsymColon}{%
     \raisebox{-0.09ex}{\text{\normalfont{:}}}}
\newcommand{\instsyml}{\subtypingycolor{\instsymColon\hspace{0.05ex}\instsymLeq}}
\newcommand{\instsymr}{\subtypingycolor{\instsymLeq\hspace{0.05ex}\instsymColon}}
\newcommand{\instsymlop}{\mathrel{\instsyml}}
\newcommand{\instsymrop}{\mathrel{\instsymr}}

\newcommand{\instjudg}[4]{\ensuremath{{#1} \entails {#2} \instsymlop {#3} \ctxout{#4}}}
\newcommand{\instjudgr}[4]{\ensuremath{{#1} \entails {#3} \instsymrop {#2} \ctxout{#4}}}

\newcommand{\chkcolor}{dBlue}
\newcommand{\syncolor}{dRed}
\newcommand{\appcolor}{dDkGreen}
\newcommand{\chk}{\mathrel{\mathcolor{\chkcolor}{\Leftarrow}}}
\newcommand{\uncoloredsyn}{{\Rightarrow}}
\newcommand{\syn}{\mathrel{\mathcolor{\syncolor}{\uncoloredsyn}}}
\newcommand{\appsep}{\;{\mathcolor{\appcolor}{\bullet}}\;}
\newcommand{\app}{\mathrel{\mathcolor{\appcolor}{{\uncoloredsyn}\hspace{-1.2ex}{\uncoloredsyn}}}}

\newcommand{\chkjudg}[4]{\ensuremath{{#1} \entails {#2} \chk {#3} \ctxout{#4}}}
\newcommand{\appjudg}[5]{\ensuremath{{#1} \entails {#3} \appsep {#2} \app {#4} \ctxout{#5}}}
\newcommand{\synjudg}[4]{\ensuremath{{#1} \entails {#2} \syn {#3} \ctxout{#4}}}

\newcommand{\declchkjudg}[3]{\ensuremath{{#1} \entails {#2} \chk {#3}}}
\newcommand{\declappjudg}[4]{\ensuremath{#1} \entails {#3} \appsep {#2}  \app {#4}}
\newcommand{\declsynjudg}[3]{\ensuremath{{#1} \entails {#2} \syn {#3}}}

\newcommand{\hyp}[2]{{#1}:{#2}}
\newcommand{\hypeq}[2]{{#1} = {#2}}
\newcommand{\tighthypeq}[2]{{#1}{=}{#2}}
\newcommand{\alltype}[1]{\All{#1}}

\newcommand{\extendssym}{\longrightarrow}
\newcommand{\extends}[2]{{#1} \extendssym {#2}}
\newcommand{\substextend}[2]{\extends{#1}{#2}}

\newcommand{\judgetp}[2]{{#1} \entails {#2}}

\newcommand{\typesize}[2]{|{#1} {\;\entails} {#2}|}

\newcommand{\judgectx}[1]{{#1}~\mathit{ctx}}

\newcommand{\ahat}{\hat{\alpha}}
\newcommand{\bhat}{\hat{\beta}}

\newcommand{\rulename}[1]{\text{\normalfont\textsf{#1}}}

\newcommand{\substextendrulename}[1]{\ensuremath{{\extendssym}{\rulename{#1}}}\xspace}
\newcommand{\substextendId}{\substextendrulename{ID}}

\newcommand{\substextendUU}{\substextendrulename{Uvar}}
\newcommand{\substextendVV}{\substextendrulename{Var}}
\newcommand{\substextendEE}{\substextendrulename{Unsolved}}
\newcommand{\substextendSolSol}{\substextendrulename{Solved}}
\newcommand{\substextendMonMon}{\substextendrulename{Marker}}

\newcommand{\substextendSolve}{\substextendrulename{Solve}}
\newcommand{\substextendAdd}{\substextendrulename{Add}}
\newcommand{\substextendAddSolved}{\substextendrulename{AddSolved}}

\newcommand{\Dsubrulename}[1]{\ensuremath{{\declsubtype}\rulename{#1}}\xspace}
\newcommand{\DsubVar}{\Dsubrulename{Var}}
\newcommand{\DsubUnit}{\Dsubrulename{Unit}}
\newcommand{\DsubArr}{\Dsubrulename{\ensuremath{\arr}}}
\newcommand{\DsubAllL}{\Dsubrulename{\ensuremath{\forall}{L}}}
\newcommand{\DsubAllR}{\Dsubrulename{\ensuremath{\forall}{R}}}

\newcommand{\Subrulename}[1]{\ensuremath{{\subtype}\rulename{#1}}\xspace}
\newcommand{\SubVar}{\Subrulename{Var}}
\newcommand{\SubUnit}{\Subrulename{Unit}}
\newcommand{\SubExvar}{\Subrulename{Exvar}}
\newcommand{\SubArr}{\Subrulename{\ensuremath{\arr}}}
\newcommand{\SubAllL}{\Subrulename{\ensuremath{\forall}{L}}}
\newcommand{\SubAllR}{\Subrulename{\ensuremath{\forall}{R}}}

\newcommand{\SubInst}[1]{\Subrulename{Instantiate{#1}}}
\newcommand{\SubInstL}{\SubInst{L}}
\newcommand{\SubInstR}{\SubInst{R}}

\newcommand{\DeclWFrulename}[1]{\ensuremath{\rulename{Decl{#1}WF}}\xspace}
\newcommand{\DeclUvarWF}{\DeclWFrulename{Uvar}}
\newcommand{\DeclUnitWF}{\DeclWFrulename{Unit}}
\newcommand{\DeclArrowWF}{\DeclWFrulename{Arrow}}
\newcommand{\DeclForallWF}{\DeclWFrulename{Forall}}

\newcommand{\WFrulename}[1]{\ensuremath{\rulename{{#1}WF}}\xspace}
\newcommand{\UvarWF}{\WFrulename{Uvar}}
\newcommand{\UnitWF}{\WFrulename{Unit}}
\newcommand{\EvarWF}{\WFrulename{Evar}}
\newcommand{\SolvedEvarWF}{\WFrulename{SolvedEvar}}
\newcommand{\ArrowWF}{\WFrulename{Arrow}}
\newcommand{\ForallWF}{\WFrulename{Forall}}

\newcommand{\CWFrulename}[1]{\ensuremath{\rulename{{#1}Ctx}}\xspace}
\newcommand{\EmptyCWF}{\CWFrulename{Empty}}
\newcommand{\UvarCWF}{\CWFrulename{Uvar}}
\newcommand{\EvarCWF}{\CWFrulename{Evar}}
\newcommand{\SolvedEvarCWF}{\CWFrulename{SolvedEvar}}
\newcommand{\VarCWF}{\CWFrulename{Var}}
\newcommand{\MarkerCWF}{\CWFrulename{Marker}}

\newcommand{\Instrulename}[1]{\ensuremath{\rulename{Inst{#1}}}\xspace}
\newcommand{\InstLrulename}[1]{\Instrulename{L{#1}}}
\newcommand{\InstRrulename}[1]{\Instrulename{R{#1}}}

\newcommand{\InstLSolve}{\InstLrulename{Solve}}
\newcommand{\InstLReach}{\InstLrulename{Reach}}
\newcommand{\InstLArr}{\InstLrulename{Arr}}
\newcommand{\InstLAllR}{\InstLrulename{AllR}}

\newcommand{\InstRSolve}{\InstRrulename{Solve}}
\newcommand{\InstRReach}{\InstRrulename{Reach}}
\newcommand{\InstRArr}{\InstRrulename{Arr}}
\newcommand{\InstRAllL}{\InstRrulename{AllL}}

\newcommand{\Decltyrulename}[1]{\ensuremath{\rulename{Decl#1}}\xspace}

\newcommand{\DeclIntrorulename}[1]{\Decltyrulename{\ensuremath{#1}I}}
\newcommand{\DeclIntroSynrulename}[1]{\Decltyrulename{\ensuremath{#1}I$\syn$}}
\newcommand{\DeclElimrulename}[1]{\Decltyrulename{\ensuremath{#1}E}}
\newcommand{\DeclApprulename}[1]{\Decltyrulename{\ensuremath{#1}App}}

\newcommand{\DeclVar}{\Decltyrulename{Var}}
\newcommand{\DeclSub}{\Decltyrulename{Sub}}
\newcommand{\DeclAnno}{\Decltyrulename{Anno}}

\newcommand{\DeclUnitIntro}{\DeclIntrorulename{\unitty}}
\newcommand{\DeclUnitIntroSyn}{\DeclIntroSynrulename{\unitty}}

\newcommand{\DeclArrIntro}{\DeclIntrorulename{\arr}}
\newcommand{\DeclArrIntroSyn}{\DeclIntroSynrulename{\arr}}
\newcommand{\DeclArrElim}{\DeclElimrulename{\arr}}

\newcommand{\DeclAllIntro}{\DeclIntrorulename{\AllSym}}

\newcommand{\DeclArrApp}{\DeclApprulename{\arr}}
\newcommand{\DeclAllApp}{\DeclApprulename{\forall}}
\newcommand{\Tyrulename}[1]{\ensuremath{\rulename{#1}}\xspace}

\newcommand{\Introrulename}[1]{\Tyrulename{\ensuremath{#1}I}}
\newcommand{\IntroSynrulename}[1]{\Tyrulename{\ensuremath{#1}I$\syn$}}
\newcommand{\Elimrulename}[1]{\Tyrulename{\ensuremath{#1}E}}
\newcommand{\Apprulename}[1]{\Tyrulename{\ensuremath{#1}App}}

\newcommand{\Var}{\Tyrulename{Var}}
\newcommand{\Sub}{\Tyrulename{Sub}}
\newcommand{\Anno}{\Tyrulename{Anno}}

\newcommand{\UnitIntro}{\Introrulename{\unitty}}
\newcommand{\UnitIntroSyn}{\IntroSynrulename{\unitty}}

\newcommand{\ArrIntro}{\Introrulename{\arr}}
\newcommand{\ArrIntroSyn}{\IntroSynrulename{\arr}}
\newcommand{\ArrElim}{\Elimrulename{\arr}}

\newcommand{\AllIntro}{\Introrulename{\AllSym}}
\newcommand{\ArrApp}{\Apprulename{\arr}}
\newcommand{\AllApp}{\Apprulename{\forall}}
\newcommand{\SolveApp}{\Apprulename{\ahat}}
\newcommand{\AssignRuleName}[1]{\ensuremath{\rulename{A#1}}\xspace}
\newcommand{\AssignIntroName}[1]{\AssignRuleName{\ensuremath{#1}I}}
\newcommand{\AssignElimName}[1]{\AssignRuleName{\ensuremath{#1}E}}
\newcommand{\AssignVar}{\AssignRuleName{Var}}
\newcommand{\AssignUnit}{\AssignRuleName{Unit}}
\newcommand{\AssignArrIntro}{\AssignIntroName{\arr}}
\newcommand{\AssignArrElim}{\AssignElimName{\arr}}
\newcommand{\AssignAllIntro}{\AssignIntroName{\forall}}
\newcommand{\AssignAllElim}{\AssignElimName{\forall}}
\newcommand{\judge}[3]{{#1} \vdash {#2} : {#3}}

\newcommand{\Fsub}{\ensuremath{{\text{F}}_{\texttt{<:}}}\xspace}

\newcommand{\Csharp}{\ensuremath{\text{\textrm{C}}^\sharp}}

\newcommand{\MLF}{\ensuremath{\mathsf{ML}^\mathsf{F}}\xspace}

\newcommand{\contextapp}[2]{{[{#1}]{#2}}}

\newcommand{\LOCALCOPY}[1]{%
          \href{papers/#1}{\bf \textcolor{dGreen}{local copy}}}

\renewcommand{\FLabel}[1]{\label{#1}}

\declaretheoremstyle[
  bodyfont=\sl
]{mytheoremstyle}

\newtheoremstyle{better}%
   {\topsep}%
   {\topsep}%
   {\upshape\slshape}%
   {}%
   {\bfseries}%
   {.}%
   {0.7em}%
   {}%
\theoremstyle{better}
\newtheorem*{lemma*}{Lemma}
\newtheorem*{definition*}{Definition}

\begin{document}
\title{%
  Complete
  and
  Easy
  Bidirectional~Typechecking for Higher-Rank Polymorphism
}

\ifnum\OPTIONConf=1     %
  \ifnum\OPTIONArxiv=0%
          \conferenceinfo{ICFP~'13}{September 25--27, 2013, Boston, MA, USA} 
          \copyrightyear{2013} 
          \copyrightdata{978-1-4503-2326-0/13/09}
          \doi{2500365.2500582}
          \exclusivelicense
  \else
        \permission{%
          This work is licensed under the \\
                \href{http://creativecommons.org}{Creative Commons}\;\href{http://creativecommons.org/licenses/by-nd/3.0/}{Attribution---No Derivative Works} License.
        }
        \conferenceinfo{ICFP '13,}{September 25--27, 2013, Boston, MA, USA.}
        \copyrightyear{2013}
        \copyrightdata{\vspace{1pt}%
              Copyright {\sf\copyright}~2013 Jana Dunfield and Neelakantan R.\ Krishnaswami%
      }
   \fi
    \authorinfo{
       Jana Dunfield
     \and
       Neelakantan R.\ Krishnaswami
   }
    {Max Planck Institute for Software Systems \\ Kaiserslautern and Saarbr\"ucken, Germany}
    {jd169@queensu.ca \and nk480@cl.cam.ac.uk}
\fi

\setlength{\pdfpageheight}{\paperheight}
\setlength{\pdfpagewidth}{\paperwidth}

\exclusivelicense                %

\newcommand{\XXXACM}[1]{\vspace{#1}}  %

\maketitle

\begin{abstract}
  Bidirectional typechecking, in which terms either synthesize a type or are checked against
  a known type, has become popular for its scalability (unlike Damas-Milner type inference,
  bidirectional typing remains decidable even for very expressive type systems),
  its error reporting, and its relative ease of implementation.
  Following design principles from proof theory,
  bidirectional typing can be applied to many type constructs.  The principles underlying
  a bidirectional approach to polymorphism, however, are less obvious.
  We give a declarative, bidirectional account of higher-rank polymorphism,
  grounded in proof theory; this calculus enjoys many properties such as $\eta$-reduction
  and predictability of annotations.
  We give an algorithm for implementing the declarative system;
  our algorithm is remarkably simple and well-behaved, despite being both sound and complete. 
\end{abstract}

{
\\[-10pt]
}

\setcounter{footnote}{0}

\section{Introduction}
\Label{sec:intro}

Bidirectional typechecking~\citep{Pierce00} has become one of the most
popular techniques for implementing typecheckers in new languages.
This technique has been used for dependent
types~\citep{Coquand96:typechecking-dependent-types,Abel08:MPC,Loeh08:Tutorial,Asperti12};
subtyping~\citep{Pierce00}; intersection, union, indexed and refinement
types~\citep{XiThesis,Davies00icfpIntersectionEffects,Dunfield04:Tridirectional};
termination checking~\citep{Abel04:termination}; higher-rank
polymorphism~\citep{PeytonJones07%
,Dunfield09%
}; refinement types for LF~\citep{LovasThesis}; contextual modal
types~\citep{Pientka08:POPL}%
; compiler intermediate
representations~\citep{Chlipala05}; and object-oriented languages
including \Csharp~\citep{Bierman07} and Scala~\citep{Odersky01:ColoredLocal}.
As can be seen, it scales well to advanced type systems;
moreover, it is easy to implement, and yields relatively high-quality
error messages~\citep{PeytonJones07}.

However, the theoretical foundation of bidirectional typechecking has
lagged behind its application. As shown by \citet{Watkins04},
bidirectional typechecking can be understood in terms of the
normalization of intuitionistic type theory, in which normal forms
correspond to the checking mode of bidirectional typechecking, and
neutral (or atomic) terms correspond to the synthesis mode.
This enables a proof of the elegant property that type annotations are
only necessary at reducible expressions, and that normal forms need no
annotations at all. The benefit of the proof-theoretic view is that it
gives a simple and easy-to-understand declarative account of where
type annotations are necessary, without reference to the details of
the typechecking algorithm.

While the proof-theoretic account of bidirectional typechecking has
been scaled up as far as type refinements and intersection and union
types~\citep{Pfenning08}, as yet there has been no completely
satisfactory account of how to extend the proof-theoretic approach
to handle polymorphism. This is especially vexing, since the ability of bidirectional
\emph{algorithms} to gracefully accommodate polymorphism (even
higher-rank polymorphism) has been one of their chief attractions.

In this paper, we extend the proof-theoretic account of bidirectional
typechecking to full higher-rank polymorphism (i.e., predicative
System F), and consequently show that bidirectional typechecking is
not merely sound with respect to the declarative semantics, but also
that it is \emph{complete}. Better still, the algorithm we give for
doing so is extraordinarily \emph{simple}.

First, as a specification of type checking, we give a declarative
bidirectional type system which guesses all quantifier
instantiations. This calculus is a small but significant
contribution of this paper, since it possesses desirable
properties, such as the preservation of typability under
$\eta$-reduction, that are missing from the type assignment version of
System F. Furthermore, we can use the bidirectional character
of our declarative calculus to show a number of refactoring theorems,
enabling us to precisely characterize what sorts of substitutions (and
reverse substitutions) preserve typability, where type annotations are
needed, and when programmers may safely delete type annotations.

Then, we give a bidirectional algorithm that always finds
corresponding instantiations. As a consequence of completeness, we can
show that our algorithm never needs explicit type applications, and
that type annotations are only required for polymorphic, reducible
expressions---which, in practice, means that only let-bindings of
functions at polymorphic type need type annotations; no other
expressions need annotations.

Our algorithm is both simple to understand and simple to implement.
The key data structure is an ordered context containing all bindings,
including type variables, term variables, and existential variables
denoting partial type information. By maintaining order, we are able
to easily manage scope information, which is particularly important in
higher-rank systems, where different quantifiers may be instantiated
with different sets of free variables. Furthermore, ordered contexts
admit a notion of \emph{extension} or \emph{information increase},
which organizes and simplifies the soundness and completeness proofs
of the algorithmic system with respect to the declarative one.

\paragraph{Contributions.}
We make the following contributions:

\XXXACM{-1pt}

\begin{itemize}
\item We give a declarative, bidirectional account of higher-rank
  polymorphism, grounded strongly in proof theory. This calculus
  has important properties (such as $\eta$-reduction) that the type
  assignment variant of System F lacks, yet is sound and complete
  (up to $\beta\eta$-equivalence) with respect to System F.
  
  As a result, we can explain where type annotations are
  needed, where they may be deleted, and why important code
  transformations are sound, all without reference to the
  implementation.

\item We give a very simple algorithm for implementing the declarative
  system. Our algorithm does not need any data structure more
  sophisticated than a list, but can still solve all of the
  problems which arise in typechecking higher-rank polymorphism
  without any need for search or backtracking.

\item We prove that our algorithm is both \emph{sound} and
  \emph{complete} with respect to our declarative specification of
  typing.  This proof is cleanly structured around \emph{context
    extension}, a relational
  notion of information increase, corresponding to the intuition
  that our algorithm progressively resolves type constraints.

  As a result of completeness, programmers may safely ``pay no
  attention to the implementor behind the curtain'', and ignore all the
  algorithmic details of unification and type inference: 
  the algorithm does exactly what the declarative
  specification says, no more and no less.
\end{itemize}

\XXXACM{-4pt}

\paragraph{Lemmas and proofs.}
Proofs of the main results, as well as statements of all lemmas (and their proofs),
can be found in the appendix, available at \href{http://www.cs.queensu.ca/~jana/papers/bidir/}{\url{www.cs.queensu.ca/~jana/papers/bidir/}}.

\section{Declarative Type System}
\Label{sec:declarative}

\begin{figure}[t]
  \XXXACM{-6pt}
  \[
    \begin{array}{llcl}
    \mbox{Terms} & e & \bnfas &
                  x
          \bnfalt \unitexp
          \bnfalt \fun{x}{e}
          \bnfalt e\;e
          \bnfalt (e : A)
    \end{array}
  \]
  \caption{Source expressions}
  \FLabel{fig:source-syntax}
  \XXXACM{-12pt}
\end{figure}

\begin{figure}[t]
\[
    \begin{array}[t]{llcl}
    \mbox{Types} & A, B, C & \bnfas &
          \unitty \bnfalt \alpha \bnfalt \alltype{\alpha}{A} \bnfalt A \arr B \\
    \mbox{Monotypes} & \tau,\sigma & \bnfas & \unitty \bnfalt \alpha \bnfalt \tau \arr \sigma \\
    \mbox{Contexts} & \Psi
                    & \bnfas & \cdot
                     \bnfalt   \Psi, \alpha
                     \bnfalt   \Psi, x : A
    \end{array}
\]
\caption{Syntax of declarative types and contexts}
\FLabel{fig:declarative-typing-syntax}
\XXXACM{-6pt}
\end{figure}

\begin{figure}[t]
    \judgbox{\judgetp{\Psi}{A}}{Under context $\Psi$, type $A$ is well-formed}
    \begin{mathpar}
        \Infer{\DeclUvarWF}
              {\alpha \in \Psi}
              {\judgetp{\Psi}{\alpha}}
        \and
        \Infer{\DeclUnitWF}
              { }
              {\judgetp{\Psi}{\unitty}}
       \XXXACM{-2pt}
        \\
        \Infer{\DeclArrowWF}
              {\judgetp{\Psi}{A} \\ \judgetp{\Psi}{B}}
              {\judgetp{\Psi}{A \arr B}}
        \and
        \Infer{\DeclForallWF}
              {\judgetp{\Psi, \alpha}{A}}
              {\judgetp{\Psi}{\alltype{\alpha}{A}}}
    \end{mathpar}

    \medskip

    \XXXACM{-4pt}
    \judgbox{\declsubjudg{\Psi}{A}{B}}{Under context $\Psi$, type $A$ is a subtype of $B$}
    \XXXACM{-2pt}
    \begin{mathpar} 
    \Infer{\DsubVar}
              {\alpha \in \Psi}
              {\declsubjudg{\Psi}{\alpha}{\alpha}}
    \and
    \Infer{\DsubUnit}
              { }
              {\declsubjudg{\Psi}{\unitty}{\unitty}}
    \XXXACM{-2pt}
    \\
    \Infer{\DsubArr}
              {\declsubjudg{\Psi}{B_1}{A_1}
                \\
                \declsubjudg{\Psi}{A_2}{B_2}
              }
              {\declsubjudg{\Psi}{A_1 \arr A_2}{B_1 \arr B_2}}
    \XXXACM{-2pt}
    \\
    \Infer{\DsubAllL}
              {\judgetp{\Psi}{\tau}
                \\
                \declsubjudg{\Psi}{[\tau/\alpha]A}{B}}
              {\declsubjudg{\Psi}{\alltype{\alpha}{A}}{B}}
    \and
    \Infer{\DsubAllR}
              {\declsubjudg{\Psi, \beta}{A}{B}}
              {\declsubjudg{\Psi}{A}{\alltype{\beta}{B}}}
    \end{mathpar}
  
  \XXXACM{-4pt}  %
  \caption{Well-formedness of types and subtyping in the declarative system}
  \FLabel{fig:declarative}
  \XXXACM{-10pt}  %
\end{figure}

\begin{figure*}[htbp]
  \judgbox{\declchkjudg{\Psi}{e}{A}}%
     {Under context $\Psi$, $e$ checks against input type $A$} \\
  \judgbox{\declsynjudg{\Psi}{e}{A}}%
     {Under context $\Psi$, $e$ synthesizes output type $A$} \\
  \judgbox{\declappjudg{\Psi}{e}{A}{C}}%
     {Under context $\Psi$, applying a function of type $A$ to $e$ synthesizes type $C$} \\
  \begin{mathpar}
     \Infer{\DeclVar}
          {(x : A) \in \Psi}
          {\declsynjudg{\Psi}{x}{A}}
     \and
     \Infer{\DeclSub}
          {\declsynjudg{\Psi}{e}{A}
            \\
            \declsubjudg{\Psi}{A}{B}
          }
          {\declchkjudg{\Psi}{e}{B}}
     \and
     \Infer{\DeclAnno}
          {\judgetp{\Psi}{A}
            \\
            \declchkjudg{\Psi}{e}{A}
          }
          {\declsynjudg{\Psi}{(e : A)}{A}}
     \vspace{-0ex}\\
     \Infer{\DeclUnitIntro}
          {}
          {\declchkjudg{\Psi}{\unitexp}{\unitty}}
     \and
     \Infer{\DeclUnitIntroSyn}
                { }
                {\declsynjudg{\Psi}{\unitexp}{\unitty}}
     \and
     \Infer{\DeclAllIntro}
           {\declchkjudg{\Psi, \alpha}{e}{A}
           }
           {\declchkjudg{\Psi}{e}{\alltype{\alpha}{A}}}
     \and
     \Infer{\DeclAllApp}
            {\judgetp{\Psi}{\tau}
             \\
             \declappjudg{\Psi}{e}{[\tau/\alpha]A}{C}}
            {\declappjudg{\Psi}{e}{\alltype{\alpha}{A}}{C}}
     \\
     \Infer{\DeclArrIntro}
          {\declchkjudg{\Psi, x : A}{e}{B}
          }
          {\declchkjudg{\Psi}{\lam{x} e}{A \arr B}}
     \and
     \Infer{\DeclArrIntroSyn}
                {\judgetp{\Psi}{\sigma \arr \tau}
                  \\
                  \declchkjudg{\Psi, x : \sigma}{e}{\tau}
                }
                {\declsynjudg{\Psi}{\lam{x} e}{\sigma \arr \tau}}
     \and
     \Infer{\DeclArrElim}
          {\declsynjudg{\Psi}{e_1}{A}
            \\
            \declappjudg{\Psi}{e_2}{A}{C}
          }
          {\declsynjudg{\Psi}{e_1\,e_2}{C}}
     \\
     ~
     \hspace{55ex}
     \Infer{\DeclArrApp}
            {\declchkjudg{\Psi}{e}{A}}
            {\declappjudg{\Psi}{e}{A \arr C}{C}}
      \end{mathpar}

  \caption{Declarative typing}
  \FLabel{fig:decl-typing}
\end{figure*}

In order to show that our algorithm is sound and complete, we need to
give a declarative type system to serve as the specification for our
algorithm. Surprisingly, it turns out that finding the correct
declarative system to use as a specification is itself an interesting
problem!

Much work on type inference for higher-rank polymorphism takes the
type assignment variant of System F as a specification
of type inference. Unfortunately, under these rules typing is not
stable under $\eta$-reductions. For example, suppose $f$ is a variable
of type $\unitty \to \alltype{\alpha}{\alpha}$. Then the term
$\fun{x}{f\;x}$ can be ascribed the type $\unitty \to \unitty$, since
the polymorphic quantifier can be instantiated to $\unitty$ between
the $f$ and the $x$.  But the $\eta$-reduct $f$ cannot be
ascribed the type $\unitty \to \unitty$, because the quantifier cannot
be instantiated until after $f$ has been applied. This is especially unfortunate
in pure languages like Haskell, where the $\eta$ law is a valid
program equality. 

Therefore, we do not use the type assignment version of System F as
our declarative specification of type checking and inference. Instead,
we give a declarative, bidirectional system as the
specification.
Traditionally, bidirectional systems are given in
terms of a \emph{checking} judgment $\declchkjudg{\Psi}{e}{A}$, which
takes a type $A$ as input and ensures that the term $e$
\emph{checks against} that type, and a \emph{synthesis} judgment
$\declsynjudg{\Psi}{e}{A}$, which takes a term $e$ and \emph{produces}
a type $A$.
This two-judgment formulation is satisfactory for simple types,
but breaks down in the presence of polymorphism.

The essential problem is as follows: the standard bidirectional rule
for checking applications $e_1\;e_2$ in non-polymorphic systems is to
synthesize type $A \arr B$ for $e_1$, and then check $e_2$ against $A$,
returning $B$ as the type.  With polymorphism, however, we may have an
application $e_1\;e_2$ in which $e_1$ synthesizes a term of
polymorphic type (e.g., $\alltype{\alpha}{\alpha \arr \alpha}$). 
Furthermore, we do not know \emph{a priori} how many quantifiers we
need to instantiate. 

To solve this problem, we turn to
\emph{focalization}~\citep{Andreoli92:focusing}, the proof-theoretic foundation
of bidirectional typechecking. In focused sequent calculi, it is
natural to give terms in \emph{spine form}
\citep{Cervesato03,Simmons12}, sequences of applications to a
head. So we view every application as really being a spine consisting
of a series of type applications followed by a term application, and
introduce an \emph{application judgment} $\declappjudg{\Psi}{e}{A}{C}$, 
which says that if a term of type $A$ is applied to argument $e$, the result
has type $C$.  Consequently, quantifiers will be instantiated exactly when
needed to reveal a function type.

The application judgment lets us suppress explicit type applications,
but to get the $\eta$ law, we need more. Recall the example with $f :
\unitty \to \alltype{\alpha}{\alpha}$.  In $\eta$-reducing
$\fun{x}{f\;x}$ to $f$, we reduce the number of applications in the
term. That is, we no longer have a syntactic position at which we can
(implicitly) instantiate polymorphic quantifiers. To handle this, we
follow \citet{Odersky96} in modeling type instantiation using
\emph{subtyping}, where subtyping is defined as a
``more-polymorphic-than'' relation that guesses type instantiations
arbitrarily deeply within types. As a result, $\unitty \to
\alltype{\alpha}{\alpha}$ is a subtype of $\unitty \to \unitty$, and
the $\eta$ law holds.

Happily, subtyping \emph{does} fit naturally into bidirectional
systems~\citep{Davies00icfpIntersectionEffects,DunfieldThesis,LovasThesis},
so we can give a declarative, bidirectional type system that guesses
type instantiations, but is otherwise entirely syntax-directed.
In particular, subsumption is confined to a single rule (which switches
from checking to synthesis), and our use of an application judgment
determines when to instantiate quantifiers.
The resulting system is very well-behaved, and ensures that the
expected typability results (such as typability being preserved by
$\eta$-reductions) continue to hold. Furthermore, our declarative
formulation makes it clear that the fundamental algorithmic problem in
extending bidirectional typechecking to polymorphism is precisely the
problem of figuring out what the missing type applications are.

Preserving the $\eta$-rule for functions comes at a cost.  The
subtyping relation induced by instantiation is undecidable for
\emph{impredicative} polymorphism~\citep{Tiuryn96,Chrzaszcz98}.  Since
we want a complete typechecking algorithm, we consequently restrict
our system to predicative polymorphism, where polymorphic quantifiers
can be instantiated only with monomorphic types. We discuss alternatives
in \Sectionref{sec:related}.

\subsection{Typing in Detail}

\paragraph{Language overview.}
In \Figureref{fig:source-syntax}, we give the grammar for our
language. We have a unit term $\unitexp$, variables $x$,
lambda-abstraction $\lam{x}e$, application $e_1\,e_2$, and type
annotation $(e : A)$.  We write $A$, $B$, $C$ for types
(\Figureref{fig:declarative-typing-syntax}):
types are the unit type $\unitty$, type variables $\alpha$, universal
quantification $\alltype{\alpha}{A}$, and functions $A \arr B$.
Monotypes $\tau$ and $\sigma$ are the same, less the universal
quantifier.  Contexts $\Psi$ are lists of type variable declarations,
and term variables $x : A$, with the expected well-formedness
condition. (We give a single-context formulation mixing type and term
hypotheses to simplify the presentation.)

\paragraph{Checking, synthesis, and application.} Our type system has
three main judgments, given in \Figureref{fig:decl-typing}. The
checking judgment $\declchkjudg{\Psi}{e}{A}$ asserts that $e$ checks
against the type $A$ in the context $\Psi$.  The synthesis
judgment $\declsynjudg{\Psi}{e}{A}$ says that we can synthesize
the type $A$ for $e$ in the context $\Psi$. Finally, an
\emph{application judgment} $\declappjudg{\Psi}{e}{A}{C}$ says that
if a (possibly polymorphic) function of type $A$ is applied to argument $e$,
then the whole application synthesizes $C$ for the whole application.

As is standard in the proof-theoretic presentations of bidirectional
typechecking, each of the introduction forms in our calculus has a
corresponding checking rule.  The \DeclUnitIntro rule says that
$\unitexp$ checks against the unit type $\unitty$.  The \DeclArrIntro
rule says that $\fun{x}{e}$ checks against the function type $A \arr B$
if $e$ checks against $B$ with the additional hypothesis that $x$
has type $A$.  The \DeclAllIntro rule says that $e$ has type
$\alltype{\alpha}{A}$ if $e$ has type $A$ in a context extended with a
fresh $\alpha$.\footnote{Note that we do not require an explicit type
  abstraction operation. As a result, an implementation needs to use
  some technique like scoped type variables~\citep{PeytonJones04} to mention
  bound type variables in annotations. This point does not matter to
  the abstract syntax, though.} Sums, products and recursive types can
be added similarly (we leave them out for simplicity).
Rule \DeclSub mediates between synthesis and
checking: it says that $e$ can be checked against $B$, if $e$
synthesizes $A$ and $A$ is a subtype of $B$ (that is, $A$ is at least
as polymorphic as $B$).

As expected, we can infer a type for a variable (the \DeclVar rule)
just by looking it up in the context. Likewise, the \DeclAnno rule
says that we can synthesize a type $A$ for a term with a type annotation $(e : A)$
just by returning that type (after checking that the term does actually 
check against $A$).

Application is a little more complex: we have to eliminate universals until we reach an arrow type.
To do this, we use an application judgment $\declappjudg{\Psi}{e}{A}{C}$,
which says that if we apply a term of type $A$ to an argument $e$,
we get something of type $C$. This judgment works by guessing
instantiations of polymorphic quantifiers in rule \DeclAllApp. Once we have
instantiated enough quantifiers to expose an arrow $A \arr C$,
we check $e$ against $A$ and return $C$ in rule \DeclArrApp.

In the following example, where we are applying some function polymorphic
in $\alpha$, \DeclAllApp instantiates the outer quantifier (to the unit type $\unitty$;
we elide the premise $\judgetp{\Psi}{\unitty}$), but leaves the inner quantifier over
$\beta$ alone.

\begin{mathdispl}
  \Infer{\DeclAllApp}
     {
       \Infer{\DeclArrApp}
           {\declchkjudg{\Psi}{x}{(\alltype{\beta} \beta{\arr}\beta)}}
           {\declappjudg{\Psi}{x}{(\alltype{\beta} \beta{\arr}\beta) \arr \unitty \arr \unitty}{\unitty \arr \unitty}}
     }
     {\declappjudg{\Psi}{x}{\big(\alltype{\alpha}  (\alltype{\beta} \beta{\arr}\beta) \arr \alpha \arr \alpha\big)}{\unitty \arr \unitty}}
\end{mathdispl}

In the minimal proof-theoretic formulation of
bidirectionality~\citep{Davies00icfpIntersectionEffects,Dunfield04:Tridirectional},
introduction forms are checked and elimination forms synthesize, full
stop.  Even $\unitexp$ cannot synthesize its type!  Actual
bidirectional typecheckers tend to take a more liberal view, adding
synthesis rules for at least some introduction forms.  To show that
our system can accommodate these kinds of extensions, we add the
\DeclUnitIntroSyn and \DeclArrIntroSyn rules, which synthesize a unit
type for $\unitexp$ and a monomorphic function type for lambda
expressions $\fun{x}{e}$. We examine other variations, including a purist
bidirectional no-inference alternative, and a liberal Damas-Milner
alternative, in \Sectionref{sec:variants}.

\paragraph{Instantiating types.}
We express the fact that one type is a polymorphic generalization
of another by means of the subtyping judgment given in
\Figureref{fig:declarative}.
One important aspect of the judgment is that types are compared relative to a
context of free variables.  This simplifies our rules, by letting us eliminate the
awkward side conditions on sets of free variables that plague many presentations.
Most of the subtyping judgment is typical: it proceeds structurally on types, with a
contravariant twist for the arrow; all the real action is contained within
the two subtyping rules for the universal quantifier.

The left rule, \DsubAllL, says that a type $\alltype{\alpha}{\!A}$ is a
subtype of $B$, if some instance $[\tau/\alpha]A$ is a subtype of
$B$.  This is what makes these rules only a declarative specification: 
\DsubAllL guesses the instantiation $\tau$ ``out of thin air'',
and so the rules do not directly yield an algorithm.

The right rule \DsubAllR is a little more subtle. It says that $A$ is a subtype
of $\alltype{\beta}{\!B}$ if we can show that $A$ is a subtype of $B$
in a context extended with $\beta$. There are two intuitions for this
rule, one semantic, the other proof-theoretic. The semantic intuition
is that since $\alltype{\beta}{\!B}$ is a subtype of $[\tau/\beta]B$ for
any $\tau$, we need $A$ to be a subtype of $[\tau/\beta]B$ for any
$\tau$. Then, if we can show that $A$ is a subtype of $B$, with a free
variable $\beta$, we can appeal to a substitution principle for subtyping
to conclude that for all $\tau$, type $A$ is a subtype of $[\tau/\beta]B$.

The proof-theoretic intuition is simpler. The rules \DsubAllL and \DsubAllR
are just the left and right rules for universal quantification in the sequent
calculus.
Type inference is a form of theorem proving, and our subtype relation
gives some of the inference rules a theorem prover may use.
Following good proof-theoretic hygiene enables us to leave the reflexivity and
transitivity rules out of the subtype relation, since they are admissible properties
(in sequent calculus terms, they are the identity and cut-admissibility properties).
The absence of these rules (particularly, the absence of transitivity), in turn,
simplifies a number of proofs.
In fact, the rules are practically syntax-directed: the only exception is
when both types are quantifiers, and either \DsubAllL or \DsubAllR could
be tried.  Since \DsubAllR is invertible, however, in practice one can apply
it eagerly.

\paragraph{Let-generalization.}
In many accounts of type inference, let-bindings are treated
specially.  For example, traditional Damas-Milner type inference only
does polymorphic generalization at let-bindings. Instead, we have
sought to avoid a special treatment of let-bindings. In logical terms,
let-bindings internalize the cut rule, and so special treatment puts
the cut-elimination property of the calculus at risk---that is, typability
may not be preserved when a let-binding is substituted
away. To make let-generalization safe, additional properties like the
principal types property are needed, a property endangered by
rich type system features like higher-rank polymorphism,
refinement types~\citep{DunfieldThesis} and GADTs~\citep{Vytiniotis10}.

To emphasize this point, we have omitted let-binding from our
formal development.  But since cut is admissible---i.e., the
substitution theorem holds---restoring let-bindings is easy, as 
long as they get no special treatment incompatible with substitution.
For example, the standard bidirectional rule for let-bindings 
is suitable:
\begin{displaymath}
  \inferrule*[]
            { \declsynjudg{\Psi}{e}{A} \\
              \declchkjudg{\Psi, x:A}{e'}{C}
            }
            { \declchkjudg{\Psi}{\Let{x}{e}{e'}}{C} 
            }
\end{displaymath}
Note the absence of generalization in this rule. 

\subsection{Bidirectional Typing and Type Assignment System F}

\begin{figure}[t]
\judgbox{\judge{\Psi}{e}{A}}{Under context $\Psi$, $e$ has type $A$}
\begin{mathpar}
\Infer{\AssignVar}
      {(x:A) \in \Psi}
      {\judge{\Psi}{x}{A}}
\and
\Infer{\AssignUnit}
      { }
      {\judge{\Psi}{\unitexp}{\unitty}}
\\
\Infer{\AssignArrIntro}
      {\judge{\Psi, x:A}{e}{B}}
      {\judge{\Psi}{\lam{x} e}{A \arr B}}
~~~
\Infer{\AssignArrElim}
      {\judge{\Psi}{e_1}{A \arr B} \\
       \judge{\Psi}{e_2}{A}}
      {\judge{\Psi}{e_1\;e_2}{B}}
\\
\Infer{\AssignAllIntro}
      {\judge{\Psi, \alpha}{e}{A}}
      {\judge{\Psi}{e}{\alltype{\alpha}{A}}}
\and
\Infer{\AssignAllElim}
      {\judge{\Psi}{e}{\alltype{\alpha}{A}} \\
       \judgetp{\Psi}{\tau}}
      {\judge{\Psi}{e}{[\tau/\alpha]A}}    
\end{mathpar}
\caption{Type assignment rules for predicative System F}
\FLabel{fig:type-assignment}
\end{figure}

Since our declarative specification is (quite consciously) not the
usual type-assignment presentation of System F, a natural question
is to ask what the relationship is. Luckily, the two systems are quite
closely related: we can show that if a term is well-typed in
our type assignment system,
it is always possible to add type annotations to make the
term well-typed in the bidirectional system;
conversely, if the bidirectional system types a term,
then some $\beta\eta$-equal term is well-typed under the type assignment system.

We formalize these properties with the following theorems, taking
$|e|$ to be the erasure of all type annotations from a term.
We give the rules for our type assignment System F in \Figureref{fig:type-assignment}. 

\begin{restatable}[Completeness of Bidirectional Typing]{theorem}{completenessbidirectional}
  \Label{thm:completeness-bidirectional}
  If $\judge{\Psi}{e}{A}$ then there exists $e'$ such that $\declsynjudg{\Psi}{e'}{A}$
  and $|e'| = e$. 
\end{restatable}
\vspace{-3pt}
\begin{restatable}[Soundness of Bidirectional Typing]{theorem}{soundnessbidirectional}
  \Label{thm:soundness-bidirectional}
  If $\declchkjudg{\Psi}{e}{A}$ then there exists $e'$ such that $\judge{\Psi}{e'}{A}$
  and $e' \beeq  |e|$.
\end{restatable}

Note that in the soundness theorem, the equality is up to $\beta$
and $\eta$.  We may need to $\eta$-expand bidirectionally-typed
terms to make them typecheck under the type assignment system, and
within the proof of soundness, we $\beta$-reduce identity coercions.

\subsection{Robustness of Typing}

Type annotations are an essential part of the bidirectional approach:
they mediate between type checking and type synthesis.
However, we want to relieve programmers from having to
write redundant type annotations, and even more importantly,
enable programmers to easily predict where type annotations
are needed.

Since our declarative system is bidirectional, the basic property is
that type annotations are required only at redexes. Additionally, these
typing rules can infer (actually, guess) all monomorphic types, so the answer to
the question of where annotations are needed is: only on bindings of
polymorphic type.\footnote{The number of
  annotations can be reduced still further; see \Sectionref{sec:variants}
  for how to infer the types of all terms typable under Damas-Milner.}  %
Where bidirectional typing really stands out is in its robustness under substitution.
We can freely substitute and ``unsubstitute'' terms:

\begin{restatable}[Substitution]{theorem}{termsubstitution}
\Label{thm:term-substitution}
  Assume $\declsynjudg{\Psi}{e}{A}$.
~\\[-3ex]
  \begin{itemize}
  \item If $\declchkjudg{\Psi, x:A}{e'}{C}$ then $\declchkjudg{\Psi}{[e/x]e'}{C}$.
  \item If $\declsynjudg{\Psi, x:A}{e'}{C}$ then $\declsynjudg{\Psi}{[e/x]e'}{C}$.
  \end{itemize}
\end{restatable}

\begin{restatable}[Inverse Substitution]{theorem}{termUnsubstitution}
\Label{thm:term-unsubstitution}
    Assume $\declchkjudg{\Psi}{e}{A}$.
    ~\\[-3ex]
    \begin{itemize}
      \item If\, $\declchkjudg{\Psi}{[(e:A)/x]e'}{C}$ then\, $\declchkjudg{\Psi, x:A}{e'}{C}$.
      \item If\, $\declsynjudg{\Psi}{[(e:A)/x]e'}{C}$ then\, $\declsynjudg{\Psi, x:A}{e'}{C}$.
    \end{itemize}
\end{restatable}

Substitution is stated in terms of synthesizing expressions, since any checking term
can be turned into a synthesizing term by adding an annotation.  Dually, inverse
substitution allows extracting any checking term into a let-binding with a type annotation.\footnote{%
The generalization of \Theoremref{thm:term-unsubstitution} to any synthesizing term---not just
$(e : A)$---does not hold.
For example, given $e = \lam{y} y$ and $e' = x$ and
$\declsynjudg{\Psi}{\lam{y} y}{\unitty \arr \unitty}$
and $\declchkjudg{\Psi}{\lam{y} y}{C_1 \arr C_2}$,
we cannot derive $\declchkjudg{\Psi, x : \unitty \arr \unitty}{x}{C_1 \arr C_2}$
unless $C_1$ and $C_2$ happen to be $\unitty$.
} %
However, doing so indiscriminately can lead to a term with many redundant annotations, and so
we also characterize when annotations can safely be removed:

\begin{restatable}[Annotation Removal]{theorem}{annotationRemoval}
\Label{thm:annotation-dropping}
  We have that:
  ~\\[-3ex]
  \begin{itemize}
  \item If\, $\declchkjudg{\Psi}{\big((\fun{x}{e}) : A\big)}{C}$ then $\declchkjudg{\Psi}{\fun{x}{e}}{C}$.
  \item If\, $\declchkjudg{\Psi}{(\unitexp : A)}{C}$ then $\declchkjudg{\Psi}{\unitexp}{C}$.
  \item If\, $\declsynjudg{\Psi}{e_1\;(e_2 : A)}{C}$ then $\declsynjudg{\Psi}{e_1\;e_2}{C}$.
  \item If\, $\declsynjudg{\Psi}{(x : A)}{A}$ then $\declsynjudg{\Psi}{x}{B}$ and $\declsubjudg{\Psi}{B}{A}$.
  \item If\, $\declsynjudg{\Psi}{\big((e_1\,e_2) : A\big)}{A}$
    \\ then $\declsynjudg{\Psi}{e_1\,e_2}{B}$
    and $\declsubjudg{\Psi}{B}{A}$. 
  \item If\, $\declsynjudg{\Psi}{\big((e : B) : A\big)}{A}$
    \\ then $\declsynjudg{\Psi}{(e : B)}{B}$
    and $\declsubjudg{\Psi}{B}{A}$. 
  \item If $\declsynjudg{\Psi}{((\fun{x}{e}):\sigma \to \tau)}{\sigma \to \tau}$ then $\declsynjudg{\Psi}{\fun{x}{e}}{\sigma \to \tau}$. 
  \end{itemize}
\end{restatable}

\noindent We can also show that the expected $\eta$-laws hold:
\begin{restatable}[Soundness of Eta]{theorem}{soundnessofeta}
~\\\Label{thm:soundness-of-eta}
  If\, $\declchkjudg{\Psi}{\fun{x}{e\;x}}{A}$ and $x \not\in \FV{e}$, then $\declchkjudg{\Psi}{e}{A}$.   
\end{restatable}

\section{Algorithmic Type System}
\Label{sec:algorithmic}

Our declarative bidirectional system is a fine specification of how typing
should behave, but it enjoys guessing entirely too much: the typing rules \DeclAllApp
and \DeclArrIntroSyn could only be implemented with the help of an oracle.  The
declarative subtyping rule \DsubAllL has the same problem.

The first step in building our \emph{algorithmic} bidirectional system will be to modify
the three oracular rules so that, instead of guessing a type, they defer the choice
by creating an existential type variable, to be solved later.
However, our existential variables are not exactly unification variables; they
are organized into  \emph{ordered algorithmic contexts} (\Sectionref{sec:alg-contexts}),
which define the variables' scope and controls the free variables of their solutions.

The algorithmic type system consists of subtyping rules (\Figureref{fig:alg-subtyping},
discussed in \Sectionref{sec:alg-subtyping}),
instantiation rules (\Figureref{fig:instantiation},
discussed in \Sectionref{sec:instantiation}),
and typing rules (\Figureref{fig:alg-typing},
discussed in \Sectionref{sec:alg-typing}).
All of the rules manipulate
the contexts in a way consistent with \emph{context extension},
a metatheoretic notion described in \Sectionref{sec:context-extension};
context extension is key in stating and proving decidability, soundness
and completeness.

\begin{figure}[t]
  \centering
  \[
      \begin{array}{llcl}
      \text{Types} & A, B, C & \bnfas &
            \unitty \bnfalt \alpha \bnfalt \ahat \bnfalt \alltype{\alpha}{A} \bnfalt A \arr B
      \\[2pt]
      \text{Monotypes} & \tau,\sigma & \bnfas &
            \unitty \bnfalt \alpha \bnfalt \ahat \bnfalt \tau \arr \sigma
      \\[4pt]
      \text{Contexts} & \Gamma, \Delta, \Theta & \bnfas &
                  \cdot
                  \bnfalt \Gamma, \alpha 
                  \bnfalt \Gamma, x:A
                  \\[1pt] &&&\!\!\!
                  \bnfalt \Gamma, \ahat
                  \bnfalt \Gamma, \hypeq{\ahat}{\tau}
                  \bnfalt \Gamma, \MonnierComma{\ahat}
      \\[2pt]
      \text{Complete Contexts}     & \Omega & \bnfas &
                  \cdot
                  \bnfalt    \Omega, \alpha
                  \bnfalt    \Omega, x:A
                  \\[1pt] &&&\!\!\!
                  \bnfalt    \Omega, \hypeq{\ahat}{\tau} 
                  \bnfalt    \Omega, \MonnierComma{\ahat}
      \end{array}
  \]
  
  \caption{Syntax of types, monotypes, and contexts in the algorithmic system}
  \FLabel{fig:alg-syntax}
\end{figure}

\begin{figure}[t]
  \judgbox{\judgetp{\Gamma}{A}}{Under context $\Gamma$, type $A$ is well-formed}
  \begin{mathpar}
      \Infer{\UvarWF}
            { }
            { \judgetp{\Gamma[\alpha]}{\alpha} }
      \and
      \Infer{\UnitWF}
            { }
            { \judgetp{\Gamma}{\unitty} }
      \and
      \Infer{\ArrowWF}
            { \judgetp{\Gamma}{A} \\ \judgetp{\Gamma}{B} }
            { \judgetp{\Gamma}{A \arr B} }
      \and
      \Infer{\ForallWF}
            { \judgetp{\Gamma, \alpha}{A} }
            { \judgetp{\Gamma}{\alltype{\alpha}{A}} }
      \vspace{-1.3ex}
      \\
      \Infer{\EvarWF}
            { }
            { \judgetp{\Gamma[\ahat]}{\ahat} }
       ~~~~
      \Infer{\SolvedEvarWF}
            { }
            { \judgetp{\Gamma[\hypeq{\ahat}{\tau}]}{\ahat} }
  \end{mathpar}
 
  \medskip
  
  \judgbox{\judgectx{\Gamma}}{Algorithmic context $\Gamma$ is well-formed}
  \begin{mathpar}
      \Infer{\EmptyCWF}
            { }
            { \judgectx{\cdot} }
      \and
      \Infer{\UvarCWF}
            { \judgectx{\Gamma}
              \\
              \alpha \notin \dom{\Gamma}}
            { \judgectx{\Gamma, \alpha} }
      \and
      \Infer{\VarCWF}
            { \judgectx{\Gamma}
              \\
              x \notin \dom{\Gamma}
               \\
              \judgetp{\Gamma}{A}
            }
            { \judgectx{\Gamma, x : A} }
      \and
      \Infer{\EvarCWF}
            { \judgectx{\Gamma}
              \\
              \ahat \notin \dom{\Gamma}}
            { \judgectx{\Gamma, \ahat} }
      \and
      \Infer{\SolvedEvarCWF}
            { \judgectx{\Gamma}
              \\
              \ahat \notin \dom{\Gamma}
              \\
              \judgetp{\Gamma}{\tau}
            }
            { \judgectx{\Gamma, \hypeq{\ahat}{\tau}} }
      \and
      \Infer{\MarkerCWF}
            { \judgectx{\Gamma}
              \\
              \MonnierComma{\ahat} \notin \Gamma
              \\
              \ahat \notin \dom{\Gamma}
            }
            { \judgectx{\Gamma, \MonnierComma{\ahat}} }
  \end{mathpar}
  \caption{Well-formedness of types and contexts in the algorithmic system}
  \FLabel{fig:algorithmic-wf}
\end{figure}

\subsection{Algorithmic Contexts}
\Label{sec:alg-contexts}

A notion of (ordered) algorithmic context is central to our approach.  Like declarative
contexts $\Psi$, algorithmic contexts $\Gamma$ (see \Figureref{fig:alg-syntax};
we also use the letters $\Delta$ and $\Theta$) contain declarations of universal
type variables $\alpha$ and term variable typings $x : A$.  Unlike declarative
contexts, algorithmic contexts also contain declarations of existential type variables
$\ahat$, which are either unsolved (and we simply write $\ahat$) or solved to some
monotype ($\hypeq{\ahat}{\tau}$).  Finally, for scoping reasons that will become
clear when we examine the rules, algorithmic contexts also
include a \emph{marker} $\MonnierComma{\ahat}$.

Complete contexts $\Omega$ are the same as contexts, except that they
cannot have unsolved variables.

The well-formedness rules for contexts (\Figureref{fig:algorithmic-wf}, bottom) do not
only prohibit duplicate declarations, but also enforce order:
if $\Gamma = (\Gamma_L, x : A, \Gamma_R)$, the type $A$ must be well-formed under $\Gamma_L$;
it cannot refer to variables $\alpha$ or $\ahat$ in $\Gamma_R$.
Similarly, if $\Gamma = (\Gamma_L, \hypeq{\ahat}{\tau}, \Gamma_R)$,
the solution type $\tau$ must be well-formed under $\Gamma_L$.
Consequently, circularity is ruled out:
$(\hypeq{\ahat}{\bhat}, \hypeq{\bhat}{\ahat})$
is not well-formed.

\paragraph{Contexts as substitutions on types.}
An algorithmic context can be viewed as a substitution for its solved
existential variables.  For example,
$\hypeq{\ahat}{\unitty}, \hypeq{\bhat}{\ahat \arr \unitty}$ 
can be applied as if it were the substitution
$\unitty/\ahat, (\ahat{\arr}\unitty) / \bhat$
(applied right to left), or the simultaneous substitution
$\unitty/\ahat, (\unitty{\arr}\unitty) / \bhat$.
We write $[\Gamma]A$ for $\Gamma$ applied as a substitution
to type $A$; this operation is defined in \Figureref{fig:substitution}.

\paragraph{Complete contexts.}
Complete contexts $\Omega$ (\Figureref{fig:alg-syntax}) have no
unsolved variables.  Therefore, applying such a context to a type
$A$ (provided it is well-formed: $\judgetp{\Omega}{A}$)
yields a type $[\Omega]A$ with no existentials.  Complete contexts are essential for stating and
proving soundness and completeness, but are not explicitly distinguished in any
of our rules.

\begin{figure}[t]
  \[
  \begin{array}[t]{l@{~~~}c@{~~~}ll}
      [\Gamma]\alpha   & = &   \alpha &
      \\{}
      [\Gamma]\unitty   & = &   \unitty &
      \\[1pt]
      \big[\Gamma[\hypeq{\ahat}{\tau}]\big] \ahat
               & = &   \big[\Gamma[\hypeq{\ahat}{\tau}]\big]\tau &
      \\[2pt]
      \big[\Gamma[\ahat]\big]\ahat   & = &   \ahat &
      \\[2pt]
      [\Gamma](A \arr B)   & = &
          ([\Gamma]A) \arr ([\Gamma]B) &
      \\{}
      [\Gamma](\alltype{\alpha} A)
         & = & 
         \alltype{\alpha} [\Gamma]A & 
  \end{array}
  \]
  \caption{Applying a context, as a substitution, to a type}
  \FLabel{fig:substitution}
\end{figure}

\paragraph{Hole notation.}
Since we will manipulate contexts not only by appending declarations (as in the declarative
system) but by inserting and replacing declarations in the middle,
a notation for contexts with a hole is useful:
\vspace{-4pt}
\begin{displ}
  $\Gamma = \Gamma_0[\Theta]$~~~means $\Gamma$ has the form $(\Gamma_L, \Theta, \Gamma_R)$
\end{displ}
For example, if $\Gamma = \Gamma_0[\bhat] = (\ahat, \bhat, x : \bhat)$,
then $\Gamma_0[\hypeq{\bhat}{\ahat}] = (\ahat, \hypeq{\bhat}{\ahat}, x : \bhat)$.
Since this notation is concise, we use it even in rules that do not replace declarations,
such as the rules for type well-formedness in \Figureref{fig:algorithmic-wf}.

Occasionally, we also need contexts with \emph{two} ordered holes:
\begin{displ}
  $\Gamma = \Gamma_0[\Theta_1][\Theta_2]$
  ~~~means
  $\Gamma$ has the form $(\Gamma_L, \Theta_1, \Gamma_M, \Theta_2, \Gamma_R)$
\end{displ}

\paragraph{Input and output contexts.}
Our declarative system used a subtyping judgment and three typing judgments:
checking, synthesis, and application.  Our algorithmic system includes similar
judgment forms, except that we replace the declarative context $\Psi$ with an
algorithmic context $\Gamma$ (the \emph{input context}), and add an \emph{output} context
$\Delta$ written after a backwards turnstile: $\subjudg{\Gamma}{A}{B}{\Delta}$ for subtyping,
$\chkjudg{\Gamma}{e}{A}{\Delta}$ for checking, and so on.
Unsolved existential variables get solved when they are compared against a type.
For example, $\ahat \subtype \beta$ would lead to replacing the unsolved declaration
$\ahat$ with $\hypeq{\ahat}{\beta}$ in the context (provided $\beta$ is declared
to the left of $\ahat$).  Input contexts thus evolve into output contexts that are ``more solved''.

The differences between the declarative and algorithmic systems,
particularly manipulations of existential variables, are most prominent in
the subtyping rules, so we discuss those first.

\subsection{Algorithmic Subtyping}
\Label{sec:alg-subtyping}

\begin{figure*}[htbp]
  \judgbox{\subjudg{\Gamma}{A}{B}{\Delta}}%
     {Under input context $\Gamma$,
       type $A$ is a subtype of $B$, with output context $\Delta$}
  \begin{mathpar}
  \Infer{\SubVar}
              { }
              {\subjudg{\Gamma[\alpha]}{\alpha}{\alpha}{\Gamma[\alpha]}}
  \and
  \Infer{\SubUnit}
            { }
            {\subjudg{\Gamma}{\unitty}{\unitty}{\Gamma}}
  \and
  \Infer{\SubExvar}
            { }
            {\subjudg{\Gamma[\ahat]}{\ahat}{\ahat}{\Gamma[\ahat]}}
  \and
  \Infer{\SubArr}
            {\subjudg{\Gamma}{B_1}{A_1}{\Theta} \\
              \subjudg{\Theta}{[\Theta]A_2}{[\Theta]B_2}{\Delta}}
            {\subjudg{\Gamma}{A_1 \arr A_2}{B_1 \arr B_2}{\Delta}}
  \\
  \Infer{\SubAllL}
            {\subjudg{\Gamma, \MonnierComma{\ahat}, \ahat}
                     {[\ahat/\alpha]A}
                     {B}
                     {\Delta, \MonnierComma{\ahat}, \Theta}}
            {\subjudg{\Gamma}{\alltype{\alpha}{A}}{B}{\Delta}}
  \and
  \Infer{\SubAllR}
            {\subjudg{\Gamma, \alpha}{A}{B}{\Delta, \alpha, \Theta}}
            {\subjudg{\Gamma}{A}{\alltype{\alpha}{B}}{\Delta}}
  \\
  \Infer{\SubInstL}
            {
              \ahat \notin \FV{%
                A}
              \\
              \instjudg{\Gamma[\ahat]}{\ahat}{%
                   A}{\Delta}
            }
            {\subjudg{\Gamma[\ahat]}{\ahat}{A}{\Delta}}
  \and
  \Infer{\SubInstR}
            {
              \ahat \notin \FV{%
                A}
              \\
              \instjudgr{\Gamma[\ahat]}{\ahat}{%
                A}{\Delta}
            }
            {\subjudg{\Gamma[\ahat]}{A}{\ahat}{\Delta}}
  \end{mathpar}  
  \caption{Algorithmic subtyping}
  \FLabel{fig:alg-subtyping}
\end{figure*}

\begin{figure*}[htbp]
      \judgbox{\instjudg{\Gamma}{\ahat}{A}{\Delta}}%
         {Under input context $\Gamma$,
           instantiate $\ahat$ such that $\ahat \subtype A$, 
           with output context $\Delta$}
      \begin{mathpar}
        \Infer{\InstLSolve}
                { \Gamma \entails \tau} %
                { \instjudg{\Gamma, \ahat, \Gamma'}
                            {\ahat}
                            {\tau}
                            {\Gamma, \hypeq{\ahat}{\tau}, \Gamma'}
                 }
        \and
        \Infer{\InstLReach}
                { }
                {\instjudg{\Gamma[\ahat][\bhat]}
                            {\ahat}
                            {\bhat}
                            {\Gamma[\ahat][\hypeq{\bhat}{\ahat}]}}
        \and
        \Infer{\InstLArr}
                {\instjudgr{\Gamma[\ahat_2, \ahat_1, \hypeq{\ahat}{\ahat_1 \arr \ahat_2}]}
                            {\ahat_1}
                            {A_1}
                            {\Theta} \\
                 \instjudg{\Theta}
                            {\ahat_2}
                            {[\Theta]A_2}
                            {\Delta}}
                {\instjudg{\Gamma[\ahat]}
                            {\ahat}
                            {A_1 \arr A_2}
                            {\Delta}}
        \and
        \Infer{\InstLAllR}
              {\instjudg{\Gamma[\ahat], \beta}{\ahat}{B}{\Delta, \beta, \Delta'}}
              {\instjudg{\Gamma[\ahat]}{\ahat}{\alltype{\beta}{B}}{\Delta}}
      \end{mathpar}    
    \\[-1.5ex]
      \judgbox{\instjudgr{\Gamma}{\ahat}{A}{\Delta}}
         {Under input context $\Gamma$,
           instantiate $\ahat$ such that $A \subtype \ahat$,
           with output context $\Delta$}
      \begin{mathpar}
        \Infer{\InstRSolve}
                { \Gamma \entails \tau}
                { \instjudgr{\Gamma, \ahat, \Gamma'}
                            {\ahat}
                            {\tau}
                            {\Gamma, \hypeq{\ahat}{\tau}, \Gamma'}
                 }
        \and
        \Infer{\InstRReach}
                { }
                {\instjudgr{\Gamma[\ahat][\bhat]}
                           {\ahat}
                           {\bhat}
                           {\Gamma[\ahat][\hypeq{\bhat}{\ahat}]}}
        \and
        \Infer{\InstRArr}
              {\instjudg{\Gamma[\ahat_2, \ahat_1, \hypeq{\ahat}{\ahat_1 \arr \ahat_2}]}
                        {\ahat_1}
                        {A_1}
                        {\Theta} \\
                 \instjudgr{\Theta}
                           {\ahat_2}
                           {[\Theta]A_2}
                           {\Delta}}
              {\instjudgr{\Gamma[\ahat]}
                         {\ahat}
                         {A_1 \arr A_2}
                         {\Delta}}
        \and 
        \Infer{\InstRAllL}
              {\instjudgr{\Gamma[\ahat], \MonnierComma{\bhat}, \bhat}{\ahat}{[\bhat/\beta]B}{\Delta, \MonnierComma{\bhat}, \Delta'}}
              {\instjudgr{\Gamma[\ahat]}{\ahat}{\alltype{\beta}{B}}{\Delta}}
      \end{mathpar}

\caption{Instantiation}
\FLabel{fig:instantiation}
\end{figure*}

\begin{figure*}[htbp]
  \judgbox{\chkjudg{\Gamma}{e}{A}{\Delta}}%
     {Under input context $\Gamma$, $e$ checks against input type $A$,
       with output context $\Delta$} \\[1ex]
  \judgbox{\synjudg{\Gamma}{e}{A}{\Delta}}%
     {Under input context $\Gamma$, $e$ synthesizes output type $A$,
       with output context $\Delta$} \\[1ex]
  \judgbox{\appjudg{\Gamma}{e}{A}{C}{\Delta}}%
     {Under input context $\Gamma$,
       applying a function of type $A$ to $e$ synthesizes type $C$,
       with output context $\Delta$} \\
  \begin{mathpar}
     \Infer{\Var}
          {(x : A) \in \Gamma}
          {\synjudg{\Gamma}{x}{A}{\Gamma}}
     \and
     \Infer{\Sub}
          {\synjudg{\Gamma}{e}{A}{\Theta}
            \\
            \subjudg{\Theta}{[\Theta]A}{[\Theta]B}{\Delta}
          }
          {\chkjudg{\Gamma}{e}{B}{\Delta}}
     \and
     \Infer{\Anno}
          {\judgetp{\Gamma}{A}
           \\
           \chkjudg{\Gamma}{e}{A}{\Delta}
          }
          {\synjudg{\Gamma}{(e : A)}{A}{\Delta}}
     \\
     \Infer{\UnitIntro}
           {}
           {\chkjudg{\Gamma}{\unitexp}{\unitty}{\Gamma}}
     \and
     \Infer{%
       {\UnitIntroSyn}}
           { }
           {{\synjudg{\Gamma}{\unitexp}{\unitty}{\Gamma}}}
     \and
     \Infer{\AllIntro}
           {\chkjudg{\Gamma, \alpha}{e}{A}{\Delta, \alpha, \Theta}
           }
           {\chkjudg{\Gamma}{e}{\alltype{\alpha}{A}}{\Delta}}
     \and
     \Infer{\AllApp}
            {\appjudg{\Gamma,\ahat}{e}{[\ahat/\alpha]A}{C}{\Delta}}
            {\appjudg{\Gamma}{e}{\alltype{\alpha}{A}}{C}{\Delta}}
     \\
\def\CompactJudgments{1}   %
     \Infer{\!\ArrIntro}
          {\chkjudg{\Gamma, x : A}{e}{B}{\Delta, x : A, \Theta}
          }
          {\chkjudg{\Gamma}{\lam{x} e}{A \arr B}{\Delta}}
     \hspace{3ex}
     \Infer{%
        {\!\ArrIntroSyn}
         }
           { \chkjudg{\Gamma, \ahat, \bhat, x : \ahat}{e}{\bhat}{\Delta, x : \ahat, \Theta}
           }
           {{\synjudg{\Gamma}{\lam{x} e}{\ahat \arr \bhat}{\Delta}}}
     \hspace{3ex}
     \Infer{\!\ArrElim}
           {\synjudg{\Gamma}{e_1}{A}{\Theta}
             \\
             \appjudg{\Theta}{e_2}{[\Theta]A}{C}{\Delta}
           }
           {\synjudg{\Gamma}{e_1\,e_2}{C}{\Delta}}
      \\
\def\CompactJudgments{0}
      \Infer{\SolveApp}
            {\chkjudg{\Gamma[\ahat_2, \ahat_1, \hypeq{\ahat}{\ahat_1 \arr \ahat_2}]}{e}{\ahat_1}{\Delta}}
            {\appjudg{\Gamma[\ahat]}{e}{\ahat}{\ahat_2}{\Delta}}
      \and
      \Infer{\ArrApp}
            {\chkjudg{\Gamma}{e}{A}{\Delta}}
            {\appjudg{\Gamma}{e}{A \arr C}{C}{\Delta}}
  \end{mathpar}
  \caption{Algorithmic typing}
  \FLabel{fig:alg-typing}
\end{figure*}

The first four subtyping rules in \Figureref{fig:alg-subtyping} do not directly manipulate
the context, but do illustrate how contexts are propagated.

Rules \SubVar and \SubUnit are reflexive rules; neither involves existential variables,
so the output context in the conclusion is the same as the input context $\Gamma$.
Rule \SubExvar concludes that any unsolved existential variable is a subtype of itself,
but this gives no clue as to how to solve that existential, so the output context is similarly
unchanged.

Rule \SubArr is a bit more interesting: it has two premises, where the first premise
has an output context $\Theta$, which is used as the input context to the
second (subtyping) premise; the second premise has output context $\Delta$,
which is the output of the conclusion.\footnote{Rule \SubArr enforces that the
function domains $B_1$, $A_1$ are compared first: $\Theta$ is an input to the
second premise.  But this is an arbitrary choice; the system would behave the
same if we chose to check the codomains first.}
Note that in \SubArr's second premise, we do not
simply check that $A_2 \subtype B_2$, but apply the first premise's output $\Theta$
to those types:
\[
\subjudg{\Theta}{[\Theta]A_2}{[\Theta]B_2}{\Delta}
\]
This maintains
a general invariant: whenever we try to derive $\subjudg{\Gamma}{A}{B}{\Delta}$,
the types $A$ and $B$ are already fully applied under $\Gamma$.  That is, they contain
no existential variables already solved in $\Gamma$.  On balance, this invariant simplifies 
the system: the extra applications of $\Theta$ in \SubArr avoid the need for
extra rules for replacing solved variables with their solutions.

All the rules discussed so far have been natural extensions of the declarative rules,
with \SubExvar being a logical way to extend reflexivity to types containing existentials.
Rule \SubAllL diverges significantly from the corresponding
declarative rule \DsubAllL.
Instead of replacing the type variable $\alpha$ with a guessed $\tau$,
rule \SubAllL replaces $\alpha$ with a new existential variable $\ahat$,
which it adds to the premise's input context:
$\subjudg{\Gamma, \MonnierComma{\ahat}, \ahat}{[\ahat/\alpha]A}{B}{\Delta, \MonnierComma{\ahat}, \Theta}$.
The peculiar-looking $\MonnierComma{\ahat}$ is a \emph{scope marker},
pronounced ``marker $\ahat$'', which will delineate existentials created by
\emph{articulation} (the step of solving $\ahat$ to $\ahat_1 \arr \ahat_2$,
discussed in the next subsection).  
The output context $(\Delta, \MonnierComma{\ahat}, \Theta)$
allows for some additional (existential) variables to appear after $\MonnierComma{\ahat}$,
in a trailing context $\Theta$.  These existential variables could mention $\ahat$,
or (if they appear between $\MonnierComma{\ahat}$ and $\ahat$)
could be mentioned \emph{by} $\ahat$; since $\ahat$ goes out of scope in
the conclusion, we drop such ``trailing existentials'' from the concluding
output context, which is simply $\Delta$.\footnote{%
   In our setting, it is safe to drop trailing existentials that are unsolved:
   such variables are unconstrained, and we can treat them as having been instantiated
   to any well-formed type, such as $\unitty$.  In a dependently typed setting, we would need
   to check that at least one solution exists.
}

Rule \SubAllR is fairly close to the declarative version, but for scoping reasons
similar to \SubAllL, it also drops $\Theta$, the part of the context to the right of
the universal type variable $\alpha$.  (Articulation makes no sense for universal variables,
so $\alpha$ can act as its own marker.)

The last two rules are essential: they derive
subtypings with an unsolved existential on one side, and an arbitrary type on
the other.  Rule \SubInstL derives $\ahat \subtype A$, and \SubInstR derives
$A \subtype \ahat$.  These rules do not directly change the output context;
they just do an ``occurs check'' $\ahat \notin \FV{A}$ to avoid circularity,
and leave all the real work to the instantiation judgment.

\subsection{Instantiation}
\Label{sec:instantiation}

Two almost-symmetric judgments instantiate unsolved existential variables:
$\instjudg{\Gamma}{\ahat}{A}{\Delta}$
and
$\instjudgr{\Gamma}{\ahat}{A}{\Delta}$.
The symbol $\instsyml$ suggests assignment of the variable to its left,
but also subtyping: the subtyping rule \SubInstL moves
from instantiation $\ahat \instsymlop A$, read ``instantiate $\ahat$ to a subtype of $A$'',
to subtyping $\ahat \subtype A$.  The symmetric judgment $A \instsymrop \ahat$
can be read ``instantiate $\ahat$ to a supertype of $A$''.

The first instantiation rule in \Figureref{fig:instantiation}, \InstLSolve, sets
$\ahat$ to $\tau$ in the output context: its conclusion is
$\instjudg{\Gamma, \ahat, \Gamma'}{\ahat}{\tau}{\Gamma, \hypeq{\ahat}{\tau}, \Gamma'}$.
The premise $\judgetp{\Gamma}{\tau}$ checks that the monotype $\tau$
is well-formed under the prefix context $\Gamma$.  
To check the soundness of this rule, we can take the conclusion $\ahat \instsymlop \tau$,
substitute our new solution for $\ahat$, and check that the resulting subtyping makes
sense.  Since $[\Gamma, \hypeq{\ahat}{\tau}, \Gamma']\ahat = \tau$, we ask whether
$\tau \subtype \tau$ makes sense, and of course it does through reflexivity.

Rule \InstLArr can be applied when the type $A$ in $\ahat \instsymlop A$ has the
form $A_1 \arr A_2$.  It follows that $\ahat$'s solution must have the form
$\cdots \arr \cdots$, so we ``articulate'' $\ahat$, giving it the solution $\ahat_1 \arr \ahat_2$
where the $\ahat_k$ are fresh existentials.
We insert their declarations just before $\ahat$---they must
be to the left of $\ahat$ so they can be mentioned in its solution, but they must be
close enough to $\ahat$ that they appear to the \emph{right} of the marker
$\MonnierComma{\ahat}$ introduced by \SubAllL.  Note that the first premise
 $A_1 \instsymrop \ahat_1$ switches to the other instantiation judgment.  Also,
the second premise $\instjudg{\Theta}{\ahat_2}{[\Theta]A_2}{\Delta}$ applies
$\Theta$ to $A_2$, to apply any solutions found in the first premise.  %

The other rules are somewhat subtle.  Rule \InstLReach derives
\[
  \instjudg{\Gamma[\ahat][\bhat]}{\ahat}{\bhat}{\Gamma[\ahat][\hypeq{\bhat}{\ahat}]}
\]
where, as explained in \Sectionref{sec:alg-contexts},
$\Gamma[\ahat][\bhat]$ denotes a context where
some unsolved existential variable $\ahat$ is declared to the left of $\bhat$.
In this situation, we cannot use \InstLSolve to set $\ahat$ to $\bhat$
because $\bhat$ is not well-formed under the part of the context to the left of $\ahat$.
Instead, we set $\bhat$ to $\ahat$.

Rule \InstLAllR is the instantiation version of \SubAllR.  Since our polymorphism
is predicative, we can't assign $\alltype{\beta} B$ to $\ahat$, but we can decompose
the quantifier in the same way that subtyping does.

The rules for the second judgment $A \instsymrop \ahat$ are similar:
\InstRSolve, \InstRReach and \InstRArr are direct analogues of the first
three $\ahat \instsymlop A$ rules, and \InstRAllL is the instantiation version of
\SubAllL.

\paragraph{Example.}  The interplay between instantiation and quantifiers
is delicate. For example, consider the problem of instantiating $\bhat$ to a
supertype of $\alltype{\alpha} \alpha$.
In this case, the type $\alltype{\alpha} \alpha$ is so polymorphic that it places no
constraints at all on $\bhat$. Therefore, it seems we are at risk of being forced
to make a necessarily incomplete choice---but the instantiation judgment's ability
to ``change its mind'' about which variable to instantiate saves the day:

\begin{mathpar}
\Infer{\InstRAllL}
      {\inferrule*[Right=\InstRReach]
             { }
             {\instjudgr{\Gamma[\bhat], \MonnierComma{\ahat}, \ahat}
                       {\bhat}
                       {\ahat}
                       {\Gamma[\bhat], \MonnierComma{\ahat}, \hypeq{\ahat}{\bhat}}
             }
      }
      {\instjudgr{{\Gamma[\bhat]}}{\bhat}{\alltype{\alpha}{\alpha}}{{\Gamma[\bhat]}}}
\end{mathpar}
Here, we introduce a new variable $\ahat$ to go under the
universal quantifier; then, instantiation applies \InstRReach
to set $\ahat$, not $\bhat$. Hence, $\bhat$ is,
correctly, \emph{not constrained} by this subtyping problem.

Thus, instantiation does not necessarily solve \emph{any} existential variable.
However, instantiation to any monotype $\tau$ will solve an existential variable---that is,
for input context $\Gamma$ and output $\Delta$,
we have $\unsolved{\Delta} < \unsolved{\Gamma}$.
This will be critical for decidability of subtyping (\Sectionref{sec:decidability-subtyping}).

\paragraph{Another example.}
\begin{figure*}[thbp]
  \centering
  \newcommand{\HiliteU}[1]{\mathcolorbox{red!15}{#1}}
  \newcommand{\HiliteV}[1]{\mathcolorbox{green!15}{#1}}
  \newcommand{\HiliteW}[1]{\mathcolorbox{blue!15}{#1}}
  \newcommand{\hGammap}{\HiliteU{\Gamma'}}
  \newcommand{\hDelta}{\HiliteV{\Delta}}
  \renewcommand{\shadebox}[1]{#1}

    \begin{mathdispl}
            \Infer{\InstRAllL}
                 {
                   \hspace{-126pt}
                   \Infer{\InstRArr}
                        {
                              \Infer{\InstLReach}
                                  {
                                       \judgetp{\overbrace{\hGammap, \MonnierComma{\bhat}}^{\hspace{-20pt}\text{context to the left of $\bhat$\hspace{-20pt}}}}
                                               {\bhat_1}
                                  }
                                  {
                                      \instjudg{\hGammap,
                                            \rput[tl](-20pt,-17pt){%
                                              \hGammap = \Gamma[\bhat_2, \bhat_1, \tighthypeq{\ahat}{\bhat_1{\arr}\bhat_2}]
                                            }
                                            \MonnierComma{\bhat}, \bhat}
                                              {\bhat_2}
                                              {\bhat}
                                              {\hGammap, \MonnierComma{\bhat}, \shadebox{\tighthypeq{\bhat}{\bhat_1}}}
                                  }
                              \\
                              \Infer{\InstRSolve}
                                     {\judgetp{\overbrace{\dots, \bhat_2}^{\hspace{-20pt}\text{context to the left of $\bhat_1$}\hspace{-20pt}}}{\bhat_2}}
                                     {
                                          \instjudgr{\hGammap, \MonnierComma{\bhat}, \tighthypeq{\bhat}{\bhat_1}}
                                                  {\bhat_1}
                                                  {\bhat_2}
                                                  {
                                                    \hDelta,
                                                    \MonnierComma{\bhat}, \tighthypeq{\bhat}{\bhat_1}}
                                    }
                        }
                        {
                           \instjudgr{\Gamma[\ahat], \shadebox{\MonnierComma{\bhat}, \bhat}}
                                     {\ahat}
                                     {\bhat{\arr}\bhat}
                                     {\hDelta,
                                       \MonnierComma{\bhat}, \tighthypeq{\bhat}{\bhat_1}}
                        }
                       \hspace{-157pt}
                 }
                 {
                   \instjudgr{%
                     \Gamma[\ahat]
                     }
                     {\ahat}
                     {
                       (\alltype{\beta} \beta{\arr}\beta)
                     }
                     {
                       \rput[bl](80pt,55pt){\hDelta = \Gamma[\bhat_2, \shadebox{\tighthypeq{\bhat_1}{\bhat_2}}, \tighthypeq{\ahat}{\bhat_1{\arr}\bhat_2}]}
                       \hDelta
                     }
                 }
    \end{mathdispl}

  \caption{Example of instantiation}
  \FLabel{fig:inst-example}
\end{figure*}

In \Figureref{fig:inst-example} we show a derivation that uses quantifier instantiation
(\InstRAllL), articulation (\InstRArr) and ``reaching'' (\InstLReach), as well as \InstRSolve.
In the output context
$\Delta = \Gamma[\bhat_2, \tighthypeq{\bhat_1}{\bhat_2}, \tighthypeq{\ahat}{\bhat_1{\arr}\bhat_2}]$
note that $\ahat$ is solved to $\bhat_1 \arr \bhat_2$,
and $\bhat_2$ is solved to $\bhat_1$.  Thus, $[\Delta]\ahat = \bhat_1{\arr}\bhat_1$, which is
a monomorphic approximation of $\alltype{\beta}\!\beta{\arr}\beta$.

\subsection{Algorithmic Typing}
\Label{sec:alg-typing}

We now turn to the typing rules in \Figureref{fig:alg-typing}.  Many of these rules follow
the declarative rules, with extra context machinery.
Rule \Var uses an assumption $x : A$ without generating any new information, so the output
context in its conclusion $\synjudg{\Gamma}{x}{A}{\Gamma}$ is just the input context.
Rule \Sub's first premise has an output context $\Theta$, used as the input context to the
second (subtyping) premise, which has output context $\Delta$, the output of the conclusion.
Rule \Anno does not directly change the context, but the derivation of its premise might
include the use of some rule that does, so we propagate the premise's output context $\Delta$
to the conclusion.

\paragraph{Unit and $\forall$.}
In the second row of typing rules, \UnitIntro and \UnitIntroSyn generate no new information
and simply propagate the input context.

\AllIntro is more interesting: Like the declarative
rule \DeclAllIntro, it adds a universal type variable $\alpha$ to the (input) context.
The output context of the premise $\chkjudg{\Gamma, \alpha}{e}{A}{\Delta, \alpha, \Theta}$
allows for some additional (existential) variables to appear after $\alpha$, in a trailing context
$\Theta$.  These existential variables could depend on $\alpha$; since $\alpha$ goes out of
scope in the conclusion, we must drop them from the concluding output context,
which is just $\Delta$: the part of the premise's output context that cannot depend on $\alpha$.

The application-judgment rule \AllApp serves a similar purpose to the subtyping rule
\SubAllL, but does \emph{not} place a marker before $\ahat$:
the variable $\ahat$ may appear in the output
type $C$, so $\ahat$ must survive in the output context $\Delta$.

\paragraph{Functions.}
In the third row of typing rules, rule \ArrIntro follows the same scheme:
the declarations $\Theta$ following $x : A$ are dropped in the conclusion's output context.

Rule \ArrIntroSyn corresponds to \DeclArrIntroSyn, one of the guessing rules, so
we create new existential variables $\ahat$ (for the function domain) and $\bhat$
(for the codomain) and check the function body against $\bhat$.  As in \AllApp, we do not
place a marker before $\ahat$, because $\ahat$ and $\bhat$ appear in the output
type ($\lam{x} e \syn \ahat \arr \bhat$).

Rule \ArrElim is the expected analogue of \DeclArrElim; like other rules with two
premises, it applies the intermediate context $\Theta$.

On the last row of typing rules, \SolveApp derives $\ahat \appsep e \app \ahat_2$
where $\ahat$ is unsolved in the input context.  Here we have an application judgment,
which is supposed to synthesize a type for an application $e_1\;e$ where $e_1$ has
type $\ahat$.  We know that $e_1$ should have function type; similarly to \InstLArr/\InstRArr,
we introduce $\ahat_1$ and $\ahat_2$ and add $\hypeq{\ahat}{\ahat_1{\arr}\ahat_2}$
to the context.  (Rule \SolveApp is the only algorithmic typing rule that
does not correspond to a declarative rule.)

Finally, rule \ArrApp is analogous to \DeclArrApp.

\section{Context Extension}%
\Label{sec:context-extension}

\begin{figure*}[htbp]
  \judgbox{\substextend{\Gamma}{\Delta}}{%
          $\Gamma$ is extended by $\Delta$%
  }
  ~\\[-5ex]
  \begin{mathpar}
    \hspace{30ex}
    \Infer{\substextendId}
        { }
        { \substextend{\emptyctx}{\emptyctx} }
    \and
    \Infer{\substextendVV}
          {\substextend{\Gamma}{\Delta}
           \\
           [\Delta]A = [\Delta]A'}
          {\substextend{\Gamma, x:A}{\Delta, x:A'}}
    \and
    \Infer{\substextendUU}
        { \substextend{\Gamma}{\Delta} }
        { \substextend{\Gamma, \alpha}{\Delta, \alpha} }
    \and
    \Infer{\substextendEE}
        { \substextend{\Gamma}{\Delta} }
        { \substextend{\Gamma, \ahat}{\Delta, \ahat} }
    ~~~~~~
    \Infer{\substextendSolSol}
        { \substextend{\Gamma}{\Delta} \\ [\Delta]\tau = [\Delta]\tau'}
        { \substextend{\Gamma, \hypeq{\ahat}{\tau}}{\Delta, \hypeq{\ahat}{\tau'}} }
    ~~~~~~
    \Infer{\substextendSolve}
        { \substextend{\Gamma}{\Delta} }
        { \substextend{\Gamma, \ahat}{\Delta, \hypeq{\ahat}{\tau}} }
    ~~~~~~
    \Infer{\substextendAdd}
        { \substextend{\Gamma}{\Delta} }
        { \substextend{\Gamma}{\Delta, \ahat} }
    \and
    \Infer{\substextendAddSolved}
        { \substextend{\Gamma}{\Delta} }
        { \substextend{\Gamma}{\Delta, \hypeq{\ahat}{\tau}} }
    \and
    \Infer{\substextendMonMon}
        { \substextend{\Gamma}{\Delta} }
        { \substextend{\Gamma, \MonnierComma{\ahat}}{\Delta, \MonnierComma{\ahat}} }
  \end{mathpar}
  
  \caption{Context extension}
  \FLabel{fig:substextend}
\end{figure*}

We motivated the algorithmic rules by saying that they evolved input
contexts to output contexts that were ``more solved''.
To state and prove the metatheoretic results of
decidability, soundness and completeness (Sections \ref{sec:decidability}--\ref{sec:completeness}),
we introduce a context extension judgment $\substextend{\Gamma}{\Delta}$.
This judgment captures a notion of information increase from an input context $\Gamma$ to
an output context $\Delta$, and relates algorithmic
contexts $\Gamma$ and $\Delta$ to completely solved extensions $\Omega$,
which correspond---via the context application described in
\Sectionref{sec:context-application}---to declarative contexts $\Psi$.

The judgment $\substextend{\Gamma}{\Delta}$ is read
``$\Gamma$ is extended by $\Delta$'' (or $\Delta$ extends $\Gamma$).
Another reading is that $\Delta$ carries at least as much information
as $\Gamma$.
A third reading is that $\substextend{\Gamma}{\Delta}$ means
that $\Gamma$ is \emph{entailed by} $\Delta$: all positive information derivable
from $\Gamma$ (say, that existential variable $\ahat$ is in scope)
can also be derived from $\Delta$ (which may have more information,
say, that $\ahat$ is equal to a particular type).  This reading is realized
by several key lemmas; for instance, extension preserves
well-formedness: if $\judgetp{\Gamma}{A}$ and
$\substextend{\Gamma}{\Delta}$, then $\judgetp{\Delta}{A}$.

The rules deriving the context extension judgment (\Figureref{fig:substextend}) say that
the empty context extends the empty context (\substextendId);
a term variable typing $x : A'$ extends $x : A$ if applying
the extending context $\Delta$ to $A$ and $A'$ yields the same type (\substextendVV);
universal type variables must match (\substextendUU);
scope markers must match (\substextendMonMon);
and, existential variables may:
  
  \begin{itemize}
  \item appear unsolved in both contexts (\substextendEE),
  \item appear solved in both contexts, if applying the extending context $\Delta$
     makes the solutions $\tau$ and $\tau'$ equal (\substextendSolSol),
  \item get solved by the extending context (\substextendSolve),
  \item be added by the extending context, either without a solution (\substextendAdd)
    or with a solution (\substextendAddSolved);
  \end{itemize}

Extension does \emph{not} allow solutions to disappear: information must increase.
It \emph{does} allow solutions to change, but only if the change preserves or increases
information.  The extension
\[
   \substextend{\big(\ahat, \hypeq{\bhat}{\ahat}\big)}{\big(\hypeq{\ahat}{\unitty}, \hypeq{\bhat}{\ahat}\big)}
\]
directly increases information about $\ahat$, and indirectly increases information
about $\bhat$.  Perhaps more interestingly, the extension
\[
   \substextend{\underbrace{\big(\hypeq{\ahat}{\unitty}, \hypeq{\bhat}{\ahat}\big)}_{\Delta}}{\underbrace{\big(\hypeq{\ahat}{\unitty}, \hypeq{\bhat}{\unitty}\big)}_{\Omega}}
\]
also holds: while the solution of $\bhat$ in $\Omega$ is different, in the sense that
$\Omega$ contains $\hypeq{\bhat}{\unitty}$ while $\Delta$ contains $\hypeq{\bhat}{\ahat}$,
applying $\Omega$ to the two solutions gives the same thing:
applying $\Omega$ to $\Delta$'s solution of $\bhat$ gives
$[\Omega]\ahat = [\Omega]\unitty = \unitty$,
while applying $\Omega$ to $\Omega$'s own solution for $\bhat$ also
gives $\unitty$, because $[\Omega]\unitty = \unitty$.

Extension is quite rigid, however, in two senses.  First, if a declaration
appears in $\Gamma$, it appears in all extensions of $\Gamma$.  Second,
\emph{extension preserves order}.  For example,
if $\bhat$ is declared after $\ahat$ in $\Gamma$, then $\bhat$ will also be declared after
$\ahat$ in every extension of $\Gamma$.  This holds for every variety of declaration.
This rigidity aids in enforcing type variable scoping and dependencies, which are
nontrivial in a setting with higher-rank polymorphism.

This combination of rigidity (in demanding that the order of declarations be preserved)
with flexibility (in how existential type variable solutions are expressed)
manages to satisfy scoping and dependency relations \emph{and} give
enough room to maneuver in the algorithm and metatheory.

\subsection{Context Application}
\Label{sec:context-application}

A complete context $\Omega$ (\Figureref{fig:alg-syntax}) has no
unsolved variables, so applying it to a (well-formed) type yields a type
$[\Omega]A$ with no existentials.
Such a type is well-formed under a \emph{declarative} context---with just
$\alpha$ and $x : A$ declarations---obtained by dropping all the existential
declarations and applying $\Omega$ to declarations $x : A$ (to yield $x : [\Omega]A$).
We can think of this context as the result of applying $\Omega$ to itself:
$[\Omega]\Omega$.

\begin{figure}
\begin{displ}
    \begin{array}[t]{l@{~~~}c@{~~~}ll}
     \contextapp{\cdot}{\cdot} & = &   \cdot &              %
    \\
     \contextapp{\Omega, \hyp{x}{A}}{(\Gamma, \hyp{x}{A_\Gamma})} & = &   \contextapp{\Omega}{\Gamma},\; \hyp{x}{\contextapp{\Omega}{A}} & 
       \mbox{if } \contextapp{\Omega}{A} = \contextapp{\Omega}{A_\Gamma}
    \\
      \contextapp{\Omega, \alpha}{(\Gamma, \alpha)} & = &   \contextapp{\Omega}{\Gamma},\; \alpha &
    \\
     \contextapp{\Omega, \hypeq{\ahat}{\tau}}{(\Gamma, \ahat)} & = &   \contextapp{\Omega}{\Gamma} &
    \\
     \contextapp{\Omega, \hypeq{\ahat}{\tau}}{(\Gamma, \hypeq{\ahat}{\tau_\Gamma})}
       & = &
       \contextapp{\Omega}{\Gamma} 
     &    \mbox{if } \contextapp{\Omega}{\tau} = \contextapp{\Omega}{\tau_\Gamma}
    \\
    \contextapp{\Omega, \hypeq{\ahat}{\tau}}{\Gamma} & = &   \contextapp{\Omega}{\Gamma}
    &    \mbox{if } \ahat \notin \dom{\Gamma}
    \\
    \contextapp{\Omega, \MonnierComma{\ahat}}{(\Gamma, \MonnierComma{\ahat})} & = &   \contextapp{\Omega}{\Gamma} &
    \end{array}
\end{displ}
\caption{Applying a complete context $\Omega$ to a context}
\FLabel{fig:context-substitution}
\end{figure}

More generally, we can apply $\Omega$ to any
context $\Gamma$ that it extends.
This operation of context application $[\Omega]\Gamma$
is given in \Figureref{fig:context-substitution}.
The application $[\Omega]\Gamma$ is defined if and only if
$\substextend{\Gamma}{\Omega}$, and applying $\Omega$ to any such $\Gamma$
yields the same declarative context $[\Omega]\Omega$:

\begin{lemma*}[Stability of Complete Contexts]%
  If $\substextend{\Gamma}{\Omega}$
  then $[\Omega]\Gamma = [\Omega]\Omega$. 
\end{lemma*}

  \section{Decidability}
\Label{sec:decidability}

Our algorithmic type system is decidable.  Since the typing rules (\Figureref{fig:alg-typing})
depend on the subtyping rules (\Figureref{fig:alg-subtyping}), which in turn depend on
the instantiation rules (\Figureref{fig:instantiation}), showing that the typing judgments (checking, synthesis
and application) are decidable requires that we show that the instantiation and subtyping judgments
are decidable.

\subsection{Decidability of Instantiation}
\Label{sec:decidability-instantiation}

As discussed in \Sectionref{sec:instantiation}, deriving
$\instjudg{\Gamma}{\ahat}{A}{\Delta}$ does not necessarily
instantiate any existential variable (unless $A$ is a monotype).
However, the instantiation rules do preserve the size of (substituted) types:

\begin{lemma*}[Instantiation Size Preservation]%
~\\
  If $\Gamma = (\Gamma_0, \ahat, \Gamma_1)$
  and $\instjudg{\Gamma}{\ahat}{A}{\Delta}$
     or $\instjudgr{\Gamma}{\ahat}{A}{\Delta}$, \\
     and $\judgetp{\Gamma}{B}$ and $\ahat \notin \FV{[\Gamma]B}$, 
     then $|[\Gamma]B| = |[\Delta]B|$, where $|C|$ is the plain size of $C$.
\end{lemma*}

Using this lemma, we can show that the type $A$ in the instantiation judgment always
get smaller, even in rule \InstLArr: the second premise applies the intermediate
context $\Theta$ to $A_2$, but the lemma tells us that this application cannot make
$A_2$ larger, and $A_2$ is smaller than the conclusion's type ($A_1 \arr A_2$).

Now we can prove decidability of instantiation, assuming that $\ahat$ is unsolved
in the input context $\Gamma$, that $A$ is well-formed under $\Gamma$,
that $A$ is fully applied ($[\Gamma]A = A$), and that $\ahat$ does not occur in
$A$.  These conditions are guaranteed when instantiation is invoked, because the typing rule
\Sub applies the input substitution, and the subtyping rules apply the substitution where
needed---in exactly one place: the second premise of \SubArr.
The proof is based on the (substituted) types in the premises being smaller
than the (substituted) type in the conclusion.

\begin{restatable}[Decidability of Instantiation]{theorem}{instantiationdecidability}
\Label{thm:decidability-of-instantiation}
~\\
    If $\Gamma = \Gamma_0[\ahat]$
    and
    $\judgetp{\Gamma}{A}$
    such that $[\Gamma]A = A$
    and $\ahat \notin \FV{A}$, then:
    \\[-3ex]
    \begin{enumerate}[(1)]
    \item Either there exists $\Delta$ such that $\instjudg{\Gamma_0[\ahat]}{\ahat}{A}{\Delta}$,
      or not.
    \item Either there exists $\Delta$ such that $\instjudgr{\Gamma_0[\ahat]}{\ahat}{A}{\Delta}$,
      or not.
    \end{enumerate}
\end{restatable}

\subsection{Decidability of Algorithmic Subtyping}
\Label{sec:decidability-subtyping}

To prove decidability of the subtyping system in \Figureref{fig:alg-subtyping},
measure judgments $\subjudg{\Gamma}{A}{B}{\Delta}$ lexicographically by
\\[-3ex]
  \begin{enumerate}[(S1)]
  \item the number of $\forall$ quantifiers in $A$ and $B$;
  \item $|\unsolved{\Gamma}|$, the number of unsolved existentials in $\Gamma$;
  \item $\typesize{\Gamma}{A} + \typesize{\Gamma}{B}$.
  \end{enumerate}

{\noindent Part} (S3) uses \emph{contextual size}, which penalizes solved variables (*):

\begin{definition*}[Contextual Size] \Label{def:typesize}
\[
    \begin{array}[t]{lcll}
      \typesize{\Gamma}{\alpha}  & = &   1
      \\
      \typesize{{\Gamma[\ahat]}}{\ahat}  & = &   1
      \\
      \typesize{{\Gamma[\hypeq{\ahat}{\tau}]}}{\ahat} & = &   1 + \typesize{\Gamma[\hypeq{\ahat}{\tau}]}{\tau}  & \text{\normalfont (*)}
      \\
      \typesize{\Gamma}{\alltype{\alpha}{A}} & = &   1 + \typesize{\Gamma, \alpha}{A}
      \\
      \typesize{\Gamma}{A \arr B} & = &   1 + \typesize{\Gamma}{A} + \typesize{\Gamma}{B}
    \end{array}
\]
\end{definition*}

For example, if $\Gamma = (\beta, \hypeq{\ahat}{\beta})$
then $\typesize{\Gamma}{\ahat}
= 1 + \typesize{\Gamma}{\beta}
= 1 + 1 = 2$, whereas the plain size of $\ahat$ is simply 1.

The connection between (S1) and (S2) may be clarified by examining rule \SubArr,
whose conclusion says that $A_1 \arr A_2$ is a subtype of $B_1 \arr B_2$.
If $A_2$ or $B_2$ is polymorphic, then the first premise on $A_1 \arr A_2$ is smaller by (S1).
Otherwise, the first premise has the same input context as the conclusion, so it has the same (S2),
but is smaller by (S3).
If $B_1$ or $A_1$ is polymorphic, then the second premise is smaller by (S1).
Otherwise, we use the property that instantiating a monotype always solves an existential:

\begin{lemma*}[Monotypes Solve Variables]%
  If $\instjudg{\Gamma}{\ahat}{\tau}{\Delta}$ or $\instjudgr{\Gamma}{\ahat}{\tau}{\Delta}$,
  then if $[\Gamma]\tau = \tau$ and $\ahat \notin \FV{[\Gamma]\tau}$,
  we have $|\unsolved{\Gamma}| = |\unsolved{\Delta}| + 1$. 
\end{lemma*}

\newcommand{\signallemma}[1]{\textbf{(\ref{#1})}\xspace}
A couple of other lemmas are worth mentioning:
  subtyping on two monotypes cannot increase the number of unsolved existentials,
  and
  applying a substitution $\Gamma$ to a type does not increase the type's size with respect to $\Gamma$.

\begin{lemma*}[Monotype Monotonicity]%
~\\
  If $\subjudg{\Gamma}{\tau_1}{\tau_2}{\Delta}$
  then
  $|\unsolved{\Delta}| \leq |\unsolved{\Gamma}|$.
\end{lemma*}

\begin{lemma*}[Substitution Decreases Size]%
~\\
  If $\judgetp{\Gamma}{A}$ then $\typesize{\Gamma}{[\Gamma]A} \leq \typesize{\Gamma}{A}$.
\end{lemma*}

\begin{restatable}[Decidability of Subtyping]{theorem}{subtypingdecidability}   
\Label{thm:subtyping-decidable}
~\\
  Given a context $\Gamma$ and types $A$, $B$ such that $\judgetp{\Gamma}{A}$ and
  $\judgetp{\Gamma}{B}$ and $[\Gamma]A = A$ and $[\Gamma]B = B$,
  it is decidable whether there exists $\Delta$ such that $\subjudg{\Gamma}{A}{B}{\Delta}$.
\end{restatable}

\subsection{Decidability of Algorithmic Typing}
\Label{sec:decidability-typing}

\begin{restatable}[Decidability of Typing]{theorem}{typingdecidable}
\Label{thm:typing-decidable}
~\\[-3ex]
  \begin{enumerate}[(i)]
      \item \emph{Synthesis:}  
        Given a context $\Gamma$
        and a term $e$,
        it is decidable whether there exist a type $A$ and a context $\Delta$
        such that \\
        $\synjudg{\Gamma}{e}{A}{\Delta}$.

      \item \emph{Checking:}
        Given a context $\Gamma$,
        a term $e$,
        and a type $B$ such that $\judgetp{\Gamma}{B}$,
        it is decidable whether there is a context $\Delta$ such that \\
        $\chkjudg{\Gamma}{e}{B}{\Delta}$.

      \item \emph{Application:}  
        Given a context $\Gamma$,
        a term $e$,
        and a type $A$ such that $\judgetp{\Gamma}{A}$,
        it is decidable whether there exist a type $C$ and a context $\Delta$ such that \\
        $\appjudg{\Gamma}{e}{A}{C}{\Delta}$.
  \end{enumerate}
\end{restatable}

    The following induction measure suffices to prove decidability:
    \[
      \left\langle
        e,~~~
        \begin{array}[c]{llll}
            \syn
          \\[0.3ex]
            \chk,
            &
               \typesize{\Gamma}{B}
          \\[0.6ex]
            \app,
            &
               \typesize{\Gamma}{A}
        \end{array}
      \right\rangle
    \]
    where $\langle \dots \rangle$ denotes lexicographic order, and
    where (when comparing two judgments typing the same term $e$)
    the synthesis judgment (top line) is considered smaller than
    the checking judgment (second line), which in turn is considered smaller than
    the application judgment (bottom line).  That is,
    $\syn \,\bigprec\, \chk \,\bigprec\, \app$.
  In \Sub, this makes the synthesis premise smaller than the checking conclusion;
  in \ArrApp and \SolveApp, this makes the checking premise smaller than the application
  conclusion.

Since we have no explicit introduction form for polymorphism, the rule \AllIntro
has the same term $e$ in its premise and conclusion, and both the premise and conclusion
are the same kind of judgment (checking).  The rule \AllApp is similar (with application judgments
in premise and conclusion).  Therefore, given two judgments on the same 
term, and that are both checking judgments or both application judgments, we
use the size of the input type expression---which \emph{does} get smaller in \AllIntro and \AllApp.

  \section{Soundness}
\Label{sec:soundness}

We want the algorithmic specifications of subtyping and typing to be sound with respect to
the declarative specifications.  Roughly, given a derivation of an algorithmic judgment
with input context $\Gamma$ and output context $\Delta$, and some complete context $\Omega$
that extends $\Delta$ (which therefore extends $\Gamma$), applying $\Omega$ throughout
the given algorithmic judgment should yield a derivable declarative judgment.
Let's make that rough outline concrete for instantiation, showing that 
the action of the instantiation rules is consistent with declarative subtyping:

\begin{restatable}[Instantiation Soundness]{theorem}{instantiationsoundness}  \Label{thm:instantiation-soundness}
   ~\\
    Given $\substextend{\Delta}{\Omega}$ and $[\Gamma]B = B$ and $\ahat \notin \FV{B}$: \\[-3ex]
        \begin{enumerate}[(1)]
            \item If $\instjudg{\Gamma}{\ahat}{B}{\Delta}$ 
              then
              $\declsubjudg{[\Omega]\Delta}{[\Omega]\ahat}{[\Omega]B}$.

            \item If $\instjudgr{\Gamma}{\ahat}{B}{\Delta}$ 
              then
              $\declsubjudg{[\Omega]\Delta}{[\Omega]B}{[\Omega]\ahat}$.
        \end{enumerate}
\end{restatable}

Note that the declarative derivation is under $[\Omega]\Delta$,
which is $\Omega$ applied to the algorithmic output context $\Delta$.

With instantiation soundness, we can prove the expected soundness property for
subtyping:

\begin{restatable}[Soundness of Algorithmic Subtyping]{theorem}{soundness} 
\Label{thm:subtyping-soundness} ~\\
  If $\subjudg{\Gamma}{A}{B}{\Delta}$
  where $[\Gamma]A = A$ and $[\Gamma]B = B$
  and $\substextend{\Delta}{\Omega}$
  then
  $\declsubjudg{\contextapp{\Omega}{\Delta}}{\contextapp{\Omega}{A}}{\contextapp{\Omega}{B}}$. 
\end{restatable}

Finally, knowing that subtyping is sound, we can prove that typing is sound:

\begin{restatable}[Soundness of Algorithmic Typing]{theorem}{typingsoundness}
\Label{thm:typing-soundness} %
Given $\substextend{\Delta}{\Omega}$: \\[-3ex]
    \begin{enumerate}[(i)]
      \item 
          If $\chkjudg{\Gamma}{e}{A}{\Delta}$
          then
          $\declchkjudg{[\Omega]\Delta}{e}{[\Omega]A}$. 
      
      \item
          If $\synjudg{\Gamma}{e}{A}{\Delta}$
          then
          $\declsynjudg{[\Omega]\Delta}{e}{[\Omega]A}$. 

      \item
          If $\appjudg{\Gamma}{e}{A}{C}{\Delta}$
          then
          $\declappjudg{[\Omega]\Delta}{e}{[\Omega]A}{[\Omega]C}$.
    \end{enumerate}
\end{restatable}

The proofs need several lemmas, including this one:

\begin{lemma*}[Typing Extension] ~\\
  If $\chkjudg{\Gamma}{e}{A}{\Delta}$
  or $\synjudg{\Gamma}{e}{A}{\Delta}$
  or $\appjudg{\Gamma}{e}{A}{C}{\Delta}$
  then
  $\substextend{\Gamma}{\Delta}$.
\end{lemma*}

  \section{Completeness}
\Label{sec:completeness}

Completeness of the algorithmic system is something like soundness in reverse:
given a declarative derivation of
$[\Omega]\Gamma \entails [\Omega]\cdots$,
we want to get an algorithmic derivation of $\Gamma \entails \cdots \ctxout{\Delta}$.

For soundness, the output context $\Delta$ such that $\substextend{\Delta}{\Omega}$
was given;
$\substextend{\Gamma}{\Omega}$ followed from
Typing Extension (the above lemma) and transitivity of extension.
For completeness, only $\Gamma$ is given, so we have
$\substextend{\Gamma}{\Omega}$ in the antecedent.  Then we might expect to show,
along with $\Gamma \entails \cdots \ctxout{\Delta}$,
that $\substextend{\Delta}{\Omega}$.  But this is not general enough:
the algorithmic rules generate fresh existential variables, so $\Delta$ may have
existentials that are not found in $\Gamma$, nor in $\Omega$.  In completeness,
we are given a declarative derivation, which contains no existentials;
the completeness proof must build up the completing context $\Omega$
along with the algorithmic derivation.
Thus, completeness will produce an $\Omega'$ which extends both the given $\Omega$
and the output context of the algorithmic derivation: $\substextend{\Omega}{\Omega'}$
and $\substextend{\Delta}{\Omega'}$.  (By transitivity, we also get $\substextend{\Gamma}{\Omega'}$.)

As with soundness, we have three main completeness results, for instantiation,
subtyping and typing.

\smallskip

\begin{restatable}[Instantiation Completeness]{theorem}{instantiationcompletes}
   \Label{thm:instantiation-completes}
   Given $\substextend{\Gamma}{\Omega}$ 
   and $A = [\Gamma]A$ 
   and $\ahat \in \unsolved{\Gamma}$  %
   and $\ahat \notin \FV{A}$:
   ~\\[-3ex]
       \begin{enumerate}[(1)]
           \item If $\declsubjudg{[\Omega]\Gamma}{[\Omega]\ahat}{[\Omega]A}$
                 then there are $\Delta$, $\Omega'$ such that
                 \\
                 $\substextend{\Omega}{\Omega'}$
                 and $\substextend{\Delta}{\Omega'}$
                 and $\instjudg{\Gamma}{\ahat}{A}{\Delta}$. 

           \item If $\declsubjudg{[\Omega]\Gamma}{[\Omega]A}{[\Omega]\ahat}$
                 then there are $\Delta$, $\Omega'$ such that
                 \\
                 $\substextend{\Omega}{\Omega'}$
                 and $\substextend{\Delta}{\Omega'}$
                 and $\instjudgr{\Gamma}{\ahat}{A}{\Delta}$. 
      \end{enumerate}                                                    
\end{restatable}

\vspace{-0.2ex}

\begin{restatable}[Generalized Completeness of Subtyping]{theorem}{completingcompleteness}
  \Label{thm:completing-completeness}
~\\
  If $\substextend{\Gamma}{\Omega}$
  and $\judgetp{\Gamma}{A}$
  and $\judgetp{\Gamma}{B}$
  and $\declsubjudg{[\Omega]\Gamma}{[\Omega]A}{[\Omega]B}$
  then
  there exist $\Delta$ and $\Omega'$ such that 
  $\substextend{\Delta}{\Omega'}$ and 
  $\substextend{\Omega}{\Omega'}$ and 
  $\subjudg{\Gamma}{[\Gamma]A}{[\Gamma]B}{\Delta}$.
\end{restatable}

\begin{restatable}[Completeness of Algorithmic Typing]{theorem}{typingcompleteness}
  \Label{thm:typing-completeness}
  ~\\
  Given $\substextend{\Gamma}{\Omega}$ and $\judgetp{\Gamma}{A}$:
  \\[-3ex]
    \begin{enumerate}[(i)]
      \item
           If $\declchkjudg{[\Omega]\Gamma}{e}{[\Omega]A}$
           \\
           then there exist $\Delta$ and $\Omega'$
           \\
           such that
           $\substextend{\Delta}{\Omega'}$ and 
           $\substextend{\Omega}{\Omega'}$ and 
           $\chkjudg{\Gamma}{e}{[\Gamma]A}{\Delta}$.
      
      \item
           If $\declsynjudg{[\Omega]\Gamma}{e}{A}$
           \\
           then there exist $\Delta$, $\Omega'$, and $A'$
           \\
           such that
           $\substextend{\Delta}{\Omega'}$ and 
           $\substextend{\Omega}{\Omega'}$ and 
           $\synjudg{\Gamma}{e}{A'}{\Delta}$ and 
           $A = [\Omega']A'$. 
      
      \item
           If $\declappjudg{[\Omega]\Gamma}{e}{[\Omega]A}{C}$
           \\
           then there exist $\Delta$, $\Omega'$, and $C'$
           \\
           such that
           $\substextend{\Delta}{\Omega'}$ and 
           $\substextend{\Omega}{\Omega'}$
           \\
           and 
           $\appjudg{\Gamma}{e}{[\Gamma]A}{C'}{\Delta}$ and 
           $C = [\Omega']C'$. 
    \end{enumerate}
\end{restatable}

\section{Design Variations}
\label{sec:variants}

The rules we give infer monomorphic types, but require annotations for
all polymorphic bindings. In this section, we consider alternatives to
this choice.

\paragraph{Eliminating type inference.}
To eliminate type inference from
the declarative system, it suffices to drop the \DeclArrIntroSyn and
\DeclUnitIntroSyn rules. The corresponding alterations to the algorithmic 
system are a little more delicate: simply deleting the \ArrIntroSyn
and \UnitIntroSyn rules breaks completeness. To see why, suppose
that we have a variable $f$ of type $\alltype{\alpha}{\alpha \to \alpha}$, 
and consider the application $f\;\unitexp$. Our algorithm will introduce a 
new existential variable $\ahat$ for $\alpha$, and then check $\unitexp$
against $\ahat$. Without the \UnitIntroSyn rule, typechecking will fail. 
To restore completeness, we need to modify these two rules. Instead of
being \emph{synthesis} rules, we will change them to
\emph{checking} rules that check values against an unknown
existential variable. 
\vspace{-4pt}
\begin{mathpar}
\Infer{\rulename{$\unitty$I$\ahat$}}
      { }
      {\chkjudg{\Gamma[\ahat]}{\unitexp}{\ahat}{\Gamma[\ahat=\unitty]}}
\and
\Infer{\rulename{$\arr$I$\ahat$}}
      {\chkjudg{{\Gamma[\ahat_2, \ahat_1, \ahat=\ahat_1 \to \ahat_2], \hyp{x}{\ahat_1}}}{e}{\ahat_2}{\Delta, \hyp{x}{\ahat_1}, \Delta'}}
      {\chkjudg{\Gamma[\ahat]}{\fun{x}{e}}{\ahat}{\Delta}}
\end{mathpar}
With these two rules replacing \UnitIntroSyn and \ArrIntroSyn, we have
a complete algorithm for the no-inference bidirectional system.

\paragraph{Full Damas-Milner type inference.}
Another alternative is to
increase the amount of type inference done. For instance, a natural
question is whether we can extend the bidirectional approach to
subsume the inference done by the algorithm of \citet{Damas82}.
This appears feasible: we can alter the \ArrIntroSyn rule to support ML-style type inference.
\begin{mathpar}
\Infer{\ArrIntroSyn$\!'$}
      {\chkjudg{\Gamma, \MonnierComma{\ahat}, \ahat, \bhat, x:\ahat}{e}{\bhat}{\Delta, \MonnierComma{\ahat}, \Delta'}
       \\
       \tau = [\Delta'](\ahat \to \bhat) \\
       \vec{\ahat} = \unsolved{\Delta'}
      }
      {\synjudg{\Gamma}{\fun{x}{e}}{\alltype{\vec{\alpha}}{[\vec{\alpha}/\vec{\ahat}]\tau}}{\Delta}}
\end{mathpar}
In this rule, we introduce a marker $\MonnierComma{\ahat}$ into the
context, and then check the function body against the type $\bhat$.
Then, our output type substitutes away all the solved
existential variables to the right of the marker $\MonnierComma{\ahat}$,
and generalizes over all of the unsolved variables to the right of the marker.
Using an ordered context gives precise control over the scope of the
existential variables, making it easy to express polymorphic generalization.

The above is only a sketch; we have not defined the corresponding declarative
system, nor proved completeness.

\section{Related Work and Discussion}
\Label{sec:related}

\subsection{Type Inference for System F}
Because type inference for System F is
undecidable~\citep{Wells99:FUndecidable}, designing type inference
algorithms for first-class polymorphism inherently involves navigating
a variety of design tradeoffs. As a result, there have been a wide
variety of proposals for extending type systems beyond the
Damas-Milner ``sweet spot''. The main tradeoff appears to be 
a ``two-out-of-three'' choice: language designers can keep
any two of: (1) the $\eta$-law for functions, (2) impredicative
instantiation, and (3) the standard type language of System F. 

As discussed in \Sectionref{sec:declarative}, for typability under
$\eta$-reductions, it is necessary for subtyping to instantiate
deeply: that is, we must allow instantiation of quantifiers to the
right of an arrow.  However, \citet{Tiuryn96} and \citet{Chrzaszcz98}
showed that the subtyping relation for impredicative System F is
undecidable. As a result, if we want $\eta$ \emph{and} a complete
algorithm, then either the polymorphic instantiations must be
predicative, or a different language of types must be used. 

\Figureref{fig:comparison} summarizes the different choices 
made by the designers of this and related systems. 

\begin{figure*}
    \begin{center}
        \begin{tabular}{lccc}
          System & $\eta$-laws? & Impredicative? & System F type language?
          \\ \hline
          \MLF & yes & yes & no
          \\ 
          FPH & no & yes & yes
          \\ 
          HML & no & yes & yes
          \\ 
          \citet{PeytonJones07} & yes & no & yes
          \\ 
          This paper & yes & no & yes
        \end{tabular}
    \end{center}
\caption{Comparison of type inference algorithms}
\FLabel{fig:comparison}
\end{figure*}

\paragraph{Impredicativity and the $\eta$-law.}
The designers of \MLF~\citep{LeBotlan03, Remy08, LeBotlan09}
chose to use a different language of types, one with a form of bounded
quantification.  This increases the expressivity of types enough to
ensure principal types, which means that (1) required annotations are
few and predictable, and (2) their system is very robust in the face
of program transformations, including $\eta$.  However, the richness of
the \MLF type structure requires a sophisticated metatheory and
correspondingly intricate implementation techniques.

\paragraph{Impredicativity and System F types.}
Much of the other work on higher-rank polymorphism avoids changing the
language of types. 

The HML system of \citet{Leijen09} and the FPH system of
\citet{Vytiniotis08} both retain the type language of
(impredicative) System F. Each of these systems gives as a specification
a slightly different extension to the declarative Damas-Milner type system,
and handle the issue of inference in slightly different
ways. 
HML is essentially a restriction of \MLF, in which the external
language of types is limited to System F, but which uses the
technology of \MLF internally, as part of type inference. FPH, on the
other hand, extends and generalizes work on boxy types~\citep{Vytiniotis06}
to control type inference. The differences in expressive power between these two
systems are subtle---roughly speaking, FPH requires slightly more
annotations, but has a less complicated specification. However, in both systems,
the same heuristic guidance to the programmer applies:
place explicit annotations on binders with fancy types.

\paragraph{The $\eta$-law and System F types.}
\citet{PeytonJones07} developed an approach for typechecking
higher-rank predicative polymorphism that is closely related to ours.
They define a bidirectional declarative system similar to our own, but
which lacks an application judgment. This complicates the
presentation of their system, forcing them to introduce an additional
grammatical category of types beyond monotypes and polytypes, and
requires many rules to carry an additional subtyping premise. Next,
they enrich the subtyping rules of \citet{Odersky96} with the
distributivity axiom of \citet{Mitchell88}, which we rejected on
ideological grounds: it is a valid coercion, but is not orthogonal (it
is a single rule mixing two different type connectives) and does not
correspond to a rule in the sequent calculus. They do not prove the
soundness and completeness of their Haskell reference implementation,
but it appears to implement behavior close to our application
judgment.

\paragraph{History of our approach.} Several of the ideas used in the present paper descend from
\citet{Dunfield09}, an approach to first-class polymorphism (including
impredicativity) also based on ordered contexts with existential
variables instantiated via subtyping.  In fact, the present work began
as an attempt to extend \citet{Dunfield09} with type-level
computation.  During that attempt, we found several shortcomings and
problems.  The most serious is that the decidability and completeness
arguments were not valid.  These problems may be fixable, but instead
we started over, reusing several of the high-level ideas in different
technical forms.

\subsection{Other Type Systems}

\citet{Pierce00} developed bidirectional typechecking for rich
subtyping, with specific techniques for instantiating polymorphism
within function application (hence, \emph{local} type inference).
Their declarative specification is more complex than ours, and their
algorithm depends on computing approximations of upper and lower
bounds on types. \emph{Colored local type inference}
\citep{Odersky01:ColoredLocal} allows different \emph{parts} of type
expressions to be propagated in different directions.  Our approach
gets a similar effect by manipulating type expressions with
existential variables.

\subsection{Our Algorithm}

One of our main contributions is our new algorithm for type inference,
which is remarkable in its simplicity. Three key ideas
underpin our algorithm.

\paragraph{Ordered contexts.}
We move away from the traditional ``bag of
constraints'' model of type inference, and instead embed existential
variables and their values directly into an ordered context.  Thus,
straightforward scoping rules control the free variables of the types
each existential variable may be instantiated with, without any need
for model-theoretic techniques like skolemization,
which fit awkwardly into a type-theoretic discipline.  Using an ordered context
permits handling quantifiers in a manner resembling the level-based generalization
mechanism of \citet{Remy92}, used also in \MLF~\citep{LeBotlan09}.

\paragraph{The instantiation judgment.}
The original inspiration
for instantiation comes from the ``greedy'' algorithm of
\citet{Cardelli93:FsubTheSystem},
which eagerly uses type information to solve existential constraints.
In that setting---a language with rather ambitious subtyping---the
greedy algorithm was incomplete: consider a function of type
$\alltype{\alpha}{\alpha \arr \alpha \arr \alpha}$ applied to a
\textit{Cat} and an \textit{Animal}; the cat will be checked against
an existential $\ahat$, which instantiates $\ahat$ to \textit{Cat},
but checking the second argument, $\textit{Animal} \subtype
\textit{Cat}$, fails.  (Reversing the order of arguments makes typing
succeed!)

In our setting, where subtyping represents the specialization order
induced by quantifier instantiation, it is possible to get a complete
algorithm, by slightly relaxing the pure greedy strategy. Rather than
eagerly setting constraints, we first look under quantifiers (in the
\InstLAllR and \InstRAllL rules) to see if there is a feasible
monotype instantiation, and we also use the the \InstLReach and
\InstRReach to set the ``wrong'' existential variable in case we need
to equate an existential variable with one to its right in the
context. Looking under quantifiers seems forced by our restriction to predicative
polymorphism, and ``reaching'' seems forced by our use of an ordered
context, but the combination of these mechanisms fortuitously
enables our algorithm to find good upper and lower monomorphic
approximations of polymorphic types.

This is surprising, since it is quite contrary to the
implementation strategy of \MLF (also used by HML and FPH).
There, the language of type constraints supports bounds on fully quantified types,
and the algorithm incrementally refines these constraints. In
contrast, we only ever create equational constraints on existentials
(bounds are not needed), and once we have a solution for an
existential, our algorithm never needs to revisit its decision.

Distinguishing instantiation as a separate judgment is new in this
paper, and beneficial: \citet{Dunfield09} baked instantiation into the
subtyping rules, resulting in a system whose direct implementation
required substantial backtracking---over a set of rules including
arbitrary application of substitutions.  We, instead, maintain an
invariant in subtyping and instantiation that the types are always
fully applied with respect to an input context, obviating the need for
explicit rules to apply substitutions.

\paragraph{Context extension.}
Finally, we introduce a context-extension judgment as the central
invariant in our correctness proofs. This permits us to state many
properties important to our algorithm abstractly, without reference to
the details of our algorithm. %

We are not the only ones to study context-based approaches to type
inference. Recently, \citet{Gundry10} recast the classic Damas-Milner
algorithm, which manipulates unstructured sets of equality
constraints, as structured constraint solving under ordered contexts.
A (semantic) notion of information increase is central to their
development, as (syntactic) context extension is to ours.  While their
formulation supports only ML-style prenex polymorphism, the ultimate
goal is a foundation for type inference for dependent types.  To some
extent, both our algorithm and theirs can be understood in terms of
the proof system of \citet{Miller92} for mixed-prefix unification. We
each restrict the unification problem, and then give a proof search
algorithm to solve the type inference problem.

\section*{Acknowledgments}
Thanks to the anonymous ICFP reviewers for their comments, which have
(we hope) led to a more correct paper.

\iffalse
    \ifnum\OPTIONLoudLabels=15
      \bibliographystyle{plainnat}%
    \else
      \bibliographystyle{plainnat}
    \fi
    \bibliography{local}
\else

\fi

\end{document}


\maketitle
        \appendix
\else
  \makeatletter
       \onecolumn

       \renewcommand{\rmdefault}{bch}
       \renewcommand{\bfdefault}{b}

       \@setfontsize{\normalsize}{10pt}{12pt}%
       
        \appendix
  \makeatother
\fi

\tableofcontents

\clearpage

\section{Declarative Subtyping}

\subsection{Properties of Well-Formedness}

\begin{restatable}[Weakening]{proposition}{propweakening}
\XLabel{prop:weakening}
            If $\judgetp{\Psi}{A}$
            then
            $\judgetp{\Psi, \Psi'}{A}$
            by a derivation of the same size.
\end{restatable}

\begin{restatable}[Substitution]{proposition}{propsubst}
\XLabel{prop:subst}
            If $\judgetp{\Psi}{A}$ and $\judgetp{\Psi, \alpha, \Psi'}{B}$
            then
            $\judgetp{\Psi,\Psi'}{[A/\alpha]B}$.
\end{restatable}

\subsection{Reflexivity}

\begin{restatable}[Reflexivity of Declarative Subtyping]{lemma}{declarativereflexivity}
\XLabel{lem:declarative-reflexivity}
  Subtyping is reflexive: if $\judgetp{\Psi}{A}$ then
        $\declsubjudg{\Psi}{A}{A}$.
\end{restatable}

\subsection{Subtyping Implies Well-Formedness}

\begin{restatable}[Well-Formedness]{lemma}{declarativewellformed}
\XLabel{lem:declarative-well-formed}
  If $\declsubjudg{\Psi}{A}{B}$ then $\judgetp{\Psi}{A}$ and $\judgetp{\Psi}{B}$.
\end{restatable}

\subsection{Substitution}

\begin{restatable}[Substitution]{lemma}{declsubsubstitution}
\XLabel{lem:decl-sub-substitution}
   If $\judgetp{\Psi}{\tau}$ and $\declsubjudg{\Psi, \alpha, \Psi'}{A}{B}$
   then $\declsubjudg{\Psi, [\tau/\alpha]\Psi'}{[\tau/\alpha]A}{[\tau/\alpha]B}$.
\end{restatable}

\subsection{Transitivity}

\begin{restatable}[Transitivity of Declarative Subtyping]{lemma}{declarativetransitivity}
\XLabel{lem:declarative-transitivity}
   If $\declsubjudg{\Psi}{A}{B}$ and $\declsubjudg{\Psi}{B}{C}$ then $\declsubjudg{\Psi}{A}{C}$.
\end{restatable}

\subsection{Invertibility of \DsubAllR}

\begin{restatable}[Invertibility]{lemma}{declinvertibility}
\XLabel{lem:decl-invertibility}
~\\
If $\Dee$ derives $\declsubjudg{\Psi}{A}{\alltype{\beta}{B}}$
then
$\Dee'$ derives $\declsubjudg{\Psi, \beta}{A}{B}$
where $\Dee' < \Dee$.
\end{restatable}

\subsection{Non-Circularity and Equality}

\begin{definition}[Subterm Occurrence]  \XLabel{def:occurrence}
 ~\\  
  Let $A \subtermof B$ iff $A$ is a subterm of $B$. \\
  Let $A \propersubtermof B$ iff $A$ is a proper subterm of $B$ (that is, $A \subtermof B$ and $A \neq B$). \\
  Let $A \occursinsidearrow B$ iff $A$ occurs in $B$ inside an arrow, that is, there exist $B_1$, $B_2$
  such that $(B_1{\arr}B_2) \subtermof B$ and $A \subtermof B_k$ for some $k \in \{1, 2\}$.
\end{definition}

\begin{restatable}[Occurrence]{lemma}{occurrencelemma}  \XLabel{lem:occurrence}
  ~\\[-10pt]
  \begin{enumerate}[(i)]
  \item  If $\declsubjudg{\Psi}{A}{\tau}$ then $\tau \notoccursinsidearrow A$.
  \item  If $\declsubjudg{\Psi}{\tau}{B}$ then $\tau \notoccursinsidearrow B$.
  \end{enumerate}
\end{restatable}

\begin{restatable}[Monotype Equality]{lemma}{declmonotypeequality}
  \XLabel{lem:decl-monotype-equality}
   If $\declsubjudg{\Psi}{\sigma}{\tau}$ then $\sigma = \tau$.
\end{restatable}

\begin{definition}[Contextual Size] \XLabel{def:typesize}
  The size of $A$ with respect to a context $\Gamma$, written $\typesize{\Gamma}{A}$, is defined by
  \begin{mathpar}
    \begin{array}{lcl}
      \typesize{\Gamma}{\alpha}  & = &   1
      \\
      \typesize{{\Gamma[\ahat]}}{\ahat}  & = &   1
      \\
      \typesize{{\Gamma[\hypeq{\ahat}{\tau}]}}{\ahat} & = &   1 + \typesize{\Gamma[\hypeq{\ahat}{\tau}]}{\tau} 
      \\
      \typesize{\Gamma}{\alltype{\alpha}{A}} & = &   1 + \typesize{\Gamma, \alpha}{A}
      \\
      \typesize{\Gamma}{A \arr B} & = &   1 + \typesize{\Gamma}{A} + \typesize{\Gamma}{B}
    \end{array}
  \end{mathpar}  
\end{definition}

\section{Type Assignment}

\begin{restatable}[Well-Formedness]{lemma}{declarativetypingwellformed}
\XLabel{lem:declarative-typing-well-formed}
~\\
  If $\declchkjudg{\Psi}{e}{A}$
  or $\declsynjudg{\Psi}{e}{A}$
  or $\declappjudg{\Psi}{e}{A}{C}$ then
  $\judgetp{\Psi}{A}$ (and in the last case, $\judgetp{\Psi}{C}$).
\end{restatable}

\begin{restatable}[Completeness of Bidirectional Typing]{theorem}{completenessbidirectional}
  \XLabel{thm:completeness-bidirectional}
~\\
  If $\judge{\Psi}{e}{A}$ then there exists $e'$ such that $\declsynjudg{\Psi}{e'}{A}$
  and $|e'| = e$. 
\end{restatable}

\begin{restatable}[Subtyping Coercion]{lemma}{subtypingcoercion}
\XLabel{lem:subtyping-coercion}
  If $\declsubjudg{\Psi}{A}{B}$ then there exists $f$ which is $\beta\eta$-equal to the identity such that 
  $\judge{\Psi}{f}{A \arr B}$. 
\end{restatable}

\begin{restatable}[Application Subtyping]{lemma}{applicationsubtyping}
\XLabel{lem:application-subtyping}
  If $\declappjudg{\Psi}{e}{A}{C}$
  then there exists $B$ such that $\declsubjudg{\Psi}{A}{B \arr C}$
  and $\declchkjudg{\Psi}{e}{B}$ by a smaller derivation.
\end{restatable}

\begin{restatable}[Soundness of Bidirectional Typing]{theorem}{soundnessbidirectional}
  \XLabel{thm:soundness-bidirectional}
  We have that:

  \begin{itemize}
  \item  If $\declchkjudg{\Psi}{e}{A}$, then there is an $e'$ such that $\judge{\Psi}{e'}{A}$
    and $e' \beeq  |e|$.
  \item  If $\declsynjudg{\Psi}{e}{A}$, then there is an $e'$ such that $\judge{\Psi}{e'}{A}$
    and $e' \beeq  |e|$.
  \end{itemize}
\end{restatable}

\section{Robustness of Typing}

\begin{restatable}[Type Substitution]{lemma}{typesubstitution}
\XLabel{lem:type-substitution}
~\\
  Assume $\judgetp{\Psi}{\tau}$.

  \begin{itemize}
  \item If\, $\declchkjudg{\Psi, \alpha, \Psi'}{e'}{C}$ then\, $\declchkjudg{\Psi, [\tau/\alpha]\Psi'}{[\tau/\alpha]e'}{[\tau/\alpha]C}$.
  \item If\, $\declsynjudg{\Psi, \alpha, \Psi'}{e'}{C}$ then\, $\declsynjudg{\Psi, [\tau/\alpha]\Psi'}{[\tau/\alpha]e'}{[\tau/\alpha]C}$.
  \item If\, $\declappjudg{\Psi, \alpha, \Psi'}{e'}{B}{C}$ then\, $\declappjudg{\Psi, [\tau/\alpha]\Psi'}{[\tau/\alpha]e'}{[\tau/\alpha]B}{[A/\alpha]C}$.
  \end{itemize}   

  Moreover, the resulting derivation contains no more applications of typing rules than the given one.
  (Internal subtyping derivations, however, may grow.)
\end{restatable}

\bigskip

\begin{definition}[Context Subtyping]
\XLabel{def:context-subtyping}
    We define the judgment $\Psi' \leq \Psi$ with the following rules:
    \\[-0.5ex]
    \begin{mathpar}
      \Infer{\CtxsubEmpty}
                { }
                {\cdot \leq \cdot}
      \and
      \Infer{\CtxsubUvar}
                {\Psi' \leq \Psi }
                {\Psi', \alpha \leq \Psi, \alpha}
      \and
      \Infer{\CtxsubVar}
                {\Psi' \leq \Psi
                  \\
                  \declsubjudg{\Psi}{A'}{A}}
                {\Psi', x:A' \leq \Psi, x:A}
    \end{mathpar}
\end{definition}

\medskip

\begin{restatable}[Subsumption]{lemma}{termsubsumption}
\XLabel{lem:term-subsumption} 
    Suppose $\Psi' \leq \Psi$. Then:

    \begin{enumerate}[(i)]
    \item If $\declchkjudg{\Psi}{e}{A}$ and $\declsubjudg{\Psi}{A}{A'}$ then $\declchkjudg{\Psi'}{e}{A'}$.
    \item If $\declsynjudg{\Psi}{e}{A}$ then 
          there exists $A'$ such that $\declsubjudg{\Psi}{A'}{A}$ and $\declsynjudg{\Psi'}{e}{A'}$.
    \item If $\declappjudg{\Psi}{e}{C}{A}$ and $\declsubjudg{\Psi}{C'}{C}$
      \\ then 
       there exists $A'$ such that $\declsubjudg{\Psi}{A'}{A}$ and $\declappjudg{\Psi'}{e}{C'}{A'}$.
    \end{enumerate}
\end{restatable}

\begin{restatable}[Substitution]{theorem}{termsubstitution}
\XLabel{thm:term-substitution}
~\\
  Assume $\declsynjudg{\Psi}{e}{A}$.
~\\[-3ex]
  \begin{enumerate}[(i)]
  \item If $\,\declchkjudg{\Psi, x:A}{e'}{C}$ then $\,\declchkjudg{\Psi}{[e/x]e'}{C}$.
  \item If $\,\declsynjudg{\Psi, x:A}{e'}{C}$ then $\,\declsynjudg{\Psi}{[e/x]e'}{C}$.
  \item If $\,\declappjudg{\Psi, x:A}{e'}{B}{C}$ then $\,\declappjudg{\Psi}{[e/x]e'}{B}{C}$.
  \end{enumerate}
\end{restatable}

\begin{restatable}[Inverse Substitution]{theorem}{termUnsubstitution}
\XLabel{thm:term-unsubstitution}
~\\
Assume $\declchkjudg{\Psi}{e}{A}$.
\begin{enumerate}[(i)]
  \item If $\,\declchkjudg{\Psi}{[(e:A)/x]e'}{C}$ then $\,\declchkjudg{\Psi, x:A}{e'}{C}$.
  \item If $\,\declsynjudg{\Psi}{[(e:A)/x]e'}{C}$ then $\,\declsynjudg{\Psi, x:A}{e'}{C}$.
  \item If $\,\declappjudg{\Psi}{[(e:A)/x]e'}{B}{C}$ then $\,\declappjudg{\Psi, x:A}{e'}{B}{C}$.
\end{enumerate}
\end{restatable}

\begin{restatable}[Annotation Removal]{theorem}{annotationRemoval}
\XLabel{thm:annotation-dropping}
  We have that:

  \begin{itemize}
    \item If\, $\declchkjudg{\Psi}{\big((\fun{x}{e}) : A\big)}{C}$ then $\declchkjudg{\Psi}{\fun{x}{e}}{C}$.
    \item If\, $\declchkjudg{\Psi}{(\unitexp : A)}{C}$ then $\declchkjudg{\Psi}{\unitexp}{C}$.
    \item If\, $\declsynjudg{\Psi}{e_1\;(e_2 : A)}{C}$ then $\declsynjudg{\Psi}{e_1\;e_2}{C}$.
    \item If\, $\declsynjudg{\Psi}{(x : A)}{A}$ then $\declsynjudg{\Psi}{x}{B}$ and $\declsubjudg{\Psi}{B}{A}$.
    \item If\, $\declsynjudg{\Psi}{\big((e_1\,e_2) : A\big)}{A}$
       then $\declsynjudg{\Psi}{e_1\,e_2}{B}$
      and $\declsubjudg{\Psi}{B}{A}$. 
    \item If\, $\declsynjudg{\Psi}{\big((e : B) : A\big)}{A}$
      then $\declsynjudg{\Psi}{(e : B)}{B}$
      and $\declsubjudg{\Psi}{B}{A}$. 
    \item If $\declsynjudg{\Psi}{\big((\fun{x}{e}) : \sigma \arr \tau\big)}{\sigma \arr \tau}$
      then $\declsynjudg{\Psi}{\fun{x}{e}}{\sigma \arr \tau}$.
  \end{itemize}
\end{restatable}

\begin{restatable}[Soundness of Eta]{theorem}{soundnessofeta}
~\\\Label{thm:soundness-of-eta}
  If\, $\declchkjudg{\Psi}{\fun{x}{e\;x}}{A}$ and $x \not\in \FV{e}$, then $\declchkjudg{\Psi}{e}{A}$.   
\end{restatable}

\section{Properties of Context Extension}

\subsection{Syntactic Properties}

\begin{restatable}[Declaration Preservation]{lemma}{declarationpreservation}
\XLabel{lem:declaration-preservation}
  If $\substextend{\Gamma}{\Delta}$,
  and $u$ is a variable or marker $\MonnierComma{\ahat}$ declared in $\Gamma$, then $u$ is declared in $\Delta$. 
\end{restatable}

\begin{restatable}[Declaration Order Preservation]{lemma}{declarationorderpreservation}
\XLabel{lem:declaration-order-preservation}
  If $\substextend{\Gamma}{\Delta}$ 
  and $u$ is declared to the left of $v$ in $\Gamma$,
  then $u$ is declared to the left of $v$ in $\Delta$. 
\end{restatable}

\begin{restatable}[Reverse Declaration Order Preservation]{lemma}{reversedeclarationorderpreservation}
\XLabel{lem:reverse-declaration-order-preservation}
  If $\substextend{\Gamma}{\Delta}$ and $u$ and $v$ are both declared in $\Gamma$
  and $u$ is declared to the left of $v$ in $\Delta$,
  then $u$ is declared to the left of $v$ in $\Gamma$. 
\end{restatable}

\begin{restatable}[Substitution Extension Invariance]{lemma}{substitutionextensioninvariance}
\XLabel{lem:subst-extension-invariance}
   If $\judgetp{\Theta}{A}$ and $\substextend{\Theta}{\Gamma}$
   then
   $[\Gamma]A = [\Gamma]([\Theta]A)$
   and $[\Gamma]A = [\Theta]([\Gamma]A)$.
\end{restatable}

\begin{restatable}[Extension Equality Preservation]{lemma}{extensionequalitypreservation}
\XLabel{lem:extension-equality-preservation}
~\\
   If $\judgetp{\Gamma}{A}$
   and $\judgetp{\Gamma}{B}$ 
   and $[\Gamma]A = [\Gamma]B$
   and $\substextend{\Gamma}{\Delta}$,
   then 
   $[\Delta]A = [\Delta]B$.   
\end{restatable}

\begin{restatable}[Reflexivity]{lemma}{substextendreflexivity}
\XLabel{lem:substextend-reflexivity}
  If $\Gamma$ is well-formed, then $\substextend{\Gamma}{\Gamma}$. 
\end{restatable}

\begin{restatable}[Transitivity]{lemma}{substextendtransitivity}
\XLabel{lem:substextend-transitivity}
  If $\substextend{\Gamma}{\Delta}$
  and $\substextend{\Delta}{\Theta}$,
  then $\substextend{\Gamma}{\Theta}$.
\end{restatable}

\begin{definition}[Softness]  \XLabel{def:soft}
  A context $\Theta$ is \emph{soft} iff it consists only of $\ahat$ and $\hypeq{\ahat}{\tau}$
  declarations.
\end{definition}

\begin{restatable}[Right Softness]{lemma}{rightsoftness}  \XLabel{lem:softness}
  If $\substextend{\Gamma}{\Delta}$ and $\Theta$ is soft
  (and $(\Delta, \Theta)$ is well-formed)
  then $\substextend{\Gamma}{\Delta, \Theta}$.
\end{restatable}

\begin{restatable}[Evar Input]{lemma}{evarinput}  \XLabel{lem:evar-input}
~\\
  If $\substextend{\Gamma, \ahat}{\Delta}$
  then $\Delta = (\Delta_0, \Delta_{\ahat}, \Theta)$ where
  $\substextend{\Gamma}{\Delta_0}$,
  and $\Delta_{\ahat}$ is either $\ahat$ or $\hypeq{\ahat}{\tau}$,
  and $\Theta$ is soft.
\end{restatable}

\begin{restatable}[Extension Order]{lemma}{extensionorder} 
\XLabel{lem:extension-order}
~\\[-10pt]
  \begin{enumerate}[(i)]
    \item
      If $\substextend{\Gamma_L, \alpha, \Gamma_R}{\Delta}$
      then $\Delta = (\Delta_L, \alpha, \Delta_R)$
      where $\substextend{\Gamma_L}{\Delta_L}$. \\
      Moreover, if $\Gamma_R$ is soft then $\Delta_R$ is soft.
      
    \item
      If $\substextend{\Gamma_L, \MonnierComma{\ahat}, \Gamma_R}{\Delta}$
      then $\Delta = (\Delta_L, \MonnierComma{\ahat}, \Delta_R)$
      where $\substextend{\Gamma_L}{\Delta_L}$. \\
      Moreover, if $\Gamma_R$ is soft then $\Delta_R$ is soft.

    \item
      If $\substextend{\Gamma_L, \ahat, \Gamma_R}{\Delta}$
      then $\Delta = \Delta_L, \Theta, \Delta_R$
      where $\substextend{\Gamma_L}{\Delta_L}$
      and $\Theta$ is either $\ahat$ or $\hypeq{\ahat}{\tau}$ for some $\tau$.

    \item
      If $\substextend{\Gamma_L, \hypeq{\ahat}{\tau}, \Gamma_R}{\Delta}$
      then $\Delta = \Delta_L, \hypeq{\ahat}{\tau'}, \Delta_R$
      where $\substextend{\Gamma_L}{\Delta_L}$
      and $[\Delta_L]\tau = [\Delta_L]\tau'$.

    \item
      If $\substextend{\Gamma_L, x : A, \Gamma_R}{\Delta}$
      then $\Delta = (\Delta_L, x : A', \Delta_R)$
      where $\substextend{\Gamma_L}{\Delta_L}$
          and $[\Delta_L]A = [\Delta_L]A'$. \\
      Moreover, $\Gamma_R$ is soft if and only if $\Delta_R$ is soft.
  \end{enumerate}
\end{restatable}

\begin{restatable}[Extension Weakening]{lemma}{extensionweakening}
  \XLabel{lem:extension-weakening}
  If $\judgetp{\Gamma}{A}$ and $\substextend{\Gamma}{\Delta}$ then $\judgetp{\Delta}{A}$. 
\end{restatable}

\begin{restatable}[Solution Admissibility for Extension]{lemma}{extensionsolve}
\XLabel{lem:extension-solve}
  If $\judgetp{\Gamma_L}{\tau}$
  then
  $\substextend{\Gamma_L, \ahat, \Gamma_R}{\Gamma_L, \hypeq{\ahat}{\tau}, \Gamma_R}$. 
\end{restatable}

\begin{restatable}[Solved Variable Addition for Extension]{lemma}{extensionaddsolve}
\XLabel{lem:extension-addsolve}
  If $\judgetp{\Gamma_L}{\tau}$
  then
  $\substextend{\Gamma_L, \Gamma_R}{\Gamma_L, \hypeq{\ahat}{\tau}, \Gamma_R}$. 
\end{restatable}

\begin{restatable}[Unsolved Variable Addition for Extension]{lemma}{extensionadd}
\XLabel{lem:extension-add}
   We have that $\substextend{\Gamma_L, \Gamma_R}{\Gamma_L, \ahat, \Gamma_R}$. 
\end{restatable}

\begin{restatable}[Parallel Admissibility]{lemma}{parallelevaradmissibility}
\XLabel{lem:parallel-admissibility} ~\\
    If $\substextend{\Gamma_L}{\Delta_L}$ and $\substextend{\Gamma_L, \Gamma_R}{\Delta_L, \Delta_R}$ then:
    
    \begin{enumerate}[(i)]
    \item $\substextend{\Gamma_L, \ahat, \Gamma_R}{\Delta_L, \ahat, \Delta_R}$
    \item If $\judgetp{\Delta_L}{\tau'}$ then 
          $\substextend{\Gamma_L, \ahat, \Gamma_R}{\Delta_L, \ahat=\tau', \Delta_R}$.
    \item If $\judgetp{\Gamma_L}{\tau}$ and $\judgetp{\Delta_L}{\tau'}$ and $[\Delta_L]\tau = [\Delta_L]\tau'$,
          then $\substextend{\Gamma_L, \ahat=\tau, \Gamma_R}{\Delta_L, \ahat=\tau', \Delta_R}$. 
    \end{enumerate}
\end{restatable}

\begin{restatable}[Parallel Extension Solution]{lemma}{parallelextensionsolve}
\XLabel{lem:parallel-extension-solve} ~\\
  If $\substextend{\Gamma_L, \ahat, \Gamma_R}{\Delta_L, \ahat=\tau', \Delta_R}$
  and $\judgetp{\Gamma_L}{\tau}$
  and $[\Delta_L]\tau = [\Delta_L]\tau'$
  then $\substextend{\Gamma_L, \ahat=\tau, \Gamma_R}{\Delta_L, \ahat=\tau', \Delta_R}$. 
\end{restatable}

\begin{restatable}[Parallel Variable Update]{lemma}{parallelextensionupdate}
\XLabel{lem:parallel-extension-update} ~\\
  If ~~~\, $\substextend{\Gamma_L, \ahat, \Gamma_R}{\Delta_L, \ahat=\tau_0, \Delta_R}$
  and $\judgetp{\Gamma_L}{\tau_1}$
  and $\judgetp{\Delta_L}{\tau_2}$
  and $[\Delta_L]\tau_0 = [\Delta_L]\tau_1 = [\Delta_L]\tau_2$
  \\
  then $\substextend{\Gamma_L, \ahat=\tau_1, \Gamma_R}{\Delta_L, \ahat=\tau_2, \Delta_R}$. 
\end{restatable}

\subsection{Instantiation Extends}

\begin{restatable}[Instantiation Extension]{lemma}{instantiationextension}   \XLabel{lem:instantiation-extension}
~\\
   If $\instjudg{\Gamma}{\ahat}{\tau}{\Delta}$ or $\instjudgr{\Gamma}{\ahat}{\tau}{\Delta}$
   then
  $\substextend{\Gamma}{\Delta}$.
\end{restatable}

\subsection{Subtyping Extends}

\begin{restatable}[Subtyping Extension]{lemma}{subtypingextension}   \XLabel{lem:subtyping-extension}
~\\
   If $\subjudg{\Gamma}{A}{B}{\Delta}$ then
   $\substextend{\Gamma}{\Delta}$.
 \end{restatable}

\section{Decidability of Instantiation}

\begin{restatable}[Left Unsolvedness Preservation]{lemma}{leftunsolvednesspreservation}
\XLabel{lem:left-unsolvedness-preservation}
~\\
  If $\instjudg{\underbrace{\Gamma_0, \ahat, \Gamma_1}_\Gamma}{\ahat}{A}{\Delta}$
     or $\instjudgr{\underbrace{\Gamma_0, \ahat, \Gamma_1}_\Gamma}{\ahat}{A}{\Delta}$,
  and $\bhat \in \unsolved{\Gamma_0}$, then $\bhat \in \unsolved{\Delta}$.
\end{restatable}

\begin{restatable}[Left Free Variable Preservation]{lemma}{leftfreevariablepreservation}
\XLabel{lem:left-free-variable-preservation}
  If $\instjudg{\overbrace{\Gamma_0, \ahat, \Gamma_1}^\Gamma}{\ahat}{A}{\Delta}$
     or $\instjudgr{\overbrace{\Gamma_0, \ahat, \Gamma_1}^\Gamma}{\ahat}{A}{\Delta}$,
     and $\judgetp{\Gamma}{B}$ and $\ahat \notin \FV{[\Gamma]B}$
     and $\bhat \in \unsolved{\Gamma_0}$ and $\bhat \notin \FV{[\Gamma]B}$,
     then $\bhat \notin \FV{[\Delta]B}$. 
\end{restatable}

\begin{restatable}[Instantiation Size Preservation]{lemma}{instantiationsizepreservation}
\XLabel{lem:instantiation-size-preservation}
  If $\instjudg{\overbrace{\Gamma_0, \ahat, \Gamma_1}^\Gamma}{\ahat}{A}{\Delta}$
     or $\instjudgr{\overbrace{\Gamma_0, \ahat, \Gamma_1}^\Gamma}{\ahat}{A}{\Delta}$,
     and $\judgetp{\Gamma}{B}$ and $\ahat \notin \FV{[\Gamma]B}$, 
     then $|[\Gamma]B| = |[\Delta]B|$, where $|C|$ is the plain size of the term $C$.
\end{restatable}

This lemma lets us show decidability by taking the size of the type argument as the induction
metric. 

\begin{restatable}[Decidability of Instantiation]{theorem}{instantiationdecidability}
\XLabel{thm:decidability-of-instantiation}
    If $\Gamma = \Gamma_0[\ahat]$
    and
    $\judgetp{\Gamma}{A}$
    such that $[\Gamma]A = A$
    and $\ahat \notin \FV{A}$, then:
    
    \begin{enumerate}[(1)]
    \item Either there exists $\Delta$ such that $\instjudg{\Gamma_0[\ahat]}{\ahat}{A}{\Delta}$,
      or not.
    \item Either there exists $\Delta$ such that $\instjudgr{\Gamma_0[\ahat]}{\ahat}{A}{\Delta}$,
      or not.
    \end{enumerate}
\end{restatable}

\section{Decidability of Algorithmic Subtyping}

\subsection{Lemmas for Decidability of Subtyping}

\begin{restatable}[Monotypes Solve Variables]{lemma}{monotypessolvevariables}
\XLabel{lem:monotypes-solve-variables}
If $\instjudg{\Gamma}{\ahat}{\tau}{\Delta}$ or $\instjudgr{\Gamma}{\ahat}{\tau}{\Delta}$, then if $[\Gamma]\tau = \tau$ and $\ahat \notin \FV{[\Gamma]\tau}$, then $|\unsolved{\Gamma}| = |\unsolved{\Delta}| + 1$. 
\end{restatable}

\begin{restatable}[Monotype Monotonicity]{lemma}{monotypemonotonicity}  \XLabel{lem:monotype-monotonicity}
  If $\subjudg{\Gamma}{\tau_1}{\tau_2}{\Delta}$
  then
  $|\unsolved{\Delta}| \leq |\unsolved{\Gamma}|$.
\end{restatable}

\begin{restatable}[Substitution Decreases Size]{lemma}{typesizesubst}
\XLabel{lem:typesize-subst}
  If $\judgetp{\Gamma}{A}$ then $\typesize{\Gamma}{[\Gamma]A} \leq \typesize{\Gamma}{A}$.
\end{restatable}

\begin{restatable}[Monotype Context Invariance]{lemma}{monotypecontextinvariance}
\XLabel{lem:monotype-context-invariance}
~\\
   If $\subjudg{\Gamma}{\tau}{\tau'}{\Delta}$
   where $[\Gamma]\tau = \tau$ and $[\Gamma]\tau' = \tau'$
   and $|\unsolved{\Gamma}| = |\unsolved{\Delta}|$
   then $\Gamma = \Delta$. 
\end{restatable}

\subsection{Decidability of Subtyping}

\begin{restatable}[Decidability of Subtyping]{theorem}{subtypingdecidability}   
\XLabel{thm:subtyping-decidable}
~\\
  Given a context $\Gamma$ and types $A$, $B$ such that $\judgetp{\Gamma}{A}$ and
  $\judgetp{\Gamma}{B}$ and $[\Gamma]A = A$ and $[\Gamma]B = B$,
  it is decidable whether there exists $\Delta$ such that $\subjudg{\Gamma}{A}{B}{\Delta}$.
\end{restatable}

\section{Decidability of Typing}

\begin{restatable}[Decidability of Typing]{theorem}{typingdecidable}
\XLabel{thm:typing-decidable}
~\\[-2ex]
  \begin{enumerate}[(i)]
      \item \emph{Synthesis:}  
        Given a context $\Gamma$
        and a term $e$,
        \\
        it is decidable whether there exist a type $A$ and a context $\Delta$ such that \\
        $\synjudg{\Gamma}{e}{A}{\Delta}$.

      \item \emph{Checking:}
        Given a context $\Gamma$,
        a term $e$,
        and a type $B$ such that $\judgetp{\Gamma}{B}$,
        \\
        it is decidable whether there is a context $\Delta$ such that \\
        $\chkjudg{\Gamma}{e}{B}{\Delta}$.

      \item \emph{Application:}  
        Given a context $\Gamma$,
        a term $e$,
        and a type $A$ such that $\judgetp{\Gamma}{A}$,
        \\
        it is decidable whether there exist a type $C$ and a context $\Delta$ such that \\
        $\appjudg{\Gamma}{e}{A}{C}{\Delta}$.
  \end{enumerate}
\end{restatable}

\section{Soundness of Subtyping}

\begin{definition}[Filling]
  The \emph{filling} of a context $\soln{\Gamma}$ solves all unsolved variables:

  \begin{displ}
        \begin{array}[t]{lcll}
            \soln{\cdot} & = &   \cdot
            \\
            \soln{\Gamma, x : A} & = &   \soln{\Gamma}, x : A
            \\
            \soln{\Gamma, \alpha} & = &   \soln{\Gamma}, \alpha
            \\
            \soln{\Gamma, \hypeq{\ahat}{\tau}} & = &   \soln{\Gamma}, \hypeq{\ahat}{\tau}
            \\
            \soln{\Gamma, \MonnierComma{\ahat}} & = &   \soln{\Gamma}, \MonnierComma{\ahat}
            \\
            \soln{\Gamma, \ahat} & = &   \soln{\Gamma}, \hypeq{\ahat}{\unitty}
        \end{array}
  \end{displ}
\end{definition}

\subsection{Lemmas for Soundness}

\begin{restatable}[Uvar Preservation]{lemma}{ctxappuvarpreservation}
\XLabel{lem:ctxapp-uvar-preservation} ~\\
  If $\alpha \in \Omega$
  and $\substextend{\Delta}{\Omega}$
  then
  $\alpha \in [\Omega]\Delta$.
\end{restatable}
\begin{proof}  By induction on $\Omega$, following the definition of context application.
\end{proof}

\begin{restatable}[Variable Preservation]{lemma}{ctxappvarpreservation}
\XLabel{lem:ctxapp-var-preservation} ~\\
  If $(x : A) \in \Delta$ or $(x : A) \in \Omega$
  and $\substextend{\Delta}{\Omega}$
  then
  $(x : [\Omega]A) \in [\Omega]\Delta$.
\end{restatable}

\begin{restatable}[Substitution Typing]{lemma}{substitutiontyping}
\XLabel{lem:substitution-typing}
  If $\judgetp{\Gamma}{A}$ then $\judgetp{\Gamma}{[\Gamma]A}$. 
\end{restatable}

\begin{restatable}[Substitution for Well-Formedness]{lemma}{completionwf}
\XLabel{lem:completion-wf}
  If $\judgetp{\Omega}{A}$
  then
  $\judgetp{[\Omega]\Omega}{[\Omega]A}$.
\end{restatable}

\begin{restatable}[Substitution Stability]{lemma}{substitutionstability}   \XLabel{lem:substitution-stability}
~\\
  For any well-formed complete context $(\Omega, \Omega_Z)$,
  if $\judgetp{\Omega}{A}$ then $[\Omega]A = [\Omega, \Omega_Z]A$.
\end{restatable}

\begin{restatable}[Context Partitioning]{lemma}{contextpartitioning}   \XLabel{lem:context-partitioning}
  ~\\
  If $\substextend{\Delta, \MonnierComma{\ahat}, \Theta}{\Omega, \MonnierComma{\ahat}, \Omega_Z}$
  then
  there is a $\Psi$ such that
  $[\Omega, \MonnierComma{\ahat}, \Omega_Z](\Delta, \MonnierComma{\ahat}, \Theta)
   =
   [\Omega]\Delta, \Psi$. 
\end{restatable}

\begin{restatable}[Softness Goes Away]{lemma}{softnessgoesaway}  \XLabel{lem:softness-goes-away}
  ~\\
  If $\substextend{\Delta, \Theta}{\Omega, \Omega_Z}$
  where $\substextend{\Delta}{\Omega}$ and $\Theta$ is soft,
  then
  $[\Omega, \Omega_Z](\Delta, \Theta) = [\Omega]\Delta$.
\end{restatable}
\begin{proof}
  By induction on $\Theta$, following the definition of $[\Omega]\Gamma$.
\end{proof}

\begin{lemma}[Filling Completes]  \XLabel{lem:soln-completes}
  If $\substextend{\Gamma}{\Omega}$
  and $(\Gamma, \Theta)$ is well-formed,
  then
  $\substextend{\Gamma, \Theta}{\Omega, \soln{\Theta}}$.
\end{lemma}
\begin{proof}
  By induction on $\Theta$, following the definition of $\soln{-}$ and applying the rules
  for $\extendssym$.
\end{proof}

\begin{restatable}[Stability of Complete Contexts]{lemma}{completesstability}
  \XLabel{lem:completes-stability}  ~\\
  If $\substextend{\Gamma}{\Omega}$
  then $[\Omega]\Gamma = [\Omega]\Omega$. 
\end{restatable}

\begin{restatable}[Finishing Types]{lemma}{finishingtypes}
  \XLabel{lem:finishing-types} ~\\
  If
  $\judgetp{\Omega}{A}$
  and $\substextend{\Omega}{\Omega'}$
  then
  $[\Omega]A = [\Omega']A$. 
\end{restatable} 

\begin{restatable}[Finishing Completions]{lemma}{finishingcompletions}
  \XLabel{lem:finishing-completions}  ~\\
  If $\substextend{\Omega}{\Omega'}$
  then
  $[\Omega]\Omega = [\Omega']\Omega'$. 
\end{restatable}

\begin{restatable}[Confluence of Completeness]{lemma}{confluenceofcompleteness}    \XLabel{lem:completes-confluence}  ~\\
  If $\substextend{\Delta_1}{\Omega}$
  and $\substextend{\Delta_2}{\Omega}$
  then
  $[\Omega]\Delta_1 = [\Omega]\Delta_2$.
\end{restatable}

\subsection{Instantiation Soundness}

\begin{restatable}[Instantiation Soundness]{theorem}{instantiationsoundness}  \XLabel{thm:instantiation-soundness}
   ~\\
    Given $\substextend{\Delta}{\Omega}$ and $[\Gamma]B = B$ and $\ahat \notin \FV{B}$: \\[-2.5ex]
        \begin{enumerate}[(1)]
            \item If $\instjudg{\Gamma}{\ahat}{B}{\Delta}$ 
              then
              $\declsubjudg{[\Omega]\Delta}{[\Omega]\ahat}{[\Omega]B}$.

            \item If $\instjudgr{\Gamma}{\ahat}{B}{\Delta}$ 
              then
              $\declsubjudg{[\Omega]\Delta}{[\Omega]B}{[\Omega]\ahat}$.
        \end{enumerate}
\end{restatable}

\subsection{Soundness of Subtyping}

\begin{restatable}[Soundness of Algorithmic Subtyping]{theorem}{soundness} 
\XLabel{thm:subtyping-soundness} ~\\
  If $\subjudg{\Gamma}{A}{B}{\Delta}$
  where $[\Gamma]A = A$ and $[\Gamma]B = B$
  and $\substextend{\Delta}{\Omega}$
  then
  $\declsubjudg{\contextapp{\Omega}{\Delta}}{\contextapp{\Omega}{A}}{\contextapp{\Omega}{B}}$. 
\end{restatable}

\begin{restatable}[Soundness, Pretty Version]{corollary}{soundnesspretty}
  If $\subjudg{\Psi}{A}{B}{\Delta}$, then $\declsubjudg{\Psi}{A}{B}$. 
\end{restatable}

\section{Typing Extension}

\begin{restatable}[Typing Extension]{lemma}{typingextension}
\XLabel{lem:typing-extension}
 ~\\
  If $\chkjudg{\Gamma}{e}{A}{\Delta}$
  or $\synjudg{\Gamma}{e}{A}{\Delta}$
  or $\appjudg{\Gamma}{e}{A}{C}{\Delta}$
  then
  $\substextend{\Gamma}{\Delta}$.
\end{restatable}

\section{Soundness of Typing}

\begin{restatable}[Soundness of Algorithmic Typing]{theorem}{typingsoundness}
\XLabel{lem:typing-soundness} %
Given $\substextend{\Delta}{\Omega}$: \\[-2.5ex]
    \begin{enumerate}[(i)]
      \item 
          If $\chkjudg{\Gamma}{e}{A}{\Delta}$
          then
          $\declchkjudg{[\Omega]\Delta}{e}{[\Omega]A}$. 
      
      \item
          If $\synjudg{\Gamma}{e}{A}{\Delta}$
          then
          $\declsynjudg{[\Omega]\Delta}{e}{[\Omega]A}$. 

      \item
          If $\appjudg{\Gamma}{e}{A}{C}{\Delta}$
          then
          $\declappjudg{[\Omega]\Delta}{e}{[\Omega]A}{[\Omega]C}$.
    \end{enumerate}
\end{restatable}

\section{Completeness of Subtyping}

\subsection{Instantiation Completeness}

\begin{restatable}[Instantiation Completeness]{theorem}{instantiationcompletes}
   \XLabel{thm:instantiation-completes}
   ~\\
   Given $\substextend{\Gamma}{\Omega}$ 
   and $A = [\Gamma]A$ 
   and $\ahat \in \unsolved{\Gamma}$  %
   and $\ahat \notin \FV{A}$:
   
       \begin{enumerate}[(1)]
           \item If $\declsubjudg{[\Omega]\Gamma}{[\Omega]\ahat}{[\Omega]A}$
                 \\
                 then there are $\Delta$, $\Omega'$ such that
                 $\substextend{\Omega}{\Omega'}$
                 and $\substextend{\Delta}{\Omega'}$
                 and $\instjudg{\Gamma}{\ahat}{A}{\Delta}$. 

           \item If $\declsubjudg{[\Omega]\Gamma}{[\Omega]A}{[\Omega]\ahat}$
                 \\
                 then there are $\Delta$, $\Omega'$ such that
                 $\substextend{\Omega}{\Omega'}$
                 and $\substextend{\Delta}{\Omega'}$
                 and $\instjudgr{\Gamma}{\ahat}{A}{\Delta}$. 
      \end{enumerate}                                                    
\end{restatable}

\subsection{Completeness of Subtyping}

\begin{restatable}[Generalized Completeness of Subtyping]{theorem}{completingcompleteness}
  \XLabel{thm:completing-completeness}
  If $\substextend{\Gamma}{\Omega}$
  and $\judgetp{\Gamma}{A}$
  and $\judgetp{\Gamma}{B}$
  and $\declsubjudg{[\Omega]\Gamma}{[\Omega]A}{[\Omega]B}$
  then
  there exist $\Delta$ and $\Omega'$ such that 
  $\substextend{\Delta}{\Omega'}$ and 
  $\substextend{\Omega}{\Omega'}$ and 
  $\subjudg{\Gamma}{[\Gamma]A}{[\Gamma]B}{\Delta}$.
\end{restatable}

\begin{restatable}[Completeness of Subtyping]{corollary}{completeness}
  If $\declsubjudg{\Psi}{A}{B}$
  then
  there is a $\Delta$ such that
  $\subjudg{\Psi}{A}{B}{\Delta}$. 
\end{restatable}

\section{Completeness of Typing}
\Label{sec:last-section-before-proofs}

\begin{restatable}[Completeness of Algorithmic Typing]{theorem}{typingcompleteness}
  \XLabel{thm:typing-completeness}
  Given $\substextend{\Gamma}{\Omega}$ and $\judgetp{\Gamma}{A}$:
  \\[-2ex]
    \begin{enumerate}[(i)]
      \item
           If $\declchkjudg{[\Omega]\Gamma}{e}{[\Omega]A}$
           \\
           then there exist $\Delta$ and $\Omega'$
           \\
           such that
           $\substextend{\Delta}{\Omega'}$ and 
           $\substextend{\Omega}{\Omega'}$ and 
           $\chkjudg{\Gamma}{e}{[\Gamma]A}{\Delta}$.
      
      \item
           If $\declsynjudg{[\Omega]\Gamma}{e}{A}$
           \\
           then there exist $\Delta$, $\Omega'$, and $A'$
           \\
           such that
           $\substextend{\Delta}{\Omega'}$ and 
           $\substextend{\Omega}{\Omega'}$ and 
           $\synjudg{\Gamma}{e}{A'}{\Delta}$ and 
           $A = [\Omega']A'$. 
      
      \item
           If $\declappjudg{[\Omega]\Gamma}{e}{[\Omega]A}{C}$
           \\
           then there exist $\Delta$, $\Omega'$, and $C'$
           \\
           such that
           $\substextend{\Delta}{\Omega'}$ and 
           $\substextend{\Omega}{\Omega'}$ and 
           $\appjudg{\Gamma}{e}{[\Gamma]A}{C'}{\Delta}$ and 
           $C = [\Omega']C'$. 
    \end{enumerate}
\end{restatable}

\clearpage

\section*{Proofs}
\addcontentsline{toc}{section}{\fontsz{14pt}{Proofs}}

\newcounter{savedsection}
\setcounter{savedsection}{\value{section}}

\renewcommand{\thesection}{\XAlph{12}$'$}

In the rest of this document, we prove the results stated above, with the same sectioning.

\section{Declarative Subtyping}

\propweakening*

\propsubst*

The proofs of these two propositions are routine inductions.

\subsection{Properties of Well-Formedness}

\subsection{Reflexivity}

\declarativereflexivity*
\begin{proof}  By induction on $A$.

  \begin{itemize}
  \ProofCaseThing{A = \unitty}  Apply rule \DsubUnit.
   
  \ProofCaseThing{A = \alpha}  Apply rule \DsubVar.

  \ProofCaseThing{A = A_1 \arr A_2}

      \begin{llproof}
      \declsubjudgPf{\Psi}{A_1}{A_1}   {By i.h.}
      \declsubjudgPf{\Psi}{A_2}{A_2}   {By i.h.}
      \declsubjudgPf{\Psi}{A_1 \arr A_2}{A_1 \arr A_2}   {By \DsubArr}
      \end{llproof}

  \ProofCaseThing{A = \alltype{\alpha}{A_0}}
  
      \begin{llproof}
      \declsubjudgPf{\Psi, \alpha}{A_0}{A_0}   {By i.h.}
      \judgetpPf{\Psi, \alpha}{\alpha}   {By \DeclUvarWF}
      \declsubjudgPf{\Psi, \alpha}{[\alpha/\alpha]A_0}{A_0}   {By def.\ of substitution}
      \declsubjudgPf{\Psi, \alpha}{\alltype{\alpha}{A_0}}{A_0}   {By \DsubAllL}
      \declsubjudgPf{\Psi}{\alltype{\alpha}{A_0}}{\alltype{\alpha}{A_0}}   {By \DsubAllR}
      \end{llproof}
      \qedhere
  \end{itemize}
\end{proof}

\subsection{Subtyping Implies Well-Formedness}

\declarativewellformed*
\begin{proof}
  By induction on the given derivation.  All 5 cases are straightforward.
\end{proof}

\subsection{Substitution}

\declsubsubstitution*
\begin{proof}
  By induction on the given derivation.

  \begin{itemize}
    \DerivationProofCase{\DsubVar}
              {\beta \in (\Psi, \alpha, \Psi')}
              {\declsubjudg{\Psi, \alpha, \Psi'}{\beta}{\beta}}

              It is given that $\Psi \entails \tau$.

              Either $\beta = \alpha$ or $\beta \neq \alpha$.  In the former case:
              We need to show $\declsubjudg{\Psi, \Psi'}{[\tau/\alpha]\alpha}{[\tau/\alpha]\alpha}$,
              that is, $\declsubjudg{\Psi, \Psi'}{\tau}{\tau}$, which follows by \Lemmaref{lem:declarative-reflexivity}.
              In the latter case: We need to show 
               $\declsubjudg{\Psi, \Psi'}{[\tau/\alpha]\beta}{[\tau/\alpha]\beta}$,
               that is,
               $\declsubjudg{\Psi, \Psi'}{\beta}{\beta}$.  Since $\beta \in (\Psi, \alpha, \Psi')$
               and $\beta \neq \alpha$, we have $\beta \in (\Psi, \Psi')$, so applying \DsubVar
               gives the result.

    \DerivationProofCase{\DsubUnit}
              { }
              {\declsubjudg{\Psi, \alpha, \Psi'}{\unitty}{\unitty}}

              For all $\tau$, substituting $[\tau/\alpha]\unitty = \unitty$, and applying
              \DsubUnit gives the result.

    \DerivationProofCase{\DsubArr}
              {\declsubjudg{\Psi, \alpha, \Psi'}{B_1}{A_1}
                \\
                \declsubjudg{\Psi, \alpha, \Psi'}{A_2}{B_2}
              }
              {\declsubjudg{\Psi, \alpha, \Psi'}{A_1 \arr A_2}{B_1 \arr B_2}}

              \begin{llproof}
                \declsubjudgPf{\Psi, \alpha, \Psi'}{B_1}{A_1}  {Subderivation}
                \declsubjudgPf{\Psi, \Psi'}{[\tau/\alpha]B_1}{[\tau/\alpha]A_1}  {By i.h.}
                \proofsep
                \declsubjudgPf{\Psi, \alpha, \Psi'}{A_2}{B_2}  {Subderivation}
                \declsubjudgPf{\Psi, \Psi'}{[\tau/\alpha]A_2}{[\tau/\alpha]B_2}  {By i.h.}
                \proofsep
                \declsubjudgPf{\Psi, \Psi'}{([\tau/\alpha]A_1) \arr ([\tau/\alpha]A_2)}
                                           {([\tau/\alpha]B_1) \arr ([\tau/\alpha]B_2)}  {By \DsubArr}
\Hand                \declsubjudgPf{\Psi, \Psi'}{[\tau/\alpha](A_1 \arr A_2)}
                                           {[\tau/\alpha](B_1 \arr B_2)}  {By definition of subst.}
              \end{llproof}

    \DerivationProofCase{\DsubAllL}
              {\judgetp{\Psi, \alpha, \Psi'}{\sigma}
                \\
                \declsubjudg{\Psi, \alpha, \Psi'}{[\sigma/\beta]A_0}{B}}
              {\declsubjudg{\Psi, \alpha, \Psi'}{\alltype{\beta}{A_0}}{B}}

              \begin{llproof}
                 \declsubjudgPf{\Psi, \alpha, \Psi'}{[\sigma/\beta]A_0}{B}  {Subderivation}
                 \declsubjudgPf{\Psi, \Psi'}{[\tau/\alpha][\sigma/\beta]A_0}{[\tau/\alpha]B}  {By i.h.}
                 \declsubjudgPf{\Psi, \Psi'}{\big[[\tau/\alpha]\sigma \,/\, \beta\big][\tau/\alpha]A_0}{[\tau/\alpha]B}  {By distributivity of substitution}
                 \proofsep
                 \judgetpPf{\Psi, \alpha, \Psi'}{\sigma}  {Premise}
                 \judgetpPf{\Psi}{\tau}  {Given}
                 \judgetpPf{\Psi, \Psi'}{[\tau/\alpha]\sigma}  {By \Propositionref{prop:subst}}
                 \proofsep
                 \declsubjudgPf{\Psi, \Psi'}{\alltype{\beta}{[\tau/\alpha]A_0}}{[\tau/\alpha]B}  {By \DsubAllL}
\Hand            \declsubjudgPf{\Psi, \Psi'}{[\tau/\alpha]\;\big(\alltype{\beta}{A_0}\big)}{[\tau/\alpha]B}  {By definition of substitution}
              \end{llproof}

    \DerivationProofCase{\DsubAllR}
              {\declsubjudg{\Psi, \alpha, \Psi', \beta}{A}{B_0}}
              {\declsubjudg{\Psi, \alpha, \Psi'}{A}{\alltype{\beta}{B_0}}}

              \begin{llproof}
                \declsubjudgPf{\Psi, \alpha, \Psi', \beta}{A}{B_0} {Subderivation}
                \declsubjudgPf{\Psi, \Psi', \beta}{[\tau/\alpha]A}{[\tau/\alpha]B_0} {By i.h.}
                \declsubjudgPf{\Psi, \Psi'}{[\tau/\alpha]A}{\alltype{\beta} [\tau/\alpha]B_0} {By \DsubAllR}
\Hand           \declsubjudgPf{\Psi, \Psi'}{[\tau/\alpha]A}{[\tau/\alpha](\alltype{\beta} B_0)} {By definition of substitution}
              \end{llproof}   \qedhere
  \end{itemize}
\end{proof}

\subsection{Transitivity}

To prove transitivity, we use a metric that adapts ideas from a
proof of cut elimination by \citet{Pfenning95}.

\declarativetransitivity*
\begin{proof}
  By induction with the following metric:
    \[
      \left\langle
        \#\forall(B),~~~
        \Dee_1 + \Dee_2
      \right\rangle
    \]
    where $\langle \dots \rangle$ denotes lexicographic order,
    the first part $\#\forall(B)$ is the number of quantifiers in $B$,
    and the second part is the (simultaneous) size of the derivations
    $\Dee_1 \derives \declsubjudg{\Psi}{A}{B}$
    and $\Dee_2 \derives \declsubjudg{\Psi}{B}{C}$.
    We need to consider the number of quantifiers first in one case:
    when \DsubAllR concluded $\Dee_1$ and \DsubAllL concluded $\Dee_2$,
    because in that case, the derivations on which the i.h.\ must be applied are
    not necessarily smaller.

  \begin{itemize}
    \DoubleDerivationProofCase
          {\DsubVar}
              {\alpha \in \Psi}
              {\declsubjudg{\Psi}{\alpha}{\alpha}}
          {\DsubVar}
              {\alpha \in \Psi}
              {\declsubjudg{\Psi}{\alpha}{\alpha}}

          Apply rule \DsubVar.

    \ProofCaseRule{{\DsubUnit} / \DsubUnit}  Similar to the {\DsubVar} / \DsubVar case.
    
    \DoubleDerivationProofCase{\DsubArr}
              {\declsubjudg{\Psi}{B_1}{A_1}
                \\
                \declsubjudg{\Psi}{A_2}{B_2}
              }
              {\declsubjudg{\Psi}{A_1 \arr A_2}{B_1 \arr B_2}}
          {\DsubArr}
              {\declsubjudg{\Psi}{C_1}{B_1}
                \\
                \declsubjudg{\Psi}{B_2}{C_2}
              }
              {\declsubjudg{\Psi}{B_1 \arr B_2}{C_1 \arr C_2}}

              By i.h. on the 3rd and 1st subderivations, $\declsubjudg{\Psi}{C_1}{A_1}$. \\
              By i.h. on the 2nd and 4th subderivations, $\declsubjudg{\Psi}{A_2}{C_2}$. \\
              By \DsubArr, $\declsubjudg{\Psi}{A_1 \arr A_2}{C_1 \arr C_2}$.
  \end{itemize}

  If \DsubAllL concluded $\Dee_1$:

  \begin{itemize}
    \DerivationProofCase{\DsubAllL}
              {\judgetp{\Psi}{\tau}
                \\
                \declsubjudg{\Psi}{[\tau/\alpha]A_0}{B}}
              {\declsubjudg{\Psi}{\alltype{\alpha}{A_0}}{B}}

              \begin{llproof}
                \judgetpPf{\Psi}{\tau}  {Premise}
                \declsubjudgPf{\Psi}{[\tau/\alpha]A_0}{B}  {Subderivation}
                \declsubjudgPf{\Psi}{B}{C}  {Given ($\Dee_2$)}
                \declsubjudgPf{\Psi}{[\tau/\alpha]A_0}{C}  {By i.h.}
\Hand            \declsubjudgPf{\Psi}{\alltype{\alpha}{A_0}}{C}  {By \DsubAllL}
              \end{llproof}
  \end{itemize}

  If \DsubAllR concluded $\Dee_2$:

  \begin{itemize}
    \DerivationProofCase{\DsubAllR}
              {\declsubjudg{\Psi, \beta}{B}{C}}
              {\declsubjudg{\Psi}{B}{\alltype{\beta}{C}}}

              \begin{llproof}
                \judgetpPf{\Psi}{\tau}  {Premise}
                \declsubjudgPf{\Psi, \beta}{B}{C}  {Subderivation}
                \declsubjudgPf{\Psi}{A}{B}  {Given ($\Dee_1$)}
                \declsubjudgPf{\Psi, \beta}{A}{B}  {By \Propositionref{prop:weakening}}
                \declsubjudgPf{\Psi, \beta}{A}{C}  {By i.h.}
\Hand            \declsubjudgPf{\Psi}{A}{\alltype{\beta} C}  {By \DsubAllL}
              \end{llproof}
  \end{itemize}

  The only remaining possible case is {\DsubAllR} / \DsubAllL.

  \begin{itemize}
    \DoubleDerivationProofCase
          {\DsubAllR}
              {\declsubjudg{\Psi, \beta}{A}{B_0}}
              {\declsubjudg{\Psi}{A}{\alltype{\beta}{B_0}}}
          {\DsubAllL}
              {\judgetp{\Psi}{\tau}
                \\
                \declsubjudg{\Psi}{[\tau/\beta]B_0}{C}}
              {\declsubjudg{\Psi}{\alltype{\beta}{B_0}}{C}}
              
              \medskip

              \begin{llproof}
                \declsubjudgPf{\Psi, \beta}{A}{B_0}   {Subderivation of $\Dee_1$}
                \judgetpPf{\Psi}{\tau}  {Premise of $\Dee_2$}
                \declsubjudgPf{\Psi}{[\tau/\beta]A}{[\tau/\beta]B_0}  {By \Lemmaref{lem:decl-sub-substitution}}
                \eqPf{[\tau/\beta]A} {A}   {$\beta$ cannot appear in $A$}
                \declsubjudgPf{\Psi}{A}{[\tau/\beta]B_0}  {By above equality}
                \declsubjudgPf{\Psi}{[\tau/\beta]B_0}{C} {Subderivation of $\Dee_2$}
\Hand           \declsubjudgPf{\Psi}{A}{C}  {By i.h. (one less $\forall$ quantifier in $B$)}
              \end{llproof}
  \qedhere
  \end{itemize}
\end{proof}

\subsection{Invertibility of \DsubAllR}

\declinvertibility*
\begin{proof}
  By induction on the given derivation $\Dee$.
  
  \begin{itemize}
      \ProofCasesRules{\DsubVar, \DsubUnit, \DsubArr}
          Impossible: the supertype cannot have the form $\alltype{\beta}{B}$.

      \DerivationProofCase{\DsubAllR}
                {\declsubjudg{\Psi, \beta}{A}{B}}
                {\declsubjudg{\Psi}{A}{\alltype{\beta}{B}}}

          The subderivation is exactly what we need, and is strictly smaller than $\Dee$.

      \DerivationProofCase{\DsubAllL}
                {\judgetp{\Psi}{\tau}
                  \\
                  \arrayenvb{
                    \Dee_0
                    \\ \declsubjudg{\Psi}{[\tau/\alpha]A_0}{\alltype{\beta}{B}}
                  }
                }
                {\declsubjudg{\Psi}{\alltype{\alpha}{A_0}}{\alltype{\beta}{B}}}

          By i.h., $\Dee_0'$ derives $\declsubjudg{\Psi, \beta}{[\tau/\alpha]A_0}{B}$
          where $\Dee_0' < \Dee_0$. \\
          By \DsubAllL, $\Dee'$ derives $\declsubjudg{\Psi, \beta}{\alltype{\alpha}{A_0}}{B}$;
          since $\Dee_0' < \Dee_0$, we have $\Dee'  < \Dee$.
  \qedhere
  \end{itemize}
\end{proof}

\subsection{Non-Circularity and Equality}

\occurrencelemma*
\begin{proof}   By induction on the given derivation.

  \begin{itemize}
    \ProofCasesRules{\DsubVar, \DsubUnit}
        (i), (ii): Here $A$ and $B$ have no subterms at all, so the result is immediate.
    
    \DerivationProofCase{\DsubArr}
          {\declsubjudg{\Psi}{B_1}{A_1}
            \\
            \declsubjudg{\Psi}{A_2}{B_2}
          }
          {\declsubjudg{\Psi}{A_1 \arr A_2}{B_1 \arr B_2}}
          
          \begin{enumerate}[(i)]
          \item Here, $A = A_1 \arr A_2$ and $\tau = B_1 \arr B_2$.
                 
                 \begin{llproof}
                     \Pf{B_1}{\notoccursinsidearrow}{A_1}    {By i.h. (ii)}
                     \Pf{B_1 \arr B_2}{\not\subtermof}{A_1}    
                                   {Suppose $B_1 \arr B_2 \subtermof A_1$.  Then $B_1 \occursinsidearrow A_1$: contradiction.}
                     \Pf{B_2}{\notoccursinsidearrow}{A_2}    {By i.h. (i)}
                     \Pf{B_1 \arr B_2}{\not\subtermof}{A_2}  
                                   {Similar}
                 \end{llproof}
                 
                 Suppose (for a contradiction) that $B_1 \arr B_2 \occursinsidearrow A_1 \arr A_2$. \\
                 $~$~~Now $B_1 \arr B_2 \subtermof A_1$ or $B_1 \arr B_2 \subtermof A_2$. \\
                 $~$~~But above, we showed that both were false: contradiction. \\
                 Therefore, $B_1 \arr B_2 \not\propersubtermof A_1 \arr A_2$. \\
                 Therefore, $B_1 \arr B_2 \notoccursinsidearrow A_1 \arr A_2$.
          
          \item Here, $A = \tau$ and $B = B_1 \arr B_2$. \\
                 Symmetric to the previous case.
          \end{enumerate}

      \DerivationProofCase{\DsubAllL}
          {\judgetp{\Psi}{\tau'}
            \\
            \declsubjudg{\Psi}{[\tau'/\alpha]A_0}{\tau}}
          {\declsubjudg{\Psi}{\alltype{\alpha}{A_0}}{\tau}}

          In part (ii), this case cannot arise, so we prove part (i).

          By i.h. (i), $\tau \notoccursinsidearrow [\tau'/\alpha]A_0$. \\
          It follows from the definition of $\occursinsidearrow$ that
          $\tau \notoccursinsidearrow \alltype{\alpha}{A_0}$.

      \DerivationProofCase{\DsubAllR}
          {\declsubjudg{\Psi, \beta}{\tau}{B_0}}
          {\declsubjudg{\Psi}{\tau}{\alltype{\beta}{B_0}}}
          
          In part (i), this case cannot arise, so we prove part (ii).
          
          Similar to the \DsubAllL case.
          \qedhere
  \end{itemize}
\end{proof}

\declmonotypeequality*
\begin{proof}
  By induction on the given derivation.

  \begin{itemize}
      \ProofCaseRule{\DsubVar}   Immediate.

      \ProofCaseRule{\DsubUnit}   Immediate.

      \DerivationProofCase{\DsubArr}
            {\declsubjudg{\Psi}{B_1}{A_1}
              \\
              \declsubjudg{\Psi}{A_2}{B_2}
            }
            {\declsubjudg{\Psi}{A_1 \arr A_2}{B_1 \arr B_2}}

            By i.h. on each subderivation, $B_1 = A_1$ and $A_2 = B_2$.
            Therefore $A_1 \arr A_2 = B_1 \arr B_2$.

       \ProofCaseRule{\DsubAllL}
            Here $\sigma = \alltype{\alpha} A_0$, which is not a monotype, so this case is impossible.

       \ProofCaseRule{\DsubAllR}
            Here $\tau = \alltype{\beta} B_0$, which is not a monotype, so this case is impossible.
  \qedhere
  \end{itemize}
\end{proof}

\section{Type Assignment}

\declarativetypingwellformed*
\begin{proof}
  By induction on the given derivation.
  
  In all cases, we apply the induction hypothesis to all subderivations.

  \begin{itemize}
  \item   In the \DeclVar and \DeclArrIntro cases, we use our standard assumption
    that every context appearing in a derivation is well-formed.
  \item  In the \DeclArrIntroSyn case, we use inversion on the $\judgetp{\Psi}{\sigma \arr \tau}$ premise.
  \item  In the \DeclAllApp case, we use the property that if $\judgetp{\Psi}{[\tau/\alpha]A_0}$ then
    $\judgetp{\Psi}{\alltype{\alpha} A_0}$.
  \item  In the \DeclAnno case, we use its premise.
  \qedhere
  \end{itemize}
\end{proof}

\completenessbidirectional*
\begin{proof}
  By induction on the derivation of $\judge{\Psi}{e}{A}$. 

    \begin{itemize}
      \DerivationProofCase{\AssignVar}
            {x:A \in \Psi}
            {\judge{\Psi}{x}{A}}

            Immediate, by rule \DeclVar. 

      \DerivationProofCase{\AssignArrIntro}
            {\judge{\Psi, x:A}{e}{B}}
            {\judge{\Psi}{\lam{x} e}{A \arr B}}

            By inversion, we have $\judge{\Psi, x:A}{e}{B}$. \\
            By induction, we have $\declsynjudg{\Psi, x:A}{e'}{B}$, where $|e'| = e$. \\ 
            By \Lemmaref{lem:declarative-reflexivity}, $\declsubjudg{\Psi}{B}{B}$. \\
            By rule \DeclSub, $\declchkjudg{\Psi, x:A}{e'}{B}$. \\ 
            By rule \DeclArrIntro, $\declchkjudg{\Psi}{\fun{x}{e'}}{A \arr B}$. \\ 
            By \Lemmaref{lem:declarative-typing-well-formed}, $\judgetp{\Psi}{A \arr B}$. \\
            By rule \DeclAnno, $\declsynjudg{\Psi}{((\fun{x}{e'}) : A \arr B)}{A \arr B}$. \\ 
            By definition, $|((\fun{x}{e'}) : A \arr B)| = |\fun{x}{e'}| = \fun{x}{|e'|} = \fun{x}{e}$. 
      
      \DerivationProofCase{\AssignArrElim}
            {\judge{\Psi}{e_1}{A \arr B} \\
             \judge{\Psi}{e_2}{A}}
            {\judge{\Psi}{e_1\;e_2}{B}}

            By induction, $\declsynjudg{\Psi}{e'_1}{A \arr B}$ and $|e'_1| = e_1$. \\
            By induction, $\declsynjudg{\Psi}{e'_2}{A}$ and $|e'_2| = e_2$. \\
            By \Lemmaref{lem:declarative-reflexivity}, $\declsubjudg{\Psi}{A}{A}$. \\
            By rule \DeclSub, $\declchkjudg{\Psi}{e'_2}{A}$. \\
            By rule \DeclArrApp, $\declappjudg{\Psi}{e'_2}{A \arr B}{B}$. \\
            By rule \DeclArrElim, $\declsynjudg{\Psi}{e'_1\;e'_2}{B}$. \\ 
            By definition, $|e'_1\;e'_2| = |e'_1|\;|e'_2| = e_1\;e_2$. 

      \DerivationProofCase{\AssignAllIntro}
            {\judge{\Psi, \alpha}{e}{A}}
            {\judge{\Psi}{e}{\alltype{\alpha}{A}}}

            By induction, $\declsynjudg{\Psi, \alpha}{e'}{A}$ where $|e'| = e$. \\ 
            By \Lemmaref{lem:declarative-reflexivity}, $\declsubjudg{\Psi, \alpha}{A}{A}$. \\
            By rule \DeclSub, $\declchkjudg{\Psi, \alpha}{e'}{A}$. \\ 
            By rule \DeclAllIntro, $\declchkjudg{\Psi}{e'}{\alltype{\alpha}{A}}$. \\ 
            By \Lemmaref{lem:declarative-typing-well-formed}, $\judgetp{\Psi}{\alltype{\alpha}{A}}$. \\
            By rule \DeclAnno, $\declsynjudg{\Psi}{(e':\alltype{\alpha}{A})}{\alltype{\alpha}{A}}$. \\ 
            By definition, $|e':\alltype{\alpha}{A}| = |e'| = e$. 

      \DerivationProofCase{\AssignAllElim}
            {\judge{\Psi}{e}{\alltype{\alpha}{A}} \\
             \judgetp{\Psi}{\tau}}
            {\judge{\Psi}{e}{[\tau/\alpha]A}}    

            By induction, $\declsynjudg{\Psi}{e'}{\alltype{\alpha}{A}}$ where $|e'| = e$. \\
            By \Lemmaref{lem:declarative-reflexivity}, $\declsubjudg{\Psi}{[\tau/\alpha]A}{[\tau/\alpha]A}$. \\
            By \DsubAllL, $\declsubjudg{\Psi}{\alltype{\alpha}{A}}{[\tau/\alpha]A}$. \\
            By rule \DeclSub, $\declchkjudg{\Psi}{e'}{[\tau/\alpha]A}$. \\
            By \Lemmaref{lem:declarative-typing-well-formed}, $\judgetp{\Psi}{[\tau/\alpha]A}$. \\
            By rule \DeclAnno, $\declchkjudg{\Psi}{(e' : [\tau/\alpha]A)}{[\tau/\alpha]A}$. \\
            By definition, $|e' : [\tau/\alpha]A| = |e'| = e$. 
    \qedhere
    \end{itemize}
\end{proof}

\subtypingcoercion*
\begin{proof}
  By induction on the derivation of $\declsubjudg{\Psi}{A}{B}$. 

\begin{itemize}
    \DerivationProofCase{\DsubVar}
              {\alpha \in \Psi}
              {\declsubjudg{\Psi}{\alpha}{\alpha}}

   Choose $f = \fun{x}{x}$. \\
   Clearly $\judge{\Psi}{\fun{x}{x}}{\alpha \arr \alpha}$. 

    \DerivationProofCase{\DsubUnit}
              { }
              {\declsubjudg{\Psi}{\unitty}{\unitty}}

   Choose $f = \fun{x}{x}$. \\
   Clearly $\judge{\Psi}{\fun{x}{x}}{\unitty \arr \unitty}$. 

    \DerivationProofCase{\DsubArr}
              {\declsubjudg{\Psi}{B_1}{A_1}
                \\
                \declsubjudg{\Psi}{A_2}{B_2}
              }
              {\declsubjudg{\Psi}{A_1 \arr A_2}{B_1 \arr B_2}}

   By induction, we have $g : B_1 \arr A_1$, which is $\beta\eta$-equal to the identity. \\
   By induction, we have $k : A_2 \arr B_2$, which is $\beta\eta$-equal to the identity. \\ 
   Let $f$ be $\lam{h} k \circ h \circ g$. \\
   It is easy to verify that $\judge{\Psi}{f}{(A_1 \arr A_2) \arr (B_1 \arr B_2)}$. \\
   Since $k$ and $g$ are identities, $f =_{\beta\eta} \fun{h}{h}$. 

    \DerivationProofCase{\DsubAllL}
              {\judgetp{\Psi}{\tau}
                \\
                \declsubjudg{\Psi}{[\tau/\alpha]A}{B}}
              {\declsubjudg{\Psi}{\alltype{\alpha}{A}}{B}}

    By induction, $g : [\tau/\alpha]A \arr B$. \\ 
    Let $f \triangleq \fun{x}{g\;x}$. \\ 
    $f$ is an eta-expansion of $g$, which is $\beta\eta$-equal to the identity. Hence $f$ is too. \\
    Also, $\fun{x}{g\;x} : (\alltype{\alpha}{A}) \arr B$, using the \DeclAllElim rule on $x$. 

    \DerivationProofCase{\DsubAllR}
              {\declsubjudg{\Psi, \beta}{A}{B}}
              {\declsubjudg{\Psi}{A}{\alltype{\beta}{B}}}
  
   By induction, we have $g$ such that $\judge{\Psi, \beta}{g}{A \arr B}$. \\ 
   Let $f \triangleq \fun{x}{g\;x}$. \\
   Use the following derivation:
   \begin{mathpar}
                               \inferrule*
                                         { 
                                           \inferrule*
                                                     { \inferrule*[Left=Weaken] 
                                                                 { \vdots }
                                                                 {\judge{\Psi, \beta}{g}{A \arr B}}
                                                     }
                                                     {\judge{\Psi, x:A, \beta}{g}{A \arr B}}
                                           \\
                                           \judge{\Psi, x:A, \beta}{x}{A} 
                                         }
                             {
                               \inferrule*
                                         {\judge{\Psi, x:A, \beta}{g\;x}{B}}
                              {\inferrule*
                                       {
                                         \judge{\Psi, x:A}{g\;x}{\alltype{\beta}{B}}
                                       }
                                       {\judge{\Psi}{\fun{x}{g\;x}}{A \arr \alltype{\beta}{B}}}
                            }}
   \qedhere
   \end{mathpar}
\end{itemize}
\end{proof}

\applicationsubtyping*
\begin{proof}
  By induction on the given derivation $\Dee$.

  \begin{itemize}
     \DerivationProofCase{\DeclArrApp}
            {\declchkjudg{\Psi}{e}{B}}
            {\declappjudg{\Psi}{e}{B \arr C}{C}}

            \begin{llproof}
\Hand              \declchkjudgPf{\Dee' \derives \Psi}{e}{B}  {Subderivation}
\Hand              \ltPf{\Dee'}{\Dee}   {$\Dee'$ is a subderivation of $\Dee$}
\Hand              \declsubjudgPf{\Psi}{\underbrace{B \arr C}_{A}}{B \arr C}  {By \Lemmaref{lem:declarative-reflexivity}}
            \end{llproof}

     \DerivationProofCase{\DeclAllApp}
            {\judgetp{\Psi}{\tau} \\ 
             \declappjudg{\Psi}{e}{[\tau/\alpha]A_0}{C}}
            {\declappjudg{\Psi}{e}{\alltype{\alpha} A_0}{C}}

            \begin{llproof}
              \judgetpPf{\Psi}{\tau}  {Subderivation}
              \declappjudgPf{\Psi}{e}{[\tau/\alpha]A_0}{C}  {Subderivation}
              \declsubjudgPf{\Psi}{[\tau/\alpha]A_0}{B \arr C}  {By i.h.}
\Hand      \declchkjudgPf{\Dee' \derives \Psi}{e}{B}  {\ditto}
\Hand      \ltPf{\Dee'}{\Dee}   {\ditto}
\Hand      \declsubjudgPf{\Psi}{\alltype{\alpha} A_0}{B \arr C}   {By \DsubAllL}
            \end{llproof}
  \qedhere
  \end{itemize}
\end{proof}

\soundnessbidirectional*
\begin{proof}
  \begin{itemize}
     \DerivationProofCase{\DeclVar}
          {(x : A) \in \Psi}
          {\declsynjudg{\Psi}{x}{A}}

      By rule \AssignVar, $\judge{\Psi}{x}{A}$. \\ 
      Note $x \beeq x$.

     \DerivationProofCase{\DeclSub}
          {\declsynjudg{\Psi}{e}{A}
            \\
            \declsubjudg{\Psi}{A}{B}
          }
          {\declchkjudg{\Psi}{e}{B}}

     By induction, $\judge{\Psi}{e'}{A}$ and $e' \beeq |e|$. \\
     By \Lemmaref{lem:subtyping-coercion}, $f : A \arr B$ such that $f \beeq id$. \\
     By \AssignArrElim, $\judge{\Psi}{f\;e'}{B}$. \\ 
     Note $f\;e' \beeq id\; e' \beeq e' \beeq |e|$. 

     \DerivationProofCase{\DeclAnno}
          {\judgetp{\Psi}{A}
            \\
            \declchkjudg{\Psi}{e}{A}
          }
          {\declsynjudg{\Psi}{(e : A)}{A}}

     By induction, $\judge{\Psi}{e'}{A}$ such that $e' \beeq |e|$. \\
     Note $e' \beeq |e| = |e:A|$. 

     \DerivationProofCase{\DeclUnitIntro}
          {}
          {\declchkjudg{\Psi}{\unitexp}{\unitty}}

      By \AssignUnit, $\judge{\Psi}{\unitexp}{\unitty}$. \\
      Note $\unitexp \beeq \unitexp$. 

     \DerivationProofCase{\DeclUnitIntroSyn}
                { }
                {\declsynjudg{\Psi}{\unitexp}{\unitty}}

      By \AssignUnit, $\judge{\Psi}{\unitexp}{\unitty}$. \\
      Note $\unitexp \beeq \unitexp$. 

     \DerivationProofCase{\DeclAllIntro}
           {\declchkjudg{\Psi, \alpha}{e}{A}
           }
           {\declchkjudg{\Psi}{e}{\alltype{\alpha}{A}}}

     By induction, $\judge{\Psi, \alpha}{e'}{A}$ such that $e' \beeq |e|$. \\ 
     By rule \AssignAllIntro, $\judge{\Psi}{e'}{\alltype{\alpha}{A}}$. 

     \DerivationProofCase{\DeclArrIntro}
          {\declchkjudg{\Psi, x : A}{e}{B}
          }
          {\declchkjudg{\Psi}{\lam{x} e}{A \arr B}}

     By induction, $\judge{\Psi, x:A}{e'}{B}$ such that $e' \beeq |e|$. \\
     By \AssignArrIntro, $\judge{\Psi}{\fun{x}{e'}}{A \arr B}$. \\ 
     Note $\fun{x}{e'} \beeq \fun{x}{|e|} = |\fun{x}{e}|$. 

     \DerivationProofCase{\DeclArrIntroSyn}
                {\judgetp{\Psi}{\sigma \arr \tau}
                  \\
                  \declchkjudg{\Psi, x : \sigma}{e}{\tau}
                }
                {\declsynjudg{\Psi}{\lam{x} e}{\sigma \arr \tau}}

     By induction, $\judge{\Psi, x:\sigma}{e'}{\tau}$ such that $e' \beeq |e|$. \\
     By \AssignArrIntro, $\judge{\Psi}{\fun{x}{e'}}{\sigma \arr \tau}$. \\ 
     Note $\fun{x}{e'} \beeq \fun{x}{|e|} = |\fun{x}{e}|$. 

     \DerivationProofCase{\DeclArrElim}
          {\declsynjudg{\Psi}{e_1}{A}
            \\
            \declappjudg{\Psi}{e_2}{A}{C}
          }
          {\declsynjudg{\Psi}{e_1\,e_2}{C}}

     By induction, $\judge{\Psi}{e'_1}{A}$ such that $e'_1 \beeq |e_1|$. \\ 
     By \Lemmaref{lem:application-subtyping}, there is a $B$ such that \\
     1. $\declsubjudg{\Psi}{A}{B \arr C}$, and \\
     2. $\declchkjudg{\Psi}{e_2}{B}$,  which is no bigger than $\declappjudg{\Psi}{e_2}{A}{C}$. \\ 
     By \Lemmaref{lem:subtyping-coercion}, we have $f$ such that $\judge{\Psi}{f}{A \arr B \arr C}$ and $f \beeq id$. \\ 
     By induction, we get $\judge{\Psi}{e'_2}{B}$ and $e'_2 \beeq |e_2|$. \\
     By \AssignArrElim twice, $\judge{\Psi}{f\;e'_1\;e'_2}{C}$. \\
     Note $f\;e'_1\;e'_2 \beeq id\;e'_1\;e'_2 \beeq e'_1\;e'_2 \beeq |e_1|\;e'_2 \beeq |e_1|\;|e_2| = |e_1\;e_2|$. 
  \qedhere
  \end{itemize}
\end{proof}

\section{Robustness of Typing}

\typesubstitution*
\begin{proof}
  By induction on the given derivation.

  In the \DeclVar case, split on whether the variable being
  typed is in $\Psi$ or $\Psi'$; the former is immediate, and in the latter, use the fact that
  $(x : C) \in \Psi'$ implies $(x : [\tau/\alpha]C) \in [\tau/\alpha]\Psi'$.

  In the \DeclSub case, use the i.h. and \Lemmaref{lem:decl-sub-substitution}.

  In the \DeclAnno case, we are substituting in the annotation in the term, as well as in the type;
  we also need \Propositionref{prop:subst}.

  In \DeclArrIntro, \DeclArrIntroSyn and \DeclAllIntro, we add to the context in the premise,
  which is why the statement is generalized for nonempty $\Psi'$.
\end{proof}

\termsubsumption*
\begin{proof}
  By mutual induction:
  in (i), by lexicographic induction on the derivation of the checking judgment, then of the subtyping judgment;
  in (ii), by induction on the derivation of the synthesis judgment;
  in (iii), by lexicographic induction on the derivation of the application judgment, then of the subtyping judgment.

  For part (i), checking:

  \begin{itemize}
     \DerivationProofCase{\DeclSub}
          {\declsynjudg{\Psi}{e}{B}
            \\
            \declsubjudg{\Psi}{B}{A}
          }
          {\declchkjudg{\Psi}{e}{A}}

          \begin{llproof}
            \declsynjudgPf{\Psi}{e}{B}   {Subderivation}
            \declsynjudgPf{\Psi'}{e}{B'}   {By i.h.}
            \declsubjudgPf{\Psi}{B'}{B}   {\ditto}
            \proofsep
            \declsubjudgPf{\Psi}{B}{A}   {Subderivation}
            \declsubjudgPf{\Psi}{A}{A'}   {Given}
            \declsubjudgPf{\Psi}{B'}{A'}  {By \Lemmaref{lem:declarative-transitivity} (twice)}
            \declsubjudgPf{\Psi'}{B'}{A'}  {By weakening}
            \proofsep
\Hand       \declchkjudgPf{\Psi'}{e}{A'}  {By \DeclSub}
          \end{llproof}

     \DerivationProofCase{\DeclUnitIntro}
          {}
          {\declchkjudg{\Psi}{\unitexp}{\unitty}}

          \begin{llproof}
              \declsynjudgPf{\Psi'}{\unitexp}{\unitty}  {By \DeclUnitIntroSyn}
              \declsubjudgPf{\Psi}{\unitty}{A'}   {Given}
              \declsubjudgPf{\Psi'}{\unitty}{A'}   {By weakening}
\Hand            \declchkjudgPf{\Psi'}{\unitexp}{A'}   {By \DeclSub}
          \end{llproof}

     \DerivationProofCase{\DeclAllIntro}
           {\declchkjudg{\Psi, \alpha}{e}{A_0}
           }
           {\declchkjudg{\Psi}{e}{\alltype{\alpha}A_0}}

           We consider cases of $\declsubjudg{\Psi}{\alltype{\alpha} A_0}{A'}$:

           \begin{itemize}
                  \DerivationProofCase{\DsubAllR}
                        {\declsubjudg{\Psi, \beta}{\alltype{\alpha} A_0}{B}}
                        {\declsubjudg{\Psi}{\alltype{\alpha} A_0}{\alltype{\beta} B}}

                        \smallskip
                        
                        \begin{llproof}
                          \declsubjudgPf{\Psi, \beta}{\alltype{\alpha} A_0}{B}  {Subderivation}
                          \declchkjudgPf{\Psi}{e}{\alltype{\alpha} A_0}   {Given}
                          \declchkjudgPf{\Psi'}{e}{B}   {By i.h. (i)}
                        \Hand  \declchkjudgPf{\Psi'}{e}{\underbrace{\alltype{\beta} B}_{A'}}   {By \DeclAllIntro}
                        \end{llproof}

                  \DerivationProofCase{\DsubAllL}
                        {\judgetp{\Psi}{\tau}
                          \\
                          \declsubjudg{\Psi}{[\tau/\alpha]A_0}{A'}}
                        {\declsubjudg{\Psi}{\alltype{\alpha} A_0}{A'}}

                        \smallskip

                        \begin{llproof}
                          \declchkjudgPf{\Psi, \alpha}{e}{A_0}   {Subderivation}
                          \declchkjudgPf{\Psi}{e}{[\tau/\alpha]A_0}   {By \Lemmaref{lem:type-substitution}}
                          \declsubjudgPf{\Psi}{[\tau/\alpha]A_0}{A'}   {Subderivation}
                        \Hand  \declchkjudgPf{\Psi'}{e}{A'}   {By i.h. (i)}
                        \end{llproof}

           \end{itemize}

     \DerivationProofCase{\DeclArrIntro}
          {\declchkjudg{\Psi, x : A_1}{e_0}{A_2}
          }
          {\declchkjudg{\Psi}{\lam{x} e_0}{A_1 \arr A_2}}

          We consider cases of $\declsubjudg{\Psi}{A_1 \arr A_2}{A'}$:

          \begin{itemize}
              \DerivationProofCase{\DsubArr}
                   {\declsubjudg{\Psi}{B_1}{A_1}
                    \\
                    \declsubjudg{\Psi}{A_2}{B_2}}
                   {\declsubjudg{\Psi}{A_1 \arr A_2}{B_1 \arr B_2}}

                   \begin{llproof}
                     \leqPf{\Psi}{\Psi'}  {Given}
                     \declsubjudgPf{\Psi}{B_1}{A_1}   {Subderivation}
                     \leqPf{\Psi', x : B_1}{\Psi, x : A_1}  {By \CtxsubVar}
                     \declchkjudgPf{\Psi', x : B_1} {e_0} {B_2}   {By i.h. (i)}
                   \Hand  \declchkjudgPf{\Psi'} {\lam{x} e_0} {B_1 \arr B_2}   {By \DeclArrIntro}
                   \end{llproof}

              \DerivationProofCase{\DsubAllR}
                    {\declsubjudg{\Psi, \beta}{A_1 \arr A_2}{B'}}
                    {\declsubjudg{\Psi}{A_1 \arr A_2}{\alltype{\beta} B'}}
                    
                    \begin{llproof}
                      \declsubjudgPf{\Psi, \beta}{A_1 \arr A_2}{B'}  {Subderivation}
                      \declchkjudgPf{\Psi, \beta}{\lam{x} e_0}{A_1 \arr A_2}  {By weakening}
                      \declchkjudgPf{\Psi', \beta}{\lam{x} e_0}{B'}  {By i.h. (i)}
                    \Hand  \declchkjudgPf{\Psi'}{\lam{x} e_0}{\alltype{\beta} B'}  {By \DeclAllIntro}
                    \end{llproof}
          \end{itemize}
  \end{itemize}

  For part (ii), synthesis:

  \begin{itemize}
     \DerivationProofCase{\DeclVar}
          {(x : A) \in \Psi}
          {\declsynjudg{\Psi}{x}{A}}

          By inversion on $\Psi' \leq \Psi$, we have $(x : A') \in \Psi'$ where $\declsubjudg{\Psi}{A'}{A}$. \\
          By \DeclVar, $\declsynjudg{\Psi'}{x}{A'}$.

     \DerivationProofCase{\DeclAnno}
          {\judgetp{\Psi}{A}
            \\
            \declchkjudg{\Psi}{e_0}{A}
          }
          {\declsynjudg{\Psi}{(e_0 : A)}{A}}

          \begin{llproof}
            \LetPf{A'}{A}  {}
            \judgetpPf{\Psi}{A}  {Subderivation}
            \judgetpPf{\Psi'}{A}  {By weakening}
            \proofsep
            \declchkjudgPf{\Psi}{e_0}{A}  {Subderivation}
            \declchkjudgPf{\Psi'}{e_0}{A}  {By i.h.}
            \proofsep
\Hand            \declsynjudgPf{\Psi'}{(e_0 : A)}{A'} {By \DeclAnno and $A' = A$}
\Hand            \declsubjudgPf{\Psi}{A'}{A}  {By \Lemmaref{lem:declarative-reflexivity}}
          \end{llproof}

     \DerivationProofCase{\DeclUnitIntroSyn}
          { }
          {\declsynjudg{\Psi}{\unitexp}{\unitty}}
          
          \begin{llproof}
            \LetPf{A'}{\unitty}   {}
\Hand        \declsynjudgPf{\Psi'}{\unitexp}{\unitty}  {By \DeclUnitIntroSyn}
\Hand        \declsubjudgPf{\Psi}{\unitty}{\unitty}  {By \DsubUnit}
          \end{llproof}

     \DerivationProofCase{\DeclArrIntroSyn}
           {\judgetp{\Psi}{\sigma \arr \tau}
             \\
             \declchkjudg{\Psi, x : \sigma}{e_0}{\tau}
           }
           {\declsynjudg{\Psi}{\lam{x} e_0}{\sigma \arr \tau}}

           \begin{llproof}
             \LetPf{A'}{\sigma \arr \tau} {}
             \leqPf{\Psi'}{\Psi}  {Given}
             \declsubjudgPf{\Psi}{\sigma}{\sigma}  {By \Lemmaref{lem:declarative-reflexivity}}
             \leqPf{\Psi', x : \sigma}  {\Psi, x : \sigma}   {By \CtxsubVar}
             \declchkjudgPf{\Psi, x : \sigma}{e_0}{\tau}  {Subderivation}
             \declsubjudgPf{\Psi}{\tau}{\tau}   {By \Lemmaref{lem:declarative-reflexivity}}
             \declchkjudgPf{\Psi', x : \sigma}{e_0}{\tau}  {By i.h. (i) with $\tau$}
\Hand             \declsubjudgPf{\Psi}{A'}{\sigma \arr \tau}    {By \Lemmaref{lem:declarative-reflexivity}}
\Hand             \declsynjudgPf{\Psi'}{\lam{x} e_0}{A'}    {By \DeclArrIntroSyn}
           \end{llproof}

     \DerivationProofCase{\DeclArrElim}
          {\declsynjudg{\Psi}{e_1}{C}
            \\
            \declappjudg{\Psi}{e_2}{C}{A}
          }
          {\declsynjudg{\Psi}{e_1\,e_2}{A}}

          \begin{llproof}
            \declsynjudgPf{\Psi}{e_1}{C}   {Subderivation}
            \declsynjudgPf{\Psi'}{e_1}{C'}   {By i.h. (ii)}
            \declsubjudgPf{\Psi}{C'}{C}   {\ditto}
            \declappjudgPf{\Psi}{e_2}{C}{A}   {Subderivation}
\Hand    \declsubjudgPf{\Psi}{A'}{A}  {By i.h. (iii)}
           \declappjudgPf{\Psi'}{e_2}{C'}{A'}  {\ditto}
\Hand    \declsynjudgPf{\Psi'}{e_1\,e_2}{A'}  {By \DeclArrElim}
          \end{llproof}
  \end{itemize}

  For part (iii), application:

  \begin{itemize}
     \DerivationProofCase{\DeclAllApp}
            {\judgetp{\Psi}{\tau} \\ 
             \declappjudg{\Psi}{e}{[\tau/\alpha]C_0}{A}}
            {\declappjudg{\Psi}{e}{\alltype{\alpha} C_0}{A}}

            \begin{llproof}
              \declsubjudgPf{\Psi}{C'}{\alltype{\alpha} C_0}   {Given}
              \declsubjudgPf{\Psi, \alpha}{C'}{C_0}  {By \Lemmaref{lem:decl-invertibility}}
              \declsubjudgPf{\Psi}{[\tau/\alpha]C'}{[\tau/\alpha]C_0}  {By \Lemmaref{lem:decl-sub-substitution}}
              \declsubjudgPf{\Psi}{C'}{[\tau/\alpha]C_0}  {$\alpha$ cannot appear in $C'$}
              \declappjudgPf{\Psi}{e}{[\tau/\alpha]C_0}{A} {Subderivation}
            \Hand  \declappjudgPf{\Psi'}{e}{C'}{A'}  {By i.h. (iii)}
            \Hand  \declsubjudgPf{\Psi'}{A'}{A}  {\ditto}
            \end{llproof}

     \DerivationProofCase{\DeclArrApp}
            {\declchkjudg{\Psi}{e}{C_0}}
            {\declappjudg{\Psi}{e}{C_0 \arr A}{A}}

            \begin{llproof}
              \declsubjudgPf{\Psi}{C'}{C_0 \arr A}    {Given}
            \end{llproof}

            \begin{itemize}
               \DerivationProofCase{\DsubArr}
                     {\declsubjudg{\Psi}{C_0}{C_1'}
                       \\
                       \declsubjudg{\Psi}{C_2'}{A}
                     }
                     {\declsubjudg{\Psi}{C_1' \arr C_2'}{C_0 \arr A}}

                     \begin{llproof}
                       \LetPf{A'}{C_2'} {}
                       \declchkjudgPf{\Psi}{e}{C_0}    {Subderivation}
                       \declsubjudgPf{\Psi}{C_0}{C_1'}   {Subderivation}
                       \declchkjudgPf{\Psi'}{e}{C_1'}   {By i.h.}
                       \declappjudgPf{\Psi'}{e}{C_1' \arr C_2'}{C_2'}  {By \DeclArrApp}
                     \Hand  \declappjudgPf{\Psi'}{e}{C_1' \arr A'}{A'}  {$A' = C_2'$}
                        \declsubjudgPf{\Psi}{C_2'}{A} {Subderivation}
                     \Hand  \declsubjudgPf{\Psi}{A'}{A} {$A' = C_2'$}
                     \end{llproof}
            
               \DerivationProofCase{\DsubAllL}
                    {\judgetp{\Psi}{\tau}
                      \\
                      \declsubjudg{\Psi}{[\tau/\beta]B}{C_0 \arr A}
                    }
                    {\declsubjudg{\Psi}{\alltype{\beta} B}{C_0 \arr A}}

                    \begin{llproof}
                      \declsubjudgPf{\Psi}{[\tau/\beta]B}{C_0 \arr A}  {Subderivation}
                      \declappjudgPf{\Psi'}{e}{[\tau/\beta]B}{A'}   {By i.h. (iii)}
                      \Hand \declsubjudgPf{\Psi}{A'}{A}   {\ditto}
                      \judgetpPf{\Psi}{\tau}   {Subderivation}
                      \judgetpPf{\Psi'}{\tau}   {By weakening}
                     \Hand  \declappjudgPf{\Psi'}{e}{\alltype{\beta} B}{A'}  {By \DeclAllApp}
                    \end{llproof}
           \qedhere
           \end{itemize}
   \end{itemize}
\end{proof}

\termsubstitution*
\begin{proof}
  By a straightforward mutual induction on the given derivation.
\end{proof}

\termUnsubstitution*
\begin{proof}
  By mutual induction on the given derivation.
  
  \begin{enumerate}[(i)]
      \item We have $\declchkjudg{\Psi}{[(e : A)/x]e'}{C}$.

          \begin{itemize}
              \DerivationProofCase{\DeclSub}
                       {\declsynjudg{\Psi}{[(e : A)/x]e'}{B}
                         \\
                         \declsubjudg{\Psi}{B}{C}
                       }
                       {\declchkjudg{\Psi}{[(e : A)/x]e'}{C}}

                  By i.h.\ (ii), $\declsynjudg{\Psi, x:A}{e'}{B}$. \\
                  By \DeclSub, $\declchkjudg{\Psi, x:A}{e'}{C}$.
                  
              \DerivationProofCase{\DeclUnitIntro}
                   {}
                   {\declchkjudg{\Psi}{\unitexp}{\underbrace{\unitty}_{C}}}
                  
                   We have $[(e : A)/x]e' = \unitexp$.  Therefore $e' = \unitexp$,
                   and the result follows by \DeclUnitIntro.

              \DerivationProofCase{\DeclAllIntro}
                    {\declchkjudg{\Psi, \alpha}{[(e : A)/x]e'}{C'}
                    }
                    {\declchkjudg{\Psi}{[(e : A)/x]e'}{\alltype{\alpha}{C'}}}
                  
                  By i.h.\ (i), $\declchkjudg{\Psi, \alpha, x:A}{e'}{C'}$. \\ 
                  By exchange, $\declchkjudg{\Psi, x:A, \alpha}{e'}{C'}$. \\ 
                  By \DeclAllIntro, $\declchkjudg{\Psi, x:A}{e'}{\alltype{\alpha}{C'}}$. 
                  
              \DerivationProofCase{\DeclArrIntro}
                   {\declchkjudg{\Psi, y : C_1}{e''}{C_2}
                   }
                   {\declchkjudg{\Psi}{\lam{y} e''}{C_1 \arr C_2}}

                  We have $[(e : A)/x]e' = \lam{y}{e''}$. \\
                  By the definition of substitution, $e' = \lam{y} e_0$ and $e'' = [(e : A)/x]e_0$.

                  \smallskip

                  \begin{llproof}
                    \declchkjudgPf{\Psi, y : C_1}{e''}{C_2}   {Subderivation}
                    \declchkjudgPf{\Psi, y : C_1}{[(e : A)/x]e_0}{C_2}  {By above equality}
                    \declchkjudgPf{\Psi, y : C_1, x:A}{e_0}{C_2} {By i.h. (i)}
                    \declchkjudgPf{\Psi, x : A, y:C_1}{e_0}{C_2} {By exchange}
                  \Hand  \declchkjudgPf{\Psi, x : A}{\underbrace{\lam{y} e_0}_{e'}}{\underbrace{C_1 \arr C_2}_{C}}  {By \DeclArrIntro}
                  \end{llproof}
          \end{itemize}

      \item We have $\declsynjudg{\Psi}{[(e:A)/x]e'}{C}$.

        \begin{itemize}
          \ProofCaseRule{$e' = x$}

              Note $[(e:A)/x]x = (e:A)$. \\
              Hence $\declsynjudg{\Psi}{(e:A)}{C}$; by inversion, $C = A$. \\
              By \Lemmaref{lem:declarative-typing-well-formed}, $\judgetp{\Psi}{C}$, which is $\judgetp{\Psi}{A}$. \\
              By \DeclAnno, $\declsynjudg{\Psi}{(e:A)}{A}$. \\
              By \DeclVar, $\declsynjudg{\Psi, x:A}{\underbrace{x}_{e'}}{A}$.

          \ProofCaseRule{$e' \neq x$}

              We now proceed by cases on the derivation of $\declsynjudg{\Psi}{[(e:A)/x]e'}{C}$.

              \begin{itemize}
                   \DerivationProofCase{\DeclVar}
                        {(y : C) \in \Psi}
                        {\declsynjudg{\Psi}{y}{C}}

                       Since $[(e:A)/x]e' = y$, we know that $e' = y$. \\ 
                       By \DeclVar, $\declsynjudg{\Psi, x:A}{y}{C}$. 

                   \DerivationProofCase{\DeclAnno}
                        {\declchkjudg{\Psi}{e''}{C}
                        }
                        {\declsynjudg{\Psi}{\underbrace{(e'' : C)}_{[(e:A)/x]e'}}{C}}

                       We know $[(e:A)/x]e' = (e'' : C)$ and $e' \neq x$. \\
                       Hence there is $e_0$ such that $e' = (e_0 : C)$ and $[(e:A)/x]e_0 = e''$.

                       \smallskip

                       \begin{llproof}
                         \declchkjudgPf{\Psi}{e''}{C}  {Subderivation}
                         \declchkjudgPf{\Psi}{[(e:A)/x]e_0}{C}  {By above equality}
                         \declchkjudgPf{\Psi, x:A}{e_0}{C} {By i.h.\ (i)}
                         \judgetpPf{\Psi, x:A}{C} {By \Lemmaref{lem:declarative-typing-well-formed}}
                         \declsynjudgPf{\Psi, x:A}{(e_0 : C)}{C} {By \DeclAnno}
                       \Hand  \declsynjudgPf{\Psi, x:A}{e'}{C} {By above equality}
                       \end{llproof}

                       \smallskip

                   \DerivationProofCase{\DeclUnitIntroSyn}
                              { }
                              {\declsynjudg{\Psi}{\unitexp}{\unitty}}

                       Since $[(e:A)/x]e' = \unitexp$, it follows that $e' = \unitexp$. \\ 
                       By \DeclUnitIntroSyn, $\declsynjudg{\Psi, x:A}{\unitexp}{\unitty}$. 

                   \DerivationProofCase{\DeclArrIntroSyn}
                              {\judgetp{\Psi}{\sigma \arr \tau}
                                \\
                                \declchkjudg{\Psi, y : \sigma}{e''}{\tau}
                              }
                              {\declsynjudg{\Psi}{\lam{y} e''}{\sigma \arr \tau}}

                       We have $[(e:A)/x]e' = \lam{y}{e''}$. \\
                       By definition of substitution,
                           there exists $e_0$ such that $e' = \lam{y}{e_0}$ and $e'' = [(e:A)/x]e_0$. \\ 
                       So $\declchkjudg{\Psi, y : \sigma}{[(e:A)/x]e_0}{\tau}$. \\ 
                       By i.h.\ (i), $\declchkjudg{\Psi, y : \sigma, x:A}{e_0}{\tau}$. \\
                       By exchange and \DeclArrIntro, $\declchkjudg{\Psi, x:A}{\lam{y}e_0}{\sigma \arr \tau}$. \\
                       Hence \DeclArrIntroSyn, $\declsynjudg{\Psi, x:A}{e'}{\sigma \arr \tau}$. 

                   \DerivationProofCase{\DeclArrElim}
                        {\declsynjudg{\Psi}{e_1}{B}
                          \\
                          \declappjudg{\Psi}{e_2}{B}{C}
                        }
                        {\declsynjudg{\Psi}{\underbrace{e_1\,e_2}_{[(e:A)/x]e'}}{C}}

                      Note that $[(e:A)/x]e' = e_1 \; e_2$. \\
                      So there exist $e'_1$, $e'_2$ such that
                      $e' = e'_1\;e'_2$ and $[(e:A)/x]e'_k = e_k$ for $k \in \{1, 2\}$. \\
                      Applying these equalities to each subderivation gives
                      \[
                         \declsynjudg{\Psi}{[(e:A)/x]e'_1}{B}
                         \AND
                         \declappjudg{\Psi}{[(e:A)/x]e'_2}{B}{C}
                      \]
                      By i.h.\ (ii) and (iii), $\declsynjudg{\Psi, x:A}{e'_1}{B}$
                         and  $\declappjudg{\Psi, x:A}{e'_2}{B}{C}$.\\ 
                      By \DeclArrElim, $\declsynjudg{\Psi, x:A}{e'_1\;e'_2}{C}$, which is
                         $\declsynjudg{\Psi, x:A}{e'}{C}$. 
              \end{itemize}
        \end{itemize}

    \item We have $\declappjudg{\Psi}{A}{[(e:A)/x]e'}{C}$. \\[-1.8em]
        \begin{itemize}
           \DerivationProofCase{\DeclAllApp}
                  {\judgetp{\Psi}{\tau} \\ 
                   \declappjudg{\Psi}{[(e:A)/x]e'}{[\tau/\alpha]B}{C}}
                  {\declappjudg{\Psi}{[(e:A)/x]e'}{\alltype{\alpha}{B}}{C} }

              Follows by i.h.\ (iii) and \DeclAllApp.

           \DerivationProofCase{\DeclArrApp}
                  {\declchkjudg{\Psi}{[(e:A)/x]e'}{B}}
                  {\declappjudg{\Psi}{[(e:A)/x]e'}{B \arr C}{C}}

              Follows by i.h.\ (i) and \DeclArrApp.
        \qedhere
        \end{itemize}
  \end{enumerate}
\end{proof}

\annotationRemoval*
\begin{proof}
  All of these follow directly from inversion and \Lemmaref{lem:term-subsumption}. The one exception is the
  third, which additionally requires a small induction on the application judgment. 
\end{proof}

\soundnessofeta*
\begin{proof}
  By induction on the derivation of $\declchkjudg{\Psi}{\fun{x}{e\;x}}{A}$. There are three
  non-impossible cases: 
  \begin{itemize}
       \DerivationProofCase{\DeclArrIntro}
          {\declchkjudg{\Psi, x : B}{e\;x}{C}
          }
          {\declchkjudg{\Psi}{\lam{x} e\;x}{B \arr C}}

        We have $\declchkjudg{\Psi, x : B}{e\;x}{C}$. \\ 
        By inversion on \DeclSub, we get $\declsynjudg{\Psi, x : B}{e\;x}{C'}$ and $\declsubjudg{\Psi}{C'}{C}$. \\ 
        By inversion on \DeclArrElim, we get $\declsynjudg{\Psi, x:B}{e}{A'}$ and $\declappjudg{\Psi, x:B}{x}{A'}{C'}$. \\ 
        By thinning, we know that $\declsynjudg{\Psi}{e}{A'}$. \\
        By \Lemmaref{lem:application-subtyping}, we get $B'$ so $\declsubjudg{\Psi, x:B}{A'}{B' \arr C'}$
        and $\declchkjudg{\Psi, x:B}{x}{B'}$. \\ 
        By inversion, we know that $\declsynjudg{\Psi, x:B}{x}{B}$ and $\declsubjudg{\Psi}{B}{B'}$. \\
        By \DsubArr, $\declsubjudg{\Psi, x:B}{B' \arr C'}{B \arr C}$. \\ 
        Hence by \Lemmaref{lem:declarative-transitivity}, $\declsubjudg{\Psi, x:B}{A'}{B \arr C}$. \\
        Hence $\declsubjudg{\Psi}{A'}{B \arr C}$. \\
        By \DeclSub, $\declchkjudg{\Psi}{e}{B \arr C}$.

       \DerivationProofCase{\DeclAllIntro}
           {\declchkjudg{\Psi, \alpha}{\fun{x}{e\;x}}{B}
           }
           {\declchkjudg{\Psi}{\fun{x}{e\;x}}{\alltype{\alpha}{B}}}

         By induction, $\declchkjudg{\Psi, \alpha}{\fun{x}{e\;x}}{B}$. \\ 
         By \DeclAllIntro, $\declchkjudg{\Psi}{\fun{x}{e\;x}}{\alltype{\alpha}{B}}$.

       \DerivationProofCase{\DeclSub}
          {\declsynjudg{\Psi}{\fun{x}{e\;x}}{B}
            \\
            \declsubjudg{\Psi}{B}{A}
          }
          {\declchkjudg{\Psi}{\fun{x}{e\;x}}{A}}

          We have $\declsynjudg{\Psi}{\fun{x} e\;x}{B}$ and $\declsubjudg{\Psi}{B}{A}$. \\ 
          By inversion on \DeclArrIntroSyn, $\declchkjudg{\Psi, x:\sigma}{e\;x}{\tau}$ and $B = \sigma \arr \tau$.\\
          By inversion on \DeclSub, we get $\declsynjudg{\Psi, x : \sigma}{e\;x}{C_2}$ and $\declsubjudg{\Psi}{C_2}{\tau}$. \\ 
          By inversion on \DeclArrElim, we get $\declsynjudg{\Psi, x:\sigma}{e}{C}$ and $\declappjudg{\Psi, x:\sigma}{x}{C}{C_2}$. \\ 
          By thinning, we know that $\declsynjudg{\Psi}{e}{C}$. \\
          By \Lemmaref{lem:application-subtyping},
          we get $C_1$ such that $\declsubjudg{\Psi, x:\sigma}{C}{C_1 \arr C_2}$
          and $\declchkjudg{\Psi, x:\sigma}{x}{C_1}$. \\ 
          By inversion on \DeclSub, $\declsynjudg{\Psi, x:\sigma}{x}{\sigma}$ and $\declsubjudg{\Psi}{\sigma}{C_1}$. \\
          By \DsubArr, $\declsubjudg{\Psi, x:\sigma}{C_1 \arr C_2}{\sigma \arr \tau}$. \\ 
          Hence by \Lemmaref{lem:declarative-transitivity}, $\declsubjudg{\Psi, x:\sigma}{C}{\sigma \arr \tau}$. \\
          Hence $\declsubjudg{\Psi}{C}{\sigma \arr \tau}$. \\
          Hence by \Lemmaref{lem:declarative-transitivity}, $\declsubjudg{\Psi}{C}{A}$. \\
          By \DeclSub, $\declchkjudg{\Psi}{e}{A}$. 
  \qedhere
  \end{itemize}
\end{proof}

\section{Properties of Context Extension}

\subsection{Syntactic Properties}

\declarationpreservation* 
\begin{proof}
  By a routine induction on $\substextend{\Gamma}{\Delta}$. 
\end{proof}

\declarationorderpreservation*
\begin{proof}
    By induction on the derivation of $\substextend{\Gamma}{\Delta}$. 
    
    \begin{itemize}
        \DerivationProofCase{\substextendId}
            { }
            { \substextend{\emptyctx}{\emptyctx} }

           This case is impossible. 

        \DerivationProofCase{\substextendVV}
              {\substextend{\Gamma}{\Delta}}
              {\substextend{\Gamma, x:A}{\Delta, x:A}}

            There are two cases, depending on whether or not $v = x$. 
            \begin{itemize}
            \item Case $v = x$: \\
              Since $u$ is declared to the left of $v$, $u$ is declared in $\Gamma$. \\
              By \Lemmaref{lem:declaration-preservation}, $u$ is declared in $\Delta$. \\
              Hence $u$ is declared to the left of $x$ in $\Delta, x:A$.

            \item Case $v \neq x$: \\
              Then $v$ is declared in $\Gamma$, and $u$ is declared to the left of $v$ in $\Gamma$.  \\
              By induction, $u$ is declared to the left of $v$ in $\Delta$. \\
              Hence $u$ is declared to the left of $v$ in $\Delta, x:A$. 
            \end{itemize}

        \DerivationProofCase{\substextendUU}
            { \substextend{\Gamma}{\Delta} }
            { \substextend{\Gamma, \alpha}{\Delta, \alpha} }

            This case is similar to the \substextendVV case. 

        \DerivationProofCase{\substextendEE}
            { \substextend{\Gamma}{\Delta} }
            { \substextend{\Gamma, \ahat}{\Delta, \ahat} }

            This case is similar to the \substextendVV case. 

        \DerivationProofCase{\substextendSolSol}
            { \substextend{\Gamma}{\Delta} \\ [\Delta]\tau = [\Delta]\tau'}
            { \substextend{\Gamma, \hypeq{\ahat}{\tau}}{\Delta, \hypeq{\ahat}{\tau'}} }

            This case is similar to the \substextendVV case. 

        \DerivationProofCase{\substextendMonMon}
            { \substextend{\Gamma}{\Delta} }
            { \substextend{\Gamma, \MonnierComma{\ahat}}{\Delta, \MonnierComma{\ahat}} }

            This case is similar to the \substextendVV case. 

        \DerivationProofCase{\substextendSolve}
            { \substextend{\Gamma}{\Delta} }
            { \substextend{\Gamma, \ahat}{\Delta, \hypeq{\ahat}{\tau}} }

            This case is similar to the \substextendVV case. 

        \DerivationProofCase{\substextendAdd}
            { \substextend{\Gamma}{\Delta} }
            { \substextend{\Gamma}{\Delta, \ahat} }

            By induction, $u$ is declared to the left of $v$ in $\Delta$. \\
            Therefore $u$ is declared to the left of $v$ in $\Delta, \ahat$.

        \DerivationProofCase{\substextendAddSolved}
            { \substextend{\Gamma}{\Delta} }
            { \substextend{\Gamma}{\Delta, \hypeq{\ahat}{\tau}} }

            By induction, $u$ is declared to the left of $v$ in $\Delta$. \\
            Therefore $u$ is declared to the left of $v$ in $\Delta, \hypeq{\ahat}{\tau}$.
    \qedhere
    \end{itemize}
\end{proof}

\reversedeclarationorderpreservation*
\begin{proof}
  It is given that $u$ and $v$ are declared in $\Gamma$.  Either $u$ is declared to the left of $v$
  in $\Gamma$, or $v$ is declared to the left of $u$.  Suppose the latter (for a contradiction).
  By \Lemmaref{lem:declaration-order-preservation}, $v$ is declared to
  the left of $u$ in $\Delta$.  But we know that $u$ is declared to the left of $v$ in $\Delta$:
  contradiction.   Therefore $u$ is declared to the left of $v$ in $\Gamma$.
\end{proof}

\substitutionextensioninvariance*
\begin{proof}
  To show that $[\Gamma]A = [\Theta][\Gamma]A$, observe that $\judgetp{\Theta}{A}$,
  and that by definition of $\substextend{\Theta}{\Gamma}$,
  every solved variable in $\Theta$ is solved in $\Gamma$.
  Therefore $[\Theta]([\Gamma]A) = [\Gamma]A$,
  since $\unsolved{[\Gamma]A}$ contains no variables that $\Theta$ solves. \\

  To show that $[\Gamma]A = [\Gamma][\Theta]A$, we proceed by induction on $\typesize{\Gamma}{A}$. 
  
  \begin{itemize}
      \DerivationProofCase{}
                          {\alpha \in \Theta}
                          {\judgetp{\Theta}{\alpha}}

          Note that $[\Gamma]\alpha = \alpha = [\Theta]\alpha$, so $[\Gamma]\alpha = [\Gamma][\Theta]\alpha$. 

      \DerivationProofCase{}
                          {\judgetp{\Theta}{A} \\ \judgetp{\Theta}{B}}
                          {\judgetp{\Theta}{A \arr B}}

          By induction, $[\Gamma]A = [\Gamma][\Theta]A$. \\
          By induction, $[\Gamma]B = [\Gamma][\Theta]B$. \\
          Then
          \begin{displaymath}
          \begin{array}{lcll}
            [\Gamma](A \arr B) & = & [\Gamma]A \arr [\Gamma]B & \mbox{By definition of substitution} \\
                              & = & [\Gamma][\Theta]A \arr [\Gamma][\Theta]B & \mbox{By induction hypothesis (twice)} \\
                              & = & [\Gamma]([\Theta]A \arr [\Theta]B) & \mbox{By definition of substitution} \\
                              & = & [\Gamma][\Theta](A \arr B) & \mbox{By definition of substitution}
          \end{array}
          \end{displaymath}                  

      \DerivationProofCase{}
                          {\judgetp{\Theta, \alpha}{A}}
                          {\judgetp{\Theta}{\alltype{\alpha}{A}}}

          By inversion, we have $\judgetp{\Theta, \alpha}{A}$. \\
          By rule \substextendUU, $\substextend{\Theta, \alpha}{\Gamma, \alpha}$. \\
          By induction, $[\Gamma, \alpha]A = [\Gamma, \alpha][\Theta, \alpha]A$. \\
          By definition, $[\Gamma]A = [\Gamma][\Theta]A$.  \\
          Then 
          \begin{displaymath}
          \begin{array}{lcll}
            [\Gamma]\alltype{\alpha}{A} & = & \alltype{\alpha}{[\Gamma]A} & \mbox{By definition} \\
                              & = & \alltype{\alpha}{[\Gamma][\Theta]A} & \mbox{By conclusion above } \\
                              & = & [\Gamma](\alltype{\alpha}{[\Theta]A}) & \mbox{By definition} \\
                              & = & [\Gamma][\Theta](\alltype{\alpha}{A}) & \mbox{By definition} \\
                              & = & [\Gamma, \alpha][\Theta, \alpha](\alltype{\alpha}{A}) & \mbox{By definition} 
          \end{array}
          \end{displaymath}                  

      \DerivationProofCase{}
                          { }
                          {\judgetp{\underbrace{\Theta_0, \ahat, \Theta_1}_{\Theta}}{\ahat}}

           Note that $[\Theta]\ahat = \ahat$. \\
           Hence $[\Gamma][\Theta]\ahat = [\Gamma]\ahat$.

      \DerivationProofCase{}
                          {}
                          {\judgetp{\Theta_0, \ahat=\tau, \Theta_1}{\ahat}}

          From $\substextend{\Theta}{\Gamma}$, 
          By a nested induction we get $\Gamma = \Gamma_0, \ahat=\tau', \Gamma_1$, 
                                   and $[\Gamma]\tau' = [\Gamma]\tau$. \\
          Note that $\typesize{\Theta}{\tau} < \typesize{\Theta}{\ahat}$. \\
          By induction, $[\Gamma]\tau = [\Gamma][\Theta]\tau$. \\
          Hence 
          \begin{displaymath}
            \begin{array}[b]{lcll}
              [\Gamma]\ahat & = & [\Gamma]\tau'        & \mbox{By definition} \\
                            & = & [\Gamma]\tau         & \mbox{From the extension judgment} \\
                            & = & [\Gamma][\Theta]\tau & \mbox{From the induction hypothesis} \\
                            & = & [\Gamma][\Theta]\ahat & \mbox{By definition}
            \end{array}
            \qedhere
          \end{displaymath}
  \end{itemize}
\end{proof}

\extensionequalitypreservation*
\begin{proof}
  By induction on the derivation of $\substextend{\Gamma}{\Delta}$. 

    \begin{itemize}
      \DerivationProofCase{\substextendId}
            { }
            { \substextend{\underbrace{\emptyctx}_{\Gamma}}{\underbrace{\emptyctx}_{\Delta}} }

            We have $[\Gamma]A = [\Gamma]B$, but $\Gamma = \Delta$, so  $[\Delta]A = [\Delta]B$.

      \DerivationProofCase{\substextendVV}
              {\substextend{\Gamma'}{\Delta'}}
              {\substextend{\Gamma', x:C}{\Delta', x:C}}

          We have $[\Gamma', x:C]A = [\Gamma', x:C]B$. \\
          By definition of substitution, $[\Gamma']A = [\Gamma']B$. \\
          By i.h., $[\Delta']A = [\Delta']B$. \\
          By definition of substitution, $[\Delta', x:C]A = [\Delta', x:C]B$. 

      \DerivationProofCase{\substextendUU}
            { \substextend{\Gamma'}{\Delta'} }
            { \substextend{\Gamma', \alpha}{\Delta', \alpha} }

          We have $[\Gamma', \alpha]A = [\Gamma', \alpha]B$. \\
          By definition of substitution, $[\Gamma']A = [\Gamma']B$. \\
          By i.h., $[\Delta']A = [\Delta']B$. \\
          By definition of substitution, $[\Delta', \alpha]A = [\Delta', \alpha]B$. 

      \DerivationProofCase{\substextendEE}
            { \substextend{\Gamma'}{\Delta'} }
            { \substextend{\Gamma', \ahat}{\Delta', \ahat} }
            
            Similar to the \substextendUU case.

      \DerivationProofCase{\substextendMonMon}
            { \substextend{\Gamma'}{\Delta'} }
            { \substextend{\Gamma', \MonnierComma{\ahat}}{\Delta', \MonnierComma{\ahat}} }

            Similar to the \substextendUU case.

      \DerivationProofCase{\substextendAdd}
            { \substextend{\Gamma}{\Delta'} }
            { \substextend{\Gamma}{\Delta', \ahat} }

          We have $[\Gamma]A = [\Gamma]B$. \\
          By i.h., $[\Delta']A = [\Delta']B$. \\
          By definition of substitution, $[\Delta', \ahat]A = [\Delta', \ahat]B$.

      \DerivationProofCase{\substextendAddSolved}
            { \substextend{\Gamma}{\Delta'} }
            { \substextend{\Gamma}{\Delta', \hypeq{\ahat}{\tau}} }

          We have $[\Gamma]A = [\Gamma]B$. \\
          By i.h., $[\Delta']A = [\Delta']B$. \\
          We implicitly assume that $\Delta$ is well-formed,
          so $\ahat \notin \dom{\Delta'}$. \\
          Since $\substextend{\Gamma}{\Delta'}$ and $\ahat \notin \dom{\Delta'}$,
          it follows that $\ahat \notin \dom{\Gamma}$. \\
          We have $\judgetp{\Gamma}{A}$ and $\judgetp{\Gamma}{B}$, so
          $\ahat \notin (\FV{A} \union \FV{B})$. \\
          Therefore, by definition of substitution, $[\Delta', \ahat=\tau]A = [\Delta', \ahat=\tau]B$.

      \DerivationProofCase{\substextendSolSol}
            { \substextend{\Gamma'}{\Delta'} \\ [\Delta']\tau = [\Delta']\tau'}
            { \substextend{\Gamma', \hypeq{\ahat}{\tau}}{\Delta', \hypeq{\ahat}{\tau'}} }

          We have $[\Gamma', \hypeq{\ahat}{\tau}]A = [\Gamma', \hypeq{\ahat}{\tau}]B$. \\
          
          By definition, $[\Gamma', \hypeq{\ahat}{\tau}]A = [\Gamma', \hypeq{\ahat}{\tau}]\tau$,
          but we implicitly assume that $\Gamma$ is well-formed, so $\ahat \notin \FV{\tau}$,
          so actually $[\Gamma', \hypeq{\ahat}{\tau}]A = [\Gamma']\tau$. \\
          Combined with similar reasoning for $B$, we get
          \[
               [\Gamma'][\tau/\ahat]A
               ~=~
               [\Gamma'][\tau/\ahat]B
          \]
          By i.h., $[\Delta'][\tau/\ahat]A = [\Delta'][\tau/\ahat]B$. \\
          By distributivity of substitution,
          $\big[[\Delta']\tau/\ahat\big][\Delta']A = \big[[\Delta']\tau/\ahat\big][\Delta']B$. \\
          Using the premise $[\Delta']\tau = [\Delta']\tau'$,
          we get $[[\Delta']\tau'/\ahat][\Delta']A = [[\Delta']\tau'/\ahat][\Delta']B$. \\
          By distributivity of substitution (in the other direction), $[\Delta'][\tau'/\ahat]A = [\Delta'][\tau'/\ahat]B$. \\
          It follows from the definition of substitution that
          $[\Delta', \hypeq{\ahat}{\tau'}]A = [\Delta', \hypeq{\ahat}{\tau'}]B$. 

      \DerivationProofCase{\substextendSolve}
            { \substextend{\Gamma'}{\Delta'} }
            { \substextend{\Gamma', \ahat}{\Delta', \hypeq{\ahat}{\tau}} }

          We have $[\Gamma', \ahat]A = [\Gamma', \ahat]B$. \\
          By definition of substitution, $[\Gamma']A = [\Gamma']B$. \\
          By i.h., $[\Delta'][\tau/\ahat]A = [\Delta'][\tau/\ahat]B$. \\
          It follows from the definition of substitution that
          $[\Delta', \hypeq{\ahat}{\tau}]A = [\Delta', \hypeq{\ahat}{\tau}]B$. 
    \qedhere
    \end{itemize}
\end{proof}

\substextendreflexivity*
\begin{proof}
  By induction on the structure of $\Gamma$. 

    \begin{itemize}
        \ProofCaseRule{$\Gamma = \cdot$} 
            Apply rule \substextendId. 

        \ProofCaseRule{$\Gamma = (\Gamma', \alpha)$}
            By i.h., $\substextend{\Gamma'}{\Gamma'}$.
            By rule \substextendUU, we get $\substextend{\Gamma', \alpha}{\Gamma', \alpha}$. 

        \ProofCaseRule{$\Gamma = (\Gamma', \ahat)$}
            By i.h., $\substextend{\Gamma'}{\Gamma'}$.
            By rule \substextendEE, we get $\substextend{\Gamma', \ahat}{\Gamma', \ahat}$. 

        \ProofCaseRule{$\Gamma = (\Gamma', \ahat=\tau)$}

            By i.h., $\substextend{\Gamma'}{\Gamma'}$. \\
            Clearly, $[\Gamma']\tau = [\Gamma']\tau$, so
            we can apply \substextendSolSol to get $\substextend{\Gamma', \ahat=\tau}{\Gamma', \ahat=\tau}$. 

        \ProofCaseRule{$\Gamma = (\Gamma', \MonnierComma{\ahat})$}
            By i.h., $\substextend{\Gamma'}{\Gamma'}$.
            By rule \substextendMonMon, we get $\substextend{\Gamma', \MonnierComma{\ahat}}{\Gamma', \MonnierComma{\ahat}}$.
    \qedhere
    \end{itemize}
\end{proof}

\substextendtransitivity*
\begin{proof}
  By induction on the derivation of $\substextend{\Delta}{\Theta}$.

  \begin{itemize}
  \ProofCaseRule{\substextendId}
  
      In this case $\Theta = \Delta$. \\
      Hence $\substextend{\Gamma}{\Delta}$ suffices. 

  \DerivationProofCase{\substextendUU}
                      {\substextend{\Delta'}{\Theta'}}
                      {\substextend{\Delta', \alpha}{\Theta', \alpha}}

      We have $\Delta = (\Delta', \alpha)$ and $\Theta = (\Theta', \alpha)$. \\
      By inversion on $\substextend{\Gamma}{\Delta}$,
      we have $\Gamma = (\Gamma', \alpha)$ and $\substextend{\Gamma'}{\Delta'}$. \\
      By i.h., $\substextend{\Gamma'}{\Theta'}$. \\
      Applying rule \substextendUU gives $\substextend{\Gamma', \alpha}{\Theta', \alpha}$. 

  \DerivationProofCase{\substextendUU}
                      {\substextend{\Delta'}{\Theta'}}
                      {\substextend{\Delta', \ahat}{\Theta', \ahat}}

      We have $\Delta = (\Delta', \ahat)$ and $\Theta = (\Theta', \ahat)$. \\
      Either of two rules could have derived $\substextend{\Gamma}{\Delta}$:

      \begin{itemize}
          \DerivationProofCase{\substextendEE}
                              {\substextend{\Gamma'}{\Delta'}}
                              {\substextend{\Gamma', \ahat}{\Delta', \ahat}}

            Here we have $\Gamma = (\Gamma', \ahat)$ and $\substextend{\Gamma'}{\Delta'}$. \\
            By i.h., $\substextend{\Gamma'}{\Theta'}$. \\
            Applying rule \substextendEE gives $\substextend{\Gamma', \ahat}{\Theta', \ahat}$. 

          \DerivationProofCase{\substextendAdd}
                              {\substextend{\Gamma}{\Delta'}}
                              {\substextend{\Gamma}{\Delta', \ahat}}

            By i.h., $\substextend{\Gamma}{\Theta'}$. \\
            By rule \substextendAdd, we get $\substextend{\Gamma}{\Theta', \ahat}$. 
      \end{itemize}

  \DerivationProofCase{\substextendSolSol}
                      {\substextend{\Delta'}{\Theta'} \\ [\Theta']\tau_1 = [\Theta']\tau_2}
                      {\substextend{\Delta', \ahat=\tau_1}{\Theta', \ahat=\tau_2}}

      In this case $\Delta = (\Delta', \hypeq{\ahat}{\tau_1})$ and $\Theta = (\Theta', \hypeq{\ahat}{\tau_2})$. \\
      One of three rules must have derived $\substextend{\Gamma}{\Delta', \ahat=\tau}$:

          \begin{itemize}
              \DerivationProofCase{\substextendSolSol}
                                  {\substextend{\Gamma'}{\Delta'} \\ [\Delta']\tau_0 = [\Delta']\tau_1}
                                  {\substextend{\Gamma', \ahat=\tau_0}{\Delta', \ahat=\tau_1}}

                    Here, $\Gamma = (\Gamma', \ahat=\tau_0)$ and $\Delta = (\Delta', \ahat=\tau_1)$. \\
                    By i.h., we have $\substextend{\Gamma'}{\Theta'}$. \\
                    The premises of the respective $\extendssym$ derivations give us
                       $[\Delta']\tau_0 = [\Delta']\tau_1$ and $[\Theta']\tau_1 = [\Theta']\tau_2$. \\
                    We know that $\judgetp{\Gamma'}{\tau_0}$ and $\judgetp{\Delta'}{\tau_1}$ and $\judgetp{\Theta'}{\tau_2}$. \\ 
                    By extension weakening (\Lemmaref{lem:extension-weakening}), $\judgetp{\Theta'}{\tau_0}$. \\ 
                    By extension weakening (\Lemmaref{lem:extension-weakening}), $\judgetp{\Theta'}{\tau_1}$. \\
                    Since $[\Delta']\tau_0 = [\Delta']\tau_1$, we know that $[\Theta'][\Delta']\tau_0 = [\Theta'][\Delta']\tau_1$. \\ 
                    By \Lemmaref{lem:subst-extension-invariance}, $[\Theta'][\Delta']\tau_0 = [\Theta']\tau_0$. \\
                    By \Lemmaref{lem:subst-extension-invariance}, $[\Theta'][\Delta']\tau_1 = [\Theta']\tau_1$. \\
                    So $[\Theta']\tau_0 = [\Theta']\tau_1$. \\ 

                    Hence by transitivity of equality, $[\Theta']\tau_0 = [\Theta']\tau_1 = [\Theta']\tau_2$. \\
                    By rule \substextendSolSol,  $\substextend{\Gamma', \ahat=\tau}{\Theta', \ahat=\tau_2}$. 

              \DerivationProofCase{\substextendAddSolved}
                                  {\substextend{\Gamma}{\Delta'}}
                                  {\substextend{\Gamma}{\Delta', \ahat=\tau_1}}

                    By induction, we have $\substextend{\Gamma}{\Theta'}$. \\
                    By rule \substextendAddSolved, we get $\substextend{\Gamma}{\Theta', \ahat=\tau_2}$. 

              \DerivationProofCase{\substextendSolve}
                                  {\substextend{\Gamma'}{\Delta'} }
                                  {\substextend{\Gamma', \ahat}{\Delta', \ahat=\tau_1}}

                    We have $\Gamma = (\Gamma', \ahat)$. \\
                    By induction, $\substextend{\Gamma'}{\Theta'}$. \\
                    By rule \substextendSolve, we get $\substextend{\Gamma', \ahat}{\Theta', \ahat=\tau_2}$.
          \end{itemize}
          
  \DerivationProofCase{\substextendMonMon}
                      {\substextend{\Delta'}{\Theta'}}
                      {\substextend{\Delta', \MonnierComma{\ahat}}{\Theta', \MonnierComma{\ahat}}}

      In this case we know $\Delta = (\Delta', \MonnierComma{\ahat})$ and $\Theta = (\Theta', \MonnierComma{\ahat})$. \\
      Since $\Delta = (\Delta', \MonnierComma{\ahat})$, only \substextendMonMon could derive
      $\substextend{\Gamma}{\Delta}$,
      so by inversion,
        $\Gamma = (\Gamma', \MonnierComma{\ahat})$
        and $\substextend{\Gamma'}{\Delta'}$. \\
      By induction, we have $\substextend{\Gamma'}{\Theta'}$. \\
      Applying rule \substextendMonMon gives 
      $\substextend{\Gamma', \MonnierComma{\ahat}}{\Theta', \MonnierComma{\ahat}}$. 

  \DerivationProofCase{\substextendAdd}
                      {\substextend{\Delta}{\Theta'}}
                      {\substextend{\Delta}{\Theta', \ahat}}

      In this case, we have $\Theta = (\Theta', \ahat)$. \\
      By induction, we get $\substextend{\Gamma}{\Theta'}$. \\
      By rule \substextendAdd, we get $\substextend{\Gamma}{\Theta', \ahat}$. 

  \DerivationProofCase{\substextendAddSolved}
                      {\substextend{\Delta}{\Theta'}}
                      {\substextend{\Delta}{\Theta', \ahat=\tau}}

      In this case, we have $\Theta = (\Theta', \ahat=\tau)$. \\
      By induction, we get $\substextend{\Gamma}{\Theta'}$. \\
      By rule \substextendAddSolved, we get $\substextend{\Gamma}{\Theta', \ahat=\tau}$. 

  \DerivationProofCase{\substextendSolve}
                      {\substextend{\Delta'}{\Theta'}}
                      {\substextend{\Delta', \ahat}{\Theta', \ahat=\tau}}

      In this case, we have $\Delta = (\Delta', \ahat)$ and $\Theta = (\Theta', \ahat=\tau)$. \\
      One of two rules could have derived $\substextend{\Gamma}{\Delta', \ahat}$:

          \begin{itemize}
              \DerivationProofCase{\substextendEE}
                                  {\substextend{\Gamma'}{\Delta'}}
                                  {\substextend{\Gamma',\ahat}{\Delta',\ahat}}

                   In this case, we have $\Gamma = (\Gamma', \ahat)$
                     and $\substextend{\Gamma'}{\Delta'}$
                     and $\substextend{\Delta'}{\Theta'}$. \\
                   By induction, we have $\substextend{\Gamma'}{\Theta'}$. \\
                   By rule \substextendSolve, we get $\substextend{\Gamma', \ahat}{\Theta', \ahat=\tau}$. 

               \DerivationProofCase{\substextendAdd}
                                   {\substextend{\Gamma}{\Delta'}}
                                   {\substextend{\Gamma}{\Delta',\ahat}}

                   In this case, we have $\substextend{\Gamma}{\Delta'}$ and $\substextend{\Delta'}{\Theta'}$. \\
                   By induction, we have $\substextend{\Gamma}{\Theta'}$. \\
                   By rule \substextendSolve, we get $\substextend{\Gamma}{\Theta', \ahat=\tau}$. 
           \qedhere
          \end{itemize}
  \end{itemize}
\end{proof}

\rightsoftness*
\begin{proof}
  By induction on $\Theta$, applying rules \substextendAdd and \substextendAddSolved as needed.
\end{proof}

\evarinput*
\begin{proof}
  By induction on the given derivation.

  \begin{itemize}
      \ProofCasesRules{\substextendId, \substextendVV, \substextendUU, \substextendSolSol, \substextendMonMon}
          \\
          Impossible: the left-hand context cannot have the form $\Gamma, \ahat$.

      \DerivationProofCase{\substextendEE}
           {\substextend{\Gamma}{\Delta_0}}
           {\substextend{\Gamma, \ahat}{\underbrace{\Delta_0, \ahat}_{\Delta}}}

           Let $\Theta = \emptyctx$, which is vacuously soft.
           Therefore $\Delta = (\Delta_0, \ahat) = (\Delta_0, \ahat, \Theta)$;
           the subderivation is the rest of the result.

      \DerivationProofCase{\substextendSolve}
          { \substextend{\Gamma}{\Delta_0} }
          { \substextend{\Gamma, \ahat}{\underbrace{\Delta_0, \hypeq{\ahat}{\tau}}_{\Delta}} }

          Let $\Theta = \emptyctx$, which is vacuously soft.
          Therefore $\Delta = (\Delta_0, \ahat) = (\Delta_0, \hypeq{\ahat}{\tau}, \Theta)$;
          the subderivation is the rest of the result.

      \DerivationProofCase{\substextendAdd}
           {\substextend{\Gamma, \ahat}{\Delta_0}}
           {\substextend{\Gamma, \ahat}{\underbrace{\Delta_0, \bhat}_{\Delta}}}

           Suppose $\bhat = \ahat$. \\
           $~$~~We have $\substextend{\Gamma, \ahat}{\Delta_0}$.  By \Lemmaref{lem:declaration-preservation},
           $\ahat$ is declared in $\Delta_0$. \\
           $~$~~But then $(\Delta_0, \bhat) = (\Delta_0, \ahat)$ with multiple $\ahat$ declarations, \\
           $~$~~which violates the implicit assumption that $\Delta$ is well-formed.  Contradiction. \\
           Therefore $\bhat \neq \ahat$.

           By i.h., $\Delta' = (\Delta_0, \Delta_{\ahat}, \Theta')$ where $\substextend{\Gamma}{\Delta_0}$
           and $\Theta'$ is soft.
           
           Let $\Theta = (\Theta', \bhat)$.
           Therefore $(\Delta', \bhat) = (\Delta_0, \Delta_{\ahat}, \Theta', \bhat)$.
           As $\Theta'$ is soft, $(\Theta', \bhat)$ is soft.
           Since $\Delta = (\Delta', \bhat)$, this gives $\Delta = (\Delta_0, \Delta_{\ahat}, \Theta)$.

      \ProofCaseRule{\substextendAddSolved}
          Similar to the case for \substextendAdd.
      \qedhere
  \end{itemize}
\end{proof}

\extensionorder*
\begin{proof}
  \begin{enumerate}[(i)]
  \item 
    By induction on the derivation of $\substextend{\Gamma_L, \alpha, \Gamma_R}{\Delta}$.

      \begin{itemize}
        \DerivationProofCase{\substextendId}
            { }
            { \substextend{\emptyctx}{\emptyctx} }

            This case is impossible since $(\Gamma_L, \alpha, \Gamma_R)$ cannot have the form $\cdot$.

       \ProofCasesRules{\substextendUU}

           We have two cases, depending on whether or not the rightmost variable is $\alpha$.

           \begin{itemize}
            \DerivationProofCase{\substextendUU}
                { \substextend{\Gamma}{\Delta'} }
                { \substextend{\Gamma, \alpha}{\Delta', \alpha} }

                Let $\Delta_L = \Delta'$, and let $\Delta_R = \cdot$ (which is soft). \\
                We have $\substextend{\Gamma}{\Delta'}$, which is
                $\substextend{\Gamma_L}{\Delta_L}$.
             
            \DerivationProofCase{\substextendUU}
                { \substextend{\Gamma_L, \alpha, \Gamma'_R}{\Delta'} }
                { \substextend{\Gamma_L, \alpha, \underbrace{\Gamma'_R, \beta}_{\Gamma_R}}
                  {\underbrace{\Delta', \beta}_{\Delta}} }

            By i.h., $\Delta' = (\Delta_L, \alpha, \Delta_R')$
              where $\substextend{\Gamma_L}{\Delta_L}$. \\
            Hence $\Delta = (\Delta_L, \alpha, \Delta_R', \beta)$. \\
            (Since $\beta \in \Gamma_R$, it cannot be the case that $\Gamma_R$ is soft.)
          \end{itemize}

    \DerivationProofCase{\substextendVV}
              {\substextend{\Gamma_L, \alpha, \Gamma'_R}{\Delta'}}
              {\substextend{\Gamma_L, \alpha, \underbrace{\Gamma'_R, x:A}_{\Gamma_R}}
                 {\underbrace{\Delta', x:A}_{\Delta}}}

            By i.h., $\Delta' = (\Delta_L, \alpha, \Delta_R')$
              where $\substextend{\Gamma_L}{\Delta_L}$. \\
            Hence $\Delta = (\Delta_L, \alpha, \Delta_R', x:A)$. \\
            (Since $x:A \in \Gamma_R$, it cannot be the case that $\Gamma_R$ is soft.)

        \DerivationProofCase{\substextendEE}
            { \substextend{\Gamma_L, \alpha, \Gamma'_R}{\Delta'} }
            { \substextend{\Gamma_L, \alpha, \underbrace{\Gamma'_R, \ahat}_{\Gamma_R}}
                {\underbrace{\Delta', \ahat}_{\Delta}} }

            By i.h., $\Delta' = (\Delta_L, \alpha, \Delta_R')$
              where $\substextend{\Gamma_L}{\Delta_L}$. \\
            Hence $\Delta = (\Delta_L, \alpha, \Delta_R', \ahat)$.  \\
            (If $\Gamma_R$ is soft, by i.h. $\Delta_R'$ is soft, so $\Delta_R = (\Delta_R', \ahat)$ is soft.)

        \DerivationProofCase{\substextendMonMon}
            { \substextend{\Gamma_L, \alpha, \Gamma'_R}{\Delta'} }
            { \substextend{\Gamma_L, \alpha,
                 \underbrace{\Gamma'_R, \MonnierComma{\bhat}}_{\Gamma'_R}}
                 {\underbrace{\Delta', \MonnierComma{\bhat}}_{\Delta}} }

            By i.h., $\Delta' = (\Delta_L, \alpha, \Delta_R')$
              where $\substextend{\Gamma_L}{\Delta_L}$. \\
            Hence $\Delta = (\Delta_L, \alpha, \Delta_R', \MonnierComma{\bhat})$.  \\
            (Since $\MonnierComma{\bhat} \in \Gamma_R$, it cannot be the case that $\Gamma_R$ is soft.)

        \DerivationProofCase{\substextendSolSol}
            { \substextend{\Gamma_L, \alpha, \Gamma'_R}{\Delta'} \\ [\Delta']\tau = [\Delta']\tau' }
            { \substextend{\Gamma_L, \alpha, \underbrace{\Gamma'_R, \hypeq{\ahat}{\tau}}_{\Gamma_R}}
                          {\underbrace{\Delta', \hypeq{\ahat}{\tau'}}_{\Delta'}} }

            By i.h., $\Delta' = (\Delta_L, \alpha, \Delta_R')$
              where $\substextend{\Gamma_L}{\Delta_L}$. \\
            Hence $\Delta = (\Delta_L, \alpha, \Delta_R', \hypeq{\ahat}{\tau'})$. \\
            (If $\Gamma_R$ is soft, by i.h. $\Delta_R'$ is soft, so $\Delta_R = (\Delta_R', \hypeq{\ahat}{\tau})$ is soft.)

        \DerivationProofCase{\substextendSolve}
            { \substextend{\Gamma_L, \alpha, \Gamma'_R}{\Delta'} }
            { \substextend{\Gamma_L, \alpha, \underbrace{\Gamma'_R, \ahat}_{\Gamma_R}}
                          {\underbrace{\Delta', \hypeq{\ahat}{\tau'}}_{\Delta}} }

            By i.h., $\Delta' = (\Delta_L, \alpha, \Delta_R')$
              where $\substextend{\Gamma_L}{\Delta_L}$. \\
            Therefore $\Delta = (\Delta_L, \alpha, \Delta_R, \hypeq{\ahat}{\tau})$. \\
            (If $\Gamma_R$ is soft, by i.h. $\Delta_R'$ is soft, so $\Delta_R = (\Delta_R', \hypeq{\ahat}{\tau})$ is soft.)

        \DerivationProofCase{\substextendAdd}
            { \substextend{\Gamma_L, \alpha, \Gamma_R}{\Delta'} }
            { \substextend{\Gamma_L, \alpha, \Gamma_R}{\underbrace{\Delta', \ahat}_{\Delta}} }

            By i.h., $\Delta' = (\Delta_L, \alpha, \Delta_R')$
              where $\substextend{\Gamma_L}{\Delta_L}$. \\
            Therefore $\Delta = (\Delta_L, \alpha, \Delta_R', \ahat)$. \\
            (If $\Gamma_R$ is soft, by i.h. $\Delta_R'$ is soft, so $\Delta_R = (\Delta_R', \ahat)$ is soft.)

        \DerivationProofCase{\substextendAddSolved}
            { \substextend{\Gamma_L, \alpha, \Gamma_R}{\Delta'} }
            { \substextend{\Gamma_L, \alpha, \Gamma_R}{\Delta', \hypeq{\ahat}{\tau}} }

            In this case, we know that  $\Delta = (\Delta', \hypeq{\ahat}{\tau})$. \\
            By i.h., $\Delta' = (\Delta_L, \alpha, \Delta_R')$
              where $\substextend{\Gamma_L}{\Delta_L}$. \\
            Hence $\Delta = (\Delta_L, \alpha, \Delta_R', \hypeq{\ahat}{\tau})$. \\
            (If $\Gamma_R$ is soft, by i.h. $\Delta_R'$ is soft, so $\Delta_R = (\Delta_R', \hypeq{\ahat}{\tau})$ is soft.)
      \end{itemize}

    \item Similar to the proof of (i), except that the \substextendMonMon and \substextendUU
      cases are swapped.

    \item Similar to (i), with $\Theta = \ahat$ in the \substextendEE case and
      $\Theta = (\hypeq{\ahat}{\tau})$ in the \substextendSolve case.

    \item Similar to (iii).

    \item Similar to (i), but using the equality premise of \substextendVV.
  \qedhere
  \end{enumerate}
\end{proof}

\extensionweakening*
\begin{proof}
  By a straightforward induction on $\judgetp{\Gamma}{A}$.

  In the \UvarWF case,
  we use \Lemmaref{lem:extension-order} (i).  In the \EvarWF case,
  use \Lemmaref{lem:extension-order} (iii).  In the \SolvedEvarWF case,
  use \Lemmaref{lem:extension-order} (iv).

  In the other cases, apply the i.h. to all subderivations, then apply the rule.
\end{proof}

\extensionsolve*
\begin{proof}
  By induction on $\Gamma_R$.

  \begin{itemize}
    \item Case $\Gamma_R = \cdot$: 
      \\
      By \Lemmaref{lem:substextend-reflexivity} (reflexivity), $\substextend{\Gamma_L}{\Gamma_L}$. \\
      Applying rule \substextendSolve gives $\substextend{\Gamma_L, \ahat}{\Gamma_L, \hypeq{\ahat}{\tau}}$. 

    \item Case $\Gamma_R = (\Gamma_R', x:A)$: 
      
      By i.h., $\substextend{\Gamma_L, \ahat, \Gamma_R'}{\Gamma_L, \hypeq{\ahat}{\tau}, \Gamma_R'}$. \\
      Applying rule \substextendVV gives
      $\substextend{\Gamma_L, \ahat, \Gamma_R', x:A}{\Gamma_L, \hypeq{\ahat}{\tau}, \Gamma_R', x:A}$. 

    \item Case $\Gamma_R = (\Gamma_R', \alpha)$:  By i.h.\ and rule \substextendUU.
    \item Case $\Gamma_R = (\Gamma_R', \bhat)$: By i.h.\ and rule \substextendAdd.
    \item Case $\Gamma_R = (\Gamma_R', \hypeq{\bhat}{\tau'})$: By i.h.\ and rule \substextendAddSolved.
    \item Case $\Gamma_R = (\Gamma_R', \MonnierComma{\bhat})$:  By i.h.\ and rule \substextendMonMon.
  \qedhere
  \end{itemize}
\end{proof}

\extensionaddsolve*
\begin{proof}
  By induction on $\Gamma_R$.
  The proof is exactly the same as the proof of \Lemmaref{lem:extension-solve},
  except that in the $\Gamma_R = \cdot$, we apply rule \substextendAddSolved instead of \substextendSolve.
\end{proof}

\extensionadd*
\begin{proof}  By induction on $\Gamma_R$.  The proof is exactly the same as the proof of
  \Lemmaref{lem:extension-solve}, except that in the $\Gamma_R = \cdot$ case, we apply
    rule \substextendAdd instead of \substextendSolve.
\end{proof}

\parallelevaradmissibility*
\begin{proof}
  By induction on $\Delta_R$.  As always, we assume that all contexts mentioned in the
  statement of the lemma are well-formed.  Hence, $\ahat \notin \dom{\Gamma_L} \union \dom{\Gamma_R}
  \union \dom{\Delta_L} \union \dom{\Delta_R}$.
  
  \begin{enumerate}[(i)]
  \item We proceed by cases of $\Delta_R$.  Observe that in all the extension rules, the right-hand context
    gets smaller, so as we enter subderivations of
    $\substextend{\Gamma_L, \Gamma_R}{\Delta_L, \Delta_R}$, the context $\Delta_R$ becomes smaller.
    
    The only tricky part of the proof is that to apply the i.h., we need $\substextend{\Gamma_L}{\Delta_L}$.
    So we need to make sure that as we drop items from the right of $\Gamma_R$ and $\Delta_R$, we don't
    go too far and start decomposing $\Gamma_L$ or $\Delta_L$!  It's easy to avoid decomposing $\Delta_L$:
    when $\Delta_R = \cdot$, we don't need to apply the i.h.\ anyway.  To avoid decomposing $\Gamma_L$,
    we need to reason by contradiction, using \Lemmaref{lem:declaration-preservation}.
    
        \begin{itemize}
        \ProofCaseRule{$\Delta_R = \cdot$} \\
            We have $\substextend{\Gamma_L}{\Delta_L}$.
            Applying \substextendEE to that derivation gives the result.
    
        \ProofCaseRule{$\Delta_R = (\Delta_R', \bhat)$}
            We have $\bhat \neq \ahat$ by the well-formedness assumption.

            The concluding rule of $\substextend{\Gamma_L, \Gamma_R}{\Delta_L, \Delta_R', \bhat}$
            must have been \substextendEE or \substextendAdd.  In both cases, the result follows by i.h.
            and applying \substextendEE or \substextendAdd.

            Note:  In \substextendAdd, the left-hand context doesn't change, so we clearly maintain
            $\substextend{\Gamma_L}{\Delta_L}$.  In \substextendEE, we can correctly apply the
            i.h.\ because $\Gamma_R \neq \cdot$.  Suppose, for
            a contradiction, that $\Gamma_R = \cdot$.  Then $\Gamma_L = (\Gamma_L', \bhat)$.
            It was given that $\substextend{\Gamma_L}{\Delta_L}$, that is, $\substextend{\Gamma_L', \bhat}{\Delta_L}$.
            By \Lemmaref{lem:declaration-preservation}, $\Delta_L$ has a declaration of $\bhat$.
            But then $\Delta = (\Delta_L, \Delta_R', \bhat)$ is not well-formed: contradiction.  Therefore
            $\Gamma_R \neq \cdot$.

        \ProofCaseRule{$\Delta_R = (\Delta_R', \hypeq{\bhat}{\tau})$}
            We have $\bhat \neq \ahat$ by the well-formedness assumption.

            The concluding rule must have been \substextendSolSol, \substextendSolve or \substextendAddSolved.
            In each case, apply the i.h.\ and then the corresponding rule.  (In \substextendSolSol and \substextendSolve,
            use \Lemmaref{lem:declaration-preservation} to show $\Gamma_R \neq \cdot$.)

        \ProofCaseRule{$\Delta_R = (\Delta_R', \alpha)$}
           The concluding rule must have been \substextendUU.  The result follows by i.h. and
           applying \substextendUU.

        \ProofCaseRule{$\Delta_R = (\Delta_R', \MonnierComma{\bhat})$}  Similar to the previous
           case, with rule \substextendMonMon.
        
        \ProofCaseRule{$\Delta_R = (\Delta_R', x : A)$}  Similar to the previous case,
           with rule \substextendVV.
        \end{itemize}

  \item Similar to part (i), except that when $\Delta_R = \cdot$, apply rule \substextendSolve.

  \item Similar to part (i), except that when $\Delta_R = \cdot$, apply rule \substextendSolSol,
     using the given equality to satisfy the second premise.
  \qedhere
  \end{enumerate}
\end{proof}

\parallelextensionsolve*
\begin{proof}
  By induction on $\Delta_R$.

  In the case where $\Delta_R = (\Delta_R', \hypeq{\ahat}{\tau'})$,
  we know that rule \substextendSolve must have concluded the derivation (we can use
  \Lemmaref{lem:declaration-preservation} to get a contradiction that rules out \substextendAddSolved);
  then we have a subderivation $\substextend{\Gamma_L}{\Delta_L}$, to which we can apply
  \substextendSolSol.
\end{proof}

\parallelextensionupdate*
\begin{proof}
  By induction on $\Delta_R$.  Similar to the proof of \Lemmaref{lem:parallel-extension-solve},
  but applying \substextendSolSol at the end.
\end{proof}

\subsection{Instantiation Extends}

\instantiationextension*
\begin{proof}
  By induction on the given instantiation derivation. 

    \begin{itemize}
      \DerivationProofCase{\InstLSolve}
              { \Gamma \entails \tau}     %
              { \instjudg{\Gamma, \ahat, \Gamma'}
                          {\ahat}
                          {\tau}
                          {\Gamma, \hypeq{\ahat}{\tau}, \Gamma'}
               }

               By \Lemmaref{lem:extension-solve},
               $\substextend{\Gamma, \ahat, \Gamma'}{\Gamma, \hypeq{\ahat}{\tau}, \Gamma'}$. 

      \DerivationProofCase{\InstLReach}
              { }
              {\instjudg{\Gamma[\ahat][\bhat]}
                          {\ahat}
                          {\bhat}
                          {\Gamma[\ahat][\hypeq{\bhat}{\ahat}]}}

          $\Gamma[\ahat][\bhat] = \Gamma_0, \ahat, \Gamma_1, \bhat, \Gamma_2$ for some $\Gamma_0, \Gamma_1, \Gamma_2$. \\
          By the definition of well-formedness, $\judgetp{\Gamma_0, \ahat, \Gamma_1}{\ahat}$. \\
          Therefore, by \Lemmaref{lem:extension-solve}, $\substextend{\Gamma_0, \ahat, \Gamma_1, \bhat, \Gamma_2}{\Gamma_0, \ahat, \Gamma_1, \hypeq{\bhat}{\ahat}, \Gamma_2}$. 

      \DerivationProofCase{\InstLArr}
              {\instjudgr{\Gamma[\ahat_2, \ahat_1, \hypeq{\ahat}{\ahat_1 \arr \ahat_2}]}
                          {\ahat_1}
                          {A_1}
                          {\Gamma'} \\
               \instjudg{\Gamma'}
                          {\ahat_2}
                          {[\Gamma']A_2}
                          {\Delta}}
              {\instjudg{\Gamma[\ahat]}
                          {\ahat}
                          {A_1 \arr A_2}
                          {\Delta}}

          By \Lemmaref{lem:extension-add}, we can insert an (unsolved) $\ahat_2$, giving
            $\substextend{\Gamma[\ahat]}{\Gamma[\ahat_2, \ahat]}$. \\
          By \Lemmaref{lem:extension-add} again, $\substextend{\Gamma[\ahat_2, \ahat]}{\Gamma[\ahat_2, \ahat_1, \ahat]}$. \\
          By \Lemmaref{lem:extension-solve}, we can solve $\ahat$, giving
            $\substextend{\Gamma[\ahat_2, \ahat_1, \ahat]}{\Gamma[\ahat_2, \ahat_1, \hypeq{\ahat}{\ahat_1 \arr \ahat_2}]}$. \\
          Then by transitivity (\Lemmaref{lem:substextend-transitivity}), $\substextend{\Gamma[\ahat]}{\Gamma[\ahat_2, \ahat_1, \hypeq{\ahat}{\ahat_1 \arr \ahat_2}]}$. \\
          By i.h.\ on the first subderivation,
            $\substextend{\Gamma[\ahat_2, \ahat_1, \hypeq{\ahat}{\ahat_1 \arr \ahat_2}]}{\Gamma'}$. \\
          By i.h.\ on the second subderivation, $\substextend{\Gamma'}{\Delta}$. \\
          By transitivity (\Lemmaref{lem:substextend-transitivity}), $\substextend{\Gamma[\ahat_2, \ahat_1, \hypeq{\ahat}{\ahat_1 \arr \ahat_2}]}{\Delta}$. \\
          By transitivity (\Lemmaref{lem:substextend-transitivity}), $\substextend{\Gamma[\ahat]}{\Delta}$. 

      \DerivationProofCase{\InstLAllR}
            {\instjudg{\Gamma[\ahat], \beta}{\ahat}{B}{\Delta, \beta, \Delta'}}
            {\instjudg{\Gamma[\ahat]}{\ahat}{\alltype{\beta}{B}}{\Delta}}

          By induction, $\substextend{\Gamma[\ahat], \beta}{\Delta, \beta, \Delta'}$. \\
          By \Lemmaref{lem:extension-order} (i),
          we have $\substextend{\Gamma[\ahat]}{\Delta}$. 

      \DerivationProofCase{\InstRSolve}
              { \Gamma \entails \tau}
              { \instjudgr{\Gamma, \ahat, \Gamma'}
                          {\ahat}
                          {\tau}
                          {\Gamma, \hypeq{\ahat}{\tau}, \Gamma'}
               }

          By \Lemmaref{lem:extension-solve}, we can solve $\ahat$, giving
          $\substextend{\Gamma, \ahat, \Gamma'}{\Gamma, \hypeq{\ahat}{\tau}, \Gamma'}$. 

      \DerivationProofCase{\InstRReach}
              { }
              {\instjudgr{\Gamma[\ahat][\bhat]}
                         {\ahat}
                         {\bhat}
                         {\Gamma[\ahat][\hypeq{\bhat}{\ahat}]}}

          $\Gamma[\ahat][\bhat] = \Gamma_0, \ahat, \Gamma_1, \bhat, \Gamma_2$ for some $\Gamma_0, \Gamma_1, \Gamma_2$. \\
          By the definition of well-formedness, $\judgetp{\Gamma_0, \ahat, \Gamma_1}{\ahat}$. \\
          Hence by \Lemmaref{lem:extension-solve}, we can solve $\bhat$, giving
          $\substextend{\Gamma_0, \ahat, \Gamma_1, \bhat, \Gamma_2}{\Gamma_0, \ahat, \Gamma_1, \hypeq{\bhat}{\ahat}, \Gamma_2}$. 

      \DerivationProofCase{\InstRArr}
            {\instjudg{\Gamma[\ahat_2, \ahat_1, \hypeq{\ahat}{\ahat_1 \arr \ahat_2}]}
                      {\ahat_1}
                      {A_1}
                      {\Gamma'} \\
               \instjudgr{\Gamma'}
                         {\ahat_2}
                         {[\Gamma']A_2}
                         {\Delta}}
            {\instjudgr{\Gamma[\ahat]}
                       {\ahat}
                       {A_1 \arr A_2}
                       {\Delta}}

          Because the contexts here are the same as in \InstLArr, this is the same as the \InstLArr case.

      \DerivationProofCase{\InstRAllL}
            {\instjudgr{\Gamma[\ahat], \MonnierComma{\bhat}, \bhat}{\ahat}{[\bhat/\beta]B}{\Delta, \MonnierComma{\bhat}, \Delta'}}
            {\instjudgr{\Gamma[\ahat]}{\ahat}{\alltype{\beta}{B}}{\Delta}}

          By i.h., $\substextend{\Gamma[\ahat], \MonnierComma{\bhat}, \bhat}{\Delta, \MonnierComma{\bhat}, \Delta'}$. \\
          By \Lemmaref{lem:extension-order} (ii), $\substextend{\Gamma[\ahat]}{\Delta}$.
    \qedhere
    \end{itemize}
\end{proof}

\subsection{Subtyping Extends}

\subtypingextension*
 \begin{proof}
  By induction on the given derivation.

  For cases \SubVar, \SubUnit, \SubExvar, we have $\Delta = \Gamma$,
  so \Lemmaref{lem:substextend-reflexivity} suffices.
  
  \begin{itemize}
    \DerivationProofCase{\SubArr}
         {\subjudg{\Gamma}{B_1}{A_1}{\Theta}
          \\
          \subjudg{\Theta}{[\Omega]A_2}{[\Omega]B_2}{\Delta}
         }
         { \subjudg{\Gamma}{A_1 \arr A_2}{B_1 \arr B_2}{\Delta} }

         By IH on each subderivation, $\substextend{\Gamma}{\Theta}$
         and $\substextend{\Theta}{\Delta}$.

         By \Lemmaref{lem:substextend-transitivity} (transitivity), %
         $\substextend{\Gamma}{\Delta}$, which was to be shown.

    \DerivationProofCase{\SubAllL}
          {\subjudg{\Gamma, \MonnierComma{\ahat}, \ahat}
                   {[\ahat/\alpha]A}
                   {B}
                   {\Delta, \MonnierComma{\ahat}, \Theta}}
          {\subjudg{\Gamma}{\alltype{\alpha}{A}}{B}{\Delta}}

          By IH, 
          $\substextend
                       {\Gamma, \MonnierComma{\ahat}, \ahat}
                       {\Delta, \MonnierComma{\ahat}, \Theta}$.
          
         By \Lemmaref{lem:extension-order} (ii) with $\Gamma_L = \Gamma$
         and $\Gamma_L' = \Delta$ and $\Gamma_R = \ahat$
         and $\Gamma_R' = \Theta$, we obtain
         \[
              \substextend{\Gamma}{\Delta}
         \]

    \DerivationProofCase{\SubAllR}
          {\subjudg{\Gamma, \beta}{A}{B}{\Delta, \beta, \Theta}}
          {\subjudg{\Gamma}{A}{\alltype{\beta}{B}}{\Delta}}

          By IH, we have
          $\substextend
                       {\Gamma, \beta}
                       {\Delta, \beta, \Theta}$.
          
          By \Lemmaref{lem:extension-order} (i), we obtain
          $
               \substextend{\Gamma}{\Delta}
          $, which was to be shown.

    \ProofCasesRules{\SubInstL, \SubInstR}
          In each of these rules, the premise has the same input and output contexts as the conclusion,
          so \Lemmaref{lem:instantiation-extension} suffices.
    \qedhere
  \end{itemize}
\end{proof}

\section{Decidability of Instantiation}

\leftunsolvednesspreservation*
\begin{proof}
  By induction on the given derivation.
  
    \begin{itemize}
        \DerivationProofCase{\InstLSolve}
                { \Gamma_0 \entails \tau}
                { \instjudg{\underbrace{\Gamma_0, \ahat, \Gamma_1}_\Gamma}
                            {\ahat}
                            {\tau}
                            {\Gamma_0, \hypeq{\ahat}{\tau}, \Gamma_1}
                 }

            Immediate, since to the left of $\ahat$, the contexts $\Delta$ and $\Gamma$ are the same.

        \DerivationProofCase{\InstLReach}
                { }
                {\instjudg{\Gamma[\ahat][\bhat]}
                            {\ahat}
                            {\bhat}
                            {\Gamma[\ahat][\hypeq{\bhat}{\ahat}]}}

            Immediate, since to the left of $\ahat$, the contexts $\Delta$ and $\Gamma$ are the same.

        \DerivationProofCase{\InstLArr}
                {\instjudgr{\Gamma[\ahat_2, \ahat_1, \hypeq{\ahat}{\ahat_1 \arr \ahat_2}]}
                            {\ahat_1}
                            {A_1}
                            {\Gamma'} \\
                 \instjudg{\Gamma'}
                            {\ahat_2}
                            {[\Gamma']A_2}
                            {\Delta}}
                {\instjudg{\Gamma[\ahat]}
                            {\ahat}
                            {A_1 \arr A_2}
                            {\Delta}}

            We have $\bhat \in \unsolved{\Gamma_0}$.  Therefore $\bhat \in \unsolved{\Gamma_0, \ahat_2}$. \\
            Clearly, $\ahat_2 \in \unsolved{\Gamma_0, \ahat_2}$. \\
            We have two subderivations:
            \begin{align*}
                \instjudgr{\Gamma_0, \ahat_2, \ahat_1, \hypeq{\ahat}{\ahat_1 \arr \ahat_2}, \Gamma_1}{\ahat_1}{A_1}{\Gamma'}
                && \text{(1)}
                \\
                \instjudg{\Gamma'}{\ahat_2}{[\Gamma']A_2}{\Delta}
                && \text{(2)}
            \end{align*}
            By induction on (1), $\bhat \in \unsolved{\Gamma'}$. \\
            Also by induction on (1), with $\ahat_2$ playing the role of $\bhat$, we get $\ahat_2 \in \unsolved{\Gamma'}$. \\
            Since $\bhat \in \Gamma_0$, it is declared to the left of $\ahat_2$
              in $\Gamma_0, \ahat_2, \ahat_1, \hypeq{\ahat}{\ahat_1 \arr \ahat_2}, \Gamma_1$. \\
            Hence by \Lemmaref{lem:declaration-order-preservation}, $\bhat$ is declared to the left of $\ahat_2$ in $\Gamma'$.
            That is, $\Gamma' = (\Gamma'_0, \ahat_2, \Gamma'_1)$, where $\bhat \in \unsolved{\Gamma'_0}$. \\
            By induction on (2), $\bhat \in \unsolved{\Delta}$.

        \DerivationProofCase{\InstLAllR}
              {\instjudg{\Gamma_0, \ahat, \Gamma_1, \beta}{\ahat}{B}{\Delta, \beta, \Delta'}}
              {\instjudg{\Gamma_0, \ahat, \Gamma_1}{\ahat}{\alltype{\beta}{B}}{\Delta}}

            We have $\bhat \in \unsolved{\Gamma_0}$. \\
            By induction, $\bhat \in \unsolved{\Delta, \beta, \Delta'}$. \\
            Note that $\bhat$ is declared to the left of $\beta$ in $\Gamma_0, \ahat, \Gamma_1, \beta$. \\
            By \Lemmaref{lem:declaration-order-preservation},
              $\bhat$ is declared to the left of $\beta$ in $(\Delta, \beta, \Delta')$,
              that is, in $\Delta$. Since $\bhat \in \unsolved{\Delta, \beta, \Delta'}$,
              we have $\bhat \in \unsolved{\Delta}$.

        \ProofCasesRules{\InstRSolve, \InstRReach}
           Similar to the \InstLSolve and \InstLReach cases.

        \DerivationProofCase{\InstRArr}
              {\instjudg{\Gamma[\ahat_2, \ahat_1, \hypeq{\ahat}{\ahat_1 \arr \ahat_2}]}
                        {\ahat_1}
                        {A_1}
                        {\Gamma'} \\
                 \instjudgr{\Gamma'}
                           {\ahat_2}
                           {[\Gamma']A_2}
                           {\Delta}}
              {\instjudgr{\Gamma[\ahat]}
                         {\ahat}
                         {A_1 \arr A_2}
                         {\Delta}}

            Similar to the \InstLArr case.

        \DerivationProofCase{\InstRAllL}
              {\instjudgr{\Gamma[\ahat], \MonnierComma{\chat}, \chat}{\ahat}{[\chat/\beta]B}{\Delta, \MonnierComma{\chat}, \Delta'}}
              {\instjudgr{\Gamma[\ahat]}{\ahat}{\alltype{\beta}{B}}{\Delta}}

            We have $\bhat \in \unsolved{\Gamma_0}$. \\
            By induction, $\bhat \in \unsolved{\Delta, \MonnierComma{\chat}, \Delta'}$. \\
            Note that $\bhat$ is declared to the left of $\MonnierComma{\chat}$ in $\Gamma_0, \ahat, \Gamma_1, \MonnierComma{\chat}, \chat$. \\
            By \Lemmaref{lem:declaration-order-preservation}, $\bhat$ is declared to the left of $\MonnierComma{\chat}$ in $\Delta, \MonnierComma{\chat}, \Delta'$. \\
            Hence $\bhat$ is declared in $\Delta$, and we know it is in $\unsolved{\Delta, \MonnierComma{\chat}, \Delta'}$,
            so $\bhat \in \unsolved{\Delta}$. 
    \qedhere
    \end{itemize}
\end{proof}

\leftfreevariablepreservation*
\begin{proof}
  By induction on the given instantiation derivation.

    \begin{itemize}
        \DerivationProofCase{\InstLSolve}
                { \Gamma_0 \entails \tau}
                { \instjudg{\underbrace{\Gamma_0, \ahat, \Gamma_1}_\Gamma}
                            {\ahat}
                            {\tau}
                            {\underbrace{\Gamma_0, \hypeq{\ahat}{\tau}, \Gamma_1}_{\Delta}}
                 }

            We have $\ahat \notin \FV{[\Gamma]B}$.  Since $\Delta$ differs from $\Gamma$
            only in $\ahat$, it must be the case that $[\Gamma]B = [\Delta]B$.
            It is given that $\bhat \notin \FV{[\Gamma]B}$, so $\bhat \notin \FV{[\Delta]B}$.

        \DerivationProofCase{\InstLReach}
                { }
                {\instjudg{\underbrace{\Gamma'[\ahat][\chat]}_{\Gamma}}
                            {\ahat}
                            {\chat}
                            {\underbrace{\Gamma'[\ahat][\hypeq{\chat}{\ahat}]}_{\Delta}}}

            Since $\Delta$ differs from $\Gamma$ only
            in solving $\chat$ to $\ahat$, applying $\Delta$ to a type will not introduce a $\bhat$.
            We have $\bhat \notin \FV{[\Gamma]B}$, so $\bhat \notin \FV{[\Delta]B}$.
        
        \DerivationProofCase{\InstRSolve}
                { \Gamma_0 \entails \tau}
                { \instjudgr{\Gamma_0, \ahat, \Gamma_1}
                            {\ahat}
                            {\tau}
                            {\Gamma_0, \hypeq{\ahat}{\tau}, \Gamma_1}
                 }
             
             Similar to the \InstLSolve case.

        \DerivationProofCase{\InstRReach}
                { }
                {\instjudgr{\Gamma'[\ahat][\chat]}
                           {\ahat}
                           {\chat}
                           {\Gamma'[\ahat][\hypeq{\chat}{\ahat}]}}

             Similar to the \InstLReach case.
        
        \DerivationProofCase{\InstLArr}
                {\instjudgr{\overbrace{\Gamma_0, \ahat_2, \ahat_1, \hypeq{\ahat}{\ahat_1 \arr \ahat_2}, \Gamma_1}^{\Gamma'}}
                            {\ahat_1}
                            {A_1}
                            {\Delta} \\
                 \instjudg{\Delta}
                            {\ahat_2}
                            {[\Delta]A_2}
                            {\Delta}}
                {\instjudg{\underbrace{\Gamma_0, \ahat, \Gamma_1}_\Gamma}
                            {\ahat}
                            {A_1 \arr A_2}
                            {\Delta}}

            We have $\judgetp{\Gamma}{B}$
            and $\ahat \notin \FV{[\Gamma]B}$
            and $\bhat \notin \FV{[\Gamma]B}$. \\
            By weakening, we get $\judgetp{\Gamma'}{B}$; since $\ahat \notin \FV{[\Gamma]B}$ and $\Gamma'$ only adds
            a solution for $\ahat$, it follows that $[\Gamma']B = [\Gamma]B$. \\
            Therefore $\ahat_1 \notin \FV{[\Gamma']B}$ and $\ahat_2 \notin \FV{[\Gamma']B}$ and $\bhat \notin \FV{[\Gamma']B}$. \\
            Since we have $\bhat \in \Gamma_0$, we also have $\bhat \in (\Gamma_0, \ahat_2)$. \\
            By induction on the first premise, $\bhat \notin \FV{[\Delta]B}$. \\
            Also by induction on the first premise, with $\ahat_2$ playing the role of $\bhat$, we have $\ahat_2 \notin \FV{[\Delta]B}$. \\
            Note that $\ahat_2 \in \unsolved{\Gamma_0, \ahat_2}$. \\
            By \Lemmaref{lem:left-unsolvedness-preservation}, $\ahat_2 \in \unsolved{\Delta}$. \\
            Therefore $\Delta$ has the form $(\Delta_0, \ahat_2, \Delta_1)$. \\
            Since $\bhat \neq \ahat_2$, we know that $\bhat$ is declared to the left of $\ahat_2$ in $\Gamma_0, \ahat_2$,
            so by \Lemmaref{lem:declaration-order-preservation}, $\bhat$ is declared to the left of $\ahat_2$ in $\Delta$.
            Hence $\bhat \in \Delta_0$. \\
            Furthermore, by \Lemmaref{lem:instantiation-extension}, we have $\substextend{\Gamma'}{\Delta}$. \\
            Then by \Lemmaref{lem:extension-weakening}, we have $\judgetp{\Delta}{B}$.
            Using induction on the second premise, $\bhat \notin \FV{[\Delta]B}$. 

        \DerivationProofCase{\InstLAllR}
              {\instjudg{\Gamma_0, \ahat, \Gamma_1, \gamma}{\ahat}{C}{\Delta, \gamma, \Delta'}}
              {\instjudg{\underbrace{\Gamma_0, \ahat, \Gamma_1}_{\Gamma}}{\ahat}{\alltype{\gamma}{C}}{\Delta}}

            We have $\judgetp{\Gamma}{B}$ and $\ahat \notin \FV{[\Gamma]B}$ and $\bhat \in \Gamma_0$ and $\bhat \notin \FV{[\Gamma]B}$. \\
            By weakening, $\judgetp{\Gamma, \gamma}{B}$; by the definition of substitution, $[\Gamma, \gamma]B = [\Gamma]B$. \\
            Substituting equals for equals, $\ahat \notin \FV{[\Gamma, \gamma]B}$ and $\bhat \notin \FV{[\Gamma, \gamma]B}$. \\
            By induction, $\bhat \notin \FV{[\Delta, \gamma, \Delta']B}$. \\
            Since $\bhat$ is declared to the left of $\gamma$ in $(\Gamma, \gamma)$, we can use
            \Lemmaref{lem:declaration-order-preservation} to show that
              $\bhat$ is declared to the left of $\gamma$ in $(\Delta, \gamma, \Delta')$,
              that is, in $\Delta$. \\
            We have $\judgetp{\Gamma}{B}$, so $\gamma \notin \FV{B}$.  Thus each free variable $u$ in $B$ is in $\Gamma$,
            to the left of $\gamma$ in $(\Gamma, \gamma)$. \\
            Therefore, by \Lemmaref{lem:declaration-order-preservation}, each free variable $u$ in $B$ is in $\Delta$. \\
            Therefore $[\Delta, \gamma, \Delta']B = [\Delta]B$. \\
            Earlier, we obtained $\bhat \notin \FV{[\Delta, \gamma, \Delta']B}$, so substituting equals for equals,
            $\bhat \notin \FV{[\Delta]B}$.

        \DerivationProofCase{\InstRArr}
              {\instjudg{\Gamma_0, \ahat_2, \ahat_1, \hypeq{\ahat}{\ahat_1 \arr \ahat_2}, \Gamma_1}
                        {\ahat_1}
                        {A_1}
                        {\Delta} \\
                 \instjudgr{\Gamma'}
                           {\ahat_2}
                           {[\Delta]A_2}
                           {\Delta}}
              {\instjudgr{\Gamma_0, \ahat, \Gamma_1}
                         {\ahat}
                         {A_1 \arr A_2}
                         {\Delta}}

            Similar to the \InstLArr case.

        \DerivationProofCase{\InstRAllL}
              {\instjudgr{\Gamma[\ahat], \MonnierComma{\chat}, \chat}{\ahat}{[\chat/\gamma]C}{\Delta, \MonnierComma{\chat}, \Delta'}}
              {\instjudgr{\Gamma[\ahat]}{\ahat}{\alltype{\gamma}{C}}{\Delta}}

             We have $\judgetp{\Gamma}{B}$ and $\ahat \notin \FV{[\Gamma]B}$ and $\bhat \in \Gamma_0$ and $\bhat \notin \FV{[\Gamma]B}$. \\
             By weakening, $\judgetp{\Gamma, \MonnierComma{\chat}, \chat}{B}$;
               by the definition of substitution, $[\Gamma, \MonnierComma{\chat}, \chat]B = [\Gamma]B$. \\
             Substituting equals for equals, $\ahat \notin \FV{[\Gamma, \MonnierComma{\chat}, \chat]B}$ and $\bhat \notin \FV{[\Gamma, \MonnierComma{\chat}, \chat]B}$. \\
             By induction, $\bhat \notin \FV{[\Delta, \MonnierComma{\chat}, \Delta']B}$. \\
             Note that $\bhat$ is declared to the left of $\MonnierComma{\chat}$ in $\Gamma, \MonnierComma{\chat}, \chat$. \\
             By \Lemmaref{lem:declaration-order-preservation}, $\bhat$ is declared to the left of $\MonnierComma{\chat}$ in $\Delta, \MonnierComma{\chat}, \Delta'$. \\
             So $\bhat$ is declared in $\Delta$. \\
             Now, note that each free variable $u$ in $B$ is in $\Gamma$, which is to the left of $\MonnierComma{\chat}$ in $\Gamma, \MonnierComma{\chat}, \chat$. \\
             Therefore, by \Lemmaref{lem:declaration-order-preservation}, each free variable $u$ in $B$ is in $\Delta$. \\
             Therefore $[\Delta, \MonnierComma{\chat}, \Delta']B = [\Delta]B$. \\
             Earlier, we obtained $\bhat \notin \FV{[\Delta, \MonnierComma{\chat}, \Delta']B}$, so substituting equals for equals,
             $\bhat \notin \FV{[\Delta]B}$.
    \qedhere
    \end{itemize}
\end{proof}

\instantiationsizepreservation*
\begin{proof}
  By induction on the given derivation.
  
  \begin{itemize}
      \DerivationProofCase{\InstLSolve}
              { \Gamma_0 \entails \tau}
              { \instjudg{\underbrace{\Gamma_0, \ahat, \Gamma_1}_\Gamma}
                          {\ahat}
                          {\tau}
                          {\Gamma_0, \hypeq{\ahat}{\tau}, \Gamma_1}
               }

              Since $\Delta$ differs from $\Gamma$ only in solving $\ahat$, and we know $\ahat \notin \FV{[\Gamma]B}$,
              we have $[\Delta]B = [\Gamma]B$; therefore $|[\Delta]B = [\Gamma]B|$.

      \DerivationProofCase{\InstLReach}
              { }
              {\instjudg{\Gamma[\ahat][\bhat]}
                          {\ahat}
                          {\bhat}
                          {\Gamma[\ahat][\hypeq{\bhat}{\ahat}]}
              }

              Here, $\Delta$ differs from $\Gamma$ only in solving $\bhat$ to $\ahat$.  However, $\ahat$ has the same
              size as $\bhat$, so even if $\bhat \in \FV{[\Gamma]B}$, we have $|[\Delta]B = [\Gamma]B|$.

      \DerivationProofCase{\InstLArr}
              {\instjudgr{\overbrace{\Gamma_0, \ahat_2, \ahat_1, \hypeq{\ahat}{\ahat_1 \arr \ahat_2}, \Gamma_1}^{\Gamma'}}
                          {\ahat_1}
                          {A_1}
                          {\Theta} \\
               \instjudg{\Theta}
                          {\ahat_2}
                          {[\Theta]A_2}
                          {\Delta}}
              {\instjudg{\underbrace{\Gamma_0, \ahat, \Gamma_1}_\Gamma}
                          {\ahat}
                          {A_1 \arr A_2}
                          {\Delta}}

      We have $\judgetp{\Gamma}{B}$ and $\ahat \notin \FV{[\Gamma]B}$.
      Since $\ahat_1, \ahat_2 \notin \dom{\Gamma}$, we have $\ahat, \ahat_1, \ahat_2 \notin \FV{[\Gamma]B}$.
      It follows that $[\Gamma']B = [\Gamma]B$. \\
      By weakening, $\judgetp{\Gamma'}{B}$. \\
      By induction on the first premise, $|[\Gamma']B| = |[\Theta]B|$. \\
      By \Lemmaref{lem:declaration-order-preservation},
        since $\ahat_2$ is declared to the left of $\ahat_1$ in $\Gamma'$,
        we have that $\ahat_2$ is declared to the left of $\ahat_1$ in $\Theta$. \\
      By \Lemmaref{lem:left-unsolvedness-preservation}, since $\ahat_2 \in \unsolved{\Gamma'}$, it is unsolved in $\Theta$:
      that is, $\Theta = (\Theta_0, \ahat_2, \Theta_1)$. \\
      By \Lemmaref{lem:instantiation-extension}, we have $\substextend{\Gamma'}{\Theta}$. \\
      By \Lemmaref{lem:extension-weakening}, $\judgetp{\Theta}{B}$. \\
      Since $\ahat_2 \notin \FV{[\Gamma']B}$,
        \Lemmaref{lem:left-free-variable-preservation} gives $\ahat_2 \notin \FV{[\Theta]B}$. \\
      By induction on the second premise, $|[\Theta]B| = |[\Delta]B|$,
        and by transitivity of equality, $|[\Gamma]B| = |[\Delta]B|$. 

      \DerivationProofCase{\InstLAllR}
            {\instjudg{\Gamma_0, \ahat, \Gamma_1, \beta}{\ahat}{A_0}{\Delta, \beta, \Delta'}}
            {\instjudg{\underbrace{\Gamma_0, \ahat, \Gamma_1}_{\Gamma}}{\ahat}{\alltype{\beta}{A_0}}{\Delta}}

      We have $\judgetp{\Gamma}{B}$ and $\ahat \notin \FV{[\Gamma]B}$. \\
      By weakening, $\judgetp{\Gamma, \beta}{B}$. \\
      From the definition of substitution, $[\Gamma]B = [\Gamma, \beta]B$.
      Hence $\ahat \notin \FV{[\Gamma, \beta]B}$. \\
      The input context of the premise is $(\Gamma_0, \ahat, \Gamma_1, \beta)$, which is $(\Gamma, \beta)$, so
      by induction, $|[\Gamma, \beta]B| = |[\Delta, \beta, \Delta']B|$. \\
      Suppose $u$ is a free variable in $B$.
      Then $u$ is declared in $\Gamma$, and so occurs before $\beta$ in $\Gamma, \beta$. \\
      By \Lemmaref{lem:declaration-order-preservation}, $u$ is declared before $\beta$ in $\Delta, \beta, \Delta'$. \\
      So every free variable $u$ in $B$ is declared in $\Delta$. \\
      Hence $[\Delta, \beta, \Delta']B = [\Delta]B$. \\
      We have $[\Gamma]B = [\Gamma, \beta]B$, so $|[\Gamma]B| = |[\Gamma, \beta]B|$;
      by transitivity of equality,  $|[\Gamma]B| = |[\Delta]B|$.

      \DerivationProofCase{\InstRSolve}
              { \Gamma_0 \entails \tau}
              { \instjudgr{\Gamma_0, \ahat, \Gamma_1}
                          {\ahat}
                          {\tau}
                          {\Gamma_0, \hypeq{\ahat}{\tau}, \Gamma_1}
              }

              Similar to the \InstLSolve case.

      \DerivationProofCase{\InstRReach}
              { }
              {\instjudgr{\Gamma[\ahat][\bhat]}
                         {\ahat}
                         {\bhat}
                         {\Gamma[\ahat][\hypeq{\bhat}{\ahat}]}}

              Similar to the \InstLReach case.

      \DerivationProofCase{\InstRArr}
            {\instjudg{\overbrace{\Gamma_0, \ahat_2, \ahat_1, \hypeq{\ahat}{\ahat_1 \arr \ahat_2}, \Gamma_1}^{\Gamma'}}
                      {\ahat_1}
                      {A_1}
                      {\Theta} \\
               \instjudgr{\Theta}
                         {\ahat_2}
                         {[\Theta]A_2}
                         {\Delta}}
            {\instjudgr{\underbrace{\Gamma_0, \ahat, \Gamma_1}_{\Gamma}}
                       {\ahat}
                       {A_1 \arr A_2}
                       {\Delta}}

           Similar to the \InstLArr case.

      \DerivationProofCase{\InstRAllL}
            {\instjudgr{\Gamma'[\ahat], \MonnierComma{\bhat}, \bhat}{\ahat}{[\bhat/\beta]A_0}{\Delta, \MonnierComma{\bhat}, \Delta'}}
            {\instjudgr{\Gamma'[\ahat]}{\ahat}{\alltype{\beta}{A_0}}{\Delta}}

          We have $\judgetp{\Gamma}{B}$ and $\ahat \notin \FV{[\Gamma]B}$. \\
          By weakening, $\judgetp{\Gamma, \MonnierComma{\bhat}, \bhat}{B}$. \\
          From the definition of substitution, $[\Gamma]B = [\Gamma, \MonnierComma{\bhat}, \bhat]B$.
          Hence $\ahat \notin \FV{[\Gamma, \MonnierComma{\bhat}, \bhat]B}$. \\
          By induction, $|[\Gamma, \MonnierComma{\bhat}, \bhat]B| = |[\Delta, \MonnierComma{\bhat}, \Delta']B|$. \\
          Suppose $u$ is a free variable in $B$. \\
          Then $u$ is declared in $\Gamma$, and so occurs before $\MonnierComma{\bhat}$ in $\Gamma, \MonnierComma{\bhat}, \bhat$. \\
          By \Lemmaref{lem:declaration-order-preservation}, $u$ is declared before $\MonnierComma{\bhat}$ in $\Delta, \MonnierComma{\bhat}, \Delta'$. \\
          So every free variable $u$ in $B$ is declared in $\Delta$. \\
          Hence $[\Delta, \MonnierComma{\bhat}, \Delta']B = [\Delta]B$. \\
          Since $[\Gamma]B = [\Gamma, \MonnierComma{\bhat}, \bhat]B$, we have
          $|[\Gamma]B| = |[\Gamma, \MonnierComma{\bhat}, \bhat]B|$;
          by transitivity of equality,
          $|[\Gamma]B| = |[\Delta]B|$.
          \qedhere
  \end{itemize}
\end{proof}

\instantiationdecidability*
\begin{proof}
  By induction on the derivation of $\judgetp{\Gamma}{A}$. 

  \begin{enumerate}[(1)]
  \item $\instjudg{\Gamma}{\ahat}{A}{\Delta}$ is decidable. 

      \begin{itemize}
       \DerivationProofCase{\UvarWF}
                           { }
                           {\judgetp{\underbrace{\Gamma_L, \ahat, \Gamma_R}_{\Gamma'[\alpha]}}{\alpha}}

          If $\alpha \in \Gamma_L$,
          then by \UvarWF we have $\judgetp{\Gamma_L}{\alpha}$,
          and by rule \InstLSolve we have a derivation. \\
          Otherwise no rule matches, and so no derivation exists. 
       
       \ProofCaseRule{\UnitWF}
          By rule \InstLSolve.

       \DerivationProofCase{\EvarWF}
                           { }
                           {\judgetp{\underbrace{\Gamma_L, \ahat, \Gamma_R}_{\Gamma}}{\bhat}}

           By inversion, we have $\bhat \in \Gamma$, and $[\Gamma]\bhat = \bhat$.
           Since $\ahat \notin \FV{[\Gamma]\bhat} = \FV{\bhat} = \{\bhat\}$,
           it follows that $\ahat \neq \bhat$:
           Either $\bhat \in \Gamma_L$ or $\bhat \in \Gamma_R$. \\
           If $\bhat \in \Gamma_L$, then we have a derivation by $\InstLSolve$. \\
           If $\bhat \in \Gamma_R$, then we have a derivation by $\InstLReach$. 

       \DerivationProofCase{\SolvedEvarWF}
                           { }
                           {\judgetp{\underbrace{\Gamma'[\hypeq{\bhat}{\tau}]}_{\Gamma}}{\bhat}}

           It is given that $[\Gamma]\bhat = \bhat$, so this case is impossible.

       \DerivationProofCase{\ArrowWF}
                           {\judgetp{\Gamma}{A_1} \\ \judgetp{\Gamma}{A_2}}
                           {\judgetp{\underbrace{\Gamma_L, \ahat, \Gamma_R}_\Gamma}{A_1 \arr A_2}}

           By assumption, $[\Gamma](A_1 \arr A_2) = A_1 \arr A_2$
           and $\ahat \notin \FV{[\Gamma](A_1 \arr A_2)}$. \\
           If $A_1 \arr A_2$ is a monotype and is well-formed under $\Gamma_L$, we can apply \InstLSolve. \\
           Otherwise, the only rule with a conclusion matching $A_1 \arr A_2$ is \InstLArr. \\
           First, consider whether $\instjudgr{\Gamma_L, \ahat_2, \ahat_1, \hypeq{\ahat}{\ahat_1 \arr \ahat_2}, \Gamma_R}{\ahat_1}{A}{-}$
           is decidable. \\[1ex]
           By definition of substitution,
             $[\Gamma](A_1 \arr A_2) = ([\Gamma]A_1) \arr ([\Gamma]A_2)$.
             Since $[\Gamma](A_1 \arr A_2) = A_1 \arr A_2$,
             we have $[\Gamma]A_1 = A_1$ and $[\Gamma]A_2 = A_2$. \\
           By weakening,
             $\judgetp{\Gamma_L, \ahat_2, \ahat_1, \hypeq{\ahat}{\ahat_1 \arr \ahat_2}, \Gamma_R}{A_1 \arr A_2}$. \\
           Since $\judgetp{\Gamma}{A_1}$ and $\judgetp{\Gamma}{A_2}$, we have
           $\ahat_1, \ahat_2 \notin \FV{A_1} \union \FV{A_2}$. \\
           Since $\ahat \notin \FV{A} \supseteq \FV{A_1}$, it follows that $[\Gamma']A_1 = A_1$. \\
           By i.h., either there exists $\Theta$ such that $\instjudgr{\Gamma_L, \ahat_2, \ahat_1, \hypeq{\ahat}{\ahat_1 \arr \ahat_2}, \Gamma_R}{\ahat_1}{A_1}{\Theta}$, or not. \\
           If not, then no derivation by \InstLArr exists. \\
           If so, then we have $\instjudg{\Gamma_L, \ahat_2, \ahat_1, \hypeq{\ahat}{\ahat_1 \arr \ahat_2}, \Gamma_R}{\ahat_1}{A_1}{\Theta}$. \\
           By \Lemmaref{lem:left-unsolvedness-preservation}, we know that $\ahat_2 \in \unsolved{\Theta}$. \\
           By \Lemmaref{lem:left-free-variable-preservation}, we know that $\ahat_2 \notin \FV{[\Theta]A_2}$. \\
           Clearly, $[\Theta]([\Theta]A_2) = [\Theta]A_2$. \\
           Hence by i.h., either there exists $\Delta$ such that $\instjudg{\Theta}{\ahat_2}{[\Theta]A_2}{\Delta}$, or not. \\
           If not, then no derivation by \InstLArr exists. \\
           If it does, then by rule \InstLArr, we have $\instjudg{\Gamma}{\ahat}{A}{\Delta}$.

      \DerivationProofCase{\ForallWF}
                          {\judgetp{\Gamma, \alpha}{A_0}}
                          {\judgetp{\Gamma}{\alltype{\alpha}{A_0}}}

           We have $\alltype{\alpha} A_0 = [\Gamma](\alltype{\alpha}A_0)$.
           By definition of substitution, $[\Gamma](\alltype{\alpha}A_0) = \alltype{\alpha}[\Gamma]A_0$, so $A_0 = [\Gamma]A_0$. \\
           By definition of substitution, $[\Gamma, \alpha]A_0 = [\Gamma]A_0$. \\
           We have $\ahat \notin \FV{[\Gamma](\alltype{\alpha} A_0)}$.  Therefore
           $\ahat \notin \FV{[\Gamma]A_0} = \FV{[\Gamma, \alpha]A_0}$. \\
           By i.h., either there exists $\Theta$
             such that $\instjudg{\Gamma, \alpha}{\ahat}{A_0}{\Theta}$, or not.
           \\[1ex]
           Suppose $\instjudg{\Gamma, \alpha}{\ahat}{A_0}{\Theta}$. \\
           By \Lemmaref{lem:instantiation-extension}, $\substextend{\Gamma}{\Theta}$; \\
           by \Lemmaref{lem:extension-order} (i), $\Theta = \Delta, \alpha, \Delta'$. \\
           Hence by rule \InstLAllR, $\instjudg{\Gamma}{\ahat}{\alltype{\alpha} A_0}{\Delta}$.
           \\[1ex]
           Suppose not. \\
           Then there is no derivation, since \InstLAllR is the only rule matching $\alltype{\alpha} A_0$. 
      \end{itemize}

  \item $\instjudgr{\Gamma}{\ahat}{A}{\Delta}$ is decidable. 
      \begin{itemize}
       \ProofCaseRule{\UvarWF}

                           Similar to the \UvarWF case in part (1), but applying rule \InstRSolve instead
                           of \InstLSolve.

       \ProofCaseRule{\UnitWF}
            Apply \InstRSolve.

       \DerivationProofCase{\EvarWF}
                           { }
                           {\judgetp{\underbrace{\Gamma_L, \ahat, \Gamma_R}_\Gamma}{\bhat}}

            Similar to the \EvarWF case in part (1), but applying
            \InstRSolve/\InstRReach instead of \InstLSolve/\InstLReach.

       \ProofCaseRule{\SolvedEvarWF} ~\\
           Impossible, for exactly the same reasons as in the \SolvedEvarWF case of part (1).

       \DerivationProofCase{\ArrowWF}
                           {\judgetp{\Gamma}{A_1} \\ \judgetp{\Gamma}{A_2}}
                           {\judgetp{\underbrace{\Gamma_L, \ahat, \Gamma_R}_\Gamma}{A_1 \arr A_2}}

              As the \ArrowWF case of part (1), except applying \InstRArr instead of \InstLArr.

      \DerivationProofCase{\ForallWF}
                          {\judgetp{\Gamma, \beta}{B}}
                          {\judgetp{\underbrace{\Gamma_L, \ahat, \Gamma_R}_\Gamma}{\alltype{\beta}{B}}}

           By assumption, $[\Gamma](\alltype{\beta}{B}) = \alltype{\beta}{B}$.
           With the definition of substitution, we get $[\Gamma]B = B$.
           Hence $[\Gamma]B = B$. \\
           Hence $[\bhat/\beta][\Gamma]B = [\bhat/\beta]B$.
           Since $\bhat$ is fresh, $[\bhat/\beta][\Gamma]B = [\Gamma][\bhat/\beta]B$. \\
           By definition of substitution, $[\Gamma, \MonnierComma{\bhat}, \bhat][\bhat/\beta]B = [\Gamma][\bhat/\beta]B$, which by transitivity of equality is $[\bhat/\beta]B$. \\
           We have $\ahat \notin \FV{[\Gamma](\alltype{\beta}{B})}$,
             so $\ahat \notin \FV{[\Gamma, \MonnierComma{\bhat}, \bhat][\bhat/\beta]B}$. \\
           Therefore, by induction, either $\instjudgr{\Gamma, \MonnierComma{\bhat}, \bhat}{\ahat}{[\bhat/\beta]B}{\Theta}$ or not.
           \\[1ex]
           Suppose $\instjudgr{\Gamma, \MonnierComma{\bhat}, \bhat}{\ahat}{[\bhat/\beta]B}{\Theta}$. \\
           By \Lemmaref{lem:instantiation-extension}, $\substextend{\Gamma, \MonnierComma{\bhat}, \bhat}{\Theta}$; \\
           by \Lemmaref{lem:extension-order} (ii), $\Theta = \Delta, \MonnierComma{\bhat}, \Delta'$. \\
           Hence by rule \InstRAllL, $\instjudgr{\Gamma}{\ahat}{\alltype{\beta}{B}}{\Delta}$.
           \\[1ex]
           Suppose not. \\
           Then there is no derivation, since \InstRAllL is the only rule matching $\alltype{\beta}{B}$. 
      \qedhere
      \end{itemize}
\end{enumerate}
\end{proof}

\section{Decidability of Algorithmic Subtyping}

\subsection{Lemmas for Decidability of Subtyping}

\monotypessolvevariables*
\begin{proof}  By induction on the given derivation. 

  \begin{itemize}
  \DerivationProofCase{\InstLSolve}
          { \Gamma_L \entails \tau}
          { \instjudg{\Gamma_L, \ahat, \Gamma_R}
                      {\ahat}
                      {\tau}
                      {\underbrace{\Gamma_L, \hypeq{\ahat}{\tau}, \Gamma_R}_{\Delta}}
           }

           It is evident that
           $|\unsolved{\Gamma_L, \ahat, \Gamma_R}| = |\unsolved{\Gamma_L, \hypeq{\ahat}{\tau}, \Gamma_R}| + 1$.

  \DerivationProofCase{\InstLReach}
          { }
          {\instjudg{\Gamma[\ahat][\bhat]}
                      {\ahat}
                      {\bhat}
                      {\Gamma[\ahat][\hypeq{\bhat}{\ahat}]}}

          Similar to the previous case.

  \DerivationProofCase{\InstLArr}
          {\instjudgr{\Gamma_0[\ahat_2, \ahat_1, \hypeq{\ahat}{\ahat_1 \arr \ahat_2}]}
                      {\ahat_1}
                      {\tau_1}
                      {\Theta} \\
           \instjudg{\Theta}
                      {\ahat_2}
                      {[\Theta]\tau_2}
                      {\Delta}}
          {\instjudg{\Gamma_0[\ahat]}
                      {\ahat}
                      {\tau_1 \arr \tau_2}
                      {\Delta}}

        \smallskip

        \begin{llproof}
            \eqPf{|\unsolved{\Gamma_0[\ahat_2, \ahat_1, \hypeq{\ahat}{\ahat_1 \arr \ahat_2}]}|}
                    {|\unsolved{\Gamma_0[\ahat]}| + 1}
                    {Immediate}
            \eqPf{|\unsolved{\Gamma_0[\ahat_2, \ahat_1, \hypeq{\ahat}{\ahat_1 \arr \ahat_2}]}|}
                   {|\unsolved{\Theta}| + 1}  {By i.h.}
            \eqPf{|\unsolved{\Gamma}|}{|\unsolved{\Theta}|}   {Subtracting $1$}
        \Hand    \continueeqPf{|\unsolved{\Delta}| + 1}   {By i.h.}
        \end{llproof}

  \DerivationProofCase{\InstLAllR}
        {\instjudg{\Gamma, \beta}{\ahat}{B}{\Delta, \beta, \Delta'}}
        {\instjudg{\Gamma}{\ahat}{\alltype{\beta}{B}}{\Delta}}

        This case is impossible, since a monotype cannot have the form $\alltype{\beta}{B}$.

  \ProofCasesRules{\InstRSolve, \InstRReach}
     Similar to the \InstLSolve and \InstLReach cases.

  \ProofCaseRule{\InstRArr}
     Similar to the \InstLArr case.

  \DerivationProofCase{\InstRAllL}
        {\instjudgr{\Gamma[\ahat], \beta}{\ahat}{B}{\Delta, \beta, \Delta'}}
        {\instjudgr{\Gamma[\ahat]}{\ahat}{\alltype{\beta}{B}}{\Delta}}

  This case is impossible, since a monotype cannot have the form $\alltype{\beta}{B}$.
  \qedhere
  \end{itemize}
\end{proof}

\monotypemonotonicity*
\begin{proof}  By induction on the given derivation.

  \begin{itemize}
    \ProofCasesRules{\SubVar, \SubExvar} ~\\
        In these rules, $\Delta = \Gamma$,
        so $\unsolved{\Delta} = \unsolved{\Gamma}$;
        therefore $|\unsolved{\Delta}| \leq |\unsolved{\Gamma}|$.

    \ProofCaseRule{\SubArr}
        We have an intermediate context $\Theta$.  
        
        By inversion, $\tau_1 = \tau_{11} \arr \tau_{12}$
        and $\tau_2 = \tau_{21} \arr \tau_{22}$.
        Therefore, we have monotypes in the first and second premises.

        By induction on the first premise,
        $|\unsolved{\Theta}| \leq |\unsolved{\Gamma}|$.
        By induction on the second premise,
        $|\unsolved{\Delta}| \leq |\unsolved{\Theta}|$.
        By transitivity of $\leq$, 
        $|\unsolved{\Delta}| \leq |\unsolved{\Gamma}|$,
        which was to be shown.

     \ProofCasesRules{\SubAllL, \SubAllR}
        We are given a derivation of subtyping on monotypes, so these
        cases are impossible.
     
     \ProofCasesRules{%
           \SubInstL, \SubInstR
        }
        The input and output contexts in the premise
        exactly match the conclusion, so the result follows
        by \Lemmaref{lem:monotypes-solve-variables}.
        \qedhere
  \end{itemize}
\end{proof}

\typesizesubst*
\begin{proof}  By induction on $\typesize{\Gamma}{A}$.
  If $A = \unitty$ or $A = \alpha$, or $A = \ahat$ and $\ahat \in \unsolved{\Gamma}$
  then $[\Gamma]A = A$. 
  Therefore, $\typesize{\Gamma}{[\Gamma]A} = \typesize{\Gamma}{A}$.

  If $A = \ahat$ and $(\hypeq{\ahat}{\tau}) \in \Gamma$, then
  by induction hypothesis, $\typesize{\Gamma}{[\Gamma]\tau} \leq \typesize{\Gamma}{\tau}$.
  Of course $\typesize{\Gamma}{\tau} \leq \typesize{\Gamma}{\tau} + 1$.
  By definition of substitution, $[\Gamma]\tau = [\Gamma]\ahat$, so
  \[
      \typesize{\Gamma}{[\Gamma]\ahat} ~\leq~ \typesize{\Gamma}{\tau} + 1
   \]
  By the definition of type size, $\typesize{\Gamma}{\ahat} = \typesize{\Gamma}{\tau} + 1$, so
  \[
      \typesize{\Gamma}{[\Gamma]\ahat} ~\leq~ \typesize{\Gamma}{\ahat}
   \]
   which was to be shown.

  If $A = A_1 \arr A_2$, the result follows via the induction hypothesis (twice).
  
  If $A = \alltype{\alpha}{A_0}$, the result follows via the induction hypothesis.
\end{proof}

\monotypecontextinvariance*
\begin{proof}
  By induction on the derivation of $\subjudg{\Gamma}{\tau}{\tau'}{\Delta}$.

  \begin{itemize}
      \ProofCasesRules{\SubVar, \SubUnit, \SubExvar}

          In these rules, the output context is the same as the input context, so
          the result is immediate.

      \DerivationProofCase{\SubArr}
                {\subjudg{\Gamma}{\tau'_1}{\tau_1}{\Theta}
                  \\
                  \subjudg{\Theta}{[\Theta]\tau_2}{[\Theta]\tau'_2}{\Delta}}
                {\subjudg{\Gamma}{\tau_1 \arr \tau_2}{\tau'_1 \arr \tau'_2}{\Delta}}

          We have that $[\Gamma](\tau_1 \arr \tau_2) = \tau_1 \arr \tau_2$.
          By definition of substitution,
          $[\Gamma]\tau_1 = \tau_1$ and $[\Gamma]\tau_2 = \tau_2$.
          Similarly, $[\Gamma]\tau_1 = \tau'_1$ and $[\Gamma]\tau_2 = \tau'_2$.
          
          By i.h., $\Theta = \Gamma$. \\
          Since $\Theta$ is predicative, $[\Theta]\tau_2$ and $[\Theta]\tau'_2$ are monotypes. \\
          Substitution is idempotent: $[\Theta][\Theta]\tau_2 = [\Theta]\tau_2$
          and $[\Theta][\Theta]\tau'_2 = [\Theta]\tau'_2$. \\
          By i.h., $\Delta = \Theta$.  Hence $\Delta = \Gamma$.  

      \ProofCasesRules{\SubAllL, \SubAllR}
          Impossible, since $\tau$ and $\tau'$ are monotypes.

      \DerivationProofCase{\SubInstL}
                {
                  \ahat \notin \FV{A}
                  \\
                  \instjudg{\Gamma_0[\ahat]}{\ahat}{A}{\Delta}
                }
                {\subjudg{\Gamma_0[\ahat]}{\ahat}{A}{\Delta}}

          By \Lemmaref{lem:monotypes-solve-variables},
          $|\unsolved{\Delta}| < |\unsolved{\Gamma_0[\ahat]}|$, but it is given that
          $|\unsolved{\Gamma_0[\ahat]}| = |\unsolved{\Delta}|$,
          so this case is impossible.

      \ProofCaseRule{\SubInstR}   Impossible, as for the \SubInstL case.
  \qedhere
  \end{itemize}
\end{proof}

\subsection{Decidability of Subtyping}

\subtypingdecidability*
\begin{proof}
  Let the judgment $\subjudg{\Gamma}{A}{B}{\Delta}$ be measured lexicographically by

  \begin{enumerate}[(S1)]
  \item the number of $\forall$ quantifiers in $A$ and $B$;
  \item $|\unsolved{\Gamma}|$, the number of unsolved existential variables in $\Gamma$;
  \item $\typesize{\Gamma}{A} + \typesize{\Gamma}{B}$.
  \end{enumerate}

  For each subtyping rule, we show that every premise is smaller than the conclusion.
  The condition that $[\Gamma]A = A$ and $[\Gamma]B = B$ is easily satisfied at each inductive
  step, using the definition of substitution.

    \begin{itemize}
      \item Rules \SubVar, \SubUnit and \SubExvar have no premises.

      \DerivationProofCase{\SubArr}
          {\subjudg{\Gamma}{B_1}{A_1}{\Theta}
            \\
            \subjudg{\Theta}{[\Theta]A_2}{[\Theta]B_2}{\Delta}
          }
          {
            \subjudg{\Gamma}{A_1 \arr A_2}{B_1 \arr B_2}{\Delta}
          }

          If $A_2$ or $B_2$ has a quantifier, then the first premise is smaller by (S1).
          Otherwise,
          the first premise shares an input context with the conclusion, so it has the same (S2).
          The types $B_1$ and $A_1$ are subterms of the conclusion's types,
          so the first premise is smaller by (S3).
          
          If $B_1$ or $A_1$ has a quantifier, then the second premise is smaller by (S1).
          Otherwise,
          by \Lemmaref{lem:monotype-monotonicity} on the first premise,
          $|\unsolved{\Theta}| \leq |\unsolved{\Gamma}|$.

          \begin{itemize}
          \item 
            If $|\unsolved{\Theta}| < |\unsolved{\Gamma}|$,
            then the second premise is smaller by (S2).
          
          \item
            If $|\unsolved{\Theta}| = |\unsolved{\Gamma}|$, we have the same (S2).

            However, by \Lemmaref{lem:monotype-context-invariance},
            $\Theta = \Gamma$, so
            $\typesize{\Theta}{[\Theta]A_2} = \typesize{\Gamma}{[\Gamma]A_2}$,
            which by \Lemmaref{lem:typesize-subst} is less than or equal to
            $\typesize{\Gamma}{A_2}$.
            
            By the same logic, $\typesize{\Theta}{[\Theta]B_2} \leq \typesize{\Gamma}{B_2}$.
            
            Therefore,
            \[
                \typesize{\Theta}{[\Theta]A_2} ~+~ \typesize{\Theta}{[\Theta]B_2}
                ~~~\leq~~~
                \typesize{\Gamma}{(A_1 \arr A_2)} ~+~ \typesize{\Gamma}{(B_1 \arr B_2)}
             \]
             and the second premise is smaller by (S3).
          \end{itemize}

      \ProofCasesRules{\SubAllL, \SubAllR}
          In each of these rules, the premise has one less quantifier than the conclusion,
          so the premise is smaller by (S1).

      \ProofCasesRules{\SubInstL, \SubInstR}
          Follows from \Theoremref{thm:decidability-of-instantiation}.
      \qedhere
  \end{itemize}
\end{proof}

\section{Decidability of Typing}

\typingdecidable*
\begin{proof}
    For rules deriving judgments of the form
    \[
        \begin{array}[t]{lll}
               \synjudg{\Gamma}{e}{-}{-}
            \\ \chkjudg{\Gamma}{e}{B}{-}
            \\ \appjudg{\Gamma}{e}{A}{-}{-}
        \end{array}
    \]
    (where we write ``$-$'' for parts of the judgments that are outputs),
    the following induction measure on such judgments is adequate to prove decidability:
    \[
      \left\langle
        e,~~~
        \begin{array}[c]{llll}
            \syn
          \\[0.3ex]
            \chk,
            &
               \typesize{\Gamma}{B}
          \\[0.6ex]
            \app,
            &
               \typesize{\Gamma}{A}
        \end{array}
      \right\rangle
    \]
    where $\langle \dots \rangle$ denotes lexicographic order, and
    where (when comparing two judgments typing terms of the same size)
    the synthesis judgment (top line) is considered smaller than
    the checking judgment (second line), which in turn is considered smaller than
    the application judgment (bottom line).  That is,
    \[
               \syn ~~\bigprec~~ \chk ~~\bigprec~~ \app
    \]
    Note that this measure only uses the input parts of the judgments, leading to a
    straightforward decidability argument.
    
    We will show that in each rule, every
    synthesis/checking/application premise is smaller than the
    conclusion.

    \begin{itemize}
      \ProofCaseRule{\Var}  No premises.

      \ProofCaseRule{\Sub}
          The first premise has the same subject term $e$ as the conclusion, but
          the judgment is smaller because the measure considers a synthesis judgment
          to be smaller than a checking judgment.

          The second premise is a subtyping judgment, which by \Theoremref{thm:subtyping-decidable}
          is decidable.

      \ProofCaseRule{\Anno}
      
          It is easy to show that the judgment $\judgetp{\Gamma}{A}$ is decidable. \\
          The second premise types $e$, but the conclusion types $(e : A)$,
          so the first part of the measure gets smaller.

      \ProofCaseRule{\UnitIntro}  No premises.

      \ProofCaseRule{\ArrIntro}  In the premise, the term is smaller.

      \ProofCaseRule{\ArrElim}  In both premises, the term is smaller.
      
      \ProofCaseRule{\AllIntro}  Both the premise and conclusion type $e$, and both are checking;
         however, $\typesize{\Gamma, \alpha}{A} < \typesize{\Gamma}{\alltype{\alpha}{A}}$,
         so the premise is smaller.

      \ProofCaseRule{\ArrApp}
         Both the premise and conclusion type $e$, but the premise is a
         checking judgment, so the premise is smaller.
      
      \ProofCaseRule{\SubstChk}  Both the premise and conclusion type $e$, and both are checking;
        however, since we can apply this rule only when $\Gamma$ has a solution for
        $\ahat$---that is, when $\Gamma = \Gamma_0[\hypeq{\ahat}{\tau}]$---we have that
        $\typesize{\Gamma}{[\Gamma]\ahat} < \typesize{\Gamma}{\ahat}$, making the last
        part of the measure smaller.

      \ProofCaseRule{\SubstApp}  Similar to \SubstChk.

      \ProofCaseRule{\AllApp}  Both the premise and conclusion type $e$, and both are application
        judgments; however, by the definition of $\typesize{\Gamma}{-}$, the size of
        the type in the premise $[\ahat/\alpha]A$ is smaller than $\alltype{\alpha}{A}$.

      \ProofCaseRule{\SolveApp}  Both the premise and conclusion type $e$, but we switch to
         checking in the premise, so the premise is smaller.
      
      \ProofCaseRule{\UnitIntroSyn}  No premises.

      \ProofCaseRule{\ArrIntroSyn}  In the premise, the term is smaller.
    \qedhere
    \end{itemize}
\end{proof}

\section{Soundness of Subtyping}

\subsection{Lemmas for Soundness}

\ctxappvarpreservation*
\begin{proof}  By mutual induction on $\Delta$ and $\Omega$.

  Suppose $(x : A) \in \Delta$.
  In the case where $\Delta = (\Delta', x : A)$ and $\Omega = (\Omega', x : A_\Omega)$,
  inversion on  $\substextend{\Delta}{\Omega}$ gives $[\Omega']A = [\Omega']A_\Omega$;
  by the definition of context application, $[\Omega', x : A_\Omega](\Delta', x : A) =
    [\Omega']\Delta', x : [\Omega']A_\Omega$, which contains $x : [\Omega']A_\Omega$,
    which is equal to $x : [\Omega']A$.  By well-formedness of $\Omega$, we know that
    $[\Omega']A = [\Omega]A$.

  Suppose $(x : A) \in \Omega$.  The reasoning is similar, because equality is symmetric.
\end{proof}

\substitutiontyping*
\begin{proof}
  By induction on $\typesize{\Gamma}{A}$ (the size of $A$ under $\Gamma$).
  
  \begin{itemize}
      \ProofCasesRules{\UvarWF, \UnitWF}
           Here $A = \alpha$ or $A = \unitty$, so applying $\Gamma$ to $A$ does not change it: $A = [\Gamma]A$.
           Since $\judgetp{\Gamma}{A}$, we have $\judgetp{\Gamma}{[\Gamma]A}$, which was to be shown.
      
      \ProofCaseRule{\EvarWF}
           In this case $A = \ahat$, but $\Gamma = \Gamma_0[\ahat]$, so applying $\Gamma$ to $A$ does not change it,
           and we proceed as in the \UnitWF case above.

      \ProofCaseRule{\SolvedEvarWF}
           In this case $A = \ahat$ and $\Gamma = \Gamma_L, \hypeq{\ahat}{\tau}, \Gamma_R$.
           Thus $[\Gamma]A = [\Gamma]\alpha = [\Gamma_L]\tau$.  We assume contexts are well-formed,
           so all free variables in $\tau$ are declared in $\Gamma_L$.  Consequently,
           $\typesize{\Gamma_L}{\tau} = \typesize{\Gamma}{\tau}$,
           which is less than $\typesize{\Gamma}{\ahat}$.  We can therefore apply the i.h. to $\tau$, yielding
           $\judgetp{\Gamma}{[\Gamma]\tau}$.  By the definition of substitution, $[\Gamma]\tau = [\Gamma]\ahat$,
           so we have $\judgetp{\Gamma}[\Gamma]\ahat$.

      \ProofCaseRule{\ArrowWF}
           In this case $A = A_1 \arr A_2$.
           By i.h., $\judgetp{\Gamma}{[\Gamma]A_1}$ and $\judgetp{\Gamma}{[\Gamma]A_2}$.
           By \ArrowWF, $\judgetp{\Gamma}{([\Gamma]A_1) \arr ([\Gamma]A_2)}$,
           which by the definition of substitution is
           $\judgetp{\Gamma}{[\Gamma](A_1 \arr A_2)}$.

      \ProofCaseRule{\ForallWF}
          In this case $A = \alltype{\alpha} A_0$.  By i.h., $\judgetp{\Gamma, \alpha}{[\Gamma, \alpha]A_0}$.
          By the definition of substitution, $[\Gamma, \alpha]A_0 = [\Gamma]A_0$,
          so by \ForallWF, $\judgetp{\Gamma}{\alltype{\alpha} [\Gamma]A_0}$,
          which by the definition of substitution is
          $\judgetp{\Gamma}{[\Gamma](\alltype{\alpha} A_0)}$.
  \qedhere
  \end{itemize}
\end{proof}

\completionwf*
\begin{proof}
    By induction on $\typesize{\Omega}{A}$, the size of $A$ under $\Omega$
    (\Definitionref{def:typesize}).
    
    We consider cases of the well-formedness rule concluding the derivation of
    $\judgetp{\Omega}{A}$.
    
    \begin{itemize}
        \DerivationProofCase{\UnitWF}
              { } 
              { \judgetp{\Omega}{\unitty} }

              \begin{llproof}
                \judgetpPf{[\Omega]\Omega} {\unitty}   {By \DeclUnitWF}
                \judgetpPf{[\Omega]\Omega} {[\Omega]\unitty}   {By definition of substitution}
              \end{llproof}

        \DerivationProofCase{\UvarWF}
              { }
              { \judgetp{\underbrace{\Omega'[\alpha]}_{\Omega}}{\alpha} }

              \begin{llproof}
                \substextendPf{\Omega}{\Omega}   {By \Lemmaref{lem:substextend-reflexivity}}
                \Pf{}{}{\alpha \in [\Omega]\Omega}  {By \Lemmaref{lem:ctxapp-uvar-preservation}}
                \judgetpPf{[\Omega]\Omega} {\alpha}   {By \DeclUvarWF}
                \judgetpPf{[\Omega]\Omega} {[\Omega]\alpha}   {By definition of substitution}
              \end{llproof}
        
        \DerivationProofCase{\SolvedEvarWF}
              { }
              { \judgetp{\underbrace{\Omega'[\hypeq{\ahat}{\tau}]}_{\Omega}}{\ahat} }

              \begin{llproof}
               \judgetpPf{\Omega}{\ahat}   {Given}
               \substextendPf{\Omega}{\Omega}    {By \Lemmaref{lem:substextend-reflexivity}}
               \judgetpPf{\Omega}{[\Omega]\ahat}   {By \Lemmaref{lem:substitution-typing}}
               \ltPf{\typesize{\Omega}{[\Omega]\ahat}}
                    {\typesize{\Omega}{\ahat}}     {Follows from definition of type size}
               \judgetpPf{[\Omega]\Omega}{[\Omega][\Omega]\ahat}   {By i.h.}
               \eqPf{[\Omega][\Omega]\ahat}{[\Omega]\ahat}   {By \Lemmaref{lem:subst-extension-invariance}}
               \judgetpPf{[\Omega]\Omega}{[\Omega]\ahat}   {Applying equality}
              \end{llproof}

        \DerivationProofCase{\EvarWF}
              { }
              { \judgetp{\underbrace{\Omega'[\ahat]}_{\Omega}}{\ahat} }
              
              Impossible: the grammar for $\Omega$ does not allow unsolved declarations.

        \DerivationProofCase{\ArrowWF}
              { \judgetp{\Omega}{A_1}
                \\
                \judgetp{\Omega}{A_2} }
              { \judgetp{\Omega}{A_1 \arr A_2} }
              
              \begin{llproof}
                \judgetpPf{\Omega}{A_1}    {Subderivation}
                \ltPf{\typesize{\Omega}{A_1}}
                   {\typesize{\Omega}{A_1 \arr A_2}} {Follows from definition of type size}
                \judgetpPf{[\Omega]\Omega}{[\Omega]A_1}   {By i.h.}
                \proofsep
                \judgetpPf{[\Omega]\Omega}{[\Omega]A_2}   {By similar reasoning on 2nd subderivation}
                \proofsep
                \judgetpPf{[\Omega]\Omega}{[\Omega]A_1 \arr [\Omega]A_2}   {By \DeclArrowWF}
                \judgetpPf{[\Omega]\Omega}{[\Omega](A_1 \arr A_2)}   {By definition of substitution}
              \end{llproof}

        \DerivationProofCase{\ForallWF}
              { \judgetp{\Omega, \alpha}{A_0} }
              { \judgetp{\Omega}{\alltype{\alpha}{A_0}} }
              
              \begin{llproof}
                \judgetpPf{\Omega, \alpha}{A_0}    {Subderivation}
                \LetPf{\Omega'}{(\Omega, \alpha)}  {}
                \ltPf{\typesize{\Omega'}{A_0}}
                   {\typesize{\Omega}{\alltype{\alpha}{A_0}}} {Follows from definition of type size}
                \judgetpPf{[\Omega'](\Omega, \alpha)}{[\Omega']A_0}   {By i.h.}
                \judgetpPf{[\Omega]\Omega, \alpha}{[\Omega']A_0}   {By definition of context application}
                \judgetpPf{[\Omega]\Omega, \alpha}{[\Omega]A_0}   {By definition of substitution}
                \judgetpPf{[\Omega]\Omega} {\alltype{\alpha} [\Omega]A_0}   {By \DeclForallWF}
                \judgetpPf{[\Omega]\Omega} {[\Omega](\alltype{\alpha} A_0)}   {By definition of substitution}
              \end{llproof}
              \qedhere
    \end{itemize}
\end{proof}

\substitutionstability*
\begin{proof}
  By induction on $\Omega_Z$.
  If $\Omega_Z = \cdot$, the result is immediate.
  Otherwise, use the i.h.\ and the fact that
  $\judgetp{\Omega}{A}$ implies $\FV{A} \sect \dom{\Omega_Z} = \emptyset$.
\end{proof}

\contextpartitioning*
\begin{proof}
  By induction on the given derivation.
  
  \begin{itemize}
      \ProofCaseRule{\substextendId}   Impossible: $\Delta, \MonnierComma{\ahat}, \Theta$ cannot have the form $\cdot$.

      \ProofCaseRule{\substextendVV}  We have $\Omega_Z = (\Omega_Z', x : A)$ and $\Theta = (\Theta', x : A')$.
        By i.h., there is $\Psi'$ such that 
        $[\Omega, \MonnierComma{\ahat}, \Omega_Z'](\Delta, \MonnierComma{\ahat}, \Theta')
       =
       [\Omega]\Delta, \Psi'$.  Then by the definition of context application,
       $[\Omega, \MonnierComma{\ahat}, \Omega_Z', x : A](\Delta, \MonnierComma{\ahat}, \Theta', x : A')
      = [\Omega]\Delta, \Psi', x : [\Omega']A$.  Let $\Psi = (\Psi', x : [\Omega']A)$.

      \ProofCaseRule{\substextendUU}  Similar to the \substextendVV case, with $\Psi = (\Psi', \alpha)$.

      \ProofCasesRules{\substextendEE, \substextendSolve, \substextendMonMon, \substextendAdd, \substextendAddSolved}
        Broadly similar to the \substextendUU case, but since the rightmost context element is soft it disappears in context application,
        so we let $\Psi = \Psi'$.
  \qedhere
  \end{itemize}
\end{proof}

\completesstability*
\begin{proof}
  By induction on the derivation of $\substextend{\Gamma}{\Omega}$. 

  \begin{itemize}

    \DerivationProofCase{\substextendId}
          { }
          {\substextend{\cdot}{\cdot}}

          In this case, $\Omega = \Gamma = \cdot$. \\
          By definition, $[\cdot]\cdot = \cdot$, which gives us the conclusion. 

    \DerivationProofCase{\substextendVV}
          {\substextend{\Gamma'}{\Omega'}
            \\
            [\Omega']A_\Gamma = [\Omega']A}
         {\substextend{\Gamma', x : A_\Gamma}{\Omega', x : A}}

         \begin{llproof}
           \eqPf {[\Omega']\Gamma'} {[\Omega']\Omega'}   {By i.h.}
           \eqPf{[\Omega']A_\Gamma} {[\Omega']A}   {Premise}
           \proofsep
           \eqPf{[\Omega]\Gamma } { [\Omega', x : A](\Gamma', x : A_\Gamma)}   {Expanding $\Omega$ and $\Gamma$}
           \continueeqPf{ [\Omega']\Gamma', x : [\Omega']A_\Gamma}  {By definition of context application}
               \trailingjust{(using $[\Omega']A_\Gamma = [\Omega']A$)}
           \continueeqPf{ [\Omega']\Omega', x : [\Omega']A}  {By above equalities}
           \continueeqPf{ [\Omega]\Omega}   {By definition of context application}
         \end{llproof}
                  
    \DerivationProofCase{\substextendUU}
          {\substextend{\Gamma'}{\Omega'}}
          {\substextend{\Gamma', \alpha}{\Omega', \alpha}}

          \begin{llproof}
            \eqPf{[\Omega]\Gamma} {[\Omega', \alpha](\Gamma', \alpha)}   {Expanding $\Omega$ and $\Gamma$}
            \continueeqPf{[\Omega']\Gamma', \alpha}  {By definition of context application}
            \continueeqPf{[\Omega']\Omega', \alpha}  {By i.h.}
            \continueeqPf{[\Omega', \alpha](\Omega', \alpha)}  {By definition of context application}
            \continueeqPf{[\Omega]\Omega} {By $\Omega = (\Omega', \alpha)$}
          \end{llproof}

    \DerivationProofCase{\substextendMonMon}
          {\substextend{\Gamma'}{\Omega'}}
          {\substextend{\Gamma', \MonnierComma{\ahat}}{\Omega', \MonnierComma{\ahat}}}

          Similar to the \substextendUU case.

    \DerivationProofCase{\substextendAddSolved}
          {\substextend{\Gamma}{\Omega'}}
          {\substextend{\Gamma}{\Omega', \hypeq{\ahat}{\tau}}}

          \begin{llproof}
            \eqPf{[\Omega]\Gamma} {[\Omega', \hypeq{\ahat}{\tau}]\Gamma}  {Expanding $\Omega$}
            \continueeqPf { [\Omega']\Gamma}  {By $\ahat \notin \dom{\Gamma}$}
            \continueeqPf { [\Omega']\Omega'}  {By i.h.}
            \continueeqPf { [\Omega', \hypeq{\ahat}{\tau}](\Omega', \hypeq{\ahat}{\tau})}
                      {By definition of context application}
            \continueeqPf { [\Omega]\Omega}   {By $\Omega = (\Omega', \hypeq{\ahat}{\tau})$}
          \end{llproof}
                   
    \DerivationProofCase{\substextendSolSol}
          {\substextend{\Gamma'}{\Omega'} \\
           [\Omega']\tau_\Gamma = [\Omega']\tau}
          {\substextend{\Gamma', \hypeq{\ahat}{\tau_\Gamma}}{\Omega', \hypeq{\ahat}{\tau}}}

          \begin{llproof}
            \eqPf{[\Omega]\Gamma} {[\Omega', \hypeq{\ahat}{\tau}](\Gamma', \hypeq{\ahat}{\tau_\Gamma})} {Expanding $\Omega$ and $\Gamma$}
                      \continueeqPf { [\Omega']\Gamma'} {By definition of context application}
                      \continueeqPf { [\Omega']\Omega'} {By i.h.}
                      \continueeqPf { [\Omega', \hypeq{\ahat}{\tau}](\Omega', \hypeq{\ahat}{\tau})}  {By definition of context application}
                      \continueeqPf { [\Omega]\Omega} {By $\Omega = (\Omega', \hypeq{\ahat}{\tau})$}
           \end{llproof}

    \DerivationProofCase{\substextendSolve}
          {\substextend{\Gamma'}{\Omega'}}
          {\substextend{\Gamma', \ahat}{\Omega', \hypeq{\ahat}{\tau}}}

          \begin{llproof}
            \eqPf{[\Omega]\Gamma} {[\Omega', \hypeq{\ahat}{\tau}](\Gamma', \ahat)}
                {Expanding $\Omega$ and $\Gamma$}
                      \continueeqPf { [\Omega']\Gamma'} {By definition of context application}
                      \continueeqPf { [\Omega']\Omega'} {By i.h.}
                      \continueeqPf { [\Omega', \hypeq{\ahat}{\tau}](\Omega', \hypeq{\ahat}{\tau})}
                           {By definition of context application}
                      \continueeqPf { [\Omega]\Omega} {By $\Omega = (\Omega', \hypeq{\ahat}{\tau})$}
            \end{llproof}

    \DerivationProofCase{\substextendEE}
        { \substextend{\Gamma}{\Delta} }
        { \substextend{\Gamma, \ahat}{\Delta, \ahat} }
    
        Impossible: $\Omega$ cannot have the form $\Delta, \ahat$.

    \DerivationProofCase{\substextendAdd}
        { \substextend{\Gamma}{\Delta} }
        { \substextend{\Gamma}{\Delta, \ahat} }

        Impossible: $\Omega$ cannot have the form $\Delta, \ahat$.
  \qedhere
  \end{itemize}
\end{proof}

\finishingtypes*
\begin{proof}
  By \Lemmaref{lem:subst-extension-invariance},
  $[\Omega']A = [\Omega'][\Omega]A$. \\
  If $\FEV{C} = \emptyset$ then $[\Omega']C = C$. \\
  Since $\Omega$ is complete and $\judgetp{\Omega}{A}$, we have
  $\FEV{[\Omega]A} = \emptyset$.  Therefore $[\Omega'][\Omega]A = [\Omega]A$.
\end{proof}

\finishingcompletions*
\begin{proof}
  By induction on the given derivation of $\substextend{\Omega}{\Omega'}$.

  Only cases \substextendId, \substextendVV, \substextendUU, \substextendSolSol,
  \substextendMonMon and \substextendAddSolved are possible.
  In all of these cases, we use the i.h. and the definition of context application;
  in cases \substextendVV and \substextendSolSol, we also use the equality in the
  premise of the respective rule.
\end{proof}

\confluenceofcompleteness*
\begin{proof} ~\\
  \begin{llproof}
    \substextendPf{\Delta_1}{\Omega}    {Given}
    \eqPf{[\Omega]\Delta_1}{[\Omega]\Omega}  {By \Lemmaref{lem:completes-stability}}
    \substextendPf{\Delta_2}{\Omega}    {Given}
    \eqPf{[\Omega]\Delta_2}{[\Omega]\Omega}  {By \Lemmaref{lem:completes-stability}}
    \eqPf{[\Omega]\Delta_1}{[\Omega]\Delta_2}  {By transitivity of equality}
  \end{llproof}
\end{proof}

\subsection{Instantiation Soundness}

\instantiationsoundness*
\begin{proof}
  By induction on the given instantiation derivation.

    \begin{enumerate}[(1)]
      \item
      \begin{itemize}
      \DerivationProofCase{\InstLSolve}
              { \Gamma_0 \entails \tau}
              { \instjudg{\underbrace{\Gamma_0, \ahat, \Gamma_1}_\Gamma}
                          {\ahat}
                          {\tau}
                         {\underbrace{\Gamma_0, \hypeq{\ahat}{\tau}, \Gamma_1}_{\Delta}}
               }

              In this case $[\Delta]\ahat = [\Delta]\tau$.
              By reflexivity of subtyping (\Lemmaref{lem:declarative-reflexivity}),
              $\declsubjudg{[\Omega]\Delta}{[\Delta]\ahat}{[\Delta]\tau}$. 

      \DerivationProofCase{\InstLReach}
              { }
              {\instjudg{\Gamma[\ahat][\bhat]}
                          {\ahat}
                          {\bhat}
                          {\underbrace{\Gamma[\ahat][\hypeq{\bhat}{\ahat}]}_{\Delta}}}

              We have $\Delta = \Gamma[\ahat][\hypeq{\bhat}{\ahat}]$.  Therefore $[\Delta]\ahat = \ahat = [\Delta]\bhat$. \\
              By reflexivity of subtyping (\Lemmaref{lem:declarative-reflexivity}),
              $\declsubjudg{[\Omega]\Delta}{[\Delta]\ahat}{[\Delta]\bhat}$.

      \DerivationProofCase{\InstLArr}
              {\instjudgr{\overbrace{\Gamma[\ahat_2, \ahat_1, \hypeq{\ahat}{\ahat_1 \arr \ahat_2}]}^{\Gamma_1}}
                          {\ahat_1}
                          {A_1}
                          {\Gamma'} \\
               \instjudg{\Gamma'}
                          {\ahat_2}
                          {[\Gamma']A_2}
                          {\Delta}}
              {\instjudg{\Gamma[\ahat]}
                          {\ahat}
                          {A_1 \arr A_2}
                          {\Delta}}

                  \begin{llproof}
                    \eqPf{[\Gamma](A_1 \arr A_2)} {[\Gamma_1](A_1 \arr A_2)}  {$\ahat \notin \FV{A_1 \arr A_2}$}
                    \notinPf{\ahat_1, \ahat_2} {\FV{A_1} \union \FV{A_2}}  {$\ahat_1, \ahat_2$ fresh}
                    \instjudgPf{\Gamma'}
                          {\ahat_2}
                          {[\Gamma']A_2}
                          {\Delta}  {Subderivation}
                     \substextendPf{\Gamma'}{\Delta}  {By \Lemmaref{lem:instantiation-extension}}
                     \substextendPf{\Delta}{\Omega}   {Given}
                     \substextendPf{\Gamma'}{\Omega}    {By  \Lemmaref{lem:substextend-transitivity}}
                     \proofsep
                     \instjudgrPf{\Gamma_1}
                          {\ahat_1}
                          {A_1}
                          {\Gamma'}      {Subderivation}
                      \declsubjudgPf{[\Omega]\Delta}{[\Omega]A_1}{[\Omega]\ahat_1}   {By i.h. and \Lemmaref{lem:completes-confluence}}
                      \proofsep
                      \instjudgPf{\Gamma'}
                           {\ahat_2}
                           {[\Gamma']A_2}
                           {\Delta}  {Subderivation}
                      \declsubjudgPf{[\Omega]\Delta}{[\Omega][\Gamma']\ahat_2}{[\Omega][\Gamma']A_2}
                              {By i.h.}
                      \substextendPf{\Gamma'}{\Omega}  {Above}
                      \declsubjudgPf{[\Omega]\Delta}{[\Omega]\ahat_2}{[\Omega]A_2}
                              {By \Lemmaref{lem:subst-extension-invariance}}
                      \proofsep
                      \declsubjudgPf{[\Omega]\Delta}{[\Omega](\ahat_1 \arr \ahat_2)}{[\Omega]A_1 \arr [\Omega]A_2} 
                                     {By \DsubArr and definition of substitution}
                  \end{llproof}
                  
                  Since $(\hypeq{\ahat}{\ahat_1 \arr \ahat_2}) \in \Gamma_1$
                  and $\substextend{\Gamma_1}{\Delta}$, we know that $[\Omega]\ahat = [\Omega](\ahat_1 \arr \ahat_2)$. \\
                  Therefore $\declsubjudg{[\Omega]\Delta}{[\Omega]\ahat}{[\Omega](A_1 \arr A_2)}$.

      \DerivationProofCase{\InstLAllR}
            {\instjudg{\Gamma[\ahat], \beta}{\ahat}{B_0}{\Delta, \beta, \Delta'}}
            {\instjudg{\Gamma[\ahat]}{\ahat}{\alltype{\beta}{B_0}}{\Delta}}

                  We have $\substextend{\Delta}{\Omega}$
                  and $[\Gamma[\ahat]](\alltype{\beta}{B_0}) = \alltype{\beta}{B_0}$
                  and $\ahat \notin \FV{\alltype{\beta}{B_0}}$. \\
                  Hence $\ahat \notin \FV{B_0}$ and by definition, $[\Gamma[\ahat], \beta]B_0 = B_0$. \\
                  By \Lemmaref{lem:soln-completes}, $\substextend{\Delta, \beta, \Delta'}{\Omega, \beta, |\Delta'|}$. \\
                  By induction, $\declsubjudg{[\Omega, \beta, |\Delta'|](\Delta, \beta, \Delta')}{[\Omega, \beta, |\Delta'|]\ahat}{[\Omega, \beta, |\Delta'|]B_0}$. \\
                  Each free variable in $\ahat$ and $B_0$ is declared in $(\Omega, \beta)$, so $\Omega, \beta, |\Delta'|$
                  behaves as $[\Omega, \beta]$ on $\ahat$ and on $B_0$, yielding
                 $\declsubjudg{[\Omega, \beta, |\Delta'|](\Delta, \beta, \Delta')}{[\Omega, \beta]\ahat}{[\Omega, \beta]B_0}$. \\
                  By \Lemmaref{lem:context-partitioning} and thinning, 
                  $\declsubjudg{[\Omega, \beta](\Delta, \beta)}{[\Omega, \beta]\ahat}{[\Omega, \beta]B_0}$. \\
                  By the definition of context application,
                    $\declsubjudg{[\Omega]\Delta, \beta}{[\Omega, \beta]\ahat}{[\Omega, \beta]B_0}$. \\
                  By the definition of substitution, $\declsubjudg{[\Omega]\Delta, \beta}{[\Omega]\ahat}{[\Omega]B_0}$. \\
                  Since $\ahat$ is declared to the left of $\beta$, we have $\beta \notin \FV{[\Omega]\ahat}$. \\
                  Applying rule \DsubAllL gives
                    $\declsubjudg{[\Omega]\Delta}{[\Omega]\ahat}{\alltype{\beta}{[\Omega]B_0}}$.
      \end{itemize}

      \item
          \begin{itemize}
          \DerivationProofCase{\InstRSolve}
                  { \Gamma_0 \entails \tau}
                  { \instjudgr{\underbrace{\Gamma_0, \ahat, \Gamma_1}_\Gamma}
                              {\ahat}
                              {\tau}
                              {\underbrace{\Gamma_0, \hypeq{\ahat}{\tau}, \Gamma_1}_{\Gamma'}}
                   }

                Similar to the \InstLSolve case.

          \DerivationProofCase{\InstRReach}
                  { }
                  {\instjudgr{\Gamma[\ahat][\bhat]}
                             {\ahat}
                             {\bhat}
                             {\underbrace{\Gamma[\ahat][\hypeq{\bhat}{\ahat}]}_{\Gamma'}}}

              Similar to the \InstLReach case.

          \DerivationProofCase{\InstRArr}
                {\instjudg{\Gamma[\ahat_2, \ahat_1, \hypeq{\ahat}{\ahat_1 \arr \ahat_2}]}
                          {\ahat_1}
                          {A_1}
                          {\Gamma'} \\
                   \instjudgr{\Gamma'}
                             {\ahat_2}
                             {[\Gamma']A_2}
                             {\Delta}}
                {\instjudgr{\Gamma[\ahat]}
                           {\ahat}
                           {A_1 \arr A_2}
                           {\Delta}}

                 Similar to the \InstLArr case.

          \DerivationProofCase{\InstRAllL}
                {\instjudgr{\Gamma[\ahat], \MonnierComma{\bhat}, \bhat}{\ahat}{[\bhat/\beta]B_0}{\Delta, \MonnierComma{\bhat}, \Delta'}}
                {\instjudgr{\Gamma[\ahat]}{\ahat}{\alltype{\beta}{B_0}}{\Delta}}

                \medskip

                \begin{llproof}
                  \eqPf{\big[\Gamma[\ahat]\big](\alltype{\beta}{B_0})} {\alltype{\beta}{B_0}} {Given}
                  \eqPf{\big[\Gamma[\ahat]\big]B_0} {B_0} {~}
                  \eqPf{\big[\Gamma[\ahat], \MonnierComma{\bhat}, \bhat\big][\bhat/\beta]B_0}
                       {[\bhat/\beta]B_0}   {~}
                  \proofsep
                  \substextendPf{\Delta}{\Omega}   {Given}
                  \substextendPf{\Delta, \MonnierComma{\bhat}, \Delta'}
                                {\Omega, \MonnierComma{\bhat}, |\Delta'|}
                                {By \Lemmaref{lem:soln-completes}}
                  \proofsep
                  \notinPf{\ahat} {\FV{\alltype{\beta}{B_0}}}  {Given}
                  \notinPf{\ahat} {\FV{B_0}}  {By definition of $\FV{-}$}
                  \proofsep
                  \decolumnizePf
                  \instjudgrPf{\Gamma[\ahat], \MonnierComma{\bhat}, \bhat}{\ahat}{[\bhat/\beta]B_0}{\Delta, \MonnierComma{\bhat}, \Delta'}  {Subderivation}
                   \declsubjudgPf{[\Omega, \MonnierComma{\bhat}, |\Delta'|](\Delta, \MonnierComma{\bhat}, \Delta')}
                                {[\Omega, \MonnierComma{\bhat}, |\Delta'|][\bhat/\beta]B_0}
                                {[\Omega, \MonnierComma{\bhat}, |\Delta'|]\ahat}
                                {By i.h.}
                   \substextendPf{\Gamma[\ahat], \MonnierComma{\bhat}, \bhat}
                                 {\Delta, \MonnierComma{\bhat}, \Delta'}
                                 {\hspace{-60pt}By \Lemmaref{lem:instantiation-extension}}
                \end{llproof} 

              By \Lemmaref{lem:declaration-order-preservation},
                $\ahat$ is declared before $\MonnierComma{\bhat}$, that is, in $\Omega$. \\
              Thus, $\big[\Omega, \MonnierComma{\bhat}, |\Delta'|\big]\ahat = [\Omega]\ahat$. \\
              By \Lemmaref{lem:evar-input}, we know that $\Delta'$ is soft, so
              by \Lemmaref{lem:softness-goes-away},
              $[\Omega, \MonnierComma{\bhat}, |\Delta'|](\Delta, \MonnierComma{\bhat}, \Delta') = 
                         [\Omega, \MonnierComma{\bhat}](\Delta, \MonnierComma{\bhat}) = 
                         [\Omega]\Delta$. \\
              Applying these equalities to the derivation above gives
              \[
                 \declsubjudg{[\Omega]\Delta}{\big[\Omega, \MonnierComma{\bhat}, |\Delta'|\big][\bhat/\beta]B_0}{[\Omega]\ahat}
              \]
              By distributivity of substitution,
              \[
                  \declsubjudg{[\Omega]\Delta}
                          {\big[[\Omega, \MonnierComma{\bhat}, |\Delta'|]\bhat/\beta\big]\big[\Omega, \MonnierComma{\bhat}, |\Delta'|\big]B_0}{[\Omega]\ahat}
              \]
              Furthermore, $[\Omega, \MonnierComma{\bhat}, |\Delta'|]B_0 = [\Omega]B_0$, since $B_0$'s free variables are either $\beta$ or in $\Omega$, giving
              \[
                   \declsubjudg{[\Omega]\Delta}{\big[[\Omega, \MonnierComma{\bhat}, |\Delta'|]\bhat/\beta\big][\Omega]B_0}{[\Omega]\ahat}
              \]
              Now apply \DsubAllL and the definition of substitution to get
                $\declsubjudg{[\Omega]\Delta}{[\Omega](\alltype{\beta}{B_0})}{[\Omega]\ahat}$.
          \qedhere
      \end{itemize}
    \end{enumerate}
\end{proof}

\subsection{Soundness of Subtyping}

\soundness*
\begin{proof} By induction on the derivation of $\subjudg{\Gamma}{A}{B}{\Delta}$.

  \begin{itemize}
     \DerivationProofCase{\SubVar}
          { }
          {\subjudg{\underbrace{\Gamma'[\alpha]}_{\Gamma}}{\alpha}{\alpha}{\underbrace{\Gamma'[\alpha]}_{\Delta}}}
          
          \begin{llproof}
            \inPf{\alpha}{\Delta}  {$\Delta = \Gamma'[\alpha]$}
            \inPf{\alpha}{[\Omega]\Delta}  {Follows from definition of context application}
            \declsubjudgPf{[\Omega]\Delta} {\alpha} {\alpha} {By \DsubVar}
            \declsubjudgPf{[\Omega]\Delta} {[\Omega]\alpha} {[\Omega]\alpha} {By def. of substitution}
          \end{llproof}

     \ProofCaseRule{\SubUnit}
          Similar to the \SubVar case, applying rule \DsubUnit instead of \DsubVar.

     \DerivationProofCase{\SubExvar}
          { }
          {\subjudg{\Gamma_L, \ahat, \Gamma_R}{\ahat}{\ahat}{\Gamma_L, \ahat, \Gamma_R}}

          \begin{llproof}
               \Pf{}{}{[\Omega]\ahat\text{~defined}}  {Follows from definition of context application}
               \judgetpPf{[\Omega]\Delta} {[\Omega]\ahat}   {Assumption that $[\Omega]\Delta$ is well-formed}
               \declsubjudgPf{[\Omega]\Delta}{[\Omega]\ahat}{[\Omega]\ahat}   {By \Lemmaref{lem:declarative-reflexivity}}
          \end{llproof}

     \DerivationProofCase{\SubArr}
          {\subjudg{\Gamma}{B_1}{A_1}{\Theta} \\
            \subjudg{\Theta}{[\Theta]A_2}{[\Theta]B_2}{\Delta}}
          {\subjudg{\Gamma}{\underbrace{A_1 \arr A_2}_{A}}{\underbrace{B_1 \arr B_2}_{B}}{\Delta}}

          \begin{llproof}
            \subjudgPf{\Gamma}{B_1}{A_1}{\Theta} {Subderivation}
            \Pf{}{}{\substextend{\Delta}{\Omega}} {Given}
            \Pf{}{}{\substextend{\Theta}{\Omega}} {By \Lemmaref{lem:substextend-transitivity}} 
            \declsubjudgPf{[\Omega]\Theta} {[\Omega]B_1} {[\Omega]A_1}  {By i.h.}
            \declsubjudgPf{[\Omega]\Delta} {[\Omega]B_1} {[\Omega]A_1}  {By \Lemmaref{lem:completes-confluence}}
            \proofsep
            \subjudgPf{\Theta}{[\Theta]A_2}{[\Theta]B_2}{\Delta} {Subderivation}
            \declsubjudgPf{[\Omega]\Delta} {[\Omega][\Theta]A_2} {[\Omega][\Theta]B_2}  {By i.h.}
            \Pf{}{}{[\Omega][\Theta]A_2 = [\Omega]A_2}  {By \Lemmaref{lem:subst-extension-invariance}}
            \Pf{}{}{[\Omega][\Theta]B_2 = [\Omega]B_2}  {By \Lemmaref{lem:subst-extension-invariance}}
            \declsubjudgPf{[\Omega]\Delta} {[\Omega]A_2} {[\Omega]B_2}  {Above equations}
            \proofsep
            \declsubjudgPf{[\Omega]\Delta} {([\Omega]A_1) \arr ([\Omega]A_2)}
                                        {([\Omega]B_1) \arr ([\Omega]B_2)}  {By \DsubArr}
            \declsubjudgPf{[\Omega]\Delta} {[\Omega](A_1 \arr A_2)}
                                        {[\Omega](B_1 \arr B_2)}  {By def. of substitution}
          \end{llproof}

     \DerivationProofCase{\SubAllL}
          {\subjudg{\Gamma, \MonnierComma{\ahat}, \ahat}
                   {[\ahat/\alpha]A_0}
                   {B}
                   {\Delta, \MonnierComma{\ahat}, \Theta}}
          {\subjudg{\Gamma}{\alltype{\alpha}{A_0}}{B}{\Delta}}

          Let $\Omega' = (\Omega, \soln{\MonnierComma{\ahat}, \Theta})$.
          
          \begin{llproof}
            \subjudgPf{\Gamma, \MonnierComma{\ahat}, \ahat}
                   {[\ahat/\alpha]A_0}
                   {B}
                   {\Delta, \MonnierComma{\ahat}, \Theta} 
                       {Subderivation}
            \proofsep
            \substextendPf{\Delta}{\Omega}   {Given}
            \substextendPf{(\Delta, \MonnierComma{\ahat}, \Theta)}{\Omega'}   {By \Lemmaref{lem:soln-completes}}
            \proofsep
            \declsubjudgPf{[\Omega'](\Delta, \MonnierComma{\ahat}, \Theta)}
                          {[\Omega'][\ahat/\alpha]A_0}
                          {[\Omega']B}
                       {By i.h.}
            \declsubjudgPf{[\Omega'](\Delta, \MonnierComma{\ahat}, \Theta)}
                          {[\Omega'][\ahat/\alpha]A_0}
                          {[\Omega]B}
                       {By $[\Omega']B =[\Omega]B$ (\Lemmaref{lem:substitution-stability})}
            \declsubjudgPf{[\Omega'](\Delta, \MonnierComma{\ahat}, \Theta)}
                          {\big[[\Omega']\ahat/\alpha\big][\Omega']A_0}
                          {[\Omega]B}
                       {By distributivity of substitution}
            \proofsep
            \judgetpPf{\Gamma, \MonnierComma{\ahat}, \ahat} {\ahat} {By \EvarWF}
            \substextendPf{\Gamma, \MonnierComma{\ahat}, \ahat}  {\Delta, \MonnierComma{\ahat}, \Theta}  {By \Lemmaref{lem:subtyping-extension}}
            \judgetpPf{\Delta, \MonnierComma{\ahat}, \Theta} {\ahat}  {By \Lemmaref{lem:extension-weakening}}
            \substextendPf{(\Delta, \MonnierComma{\ahat}, \Theta)}{\Omega'}   {Above}
            \judgetpPf{[\Omega']\Omega'} {[\Omega']\ahat}   {By \Lemmaref{lem:completion-wf}}
            \judgetpPf{[\Omega'](\Delta, \MonnierComma{\ahat}, \Theta)} {[\Omega']\ahat}   {By \Lemmaref{lem:completes-stability}}
            \proofsep
            \declsubjudgPf{[\Omega'](\Delta, \MonnierComma{\ahat}, \Theta)}
                          {\alltype{\alpha}{[\Omega']A_0}}
                          {[\Omega]B}   {By \DsubAllL}
            \declsubjudgPf{[\Omega'](\Delta, \MonnierComma{\ahat}, \Theta)}
                          {\alltype{\alpha}{[\Omega, \alpha]A_0}}
                          {[\Omega]B}   {By \Lemmaref{lem:substitution-stability}}
            \declsubjudgPf{[\Omega]\Delta}
                          {\alltype{\alpha}{[\Omega, \alpha]A_0}}
                          {[\Omega]B}   {By \Lemmaref{lem:context-partitioning} and thinning}
            \declsubjudgPf{[\Omega]\Delta}
                          {\alltype{\alpha}{[\Omega]A_0}}
                          {[\Omega]B}   {By def.\ of substitution}
            \declsubjudgPf{[\Omega]\Delta}
                          {[\Omega](\alltype{\alpha}{A_0})}
                          {[\Omega]B}   {By def.\ of substitution}
          \end{llproof}

     \DerivationProofCase{\SubAllR}
           {\subjudg{\Gamma, \alpha}{A}{B_0}{\Delta, \alpha, \Theta}}
           {\subjudg{\Gamma}{A}{\alltype{\alpha}{B_0}}{\Delta}}

           \begin{llproof}
             \subjudgPf{\Gamma, \alpha}{A}{B_0}{\Delta, \alpha, \Theta}  {Subderivation}
             \LetPf{\Omega_Z}{\soln{\Theta}}   {}
             \LetPf{\Omega'}{(\Omega, \alpha, \Omega_Z)} {}
             \substextendPf{(\Delta, \alpha, \Theta)}{\Omega'}   {By \Lemmaref{lem:soln-completes}}
             \declsubjudgPf{[\Omega'](\Delta, \alpha, \Theta)} {[\Omega']A} {[\Omega']B_0}  {By i.h.}
             \declsubjudgPf{[\Omega, \alpha](\Delta, \alpha)} {[\Omega, \alpha]A} {[\Omega, \alpha]B_0}  {By \Lemmaref{lem:substitution-stability}}
             \declsubjudgPf{[\Omega, \alpha](\Delta, \alpha)} {[\Omega]A} {[\Omega]B_0}  {By def.\ of substitution}
             \declsubjudgPf{[\Omega]\Delta} {[\Omega]A} {\alltype{\alpha}{[\Omega]B_0}}  {By \DsubAllR}
             \declsubjudgPf{[\Omega]\Delta} {[\Omega]A} {[\Omega](\alltype{\alpha}{B_0})}  {By def.\ of substitution}
           \end{llproof}

     \DerivationProofCase{\SubInstL}
          {
            \ahat \notin \FV{B}
            \\
            \instjudg{\Gamma}{\ahat}{B}{\Delta}
          }
          {\subjudg{\underbrace{\Gamma}_{\Gamma_0[\ahat]}}{\ahat}{B}{\Delta}}

          \begin{llproof}
            \instjudgPf{\Gamma}{\ahat}{B}{\Delta}   {Subderivation}
            \declsubjudgPf{[\Omega]\Delta}{[\Omega]\ahat}{[\Omega]B}   {By \Theoremref{thm:instantiation-soundness}}
          \end{llproof}

     \ProofCaseRule{\SubInstR}   Similar to the case for \SubInstL.
     \qedhere
  \end{itemize}
\end{proof}

\soundnesspretty*
\begin{proof}
  By reflexivity (\Lemmaref{lem:substextend-reflexivity}), $\substextend{\Psi}{\Psi}$. \\ 
  Since $\Psi$ has no existential variables, it is a complete context $\Omega$. \\ 
  By \Theoremref{thm:subtyping-soundness}, $\declsubjudg{[\Psi]\Psi}{[\Psi]A}{[\Psi]B}$. \\ 
  Since $\Psi$ has no existential variables, $[\Psi]\Psi = \Psi$, and $[\Psi]A = A$, and $[\Psi]B = B$. \\
  Therefore $\declsubjudg{\Psi}{A}{B}$.
\end{proof}

\section{Typing Extension}

\typingextension*
\begin{proof}
  By induction on the given derivation.
  
  \begin{itemize}
       \ProofCasesRules{\Var, \UnitIntro, \UnitIntroSyn}

          Since $\Delta = \Gamma$, the result follows by \Lemmaref{lem:substextend-reflexivity}.

     \DerivationProofCase{\Sub}
          {\synjudg{\Gamma}{e}{B}{\Theta}
            \\
            \subjudg{\Theta}{[\Theta]B}{[\Theta]A}{\Delta}
          }
          {\chkjudg{\Gamma}{e}{A}{\Delta}}

          \begin{llproof}
            \substextendPf{\Gamma}{\Theta}   {By i.h.}
            \substextendPf{\Theta}{\Delta}   {By \Lemmaref{lem:subtyping-extension}}
\Hand        \substextendPf{\Gamma}{\Delta}  {By \Lemmaref{lem:substextend-transitivity}}
          \end{llproof}

     \DerivationProofCase{\Anno}
          {\judgetp{\Gamma}{A}
           \\
           \chkjudg{\Gamma}{e}{A}{\Delta}
          }
          {\synjudg{\Gamma}{(e : A)}{A}{\Delta}}
          
          \begin{llproof}
            \Hand \substextendPf{\Gamma}{\Delta}  {By i.h.}
          \end{llproof}

     \DerivationProofCase{\AllIntro}
           {\chkjudg{\Gamma, \alpha}{e}{A_0}{\Delta, \alpha, \Theta}
           }
           {\chkjudg{\Gamma}{e}{\alltype{\alpha}{A_0}}{\Delta}}

          \begin{llproof}
            \substextendPf{\Gamma, \alpha}{\Delta, \alpha, \Theta}  {By i.h.}
\Hand       \substextendPf{\Gamma}{\Delta}   {By \Lemmaref{lem:extension-order} (i)}
          \end{llproof}

     \DerivationProofCase{\AllApp}
            {\appjudg{\Gamma, \ahat}{e}{[\ahat/\alpha]A_0}{C}{\Delta}}
            {\appjudg{\Gamma}{e}{\alltype{\alpha}{A_0}}{C}{\Delta}}

          \begin{llproof}
             \substextendPf{\Gamma, \ahat}{\Delta}  {By i.h.}
             \substextendPf{\Gamma}{\Gamma, \ahat}  {By \substextendAdd}
\Hand        \substextendPf{\Gamma}{\Delta}  {By \Lemmaref{lem:substextend-transitivity}}
          \end{llproof}

     \DerivationProofCase{\ArrIntro}
          {\chkjudg{\Gamma, x : A_1}{e}{A_2}{\Delta, x : A_1, \Theta}
          }
          {\chkjudg{\Gamma}{\lam{x} e}{A_1 \arr A_2}{\Delta}}

          \begin{llproof}
            \substextendPf{\Gamma, x : A_1}{\Delta, x : A_1, \Theta}  {By i.h.}
\Hand       \substextendPf{\Gamma}{\Delta}   {By \Lemmaref{lem:extension-order} (v)}
          \end{llproof}

     \DerivationProofCase{\ArrElim}
           {\synjudg{\Gamma}{e_1}{B}{\Theta}
             \\
             \appjudg{\Theta}{e_2}{[\Theta]B}{A}{\Delta}
           }
           {\synjudg{\Gamma}{e_1\,e_2}{A}{\Delta}}

           By the i.h. on each premise, then \Lemmaref{lem:substextend-transitivity}.

      \DerivationProofCase{\ArrIntroSyn}
           { \chkjudg{\Gamma, \ahat, \bhat, x : \ahat}{e}{\bhat}{\Delta, x : \ahat, \Theta}
           }
           {{\synjudg{\Gamma}{\lam{x} e}{\ahat \arr \bhat}{\Delta}}}

          \begin{llproof}
            \substextendPf{\Gamma, \ahat, \bhat, x : \ahat}{\Delta, x : \ahat, \Theta}  {By i.h.}
            \substextendPf{\Gamma, \ahat, \bhat}{\Delta}   {By \Lemmaref{lem:extension-order} (v)}
            \substextendPf{\Gamma}{\Gamma, \ahat, \bhat}   {By \substextendAdd (twice)}
\Hand       \substextendPf{\Gamma}{\Delta}   {By \Lemmaref{lem:substextend-transitivity}}
          \end{llproof}
           
      \DerivationProofCase{\ArrApp}
            {\chkjudg{\Gamma}{e}{A}{\Delta}}
            {\appjudg{\Gamma}{e}{A \arr C}{C}{\Delta}}

            \begin{llproof}
\Hand              \substextendPf{\Gamma}{\Delta}   {By i.h.}
            \end{llproof}

      \DerivationProofCase{\SolveApp}
            {\chkjudg{\Gamma[\ahat_2, \ahat_1, \hypeq{\ahat}{\ahat_1 \arr \ahat_2}]}{e}{\ahat_1}{\Delta}}
            {\appjudg{\Gamma[\ahat]}{e}{\ahat}{\ahat_2}{\Delta}}

          \begin{llproof}
            \substextendPf{\Gamma[\ahat_2, \ahat_1, \hypeq{\ahat}{\ahat_1 \arr \ahat_2}]}{\Delta}  {By i.h.}
            \decolumnizePf
            \substextendPf{\Gamma[\ahat]}
                          {\Gamma[\ahat_2, \ahat_1, \hypeq{\ahat}{\ahat_1 \arr \ahat_2}]}
                          {By \Lemmaref{lem:extension-addsolve}}
                          \trailingjust{then \Lemmaref{lem:parallel-admissibility} (ii)}
\Hand       \substextendPf{\Gamma}{\Delta}   {By \Lemmaref{lem:substextend-transitivity}   \qedhere}
          \end{llproof}
            
  \end{itemize}
\end{proof}

\section{Soundness of Typing}

\typingsoundness*
\begin{proof}
  By induction on the given algorithmic typing derivation.

  \begin{itemize}
     \DerivationProofCase{\Var}
          {(x : A) \in \Gamma}
          {\synjudg{\Gamma}{x}{A}{\Gamma}}
          
          \begin{llproof}
            \inPf{(x : A)}{\Gamma}  {Premise}
            \inPf{(x : A)}{\Delta}  {By $\Gamma = \Delta$}
            \substextendPf{\Delta}{\Omega}  {Given}
            \inPf{(x : [\Omega]A)}{[\Omega]\Gamma}  {By \Lemmaref{lem:ctxapp-var-preservation}}
\Hand            \declsynjudgPf{[\Omega]\Gamma}{x}{[\Omega]A} {By \DeclVar}
          \end{llproof}

     \DerivationProofCase{\Sub}
          {\synjudg{\Gamma}{e}{A}{\Theta}
            \\
            \subjudg{\Theta}{[\Theta]A}{[\Theta]B}{\Delta}
          }
          {\chkjudg{\Gamma}{e}{B}{\Delta}}

          \begin{llproof}
            \synjudgPf{\Gamma}{e}{A}{\Theta}  {Subderivation}
            \subjudgPf{\Theta}{[\Theta]A}{[\Theta]B}{\Delta}  {Subderivation}
            \substextendPf{\Theta}{\Delta}  {By \Lemmaref{lem:typing-extension}}
            \substextendPf{\Delta}{\Omega}  {Given}
            \substextendPf{\Theta}{\Omega}  {By \Lemmaref{lem:substextend-transitivity}}
            \declsynjudgPf{[\Omega]\Theta} {e} {[\Omega]A}  {By i.h.}
            \eqPf{[\Omega]\Theta} {[\Omega]\Delta}   {By \Lemmaref{lem:completes-confluence}}
            \declsynjudgPf{[\Omega]\Delta} {e} {[\Omega]A}  {By above equalities}
            \proofsep
            \subjudgPf{\Theta}{[\Theta]A}{[\Theta]B}{\Delta}  {Subderivation}
            \declsubjudgPf{[\Omega]\Delta} {[\Omega][\Theta]A} {[\Omega][\Theta]B}  {By \Theoremref{thm:subtyping-soundness}}
            \eqPf{[\Omega][\Theta]A} {[\Omega]A}   {By \Lemmaref{lem:subst-extension-invariance}}
            \eqPf{[\Omega][\Theta]B} {[\Omega]B}   {By \Lemmaref{lem:subst-extension-invariance}}
            \declsubjudgPf{[\Omega]\Delta} {[\Omega]A} {[\Omega]B}  {By above equalities}
\Hand       \declchkjudgPf{[\Omega]\Delta} {e} {[\Omega]B}  {By \DeclSub}
          \end{llproof}

     \DerivationProofCase{\Anno}
          {\judgetp{\Gamma}{A}
           \\
           \chkjudg{\Gamma}{e_0}{A}{\Delta}
          }
          {\synjudg{\Gamma}{(e_0 : A)}{A}{\Delta}}
          
          \begin{llproof}
            \chkjudgPf{\Gamma}{e_0}{A}{\Delta}  {Subderivation}
            \declchkjudgPf{[\Omega]\Delta} {e_0} {[\Omega]A}  {By i.h.}
            \proofsep
            \judgetpPf{\Gamma}{A}  {Subderivation}
            \substextendPf{\Gamma}{\Delta}  {By \Lemmaref{lem:typing-extension}}
            \substextendPf{\Delta}{\Omega}  {Given}
            \substextendPf{\Gamma}{\Omega}  {By \Lemmaref{lem:substextend-transitivity}}
            \judgetpPf{\Omega} {A}   {By \Lemmaref{lem:extension-weakening}}
            \judgetpPf{[\Omega]\Omega} {[\Omega]A}   {By \Lemmaref{lem:completion-wf}}
            \eqPf{[\Omega]\Delta}{[\Omega]\Omega}  {By \Lemmaref{lem:completes-stability}}
            \judgetpPf{[\Omega]\Delta} {[\Omega]A}   {By above equality}
            \proofsep
            \declsynjudgPf{[\Omega]\Delta} {(e_0 : [\Omega]A)} {[\Omega]A}  {By \DeclAnno}
            \Pf{}{}{A\text{~contains no existential variables}}   {Assumption about source programs}
            \eqPf{[\Omega]A} {A}  {From definition of substitution}
\Hand     \declsynjudgPf{[\Omega]\Delta} {(e_0 : A)} {[\Omega]A}  {By above equality}
          \end{llproof}

     \DerivationProofCase{\UnitIntro}
          {}
          {\chkjudg{\Gamma}{\unitexp}{\unitty}{\underbrace{\Gamma}_{\Delta}}}

          \begin{llproof}
            \declchkjudgPf{[\Omega]\Delta}{\unitexp}{\unitty}   {By \DeclUnitIntro}
\Hand    \declchkjudgPf{[\Omega]\Delta} {\unitexp} {[\Omega]\unitty}  {By definition of substitution}
          \end{llproof}

     \DerivationProofCase{\ArrIntro}
          {\chkjudg{\Gamma, x : A_1}{e_0}{A_2}{\Delta, x : A_1, \Theta}
          }
          {\chkjudg{\Gamma}{\lam{x} e}{A_1 \arr A_2}{\Delta}}

          \begin{llproof}
            \substextendPf{\Delta}{\Omega}  {Given}
            \substextendPf{\Delta, x : A_1}{\Omega, x : [\Omega]A_1}  {By \substextendVV}
            \substextendPf{\Gamma, x : A_1}{\Delta, x : A_1, \Theta}  {By \Lemmaref{lem:typing-extension}}
            \Pf{}{}{\Theta\text{~is soft}}   {By \Lemmaref{lem:extension-order} (v)}
                                            \trailingjust{(with $\Gamma_R = \cdot$, which is soft)}
            \substextendPf{\underbrace{\Delta, x : A_1, \Theta}_{\Delta'}}{\underbrace{\Omega, x : [\Omega]A_1, \soln{\Theta}}_{\Omega'}}  {By \Lemmaref{lem:soln-completes}}
            \proofsep
            \chkjudgPf{\Gamma, x : A_1}{e_0}{A_2}{\Delta'}  {Subderivation}
            \proofsep
            \declchkjudgPf{[\Omega']\Delta'} {e_0} {[\Omega']A_2}  {By i.h.}
            \eqPf{[\Omega']A_2}{[\Omega]A_2}   {By \Lemmaref{lem:substitution-stability}}
            \declchkjudgPf{[\Omega']\Delta'} {e_0} {[\Omega]A_2} {By above equality}
            \proofsep
            \substextendPf{\underbrace{\Delta, x : A_1, \Theta}_{\Delta'}}{\underbrace{\Omega, x : [\Omega]A_1, \soln{\Theta}}_{\Omega'}}  {Above}
            \Pf{}{}{\Theta\text{~is soft}}   {Above}
            \eqPf{[\Omega']\Delta'} {[\Omega]\Delta, x : [\Omega]A_1}   {By \Lemmaref{lem:softness-goes-away}}
            \declchkjudgPf{[\Omega]\Delta, x : [\Omega]A_1} {e_0} {[\Omega]A_2} {By above equality}
            \decolumnizePf
            \declchkjudgPf{[\Omega]\Delta} {\lam{x} e_0} {([\Omega]A_1) \arr ([\Omega]A_2)}  {By \DeclArrIntro}
\Hand       \declchkjudgPf{[\Omega]\Delta} {\lam{x} e_0} {[\Omega](A_1 \arr A_2)}  {By definition of substitution}
          \end{llproof}

     \DerivationProofCase{\ArrElim}
          {\synjudg{\Gamma}{e_1}{A_1}{\Theta}
            \\
            \appjudg{\Theta}{e_2}{A_1}{A_2}{\Delta}
          }
          {\synjudg{\Gamma}{e_1\,e_2}{A_2}{\Delta}}

          \begin{llproof}
            \synjudgPf{\Gamma}{e_1}{A_1}{\Theta}  {Subderivation}
            \subjudgPf{\Theta}{A_1}{B}{\Delta}  {Subderivation}
            \substextendPf{\Theta}{\Delta}  {By \Lemmaref{lem:typing-extension}}
            \substextendPf{\Delta}{\Omega}  {Given}
            \substextendPf{\Theta}{\Omega}  {By \Lemmaref{lem:substextend-transitivity}}
            \declsynjudgPf{[\Omega]\Theta} {e_1} {[\Omega]A_1}  {By i.h.}
            \eqPf{[\Omega]\Theta} {[\Omega]\Delta}   {By \Lemmaref{lem:completes-confluence}}
            \declsynjudgPf{[\Omega]\Delta} {e_1} {[\Omega]A_1}  {By above equality}
            \proofsep
            \appjudgPf{\Theta}{e_2}{A_1}{A_2}{\Delta}  {Subderivation}
            \substextendPf{\Delta}{\Omega}  {Given}
            \declappjudgPf{[\Omega]\Delta}{e_2}{[\Omega]A_1}{[\Omega]A_2}  {By i.h.}
\Hand       \declsynjudgPf{[\Omega]\Delta} {e_1 e_2} {[\Omega]A_2}  {By \DeclArrElim}
          \end{llproof}

     \DerivationProofCase{\AllIntro}
           {\chkjudg{\Gamma, \alpha}{e}{A_0}{\Delta, \alpha, \Theta}
           }
           {\chkjudg{\Gamma}{e}{\alltype{\alpha}{A_0}}{\Delta}}

          (Similar to \ArrIntro, using a different subpart of \Lemmaref{lem:extension-order}
          and applying \DeclAllIntro; written out anyway.)

          \begin{llproof}
            \substextendPf{\Delta}{\Omega}  {Given}
            \substextendPf{\Delta, \alpha}{\Omega, \alpha}  {By \substextendUU}
            \substextendPf{\Gamma, \alpha}{\Delta, \alpha, \Theta}  {By \Lemmaref{lem:typing-extension}}
            \Pf{}{}{\Theta\text{~is soft}}   {By \Lemmaref{lem:extension-order} (i) (with $\Gamma_R = \cdot$, which is soft)}
            \substextendPf{\underbrace{\Delta, \alpha, \Theta}_{\Delta'}}{\underbrace{\Omega, \alpha, \soln{\Theta}}_{\Omega'}}  {By \Lemmaref{lem:soln-completes}}
            \proofsep
            \chkjudgPf{\Gamma, \alpha}{e}{A_0}{\Delta'}  {Subderivation}
            \proofsep
            \declchkjudgPf{[\Omega']\Delta'} {e} {[\Omega']A_0}  {By i.h.}
            \eqPf{[\Omega']A_0}{[\Omega]A_0}   {By \Lemmaref{lem:substitution-stability}}
            \declchkjudgPf{[\Omega']\Delta'} {e} {[\Omega]A_0} {By above equality}
            \proofsep
            \substextendPf{\underbrace{\Delta, \alpha, \Theta}_{\Delta'}}{\underbrace{\Omega, \alpha, \soln{\Theta}}_{\Omega'}}  {Above}
            \Pf{}{}{\Theta\text{~is soft}}   {Above}
            \eqPf{[\Omega']\Delta'} {[\Omega]\Delta, \alpha}   {By \Lemmaref{lem:softness-goes-away}}
            \declchkjudgPf{[\Omega]\Delta, \alpha} {e} {[\Omega]A_0} {By above equality}
            \proofsep
            \declchkjudgPf{[\Omega]\Delta} {e} {\alltype{\alpha} [\Omega]A_0}  {By \DeclAllIntro}
\Hand       \declchkjudgPf{[\Omega]\Delta} {e} {[\Omega](\alltype{\alpha} A_0)}  {By definition of substitution}
          \end{llproof}

     \DerivationProofCase{\AllApp}
            {\appjudg{\Gamma,\ahat}{e}{[\ahat/\alpha]A_0}{C}{\Delta}}
            {\appjudg{\Gamma}{e}{\alltype{\alpha}{A_0}}{C}{\Delta}}

            \begin{llproof}
              \appjudgPf{\Gamma, \ahat}{e}{[\ahat/\alpha]A_0}{C}{\Delta}  {Subderivation}
              \substextendPf{\Delta}{\Omega}   {Given}
              \declappjudgPf{[\Omega]\Delta}{e}{[\Omega][\ahat/\alpha]A_0}{[\Omega]C}  {By i.h.}
              \declappjudgPf{[\Omega]\Delta}{e}{\big[[\Omega]\ahat\,/\,\alpha\big]\,[\Omega]A_0}{[\Omega]C}  {By distributivity of substitution}
              \proofsep
              \substextendPf{\Gamma, \ahat} {\Delta}   {By \Lemmaref{lem:typing-extension}}
              \substextendPf{\Gamma, \ahat} {\Omega}   {By \Lemmaref{lem:substextend-transitivity}}
              \judgetpPf{\Gamma, \ahat} {\ahat}   {By \EvarWF}
              \judgetpPf{\Omega} {\ahat}    {By \Lemmaref{lem:extension-weakening}}
              \judgetpPf{[\Omega]\Omega} {[\Omega]\ahat}   {By \Lemmaref{lem:completion-wf}}
              \eqPf{[\Omega]\Omega} {[\Omega]\Delta}   {By \Lemmaref{lem:completes-stability}}
              \judgetpPf{[\Omega]\Delta} {[\Omega]\ahat}   {By above equality}
              \proofsep
              \declappjudgPf{[\Omega]\Delta}{e}{\alltype{\alpha} [\Omega]A_0}{[\Omega]C}   {By \DeclAllApp}
\Hand              \declappjudgPf{[\Omega]\Delta}{e}{[\Omega](\alltype{\alpha} A_0)}{[\Omega]C}   {By definition of substitution}
            \end{llproof}

     \DerivationProofCase{\ArrApp}
            {\chkjudg{\Gamma}{e}{B}{\Delta}}
            {\appjudg{\Gamma}{e}{B \arr C}{C}{\Delta}}
            
            \begin{llproof}
              \chkjudgPf{\Gamma}{e}{B}{\Delta}  {Subderivation}
              \substextendPf{\Delta}{\Omega}   {Given}
              \declchkjudgPf{[\Omega]\Delta}{e}{[\Omega]B}   {By i.h.}
              \declappjudgPf{[\Omega]\Delta}{e}{([\Omega]B) \arr ([\Omega]C)}{[\Omega]C}   {By \DeclArrApp}
\Hand              \declappjudgPf{[\Omega]\Delta}{e}{[\Omega](B \arr C)}{[\Omega]C}   {By definition of substitution}
            \end{llproof}

      \DerivationProofCase{\SolveApp}
            {\chkjudg{\Gamma_0[\ahat_2, \ahat_1, \hypeq{\ahat}{\ahat_1 \arr \ahat_2}]}{e}{\ahat_1}{\Delta}}
            {\appjudg{\underbrace{\Gamma_0[\ahat]}_{\Gamma}}{e}{\ahat}{\ahat_2}{\Delta}}

            \begin{llproof}
                \chkjudgPf{\overbrace{\Gamma_0[\ahat_2, \ahat_1, \hypeq{\ahat}{\ahat_1 \arr \ahat_2}]}^{\Gamma'}}{e}{\ahat_1}{\Delta}   {Subderivation}
                \substextendPf{\Delta}{\Omega}   {Given}
                \declchkjudgPf{[\Omega]\Delta}{e}{[\Omega]\ahat_1}   {By i.h.}
                \declappjudgPf{[\Omega]\Delta}{e}{([\Omega]\ahat_1) \arr ([\Omega]\ahat_2)}{[\Omega]\ahat_2}    {By \DeclArrApp}
                \decolumnizePf
                \substextendPf{\Gamma'}{\Delta}   {By \Lemmaref{lem:typing-extension}}
                \substextendPf{\Delta}{\Omega}   {Given}
                \substextendPf{\Gamma'}{\Omega}   {By \Lemmaref{lem:substextend-transitivity}}
                \eqPf{[\Gamma']\ahat}  {[\Gamma'](\ahat_1 \arr \ahat_2)}   {By definition of $[\Gamma'](-)$}
                \eqPf{[\Omega][\Gamma']\ahat}  {[\Omega][\Gamma'](\ahat_1 \arr \ahat_2)}   {Applying $\Omega$ to both sides}
                \eqPf{[\Omega]\ahat}  {[\Omega](\ahat_1 \arr \ahat_2)}   {By \Lemmaref{lem:subst-extension-invariance}, twice}
                \eqPf{~}  {([\Omega]\ahat_1) \arr ([\Omega]\ahat_2)}   {By definition of substitution}
                \proofsep
\Hand     \declappjudgPf{[\Omega]\Delta}{e}{[\Omega]\ahat}{[\Omega]\ahat_2}    {By above equality}
            \end{llproof}

               \DerivationProofCase{\UnitIntroSyn}
                    { }
                    {{\synjudg{\Gamma}{\unitexp}{\unitty}{\underbrace{\Gamma}_{\Delta}}}}

                    \begin{llproof}
                      \Hand \declsynjudgPf{[\Omega]\Delta} {\unitexp} {[\Omega]\unitty}  {By \DeclUnitIntroSyn and definition of substitution}
                    \end{llproof}

               \DerivationProofCase{\ArrIntroSyn}
                    { \chkjudg{\Gamma, \ahat, \bhat, x : \ahat}{e_0}{\bhat}{\Delta, x : \ahat, \Theta}
                }
                    {{\synjudg{\Gamma}{\lam{x} e_0}{\ahat \arr \bhat}{\Delta}}}

                  \begin{llproof}
                    \substextendPf{\Gamma, \ahat, \bhat, x : \ahat}{\Delta, x : \ahat, \Theta}  {By \Lemmaref{lem:typing-extension}}
                    \Pf{}{}{\Theta\text{~is soft}}   {By \Lemmaref{lem:extension-order} (v) (with $\Gamma_R = \cdot$, which is soft)}
                    \substextendPf{\Gamma, \ahat, \bhat}{\Delta}   {\ditto}
                    \proofsep
                    \substextendPf{\Delta}{\Omega}  {Given}
                    \substextendPf{\Delta, x : \ahat}{\Omega, x : [\Omega]\ahat}  {By \substextendVV}
                    \substextendPf{\underbrace{\Delta, x : \ahat, \Theta}_{\Delta'}}{\underbrace{\Omega, x : [\Omega]\ahat, \soln{\Theta}}_{\Omega'}}  {By \Lemmaref{lem:soln-completes}}
                    \decolumnizePf
                    \chkjudgPf{\Gamma, \ahat, \bhat, x : \ahat}{e}{\bhat}{\Delta, x : \ahat, \Theta}  {Subderivation}
                    \proofsep
                    \declchkjudgPf{[\Omega']\Delta'}
                                 {e_0} {[\Omega']\bhat}  {By i.h.}
                    \eqPf{[\Omega']\bhat}{\big[\Omega, x : [\Omega]\ahat\big]\bhat}   {By \Lemmaref{lem:substitution-stability}}
                    \continueeqPf{[\Omega]\bhat}   {By definition of substitution}
                    \eqPf{[\Omega']\Delta'}{\big[\Omega, x : [\Omega]\ahat\big]\big(\Delta, x : \ahat\big)} 
                             {By \Lemmaref{lem:softness-goes-away}}
                    \continueeqPf{[\Omega]\Delta, x : [\Omega]\ahat}    {By definition of context substitution}
                    \declchkjudgPf{[\Omega]\Delta, x : [\Omega]\ahat}
                                 {e_0} {[\Omega]\bhat} {By above equalities}
                    \proofsep
                    \substextendPf{\Gamma, \ahat, \bhat}{\Delta}  {Above}
                    \substextendPf{\Gamma, \ahat, \bhat}{\Omega}   {By \Lemmaref{lem:substextend-transitivity}}
                    \judgetpPf{\Gamma, \ahat, \bhat} {\ahat}   {By \EvarWF}
                    \judgetpPf{\Omega} {\ahat}   {By \Lemmaref{lem:extension-weakening}}
                    \judgetpPf{[\Omega]\Delta}
                                {[\Omega]\ahat}   {By \Lemmaref{lem:completion-wf}}
                                \trailingjust{and \Lemmaref{lem:completes-stability}}
                    \proofsep
                    \judgetpPf{[\Omega]\Delta}
                                {[\Omega]\bhat}   {By similar reasoning}
                    \decolumnizePf
                    \judgetpPf{[\Omega]\Delta}
                                {([\Omega]\ahat) \arr ([\Omega]\bhat)}   {By \DeclArrowWF}
                    \Pf{}{}{\text{$[\Omega]\ahat$, $[\Omega]\bhat$ monotypes}}   {$\Omega$ predicative}
                    \proofsep
                    \declsynjudgPf{[\Omega]\Delta} {\lam{x} e_0} {([\Omega]\ahat) \arr ([\Omega]\bhat)}  {By \DeclArrIntroSyn}
        \Hand       \declsynjudgPf{[\Omega]\Delta} {\lam{x} e_0} {[\Omega](\ahat \arr \bhat)}  {By definition of substitution%
    \qedhere}
                  \end{llproof}                    
      \end{itemize}
\end{proof}

\clearpage
\section{Completeness}

\subsection{Instantiation Completeness}

\instantiationcompletes*
\begin{proof}
  By mutual induction on the given declarative subtyping derivation.
  
  \begin{enumerate}[(1)]
  \item
    We have $\declsubjudg{[\Omega]\Gamma}{[\Omega]\ahat}{[\Omega]A}$. We now case-analyze the shape of $A$. 
    \begin{itemize}
      \ProofCaseRule{$A = \bhat$}

          It is given that $\ahat \notin \FV{\bhat}$, so $\ahat \neq \bhat$. \\
          Since $A = \bhat$, we have $\declsubjudg{[\Omega]\Gamma}{[\Omega]\ahat}{[\Omega]\bhat}$. \\ 
          Since $\Omega$ is predicative, $[\Omega]\ahat = \tau_1$ and $[\Omega]\bhat = \tau_2$,
          so we have $\declsubjudg{[\Omega]\Gamma}{\tau_1}{\tau_2}$. \\
          By \Lemmaref{lem:decl-monotype-equality},
             $\tau_1 = \tau_2$. \\ 
          We have $A = \bhat$ and $[\Gamma]A = A$, so $[\Gamma]\bhat = \bhat$.  Thus $\bhat \in \unsolved{\Gamma}$. \\ 
          Let $\Omega'$ be $\Omega$.  By \Lemmaref{lem:substextend-reflexivity}, $\substextend{\Omega}{\Omega}$. \\
          Now consider whether $\ahat$ is declared to the left of $\bhat$, or vice versa. 

          \begin{itemize}
            \ProofCaseRule{$\Gamma = (\Gamma_0, \ahat, \Gamma_1, \bhat, \Gamma_2)$}

                Let $\Delta$ be $\Gamma_0, \ahat, \Gamma_1, \hypeq{\bhat}{\ahat}, \Gamma_2$. \\ 
                By rule \InstLReach, $\instjudg{\Gamma}{\ahat}{\bhat}{\Delta}$. \\
                It remains to show that $\substextend{\Delta}{\Omega}$. \\ 
                We have $[\Omega]\ahat = [\Omega]\bhat$.
                Then by \Lemmaref{lem:parallel-extension-solve}, $\substextend{\Delta}{\Omega}$. 

            \ProofCaseRule{$(\Gamma = \Gamma_0, \bhat, \Gamma_1, \ahat, \Gamma_2)$}

                Let $\Delta$ be $\Gamma_0, \bhat, \Gamma_1, \hypeq{\ahat}{\bhat}, \Gamma_2$. \\ 
                By rule \InstLSolve, $\instjudg{\Gamma}{\ahat}{\bhat}{\Delta}$. \\
                It remains to show that $\substextend{\Delta}{\Omega}$. \\ 
                We have $[\Omega]\bhat = [\Omega]\ahat$.
                Then by \Lemmaref{lem:parallel-extension-solve}, $\substextend{\Delta}{\Omega}$. 
          \end{itemize}

      \ProofCaseRule{$A = \alpha$}

          Since $A = \alpha$, we have $\declsubjudg{[\Omega]\Gamma}{[\Omega]\ahat}{[\Omega]\alpha}$. \\
          Since $[\Omega]\alpha = \alpha$, we have $\declsubjudg{[\Omega]\Gamma}{[\Omega]\ahat}{\alpha}$. \\ 
          By inversion, \DsubVar was used, so $[\Omega]\ahat = \alpha$; therefore,
          since $\Omega$ is well-formed, $\alpha$ is declared to the left of $\ahat$ in $\Omega$. \\ 
          We have $\substextend{\Gamma}{\Omega}$. \\
          By \Lemmaref{lem:reverse-declaration-order-preservation}, we know that $\alpha$ is declared to the left of $\ahat$ in $\Gamma$; that is, $\Gamma = \Gamma_0[\alpha][\ahat]$. \\
          Let $\Delta = \Gamma_0[\alpha][\hypeq{\ahat}{\alpha}]$ and $\Omega' = \Omega$. \\
          By \InstLSolve, $\instjudg{\Gamma_0[\alpha][\ahat]}{\ahat}{\alpha}{\Delta}$. \\
          By \Lemmaref{lem:parallel-extension-solve}, $\substextend{\Gamma_0[\alpha][\ahat=\alpha]}{\Omega}$.

      \ProofCaseRule{$A = A_1 \arr A_2$}

          By the definition of substitution, $[\Omega]A = ([\Omega]A_1) \arr ([\Omega]A_2)$. \\ 
          Therefore $\declsubjudg{[\Omega]\Gamma}{[\Omega]\ahat}{([\Omega]A_1) \arr ([\Omega]A_2)}$. \\ 
          Since we have an arrow as the supertype, only \DsubAllL or \DsubArr could have been used,
          and the subtype $[\Omega]\ahat$ must be either a quantifier or an arrow.
          But $\Omega$ is predicative, so $[\Omega]\ahat$ cannot be a quantifier.
          Therefore, it is an arrow: $[\Omega]\ahat = \tau_1 \arr \tau_2$, and \DsubArr concluded the derivation.
          Inverting \DsubArr gives
          $\declsubjudg{[\Omega]\Gamma}{[\Omega]A_2}{\tau_2}$ and 
          $\declsubjudg{[\Omega]\Gamma}{\tau_1}{[\Omega]A_1}$. \\

          Since $\ahat \in \unsolved{\Gamma}$, we know that $\Gamma$ has the form $\Gamma_0[\ahat]$. \\
          By \Lemmaref{lem:extension-add} twice, inserting unsolved variables $\ahat_2$ and $\ahat_1$ into
          the middle of the context extends it, that is: $\substextend{\Gamma_0[\ahat]}{\Gamma_0[\ahat_2, \ahat_1, \ahat]}$. \\
          Clearly, $\ahat_1 \arr \ahat_2$ is well-formed in $(\dots, \ahat_2, \ahat_1)$, so by 
          \Lemmaref{lem:extension-solve}, solving $\ahat$ extends the context:
          $\substextend{\Gamma_0[\ahat_2, \ahat_1, \ahat]}{\Gamma_0[\ahat_2, \ahat_1, \hypeq{\ahat}{\ahat_1 \arr \ahat_2}]}$.
          Then by \Lemmaref{lem:substextend-transitivity}, $\substextend{\Gamma_0[\ahat]}{\Gamma_0[\ahat_2, \ahat_1, \hypeq{\ahat}{\ahat_1 \arr \ahat_2}]}$.

          Since $\ahat \in \unsolved{\Gamma}$ and $\substextend{\Gamma}{\Omega}$, we know that
          $\Omega$ has the form $\Omega_0[\hypeq{\ahat}{\tau_0}]$.
          To show that we can extend this context,
          we apply \Lemmaref{lem:extension-addsolve} twice to introduce 
          $\hypeq{\ahat_2}{\tau_2}$ and  $\hypeq{\ahat_1}{\tau_1}$,
          and then \Lemmaref{lem:extension-solve} to overwrite $\tau_0$:
          \[
              \substextend{\underbrace{\Omega_0[\hypeq{\ahat}{\tau_0}]}_{\Omega}}
                          {\Omega_0[\hypeq{\ahat_2}{\tau_2}, \hypeq{\ahat_1}{\tau_1}, \hypeq{\ahat}{\ahat_1 \arr \ahat_2}]}
          \]
          We have $\substextend{\Gamma}{\Omega}$, that is,
          \[
                  \substextend{\Gamma_0[\ahat]}{\Omega_0[\hypeq{\ahat}{\tau_0}]}
          \]
          By \Lemmaref{lem:parallel-admissibility} (i) twice, inserting unsolved variables $\ahat_2$ and $\ahat_1$
          on both contexts in the above extension preserves extension:

          {\small
              \begin{llproof}
                      \substextendPf{\Gamma_0[\ahat_2, \ahat_1, \ahat]}
                                  {\Omega_0[\hypeq{\ahat_2}{\tau_2}, \hypeq{\ahat_1}{\tau_1}, \hypeq{\ahat}{\tau_0}]}
                                  {By \Lemmaref{lem:parallel-admissibility} (ii) twice}
                      \substextendPf{\underbrace{\Gamma_0[\ahat_2, \ahat_1, \hypeq{\ahat}{\ahat_1{\arr}\ahat_2}]}_{\Gamma_1}}
                                  {\underbrace{\Omega_0[\hypeq{\ahat_2}{\tau_2}, \hypeq{\ahat_1}{\tau_1}, \hypeq{\ahat}{\ahat_1 {\arr} \ahat_2}]}_{\Omega_1}}
                                  {By \Lemmaref{lem:parallel-extension-update}}
              \end{llproof}
          }

          Since $\ahat \notin \FV{A}$, it follows that $[\Gamma_1]A = [\Gamma]A = A$. \\ 
          Therefore $\ahat_1 \notin \FV{A_1}$ and $\ahat_1, \ahat_2 \notin \FV{A_2}$. \\ 
          By \Lemmaref{lem:finishing-completions} and \Lemmaref{lem:finishing-types}, $[\Omega_1]\Gamma_1 = [\Omega]\Gamma$ and $[\Omega_1]\ahat_1 = \tau_1$.  \\ 
          By i.h., there are $\Delta_2$ and $\Omega_2$ such that 
          $\instjudgr{\Gamma_1}{\ahat_1}{A_1}{\Delta_2}$ and $\substextend{\Delta_2}{\Omega_2}$ and $\substextend{\Omega_1}{\Omega_2}$. 

          Next, note that $[\Delta_2][\Delta_2]A_2 = [\Delta_2]A_2$. \\ 
          By \Lemmaref{lem:left-unsolvedness-preservation}, we know that $\ahat_2 \in \unsolved{\Delta_2}$. \\ 
          By \Lemmaref{lem:left-free-variable-preservation}, we know that $\ahat_2 \notin \FV{[\Delta_2]A_2}$. \\ 
          By \Lemmaref{lem:substextend-transitivity}, $\substextend{\Omega}{\Omega_2}$. \\
          We know $[\Omega_2]\Delta_2 = [\Omega]\Gamma$ because: 
          \begin{mathpar}
            \begin{array}{lcll}
              [\Omega_2]\Delta_2 
              & = & [\Omega_2]\Omega_2 & \text{By \Lemmaref{lem:completes-stability}} \\ 
              & = & [\Omega]\Omega & \text{By \Lemmaref{lem:finishing-completions}} \\
              & = & [\Omega]\Gamma & \text{By \Lemmaref{lem:completes-stability}}
            \end{array}
          \end{mathpar}
          By \Lemmaref{lem:finishing-types}, we know that $[\Omega_2]\ahat_2 = [\Omega_1]\ahat_2 = \tau_2$. \\ 
          By \Lemmaref{lem:finishing-types}, we know that $[\Omega_2]A_2 = [\Omega]A_2$. \\ 
          Hence we know that $\declsubjudg{[\Omega_2]\Delta_2}{[\Omega_2]\ahat_2}{[\Omega_2]A_2}$. \\ 
          By i.h., we have $\Delta$ and $\Omega'$ such that 
          $\instjudg{\Delta_2}{\ahat_2}{[\Delta_2]A_2}{\Delta}$ and $\substextend{\Omega_2}{\Omega'}$ and $\substextend{\Delta}{\Omega'}$.  \\
          By rule \InstLArr, $\instjudg{\Gamma}{\ahat}{A}{\Delta}$. \\ 
          By \Lemmaref{lem:substextend-transitivity}, $\substextend{\Omega}{\Omega'}$. 

      \ProofCaseRule{$A = \unitty$}

          We have $A = \unitty$, so $\declsubjudg{[\Omega]\Gamma}{[\Omega]\ahat}{[\Omega]\unitty}$. \\ 
          Since $[\Omega]\unitty = \unitty$, we have $\declsubjudg{[\Omega]\Gamma}{[\Omega]\ahat}{\unitty}$. \\ 
          The only declarative subtyping rules that can have $\unitty$ as the supertype in the conclusion are
            \DsubAllL and \DsubUnit.
          However, since $\Omega$ is predicative, $[\Omega]\ahat$ cannot be a quantifier, so \DsubAllL cannot
          have been used.  Hence \DsubUnit was used and $[\Omega]\ahat = \unitty$. \\ 
          Let $\Delta = \Gamma[\hypeq{\ahat}{\unitty}]$ and $\Omega' = \Omega$. \\
          By \InstLSolve, $\instjudg{\Gamma[\ahat]}{\ahat}{\unitty}{\Delta}$. \\
          By \Lemmaref{lem:parallel-extension-solve}, $\substextend{\Gamma[\ahat=\unitty]}{\Omega}$.

      \ProofCaseRule{$A = \alltype{\beta}{B}$}

          We have $\declsubjudg{[\Omega]\Gamma}{[\Omega]\ahat}{[\Omega](\alltype{\beta}{B})}$. \\
          By definition of substitution, $[\Omega](\alltype{\beta}{B}) = \alltype{\beta}{[\Omega]B}$, so
          we have $\declsubjudg{[\Omega]\Gamma}{[\Omega]\ahat}{\alltype{\beta}{[\Omega]B}}$.
          The only declarative subtyping rules that can have a quantifier as supertype are
             \DsubAllL and \DsubAllR.  However, since $\Omega$ is predicative, $[\Omega]\ahat$ cannot be
             a quantifier, so \DsubAllL cannot have been used.  Hence \DsubAllR was used,
             and we have a subderivation of
             $\declsubjudg{[\Omega]\Gamma, \beta}{[\Omega]\ahat}{[\Omega]B}$. \\
          Let $\Omega_1 = (\Omega, \beta)$ and $\Gamma_1 = (\Gamma, \beta)$. \\
          By \substextendUU, $\substextend{\Gamma_1}{\Omega_1}$. \\
          By the definition of substitution, $[\Omega_1]B = [\Omega]B$
          and $[\Omega_1]\ahat = [\Omega]\ahat$. \\
          Note that $[\Omega_1]\Gamma_1 = [\Omega]\Gamma, \beta$. \\
          Since $\ahat \in \unsolved{\Gamma}$, we have $\ahat \in \unsolved{\Gamma_1}$. \\
          Since $\ahat \notin \FV{A}$ and $A = \alltype{\beta}{B}$, we have $\ahat \notin \FV{B}$. \\
          By i.h., there are $\Omega_2$ and $\Delta_2$ such that
          $\instjudg{\Gamma, \beta}{\ahat}{B}{\Delta_2}$ and 
          $\substextend{\Delta_2}{\Omega_2}$ and $\substextend{\Omega_1}{\Omega_2}$. \\
          By \Lemmaref{lem:instantiation-extension}, $\substextend{\Gamma_1}{\Delta_2}$,
          that is, $\substextend{\Gamma, \beta}{\Delta_2}$. \\
          Therefore by \Lemmaref{lem:extension-order},
              $\Delta_2 = (\Delta', \beta, \Omega'')$ where $\substextend{\Gamma}{\Delta'}$. \\
          By equality, we know $\substextend{\Delta', \beta, \Delta''}{\Omega_2}$. \\
          By \Lemmaref{lem:extension-order},
             $\Omega_2 = (\Omega', \beta, \Omega'')$ where \Hand $\substextend{\Delta'}{\Omega'}$. \\
          We have $\substextend{\Omega_1}{\Omega_2}$, that is,
          $\substextend{\Omega, \beta}{\Omega', \beta, \Omega''}$,
          so \Lemmaref{lem:extension-order} gives \Hand $\substextend{\Omega}{\Omega'}$. \\
          By rule \InstLAllR, $\instjudg{\Gamma}{\ahat}{\alltype{\beta}{B}}{\Delta'}$. 
    \end{itemize}
        
  \item $\declsubjudg{[\Omega]\Gamma}{[\Omega]A}{[\Omega]\ahat}$

    These cases are mostly symmetric.  The one exception is the one connective that is not treated symmetrically
    in the declarative subtyping rules:
    \begin{itemize}
      \ProofCaseRule{$A = \alltype{\alpha}{B}$}

      Since $A = \alltype{\alpha}{B}$, we have
      $\declsubjudg{[\Omega]\Gamma}{[\Omega]\alltype{\beta}{B}}{[\Omega]\ahat}$. \\
      By symmetric reasoning to the previous case (the last case of part (1) above), \DsubAllL must have been
      used, with a subderivation of
      $\declsubjudg{[\Omega]\Gamma}{[\Omega]\ahat}{[\tau/\beta][\Omega]B}$. \\
      Since $\judgetp{[\Omega]\Gamma}{\tau}$, the type $\tau$ has no existential variables and is
      therefore invariant under substitution: $\tau = [\Omega]\tau$.
      Therefore $\big[\tau/\beta\big]\big[\Omega\big]B = \big[[\Omega]\tau/\beta\big]\big[\Omega\big]B$. \\
      By distributivity of substitution, this is $\big[\Omega\big][\tau/\beta]B$.
      Interposing $\bhat$, this is equal to $[\Omega][\tau/\bhat][\bhat/\beta]B$.
      Therefore
      $\declsubjudg{[\Omega]\Gamma}{[\Omega]\ahat}{[\Omega][\tau/\bhat][\bhat/\beta]B}$. \\
      Let $\Omega_1$ be $\Omega, \MonnierComma{\bhat}, \hypeq{\bhat}{\tau}$
      and let $\Gamma_1$ be $\Gamma, \MonnierComma{\bhat}, \bhat$.

      \begin{itemize}
          \item 
            By the definition of context application, $[\Omega_1]\Gamma_1 = [\Omega]\Gamma$.

          \item
            From the definition of substitution, $[\Omega_1]\ahat = [\Omega]\ahat$.

          \item
            It follows from the definition of substitution that $[\Omega][\tau/\bhat]C = [\Omega_1]C$ for all $C$.
            Therefore $[\Omega][\tau/\bhat][\bhat/\beta]B = [\Omega_1][\bhat/\beta]B$.
      \end{itemize}

      Applying these three equalities, $\declsubjudg{[\Omega_1]\Gamma_1}{[\Omega_1]\ahat}{[\Omega_1][\bhat/\beta]B}$. \\
      By the definition of substitution, $[\Gamma, \MonnierComma{\bhat}, \bhat]B = [\Gamma]B = B$, so $\ahat \notin \FV{[\Gamma_1]B}$. \\
      Since $\ahat \in \unsolved{\Gamma}$, we have $\ahat \in \unsolved{\Gamma_1}$. \\

      By i.h., there exist $\Delta_2$ and $\Omega_2$ such that
        $\instjudgr{\Gamma_1}{\ahat}{B}{\Delta_2}$
        and $\substextend{\Omega_1}{\Omega_2}$
        and $\substextend{\Delta_2}{\Omega_2}$. \\      
      By \Lemmaref{lem:instantiation-extension}, $\substextend{\Gamma_1}{\Delta_2}$,
        which is, $\substextend{\Gamma, \MonnierComma{\bhat}, \bhat}{\Delta_2}$. \\
      By \Lemmaref{lem:extension-order}, $\Delta_2 = (\Delta', \MonnierComma{\bhat}, \Delta'')$ and $\substextend{\Gamma}{\Delta'}$. \\
      By equality, $\substextend{\Delta', \MonnierComma{\bhat}, \Delta''}{\Omega_2}$. \\
      By \Lemmaref{lem:extension-order}, $\Omega_2 = (\Omega', \MonnierComma{\bhat}, \Omega'')$ and \Hand $\substextend{\Delta'}{\Omega'}$. \\
      By equality, $\substextend{\Omega, \MonnierComma{\bhat}, \hypeq{\bhat}{\tau}}{\Omega', \MonnierComma{\bhat}, \Omega''}$. \\
      \Hand By \Lemmaref{lem:extension-order}, $\substextend{\Omega}{\Omega'}$. \\
      By \InstRAllL, $\instjudgr{\Gamma}{\ahat}{\alltype{\beta}{B}}{\Delta'}$. 
    \qedhere
    \end{itemize}
  \end{enumerate}
\end{proof}

\clearpage

\subsection{Completeness of Subtyping}

\completingcompleteness*
\begin{proof}
  By induction on the derivation of $\declsubjudg{[\Omega]\Gamma}{[\Omega]A}{[\Omega]B}$. 
  
  \newcommand{\CASEhandled}[1]{\textcolor{dDkGreen}{\textbf{#1}}}
  \newcommand{\CASEhandledsymm}[1]{\textcolor{dBlue}{\textrm{#1}}}
  \newcommand{\CASEtobeproved}[1]{\textcolor{red}{\textbf{#1}}}
  \newcommand{\CASEimpossible}{\textcolor{dRed}{\textsl{impossible}}}
  \newcommand{\MAINLABEL}[1]{\mathsz{14pt}{#1}}
  \newcommand{\CaseBpoly}{\CASEhandled{1 (B poly)}}

  We distinguish cases of $[\Gamma]B$ and $[\Gamma]A$ that are
  \CASEimpossible, \CASEhandled{fully written out}, and
  \CASEhandledsymm{similar to fully-written-out cases}.

  \begin{displ}
    
    \begin{array}[t]{rccccccc}
      &\multicolumn{7}{c}{\MAINLABEL{[\Gamma]B}}
      \\[5pt]
      &
        & {\alltype{\beta} B'} &   \unitty  &   \alpha   &   \bhat   &    B_1 \arr B_2 &
      \\\cmidrule[1pt]{3-7}
      &\alltype{\alpha} A'
                &   \CaseBpoly
                &   \CASEhandled{2.Poly}
                &   \CASEhandled{2.Poly}
                &   \CASEhandled{2.Poly}
                &   \CASEhandled{2.Poly}
      \\\cmidrule{3-7}
      &\unitty
                &   \CaseBpoly
                &   \CASEhandled{2.Units}
                &   \CASEimpossible
                &   \CASEhandledsymm{2.BEx.Unit}
                &   \CASEimpossible
      \\\cmidrule{3-7}
      \MAINLABEL{[\Gamma]A}
      & \alpha
                &   \CaseBpoly
                &   \CASEimpossible
                &   \CASEhandled{2.Uvars}
                &   \CASEhandledsymm{2.BEx.Uvar}
                &   \CASEimpossible
      \\\cmidrule{3-7}
      & \ahat
                &   \CaseBpoly
                &   \CASEhandled{2.AEx.Unit}
                &   \CASEhandledsymm{2.AEx.Uvar}
                &
                    \begin{tabular}[c]{c}
                      \CASEhandled{2.AEx.SameEx}
                      \\
                      \CASEhandled{2.AEx.OtherEx}
                    \end{tabular}
                &   \CASEhandled{2.AEx.Arrow}
      \\\cmidrule{3-7}
      & A_1 \arr A_2
                &   \CaseBpoly
                &   \CASEimpossible
                &   \CASEimpossible
                &   \CASEhandledsymm{2.BEx.Arrow}
                &   \CASEhandled{2.Arrows}
    \end{array}
  \end{displ}

  The impossibility of the ``\CASEimpossible'' entries follows from inspection of the
  declarative subtyping rules.

  \medskip

  We first split on $[\Gamma]B$.
  
  \begin{itemize}
  \item \textbf{Case \CaseBpoly: $[\Gamma]B$ polymorphic:}  $[\Gamma]B = \alltype{\beta} B'$:
    
    \begin{llproof}
      \eqPf{B} {\alltype{\beta} B_0}   {$\Gamma$ predicative}
      \eqPf{B'} {[\Gamma]B_0}   {$\Gamma$ predicative}
      \eqPf{[\Omega]B} {[\Omega](\alltype{\beta} B_0)}   {Applying $\Omega$ to both sides}
      \continueeqPf {\alltype{\beta} [\Omega]B_0}   {By definition of substitution}
      $\Dee \derives~$ \declsubjudgPf{[\Omega]\Gamma} {[\Omega]A} {[\Omega]B}   {Given}
      $\Dee \derives~$ \declsubjudgPf{[\Omega]\Gamma} {[\Omega]A} {\alltype{\beta} [\Omega]B_0}   {By above equality}
      $\Dee' \derives~$ \declsubjudgPf{[\Omega]\Gamma, \beta}
                    {[\Omega]A}
                    {[\Omega]B_0}
            {By \Lemmaref{lem:decl-invertibility}}
      \ltPf{\Dee'} {\Dee}   {\ditto}
      $\Dee' \derives~$ \declsubjudgPf{[\Omega, \beta](\Gamma, \beta)}
                    {[\Omega, \beta]A}
                    {[\Omega, \beta]B_0}
            {By definitions of substitution}
      \subjudgPf{\Gamma, \beta}{[\Gamma, \beta]A}{[\Gamma, \beta]B_0} {\Delta'}  {By i.h.}
      \substextendPf{\Delta'}{\Omega_0'}   {\ditto}
      \substextendPf{\Omega, \beta}{\Omega_0'}   {\ditto}
      \subjudgPf{\Gamma, \beta}{[\Gamma]A}{[\Gamma]B_0} {\Delta'}  {By definition of substitution}
      \proofsep
      \substextendPf{\Gamma, \beta} {\Delta'}  {By \Lemmaref{lem:instantiation-extension}}
      \eqPf{\Delta'}{\Delta, \beta, \Theta}  {By \Lemmaref{lem:extension-order} (i)}
      \substextendPf{\Gamma}{\Delta}  {\ditto}
      \substextendPf{\Delta, \beta, \Theta}{\Omega_0'}   {By $\substextend{\Delta'}{\Omega_0'}$ and above equality}
      \eqPf{\Omega_0'}{\Omega', \beta, \Omega_R}  {By \Lemmaref{lem:extension-order} (i)}
\Hand      \substextendPf{\Delta}{\Omega'}  {\ditto}
      \proofsep
      \subjudgPf{\Gamma, \beta}{[\Gamma]A}{[\Gamma]B_0} {\Delta, \beta, \Theta}  {By above equality}
      \substextendPf{\Omega, \beta}{\Omega', \beta, \Omega_R}   {By above equality}
\Hand      \substextendPf{\Omega}{\Omega'}   {By \Lemmaref{lem:substextend-transitivity}}
      \proofsep
      \subjudgPf{\Gamma}{[\Gamma]A}{\alltype{\beta}{[\Gamma]B_0}}{\Delta}  {By \SubAllR}
\Hand      \subjudgPf{\Gamma}{[\Gamma]A}{\alltype{\beta} B'}{\Delta}  {By above equality}
    \end{llproof}

\clearpage
  \item \textbf{Cases 2.*: $[\Gamma]B$ not polymorphic:}

    We split on the form of $[\Gamma]A$.

    \begin{itemize}
    \item \textbf{Case \CASEhandled{2.Poly}: $[\Gamma]A$ is polymorphic:} $[\Gamma]A = \alltype{\alpha} A'$:

      \medskip
      
      \begin{llproof}
        \eqPf{A} {\alltype{\alpha} A_0}   {$\Gamma$ predicative}
        \eqPf{A'} {[\Gamma]A_0}   {$\Gamma$ predicative}
        \eqPf{[\Omega]A} {[\Omega](\alltype{\alpha} A_0)}   {Applying $\Omega$ to both sides}
        \eqPf{[\Omega]A} {\alltype{\alpha} [\Omega]A_0}   {By definition of substitution}
        \declsubjudgPf{[\Omega]\Gamma} {[\Omega]A} {[\Omega]B}   {Given}
        \declsubjudgPf{[\Omega]\Gamma} {\alltype{\alpha} [\Omega]A_0} {[\Omega]B}   {By above equality}
        \neqPf{[\Gamma]B} {(\alltype{\beta}{\cdots})}   {We are in the ``$[\Gamma]B$ not polymorphic'' subcase}
        \neqPf{B} {(\alltype{\beta}{\ldots})}   {$\Gamma$ predicative}
        \declsubjudgPf{[\Omega]\Gamma} {[\tau/\alpha][\Omega]A_0} {[\Omega]B}   {By inversion on \DsubAllL}
        \judgetpPf{[\Omega]\Gamma} {\tau}   {\ditto}
        \proofsep
        \substextendPf{\Gamma}{\Omega}   {Given}
        \substextendPf{\Gamma, \MonnierComma{\ahat}}{\Omega, \MonnierComma{\ahat}}
                   {By \substextendMonMon}
        \substextendPf{\Gamma, \MonnierComma{\ahat}, \ahat}{\underbrace{\Omega, \MonnierComma{\ahat}, \hypeq{\ahat}{\tau}}_{\Omega_0}}
                   {By \substextendSolve}
        \proofsep
        \eqPf{[\Omega]\Gamma} {[\Omega_0](\Gamma, \MonnierComma{\ahat}, \ahat)}  {By definition of context application (lines 16, 13)}
        \decolumnizePf
        \proofsep
        \declsubjudgPf{[\Omega]\Gamma}  {[\tau/\alpha][\Omega]A_0}  {[\Omega]B}    {Above}
        \declsubjudgPf{[\Omega_0](\Gamma, \MonnierComma{\ahat}, \ahat)}
                      {[\tau/\alpha][\Omega]A_0}
                      {[\Omega]B}
                      {By above equality}
        \declsubjudgPf{[\Omega_0](\Gamma, \MonnierComma{\ahat}, \ahat)}
                      {\big[[\Omega_0]\ahat / \alpha\big][\Omega]A_0}
                      {[\Omega]B}
                      {By definition of substitution}
        \declsubjudgPf{[\Omega_0](\Gamma, \MonnierComma{\ahat}, \ahat)}
                      {\big[[\Omega_0]\ahat / \alpha\big][\Omega_0]A_0}
                      {[\Omega_0]B}
                      {By definition of substitution}
        \declsubjudgPf{[\Omega_0](\Gamma, \MonnierComma{\ahat}, \ahat)}
                      {[\Omega_0][\ahat / \alpha]A_0}
                      {[\Omega_0]B}
                      {By distributivity of substitution}
        \proofsep
        \subjudgPf{\Gamma, \MonnierComma{\ahat}, \ahat} {[\Gamma, \MonnierComma{\ahat}, \ahat][\ahat / \alpha]A_0} {[\Gamma, \MonnierComma{\ahat}, \ahat]B} {\Delta_0}   {By i.h.}
        \substextendPf{\Delta_0} {\Omega''}    {\ditto}
        \substextendPf{\Omega_0} {\Omega''}    {\ditto}
        \subjudgPf{\Gamma, \MonnierComma{\ahat}, \ahat} {[\Gamma][\ahat / \alpha]A_0} {[\Gamma]B} {\Delta_0}   {By definition of substitution}
        \substextendPf{\Gamma, \MonnierComma{\ahat}, \ahat} {\Delta_0}   {By \Lemmaref{lem:subtyping-extension}}
        \eqPf{\Delta_0}{(\Delta, \MonnierComma{\ahat}, \Theta)}  {By \Lemmaref{lem:extension-order} (ii)}
        \substextendPf{\Gamma}{\Delta}   {\ditto}
        \eqPf{\Omega''}{(\Omega', \MonnierComma{\ahat}, \Omega_Z)}  {By \Lemmaref{lem:extension-order} (ii)}
\Hand        \substextendPf{\Delta}{\Omega'}   {\ditto}
        \substextendPf{\Omega_0} {\Omega''}    {Above}
        \substextendPf{\Omega, \MonnierComma{\ahat}, \hypeq{\ahat}{\tau}} {\Omega', \MonnierComma{\ahat}, \Omega_Z}    {By above equalities}
\Hand        \substextendPf{\Omega} {\Omega'}    {By \Lemmaref{lem:extension-order} (ii)}
        \proofsep
        \subjudgPf{\Gamma, \MonnierComma{\ahat}, \ahat}
                  {[\Gamma][\ahat/\alpha]A_0}
                  {[\Gamma]B}
                  {\Delta, \MonnierComma{\ahat}, \Theta}
                  {By above equality $\Delta_0 = (\Delta, \MonnierComma{\ahat}, \Theta)$}
        \subjudgPf{\Gamma, \MonnierComma{\ahat}, \ahat}
                  {[\ahat/\alpha][\Gamma]A_0}
                  {[\Gamma]B}
                  {\Delta, \MonnierComma{\ahat}, \Theta}
                  {By def.\ of subst.\ ($[\Gamma]\ahat = \ahat$ and $[\Gamma]\alpha = \alpha$)}
         \subjudgPf{\Gamma} {\alltype{\alpha}{[\Gamma]A_0}} {[\Gamma]B} {\Delta}   {By \SubAllL}
     \Hand      \subjudgPf{\Gamma} {\alltype{\alpha}{A'}} {[\Gamma]B} {\Delta}   {By above equality}
      \end{llproof}

      \smallskip

    \item \textbf{Case \CASEhandled{2.AEx}: $A$ is an existential variable} $[\Gamma]A = \ahat$:

      We split on the form of $[\Gamma]B$.
      \begin{itemize}
      \item \textbf{Case \CASEhandled{2.AEx.SameEx}: $[\Gamma]B$ is the same existential variable}  $[\Gamma]B = \ahat$:

        \smallskip

        \begin{llproof}
            \subjudgPf{\Gamma} {\ahat} {\ahat} {\Gamma}    {By \SubExvar}
\Hand         \subjudgPf{\Gamma} {[\Gamma]A} {[\Gamma]B} {\Gamma}    {By $[\Gamma]A = [\Gamma]B = \ahat$}
\Hand          \substextendPf{\Delta} {\Omega}   {$\Delta = \Gamma$}
\Hand          \substextendPf{\Omega} {\Omega'}   {By \Lemmaref{lem:substextend-reflexivity} and $\Omega' = \Omega$}
        \end{llproof}

      \item \textbf{Case \CASEhandled{2.AEx.OtherEx}: $[\Gamma]B$ is a different existential variable}  $[\Gamma]B = \bhat$ where $\bhat \neq \ahat$:

        Either $\ahat \in \FV{[\Gamma]\bhat}$, or $\ahat \notin \FV{[\Gamma]\bhat}$.

        \begin{itemize}
            \item $\ahat \in \FV{[\Gamma]\bhat}$: \\
                  We have $\ahat \subtermof [\Gamma]\bhat$.  \\
                  Therefore $\ahat = [\Gamma]\bhat$, or $\ahat \propersubtermof [\Gamma]\bhat$. \\
                  But we are in Case 2.AEx.\textbf{Other}Ex, so the former is impossible. \\
                  Therefore, $\ahat \propersubtermof [\Gamma]\bhat$. \\
                  Since $\Gamma$ is predicative, $[\Gamma]\bhat$ cannot have the form $\alltype{\beta} \cdots$,
                  so the only way that $\ahat$ can be a proper subterm of $[\Gamma]\bhat$ is if $[\Gamma]\bhat$
                  has the form $B_1 \arr B_2$ such that $\ahat$ is a subterm of $B_1$ or $B_2$, that is:
                  $\ahat \occursinsidearrow [\Gamma]\bhat$. \\
                  Then by a property of substitution, $[\Omega]\ahat \occursinsidearrow [\Omega][\Gamma]\bhat$. \\
                  By \Lemmaref{lem:subst-extension-invariance}, $[\Omega][\Gamma]\bhat = [\Omega]\bhat$, so
                   $[\Omega]\ahat \occursinsidearrow [\Omega]\bhat$. \\
                  We have $\declsubjudg{[\Omega]\Gamma}{[\Omega]\ahat}{[\Omega]\bhat}$,
                  and we know that $[\Omega]\ahat$ is a monotype, so we can use \Lemmaref{lem:occurrence} (ii)
                  to show that $[\Omega]\ahat \notoccursinsidearrow [\Omega]\bhat$, a contradiction.

            \item $\ahat \notin \FV{[\Gamma]\bhat}$:

                  \begin{llproof}
                    \instjudgPf{\Gamma} {\ahat} {[\Gamma]\bhat} {\Delta}   {By \Theoremref{thm:instantiation-completes} (1)}
    \Hand                \subjudgPf{\Gamma} {\ahat} {\bhat} {\Delta}    {By \SubInstL}
          \Hand     \substextendPf{\Delta}{\Omega'} {\ditto}
          \Hand     \substextendPf{\Omega}{\Omega'} {\ditto}
                  \end{llproof}
      \end{itemize}

      \item \textbf{Case \CASEhandled{2.AEx.Unit}:}  $[\Gamma]B = \unitty$:
        
        \begin{llproof}
          \substextendPf{\Gamma}{\Omega}   {Given}
          \eqPf{\unitty}{[\Omega]\unitty}   {By definition of substitution}
          \notinPf{\ahat} {\FV{\unitty}}   {By definition of $\FV{-}$}
          \declsubjudgPf{[\Omega]\Gamma}  {[\Omega]\ahat}  {[\Omega]\unitty}  {Given}
          \proofsep
          \instjudgPf{\Gamma}  {\ahat}  {\unitty}  {\Delta}  {By \Theoremref{thm:instantiation-completes} (1)}
\Hand          \substextendPf{\Omega}{\Omega'}    {\ditto}
\Hand          \substextendPf{\Delta}{\Omega'}    {\ditto}
          \proofsep
          \eqPf{\unitty}{[\Gamma]\unitty}   {By definition of substitution}
          \notinPf{\ahat} {\FV{\unitty}}   {By definition of $\FV{-}$}
          \proofsep
\Hand          \subjudgPf{\Gamma}  {\ahat}  {\unitty}  {\Delta}   {By \SubInstL}
        \end{llproof}

      \item \textbf{Case \CASEhandledsymm{2.AEx.Uvar}:}  $[\Gamma]B = \beta$:

          Similar to Case 2.AEx.Unit, using $\beta = [\Omega]\beta = [\Gamma]\beta$ and $\ahat \notin \FV{\beta}$.

      \item \textbf{Case \CASEhandled{2.AEx.Arrow}:}  $[\Gamma]B = B_1 \arr B_2$:
        
                   Since $[\Gamma]B$ is an arrow, it cannot be exactly $\ahat$.
                   
                   Suppose, for a contradiction, that $\ahat \in \FV{[\Gamma]B}$.
                   
                       \begin{llproof}
                         \subtermofPf {\ahat} {[\Gamma]B}   {$\ahat \in \FV{[\Gamma]B}$}
                         \subtermofPf {[\Omega]\ahat} {[\Omega][\Gamma]B}   {By a property of substitution}
                         \proofsep
                         \substextendPf {\Gamma}{\Omega}   {Given}
                         \eqPf {[\Omega][\Gamma]B} {[\Omega]B}   {By \Lemmaref{lem:subst-extension-invariance}}
                         \proofsep
                         \subtermofPf {[\Omega]\ahat} {[\Omega]B}   {By above equality}
                         \proofsep
                         \neqPf {[\Gamma]B} {\ahat}      {Given (2.AEx.Arrow)}
                         \neqPf {[\Omega][\Gamma]B} {[\Omega]\ahat}      {By a property of substitution}
                         \neqPf {[\Omega]B} {[\Omega]\ahat}      {By \Lemmaref{lem:subst-extension-invariance}}
                         \proofsep
                         \propersubtermofPf {[\Omega]\ahat} {[\Omega]B}   {Follows from $\subtermofsym$ and $\neq$}
                         \occursinsidearrowPf {[\Omega]\ahat} {[\Omega]B}   {$[\Omega]A$ has the form $\cdots \arr \cdots$}
                         \declsubjudgPf{[\Omega]\Gamma} {[\Omega]\ahat} {[\Omega]B}   {Given}
                         \Pf{}{}{[\Omega]B\text{~is a monotype}}  {$\Omega$ is predicative}
                         \notoccursinsidearrowPf {[\Omega]\ahat} {[\Omega]B}   {By \Lemmaref{lem:occurrence} (ii)}
                       \contraPf{\ahat \notin \FV{[\Gamma]B}}
                      \end{llproof}

                   \begin{llproof}
                     \instjudgPf{\Gamma} {\ahat} {[\Gamma]B} {\Delta}   {By \Theoremref{thm:instantiation-completes} (1)}
\Hand             \substextendPf{\Delta} {\Omega'}   {\ditto}
\Hand             \substextendPf{\Omega} {\Omega'}   {\ditto}
\Hand             \subjudgPf{\Gamma} {\ahat} {\underbrace{[\Gamma]B}_{B_1 \arr B_2}} {\Delta}   {By \SubInstL}
                   \end{llproof}
    \end{itemize}

    \item \textbf{Case \CASEhandledsymm{2.BEx}: $[\Gamma]A$ is not polymorphic and $[\Gamma]B$ is an existential variable}:
         $[\Gamma]B = \bhat$

         We split on the form of $[\Gamma]A$.

            \begin{itemize}
            \item \textbf{Case \CASEhandledsymm{2.BEx.Unit}} ($[\Gamma]A = \unitty$), \\
                  \textbf{Case \CASEhandledsymm{2.BEx.Uvar}} ($[\Gamma]A = \alpha$), \\
                  \textbf{Case \CASEhandledsymm{2.BEx.Arrow}} ($[\Gamma]A = A_1 \arr A_2$): \\
                Similar to Cases \CASEhandled{2.AEx.Unit}, \CASEhandled{2.AEx.Uvar} and \CASEhandled{2.AEx.Arrow},
                but using part (2) of \Theoremref{thm:instantiation-completes} instead of part (1),
                and applying \SubInstR instead of \SubInstL as the final step.
            \end{itemize}

    \item \textbf{Case \CASEhandled{2.Units}}: $[\Gamma]A = [\Gamma]B = \unitty$:

      \begin{llproof}
\Hand        \subjudgPf{\Gamma}{\unitty}{\unitty}{\Gamma}   {By \SubUnit}
          \substextendPf{\Gamma} {\Omega}   {Given}
\Hand          \substextendPf{\Delta} {\Omega}   {$\Delta = \Gamma$}
\Hand          \substextendPf{\Omega} {\Omega'}   {By \Lemmaref{lem:substextend-reflexivity} and $\Omega' = \Omega$}
      \end{llproof}

    \item \textbf{Case \CASEhandled{2.Uvars}}: $[\Gamma]A = [\Gamma]B = \alpha$:      

      \begin{llproof}
         \Pf{}{}{\alpha \in \Omega}  {By inversion on \DsubVar}
         \substextendPf{\Gamma}{\Omega}   {Given}
         \Pf{}{}{\alpha \in \Gamma} {By \Lemmaref{lem:extension-order}}
\Hand         \subjudgPf{\Gamma}{\alpha}{\alpha}{\Gamma}  {By \SubVar}
\Hand         \substextendPf{\Delta} {\Omega}   {$\Delta = \Gamma$}
\Hand         \substextendPf{\Omega} {\Omega'}   {By \Lemmaref{lem:substextend-reflexivity} and $\Omega' = \Omega$}        
      \end{llproof}

    \item \textbf{Case \CASEhandled{2.Arrows}}: $A = A_1 \arr A_2$ and $B = B_1 \arr B_2$:

      Only rule \DsubArr could have been used.

      \begin{llproof}
        \declsubjudgPf{[\Omega]\Gamma}  {[\Omega]B_1}  {[\Omega]A_1}    {Subderivation}
        \subjudgPf{\Gamma}  {[\Gamma]B_1}  {[\Gamma]A_1}  {\Theta}    {By i.h.}
        \substextendPf{\Theta} {\Omega_0}    {\ditto}
        \substextendPf{\Omega} {\Omega_0}    {\ditto}
        \substextendPf{\Gamma}{\Omega}   {Given}
        \substextendPf{\Gamma}{\Omega_0}   {By \Lemmaref{lem:substextend-transitivity}}
        \proofsep
        \substextendPf{\Theta}{\Omega_0}   {Above}
        \proofsep
        \eqPf{[\Omega]\Gamma}{[\Omega]\Theta}  {By \Lemmaref{lem:completes-confluence}}
        \proofsep
        \declsubjudgPf{[\Omega]\Gamma}  {[\Omega]A_2}  {[\Omega]B_2}    {Subderivation}
        \declsubjudgPf{[\Omega]\Theta}  {[\Omega]A_2}  {[\Omega]B_2}    {By above equality}
        \proofsep
        \eqPf{[\Omega]A_2} {[\Omega][\Gamma]A_2}  {By \Lemmaref{lem:subst-extension-invariance}}
        \eqPf{[\Omega]B_2} {[\Omega][\Gamma]B_2}  {By \Lemmaref{lem:subst-extension-invariance}}
        \proofsep
        \declsubjudgPf{[\Omega]\Theta}  {[\Omega][\Gamma]A_2}  {[\Omega][\Gamma]B_2}    {By above equalities}
        \subjudgPf{\Theta}  {[\Theta][\Gamma]A_2}  {[\Theta][\Gamma]B_2}  {\Delta}    {By i.h.}
\Hand   \substextendPf{\Delta} {\Omega'}    {\ditto}
        \substextendPf{\Omega_0} {\Omega'}    {\ditto}
        \decolumnizePf
        \subjudgPf{\Gamma}{([\Gamma]A_1) \arr ([\Gamma]A_2)}{([\Gamma]B_1) \arr ([\Gamma]B_2)}{\Delta} {By \SubArr}
\Hand   \subjudgPf{\Gamma}{[\Gamma](A_1 \arr A_2)}{[\Gamma](B_1 \arr B_2)}{\Delta} {By definition of substitution}
\Hand   \substextendPf{\Omega} {\Omega'}  {By \Lemmaref{lem:substextend-transitivity}}
      \end{llproof}%
    \qedhere
    \end{itemize}
  \end{itemize}
\end{proof}

\completeness*
\begin{proof}
  Let $\Omega = \Psi$ and $\Gamma = \Psi$. \\
  By \Lemmaref{lem:substextend-reflexivity}, $\substextend{\Psi}{\Psi}$, so $\substextend{\Gamma}{\Omega}$. \\
  By \Lemmaref{lem:declarative-well-formed}, $\judgetp{\Psi}{A}$ and $\judgetp{\Psi}{B}$; since $\Gamma = \Psi$,
  we have
  $\judgetp{\Gamma}{A}$ and $\judgetp{\Gamma}{B}$.
  \\
  By \Theoremref{thm:completing-completeness},
  there exists $\Delta$ such that $\subjudg{\Gamma}{[\Gamma]A}{[\Gamma]B}{\Delta}$.
  \\
  Since $\Gamma = \Psi$ and $\Psi$ is a declarative context with no existentials, $[\Psi]C = C$ for all $C$,
  so we actually have $\subjudg{\Psi}{A}{B}{\Delta}$, which was to be shown.
\end{proof}

\clearpage
\section{Completeness of Typing}

\typingcompleteness*
\begin{proof}
  By induction on the given declarative derivation.

  \begin{itemize}
     \DerivationProofCase{\DeclVar}
          {(x : A) \in [\Omega]\Gamma}
          {\declsynjudg{[\Omega]\Gamma}{x}{A}}
          
          \begin{llproof}
            \inPf{(x : A)} {[\Omega]\Gamma}  {Premise}
            \substextendPf{\Gamma}{\Omega}   {Given}
            \inPf{(x : A')}{\Gamma\text{~where $[\Omega]A' = [\Omega]A$}}   {From definition of context application}
            \LetPf{\Delta}{\Gamma} {}
            \LetPf{\Omega'}{\Omega} {}
\Hand       \substextendPf{\Gamma}{\Omega}   {Given}
\Hand       \substextendPf{\Omega}{\Omega}   {By \Lemmaref{lem:substextend-reflexivity}}
\Hand       \synjudgPf{\Gamma}{x}{A'}{\Gamma}   {By \Var}
                   \eqPf{[\Omega]A'}{[\Omega]A}   {Above}
\Hand              \continueeqPf{A}   {$\FEV{A} = \emptyset$}
          \end{llproof}

     \DerivationProofCase{\DeclSub}
          {\declsynjudg{[\Omega]\Gamma}{e}{B}
            \\
            \declsubjudg{[\Omega]\Gamma}{B}{[\Omega]A}
          }
          {\declchkjudg{[\Omega]\Gamma}{e}{[\Omega]A}}

          \begin{llproof}
            \declsynjudgPf{[\Omega]\Gamma}{e}{B}   {Subderivation}
            \synjudgPf{\Gamma}{e}{B'}{\Theta}      {By i.h.}
            \eqPf{B}{[\Omega]B'}   {\ditto}
            \substextendPf{\Theta}{\Omega_0}   {\ditto}
            \substextendPf{\Omega}{\Omega_0}   {\ditto}
            \proofsep
            \substextendPf{\Gamma}{\Omega}   {Given}
            \substextendPf{\Gamma}{\Omega_0}   {By \Lemmaref{lem:substextend-transitivity}}
            \declsubjudgPf{[\Omega]\Gamma}{B}{[\Omega]A}  {Subderivation}
            \eqPf{[\Omega]\Gamma} {[\Omega]\Theta}   {By \Lemmaref{lem:completes-confluence}}
            \declsubjudgPf{[\Omega]\Theta}{B}{[\Omega]A}  {By above equalities}
            \substextendPf{\Theta}{\Omega_0}  {Above}
            \subjudgPf{\Theta}{[\Theta]B'}{[\Theta]A}{\Delta}  {By \Theoremref{thm:completing-completeness}}
            \substextendPf{\Delta}{\Omega'}   {\ditto}
            \substextendPf{\Omega_0}{\Omega'}   {\ditto}
\Hand   \substextendPf{\Delta}{\Omega'}   {By \Lemmaref{lem:substextend-transitivity}}
\Hand   \substextendPf{\Omega}{\Omega'}   {By \Lemmaref{lem:substextend-transitivity}}
            \proofsep
\Hand   \chkjudgPf{\Gamma}{e}{A}{\Delta}  {By \Sub}
          \end{llproof}

     \DerivationProofCase{\DeclAnno}
          {\judgetp{[\Omega]\Gamma}{A}
            \\
            \declchkjudg{[\Omega]\Gamma}{e_0}{A}
          }
          {\declsynjudg{[\Omega]\Gamma}{(e_0 : A)}{A}}

          \begin{llproof}
            \eqPf{A} {[\Omega]A}   {Source type annotations cannot contain evars}
            \continueeqPf           {[\Gamma]A}   {\ditto}
            \declchkjudgPf{[\Omega]\Gamma}{e_0}{A}   {Subderivation}
            \declchkjudgPf{[\Omega]\Gamma}{e_0}{[\Omega]A}   {By above equality}
            \chkjudgPf{\Gamma}{e_0}{[\Gamma]A}{\Delta}   {By i.h.}
\Hand       \substextendPf{\Delta}{\Omega}   {\ditto}
\Hand       \substextendPf{\Omega}{\Omega'}   {\ditto}
            \proofsep
            \judgetpPf{\Gamma}{A}   {Given}
            \proofsep
            \synjudgPf{\Gamma}{(e_0 : A)}{A}{\Delta}   {By \Anno}
            \eqPf{A} {[\Omega']A}   {Source type annotations cannot contain evars}
\Hand       \synjudgPf{\Gamma}{(e_0 : [\Omega']A)}{[\Omega']A}{\Delta}   {By above equality}
          \end{llproof}

     \DerivationProofCase{\DeclUnitIntro}
          {}
          {\declchkjudg{[\Omega]\Gamma}{\unitexp}{\unitty}}

          We have $[\Omega]A = \unitty$.  Either $[\Gamma]A = \unitty$ or $[\Gamma]A = \ahat \in \unsolved{\Gamma}$.

          In the former case:

          \begin{llproof}
            \LetPf{\Delta}{\Gamma} {}
            \LetPf{\Omega'}{\Omega} {}
\Hand       \substextendPf{\Gamma}{\Omega}   {Given}
\Hand       \substextendPf{\Omega}{\Omega'}   {By \Lemmaref{lem:substextend-reflexivity}}
             \chkjudgPf{\Gamma}{\unitexp}{\unitty}{\Gamma}   {By \UnitIntro}
\Hand       \chkjudgPf{\Gamma}{\unitexp}{[\Gamma]\unitty}{\Gamma}   {$\unitty = [\Gamma]\unitty$}
          \end{llproof}

          In the latter case:

          \begin{llproof}
             \synjudgPf{\Gamma}{\unitexp}{\unitty}{\Gamma}   {By \UnitIntroSyn}
             \declsubjudgPf{[\Omega]\Gamma}{\unitty}{\unitty}   {By \DsubUnit}
             \proofsep
             \eqPf{\unitty} {[\Omega]\unitty}   {By definition of substitution}
             \continueeqPf {[\Omega][\Gamma]\ahat}   {By $[\Omega]A = \unitty$}
             \continueeqPf {[\Omega]\ahat}   {By \Lemmaref{lem:subst-extension-invariance}}
             \proofsep
             \declsubjudgPf{[\Omega]\Gamma}{[\Omega]\unitty}{[\Omega]\ahat}   {By above equalities}
             \subjudgPf{\Gamma}{\unitty}{\ahat}{\Delta}   {By \Theoremref{thm:instantiation-completes} (1)}
             \eqPf{\unitty}{[\Gamma]\unitty}  {By definition of substitution}
             \eqPf{\ahat}{[\Gamma]\ahat}  {$\ahat \in \unsolved{\Gamma}$}
             \subjudgPf{\Gamma}{[\Gamma]\unitty}{[\Gamma]\ahat}{\Delta}   {By above equalities}
\Hand  \substextendPf{\Omega}{\Omega'} {\ditto}
\Hand  \substextendPf{\Delta}{\Omega'} {\ditto}
             \chkjudgPf{\Gamma}{\unitexp}{\ahat}{\Delta}   {By \Sub}
\Hand        \chkjudgPf{\Gamma}{\unitexp}{[\Gamma]A}{\Delta}   {By $[\Gamma]A = \ahat$}
          \end{llproof}

     \DerivationProofCase{\DeclAllIntro}
           {\declchkjudg{[\Omega]\Gamma, \alpha}{e}{A_0}
           }
           {\declchkjudg{[\Omega]\Gamma}{{e}}{\alltype{\alpha}{A_0}}}

           \begin{llproof}
             \eqPf{[\Omega]A}{\alltype{\alpha}{A_0}}   {Given}
             \continueeqPf{\alltype{\alpha}{[\Omega]A'}}  {By def. of subst. and predicativity of $\Omega$}
             \eqPf{A_0}{[\Omega]A'}   {Follows from above equality}
             \declchkjudgPf{[\Omega]\Gamma, \alpha}{e}{[\Omega]A'}  {Subderivation and above equality}
             \proofsep
             \substextendPf{\Gamma}{\Omega}  {Given}
             \substextendPf{\Gamma, \alpha}{\Omega, \alpha}  {By \substextendUU}
             \proofsep
             \eqPf{[\Omega]\Gamma, \alpha}{[\Omega, \alpha](\Gamma, \alpha)}  {By definition of context substitution}
             \declchkjudgPf{[\Omega, \alpha](\Gamma, \alpha)}{e}{[\Omega]A'}  {By above equality}
             \declchkjudgPf{[\Omega, \alpha](\Gamma, \alpha)}{e}{[\Omega, \alpha]A'}  {By definition of substitution}
           \end{llproof}

           \begin{llproof}
             \chkjudgPf{\Gamma, \alpha}{e}{[\Gamma, \alpha]A'}{\Delta'}  {By i.h.}
             \substextendPf{\Delta'}{\Omega_0'}   {\ditto}
             \substextendPf{\Omega, \alpha}{\Omega_0'}   {\ditto}
             \substextendPf{\Gamma, \alpha}{\Delta'}  {By \Lemmaref{lem:typing-extension}}
             \eqPf{\Delta'} {\Delta, \alpha, \Theta}   {By \Lemmaref{lem:extension-order} (i)}
             \substextendPf{\Delta, \alpha, \Theta}{\Omega_0'}   {By above equality}
             \eqPf{\Omega_0'} {\Omega', \alpha, \Omega_Z}   {By \Lemmaref{lem:extension-order} (i)}
\Hand        \substextendPf{\Delta} {\Omega'}   {\ditto}
\Hand        \substextendPf{\Omega}{\Omega'}  {By \Lemmaref{lem:extension-order} on $\substextend{\Omega, \alpha}{\Omega_0'}$}
             \proofsep
             \chkjudgPf{\Gamma, \alpha}{e}{[\Gamma, \alpha]A'}{\Delta, \alpha, \Theta}   {By above equality}
             \chkjudgPf{\Gamma, \alpha}{e}{[\Gamma]A'}{\Delta, \alpha, \Theta}   {By definition of substitution}
             \chkjudgPf{\Gamma}{e}{\alltype{\alpha}{[\Gamma]A'}}{\Delta}   {By \AllIntro}
\Hand        \chkjudgPf{\Gamma}{e}{[\Gamma](\alltype{\alpha}{A'})}{\Delta}   {By definition of substitution}
           \end{llproof}

     \DerivationProofCase{\DeclAllApp}
            {\judgetp{[\Omega]\Gamma}{\tau}
              \\ 
             \declappjudg{[\Omega]\Gamma}{e}{[\tau/\alpha]A_0}{C}}
            {\declappjudg{[\Omega]\Gamma}{e}{\underbrace{\alltype{\alpha}A_0}_{[\Omega]A}}{C}}

            \begin{llproof}
              \judgetpPf{[\Omega]\Gamma}{\tau}   {Subderivation}
              \proofsep
              \eqPf{[\Omega]A}{\alltype{\alpha} A_0}   {Given}
              \continueeqPf{\alltype{\alpha} [\Omega]A'}  {By def. of subst. and predicativity of $\Omega$}
              \declappjudgPf{[\Omega]\Gamma}{e}{[\tau/\alpha][\Omega]A'}{C}  {Subderivation and above equality}
              \substextendPf{\Gamma}{\Omega}   {Given}
              \substextendPf{\Gamma, \ahat}{\Omega, \hypeq{\ahat}{\tau}}  {By \substextendSolve}
              \proofsep
              \eqPf{[\Omega]\Gamma} {[\Omega, \hypeq{\ahat}{\tau}](\Gamma, \ahat)}  {By definition of context application}
              \declappjudgPf{[\Omega, \hypeq{\ahat}{\tau}](\Gamma, \ahat)}{e}{[\tau/\alpha][\Omega]A'}{C}  {By above equality}
              \declappjudgPf{[\Omega, \hypeq{\ahat}{\tau}](\Gamma, \ahat)}{e}{[\tau/\alpha][\Omega, \hypeq{\ahat}{\tau}]A'}{C}  {By def.\ of subst.}
              \eqPf{\big(\big[[\Omega]\tau/\alpha\big]\big[\Omega, \hypeq{\ahat}{\tau}\big]A'\big)} {\big([\Omega, \hypeq{\ahat}{\tau}][\ahat/\alpha]A'\big)}  {By distributivity of substitution}
              \eqPf{\tau}{[\Omega]\tau}   {$\FEV{\tau} = \emptyset$}
              \eqPf{\big(\big[\tau/\alpha\big]\big[\Omega, \hypeq{\ahat}{\tau}\big]A'\big)} {\big([\Omega, \hypeq{\ahat}{\tau}][\ahat/\alpha]A'\big)}  {By above equality}
              \declappjudgPf{[\Omega, \hypeq{\ahat}{\tau}](\Gamma, \ahat)}{e}{[\Omega, \hypeq{\ahat}{\tau}][\ahat/\alpha]A'}{C}  {By above equality}
            \end{llproof}

            \begin{llproof}
              \appjudgPf {\Gamma, \ahat} {e} {[\ahat/\alpha]A'} {C'} {\Delta}   {By i.h.}
\Hand         \eqPf{C}{[\Omega]C'}   {\ditto}
\Hand         \substextendPf{\Delta}{\Omega'}   {\ditto}
\Hand         \substextendPf{\Omega}{\Omega'}   {\ditto}
              \proofsep
\Hand         \appjudgPf{\Gamma}{e}{\alltype{\alpha} A'} {C'} {\Delta}   {By \AllApp}
            \end{llproof}

     \DerivationProofCase{\DeclArrIntro}
          {\declchkjudg{[\Omega]\Gamma, x : A_1'}{e_0}{A_2'}
          }
          {\declchkjudg{[\Omega]\Gamma}{\lam{x} e_0}{A_1' \arr A_2'}}

          We have $[\Omega]A = A_1' \arr A_2'$.  Either $[\Gamma]A = A_1 \arr A_2$
          where $A_1' = [\Omega]A_1$ and $A_2' = [\Omega]A_2$---or
          $[\Gamma]A = \ahat$ and $[\Omega]\ahat = A_1' \arr A_2'$.

          In the former case:

          \begin{llproof}
            \declchkjudgPf{[\Omega]\Gamma, x : A_1'}{e_0}{A_2'}   {Subderivation}
            \proofsep
            \eqPf{A_1'}  {[\Omega]A_1}   {Known in this subcase}
            \continueeqPf {[\Omega][\Gamma]A_1}   {By \Lemmaref{lem:subst-extension-invariance}}
            \eqPf{[\Omega]A_1'} {[\Omega][\Omega][\Gamma]A_1}   {Applying $\Omega$ on both sides}
            \continueeqPf {[\Omega][\Gamma]A_1}   {By idempotence of substitution}
            \proofsep
            \eqPf{[\Omega]\Gamma, x : A_1'}  {[\Omega, x : A_1'](\Gamma, x : [\Gamma]A_1)}   {By definition of context application}
            \proofsep
            \declchkjudgPf{[\Omega, x : A_1'](\Gamma, x : [\Gamma]A_1)}{e_0}{A_2'}   {By above equality}
            \proofsep
            \substextendPf{\Gamma}{\Omega}   {Given}
            \substextendPf{\Gamma, x : [\Gamma]A_1}{\Omega, x : A_1'}   {By \substextendVV}
            \proofsep
            \chkjudgPf{\Gamma, x : [\Gamma]A_1}{e_0}{A_2}{\Delta'}   {By i.h.}
            \substextendPf{\Delta'}{\Omega_0'}   {\ditto}
            \substextendPf{\Omega, x : A_1'}{\Omega_0'}   {\ditto}
            \eqPf{\Omega_0'} {\Omega', x : A_1', \Theta}   {By \Lemmaref{lem:extension-order} (v)}
\Hand       \substextendPf{\Omega} {\Omega'}  {\ditto}
            \proofsep
            \substextendPf{\Gamma, x : [\Gamma]A_1}{\Delta'}   {By \Lemmaref{lem:typing-extension}}
            \eqPf{\Delta'}{\Delta, x : \cdots, \Theta}   {By \Lemmaref{lem:extension-order} (v)}
            \substextendPf{\Delta, x : \cdots, \Theta}{\Omega', x : A_1', \Theta}  {By above equalities}
\Hand       \substextendPf{\Delta}{\Omega'}  {By \Lemmaref{lem:extension-order} (v)}
            \decolumnizePf
            \chkjudgPf{\Gamma, x : [\Gamma]A_1}{e_0}{[\Gamma]A_2}{\Delta, \alpha, \Theta}   {By above equality}
            \chkjudgPf{\Gamma}{\lam{x} e_0}{([\Gamma]A_1) \arr ([\Gamma]A_2)}{\Delta}  {By \ArrIntro}
\Hand      \chkjudgPf{\Gamma}{\lam{x} e_0}{[\Gamma](A_1 \arr A_2)}{\Delta}  {By definition of substitution}
          \end{llproof}
          
          In the latter case:
          
          \begin{llproof}
            \eqPf{[\Omega]\ahat}{A_1' \arr A_2'}  {Known in this subcase}
            \declchkjudgPf{[\Omega]\Gamma, x : A_1'}{e_0}{A_2'}   {Subderivation}
            \substextendPf{\Gamma}{\Omega}   {Given}
            \substextendPf{\Gamma, \ahat, \bhat} {\Omega, \hypeq{\ahat}{A_1'}, \hypeq{\bhat}{A_2'}}
                      {By \substextendSolve twice}
            \eqPf{[\Omega]\ahat} {[\Omega]A_1'}   {By definition of substitution}
            \substextendPf{\Gamma, \ahat, \bhat, x : \ahat} {\Omega, \hypeq{\ahat}{A_1'}, \hypeq{\bhat}{A_2'}, x : A_1'}
                      {By \substextendVV}
            \eqPf{[\Omega]\Gamma, x : A_1'}
                   {\big[\Omega, \hypeq{\ahat}{A_1'}, \hypeq{\bhat}{A_2'}, x : A_1'\big]%
                     \big(\Gamma, \ahat, \bhat, x : \ahat\big)}
                   {By definition of context application}
            \LetPf{\Omega_0}{(\Omega, \hypeq{\ahat}{A_1'}, \hypeq{\bhat}{A_2'}, x : A_1')}   {}
            \declchkjudgPf{[\Omega_0](\Gamma, \ahat, \bhat, x : \ahat)}{e_0}{A_2'}   {By above equality}
            \chkjudgPf{\Gamma, \ahat, \bhat, x : \ahat} {e_0} {\bhat} {\Delta'}     {By i.h. with $\Omega_0$}
            \substextendPf{\Delta'} {\Omega_0'}     {\ditto}
            \substextendPf{\Omega_0} {\Omega_0'}     {\ditto}
          \end{llproof}

          \begin{llproof}
            \substextendPf{\Gamma, \ahat, \bhat, x : \ahat}{\Delta'}     {By \Lemmaref{lem:typing-extension}}
            \eqPf{\Delta'}{\Delta, x : \ahat, \Theta}   {By \Lemmaref{lem:extension-order} (v)}
            \substextendPf{\Delta, x : \ahat, \Theta} {\Omega_0'}   {By above equality}
            \eqPf{\Omega_0'} {\Omega'', x : \cdots, \Omega_Z}     {By \Lemmaref{lem:typing-extension}}
\Hand       \substextendPf{\Delta}{\Omega''}     {\ditto}
            \substextendPf{\Gamma, \ahat, \bhat}{\Delta}   {\ditto}
            \substextendPf{\Omega_0} {\underbrace{\Omega'', x : \cdots, \Omega_Z}_{\Omega_0'}}     {By above equality}
            \substextendPf{\Omega, \hypeq{\ahat}{A_1'}, \hypeq{\bhat}{A_2'}, x : A_1'}
                          {\Omega'', x : \cdots, \Omega_Z}
                          {By def.\ of $\Omega_0$}
            \eqPf{\Omega''}
                      {\Omega', \hypeq{\ahat}{\dots}, \dots}
                      {By \Lemmaref{lem:extension-order} (iii)}
\Hand       \substextendPf{\Omega}{\Omega'}  {\ditto}
            \decolumnizePf
            \chkjudgPf{\Gamma, \ahat, \bhat, x : \ahat} {e_0} {\bhat} {\Delta, x : \ahat, \Theta}     {By above equality}
            \chkjudgPf{\Gamma} {\lam{x} e_0} {\ahat \arr \bhat} {\Delta}     {By \ArrIntroSyn}
            \proofsep
            \eqPf{[\Gamma]\ahat}{\ahat}    {By definition of substitution}
            \eqPf{[\Gamma]\bhat}{\bhat}    {By definition of substitution}
            \chkjudgPf{\Gamma}{\lam{x} e_0}{([\Gamma]\ahat) \arr ([\Gamma]\bhat)}{\Delta} {By above equalities}
\Hand       \chkjudgPf{\Gamma}{\lam{x} e_0}{[\Gamma](\ahat \arr \bhat)}{\Delta} {By definition of substitution}
          \end{llproof}

     \DerivationProofCase{\DeclArrElim}
          {\declsynjudg{[\Omega]\Gamma}{e_1}{B}
            \\
            \declappjudg{[\Omega]\Gamma}{e_2}{B}{A}
          }
          {\declsynjudg{[\Omega]\Gamma}{e_1\,e_2}{A}}

          \begin{llproof}
            \declsynjudgPf{[\Omega]\Gamma}{e_1}{B}   {Subderivation}
            \substextendPf{\Gamma}{\Omega}  {Given}
            \synjudgPf{\Gamma} {e_1} {B'} {\Theta}   {By i.h.}
            \eqPf{B} {[\Omega]B'}  {\ditto}
            \substextendPf{\Theta} {\Omega_0'}  {\ditto}
            \substextendPf{\Omega} {\Omega_0'}  {\ditto}
            \proofsep
            \declappjudgPf{[\Omega]\Gamma}{e_2}{B}{A}  {Subderivation}
            \declappjudgPf{[\Omega]\Gamma}{e_2}{[\Omega]B'}{A}  {By above equality}
            \substextendPf{\Gamma}{\Omega_0'}  {By \Lemmaref{lem:substextend-transitivity}}
            \eqPf{[\Omega]\Gamma} {[\Omega]\Omega}   {By \Lemmaref{lem:completes-stability}}
            \continueeqPf {[\Omega_0']\Omega_0'}   {By \Lemmaref{lem:finishing-completions}}
            \continueeqPf {[\Omega_0']\Gamma}   {By \Lemmaref{lem:completes-stability}}
            \continueeqPf {[\Omega_0']\Theta}   {By \Lemmaref{lem:completes-confluence}}
            \declappjudgPf{[\Omega_0']\Theta}{e_2}{[\Omega]B'}{A}  {By above equality}
            \eqPf{[\Omega]B'}{[\Omega_0']B'}   {By \Lemmaref{lem:finishing-types}}
            \eqPf{[\Omega_0']B'}{[\Omega_0'][\Theta]B'}   {By \Lemmaref{lem:subst-extension-invariance}}
            \declappjudgPf{[\Omega_0']\Theta}{e_2}{[\Omega][\Theta]B'}{A}  {By above equalities}
            \proofsep
            \appjudgPf{\Theta}{e_2}{[\Theta]B'}{A'}{\Delta}  {By i.h. with $\Omega_0'$}
\Hand            \eqPf{A} {[\Omega]A'}  {\ditto}
\Hand            \substextendPf{\Delta}{\Omega'} {\ditto}
            \substextendPf{\Omega_0'}{\Omega'} {\ditto}
            \substextendPf{\Omega}{\Omega'} {By \Lemmaref{lem:substextend-transitivity}}
\Hand            \synjudgPf{\Gamma}{e_1\,e_2}{A'}{\Delta}  {By \ArrElim}
          \end{llproof}

\clearpage
      \DerivationProofCase{\DeclArrApp}
            {\declchkjudg{[\Omega]\Gamma}{e}{B}}
            {\declappjudg{[\Omega]\Gamma}{e}{\underbrace{B \arr C}_{[\Omega]A}}{C}}

          We have $[\Omega]A = B \arr C$.  Either $[\Gamma]A = B_0 \arr C_0$
          where $B = [\Omega]B_0$ and $C = [\Omega]C_0$---or
          $[\Gamma]A = \ahat$ where $\ahat \in \unsolved{\Gamma}$
          and $[\Omega]\ahat = B \arr C$.

          In the former case:

          \begin{llproof}
            \declchkjudgPf{[\Omega]\Gamma}{e}{B}  {Subderivation}
            \eqPf{B} {[\Omega]B_0}  {Known in this subcase}
            \proofsep
            \substextendPf{\Gamma}{\Omega}   {Given}
            \proofsep
            \chkjudgPf{\Gamma}{e}{[\Gamma]B_0}{\Delta}   {By i.h.}
            \appjudgPf{\Gamma}{e}{([\Gamma]B_0) \arr ([\Gamma]C_0)}{[\Gamma]C_0}{\Delta}  {By \ArrApp}
\Hand       \substextendPf{\Delta}{\Omega'}   {\ditto}
\Hand       \substextendPf{\Omega}{\Omega'}   {\ditto}
            \LetPf{C'} {[\Gamma]C_0}  {}
            \eqPf{C}{[\Omega]C_0}  {Known in this subcase}
            \continueeqPf {[\Omega][\Gamma]C_0}   {By \Lemmaref{lem:subst-extension-invariance}}
\Hand       \continueeqPf {[\Omega]C'}   {$[\Gamma]C_0 = C'$}
\Hand       \appjudgPf{\Gamma}{e}{[\Gamma](B_0 \arr C_0)}{[\Gamma]C_0}{\Delta}  {By definition of substitution}
          \end{llproof}
          
          In the latter case, $\ahat \in \unsolved{\Gamma}$,
          so the context $\Gamma$ must have the form $\Gamma_0[\ahat]$.
            
            \begin{llproof}
              \substextendPf{\Gamma}{\Omega}   {Given}
              \substextendPf{\Gamma_0[\ahat]}{\Omega}   {$\Gamma = \Gamma_0[\ahat]$}
              \eqPf{[\Omega]A}  {B \arr C}   {Above}
              \eqPf{[\Omega]\ahat}  {B \arr C}   {$A = \ahat$}
              \eqPf{\Omega} {\Omega_0[\hypeq{\ahat}{A_0}]
                             \AND   [\Omega]A_0 = B \arr C}    {Follows from $[\Omega]\ahat = B \arr C$}
              \LetPf{\Gamma'}{\Gamma_0[\ahat_2, \ahat_1, \hypeq{\ahat}{\ahat_1 \arr \ahat_2}]}   {}
              \LetPf{\Omega_0'}{\Omega_0[\hypeq{\ahat_2}{[\Omega]C}, \hypeq{\ahat_1}{[\Omega]B}, \hypeq{\ahat}{\ahat_1 \arr \ahat_2}]}   {}
              \decolumnizePf
              \substextendPf{\Gamma'}{\Omega_0'}    {By \Lemmaref{lem:parallel-admissibility} (ii) twice}
              \proofsep
              \declchkjudgPf{[\Omega]\Gamma}{e}{B}   {Subderivation}
              \proofsep
              \substextendPf{\Omega}{\Omega_0'}   {By \Lemmaref{lem:extension-addsolve}}
                       \trailingjust{then \Lemmaref{lem:parallel-admissibility} (iii)}
              \eqPf{[\Omega]\Gamma} {[\Omega]\Omega}   {By \Lemmaref{lem:completes-stability}}
              \continueeqPf {[\Omega_0']\Omega_0'}   {By \Lemmaref{lem:finishing-completions}}
              \continueeqPf {[\Omega_0']\Gamma'}   {By \Lemmaref{lem:completes-confluence}}
              \proofsep
              \eqPf{B} {[\Omega_0']\ahat_1}  {By definition of $\Omega_0'$}
              \declchkjudgPf{[\Omega_0']\Gamma'}{e}{[\Omega_0']\ahat_1}   {By above equalities}
              \proofsep
              \chkjudgPf{\Gamma'}{e}{[\Gamma']\ahat_1}{\Delta}     {By i.h.}
\Hand         \substextendPf{\Delta}{\Omega'}   {\ditto}
              \substextendPf{\Omega_0'}{\Omega'}   {\ditto}              
\Hand         \substextendPf{\Omega}{\Omega'}   {By \Lemmaref{lem:substextend-transitivity}}
              \proofsep
              \eqPf {[\Gamma']\ahat_1} {\ahat_1}   {$\ahat_1 \in \unsolved{\Gamma'}$}
              \chkjudgPf{\Gamma'}{e}{\ahat_1}{\Delta}   {By above equality}
            \end{llproof}
            
            \begin{llproof}
              \appjudgPf{\Gamma} {e} {\ahat} {\ahat_2}  {\Delta}  {By \SolveApp}
              \LetPf{C'} {\ahat_2} {}
              \eqPf{C}{[\Omega_0']\ahat_2}  {By definition of $\Omega_0'$}
              \continueeqPf{[\Omega']\ahat_2}  {By \Lemmaref{lem:finishing-types}}
\Hand         \continueeqPf{[\Omega']C'}  {By above equality}
\Hand         \appjudgPf{\Gamma}{e}{[\Gamma]A}{C'}{\Delta}  {$\ahat = [\Gamma]A$ and $\ahat_2 = C'$}
            \end{llproof}

          \DerivationProofCase{\DeclUnitIntroSyn}
                { }
                {\declsynjudg{[\Omega]\Gamma}{\unitexp}{\unitty}}

                \begin{llproof}
                  \eqPf {\unitty} {A}   {Given}
                  \synjudgPf{\Gamma} {\unitexp} {\unitty}  {\Gamma}   {By \UnitIntroSyn}
                  \LetPf{\Delta}{\Gamma}  {}
                  \LetPf{\Omega'}{\Omega}  {}
                  \substextendPf{\Gamma}{\Omega}  {Given}
\Hand             \substextendPf{\Delta}{\Omega}  {By above equality}
\Hand             \substextendPf{\Omega}{\Omega'}  {By \Lemmaref{lem:substextend-reflexivity}}
                  \LetPf{A'}{\unitty}  {}
\Hand             \synjudgPf{\Gamma} {\unitexp} {A'}  {\Delta}   {By above equalities}
\Hand             \eqPf {\unitty}  {[\Omega]A'}  {By definition of substitution}
                \end{llproof}

                \bigskip

          \DerivationProofCase{\DeclArrIntroSyn}
                {\judgetp{[\Omega]\Gamma}{\sigma \arr \tau}
                 \\
                 \declchkjudg{[\Omega]\Gamma, x : \sigma}{e_0}{\tau}
                }
                {\declsynjudg{[\Omega]\Gamma}{\lam{x} e_0}{\sigma \arr \tau}}

                \medskip

                \begin{llproof}
                  \eqPf{(\sigma \arr \tau)}{A}    {Given}
                  \declchkjudgPf{[\Omega]\Gamma, x : \sigma}{e_0}{\tau}   {Subderivation}
                  \proofsep
                  \LetPf{\Gamma'} {(\Gamma, \ahat, \bhat, x : \ahat)}  {}
                  \LetPf{\Omega_0} {(\Omega, \hypeq{\ahat}{\sigma}, \hypeq{\bhat}{\tau}, x : \sigma)}   {}
                  \proofsep
                  \substextendPf{\Gamma}{\Omega}   {Given}
                  \substextendPf{\Gamma'}{\Omega_0}   {By \substextendSolve twice, then \substextendVV}
                  \proofsep
                  \eqPf{[\Omega_0]\Gamma'} {\big([\Omega]\Gamma, x : \sigma\big)}   {By definition of context application}
                  \eqPf{\tau}{[\Omega_0]\bhat}{By definition of $\Omega_0$}
                  \declchkjudgPf{[\Omega_0]\Gamma'}{e_0}{[\Omega_0]\bhat}   {By above equalities}
                  \proofsep
                  \chkjudgPf{\Gamma'}{e_0}{\bhat} {\Delta'}   {By i.h.}
                  \substextendPf{\Delta'}{\Omega_0'}    {\ditto}
                  \substextendPf{\Omega_0}{\Omega_0'}    {\ditto}
                  \proofsep
                  \eqPf{\Delta'}{(\Delta, x : \ahat, \Theta)}  {By \Lemmaref{lem:extension-order} (v)}
                  \chkjudgPf{\Gamma, \ahat, \bhat, x : \ahat}{e_0}{\bhat} {\Delta, x : \ahat, \Theta}   {By above equalities}
                  \substextendPf{(\Delta, x : \ahat, \Theta)} {\Omega_0'}  {By above equality}
                  \eqPf{\Omega_0'}{\Omega', x : \sigma, \Omega_Z}   {By \Lemmaref{lem:extension-order} (v)}
\Hand             \substextendPf{\Delta} {\Omega'}  {\ditto}
                  \synjudgPf{\Gamma}{\lam{x} e_0}{\ahat \arr \bhat}{\Delta}   {By \ArrIntroSyn}
               \end{llproof}

               \begin{llproof}
                  \LetPf{A'} {(\ahat \arr \bhat)}  {}
\Hand              \synjudgPf{\Gamma}{\lam{x} e_0}{A'}{\Delta}   {By above equality}
                  \eqPf{\sigma \arr \tau}{([\Omega_0]\ahat) \arr ([\Omega_0]\bhat)}  {By definition of $\Omega_0$}
                  \eqPf{\sigma \arr \tau}{[\Omega_0](\ahat \arr \bhat)}  {By definition of substitution}
                  \eqPf{A} {[\Omega_0]A'}   {By above equalities}
\Hand            \eqPf{A} {[\Omega']A'}   {By \Lemmaref{lem:finishing-types}}
                  \proofsep
                  \substextendPf{\Gamma'}{\Delta'}   {By \Lemmaref{lem:typing-extension}}
\Hand             \substextendPf{\Omega}{\Omega'}  {By \Lemmaref{lem:substextend-transitivity}     \qedhere}
                \end{llproof}
  \end{itemize}
\end{proof}

\ifnum\OPTIONLoudLabels=1
  \bibliographystyle{plainnatlocalcopy}
\else
  \bibliographystyle{plainnat}
\fi
\bibliography{local}